\pdfoutput=1

\documentclass[a4paper, 11pt, twoside, openright, english]{memoir}

\usepackage{mypreamble}

\makeatletter

\title{Dualities in 2+1 Dimensions}
\author{Carl Turner}
\def\@titlefooter{XIV Modave Summer School in Mathematical Physics}
\def\@supertitle{Lectures on}
\date{2018}

\def\@secondpageinfo{
  Dr Carl Turner \\
  Junior Research Fellow \\
  Gonville \& Caius College -- University of Cambridge \\[2ex]
  \href{mailto:C.P.Turner@damtp.cam.ac.uk}{C.P.Turner@damtp.cam.ac.uk}
}

\makeatother

\begin{document}

\frontmatter
\include{titlepage}
\begin{abstract}
  \noindent
  In this course, we will discuss dualities in 2+1 dimensions. We begin by briefly reviewing the physics of 1+1d T-duality and bosonization, and discuss flux attachment and the dual photon in 2+1d. Then, we introduce 2+1d particle-vortex duality, bosonization and their associated web of dualities. This includes discussing more elaborate Abelian dualities, including self-dual theories and quiver theories. Next, we move on to non-Abelian physics, discussing QCD in three dimensions and more. This done, we will discuss the relation of these dualities to SUSY dualities, lattice physics, and large $N$ physics; and finally, we present the application of one of the Abelian dualities to quantum Hall physics.
  
  These lecture notes are expanded from lectures delivered at \textit{XIV Modave Summer School in Mathematical Physics} in September 2018. Please 
  \href{mailto:C.P.Turner@damtp.cam.ac.uk}{email the author} with any comments or corrections.
\end{abstract}

\tableofcontents

\chapter*{Conventions} \label{conventions}
\addcontentsline{toc}{chapter}{Conventions}

\paragraph{Spacetime Conventions}

We use the metric convention $(+--)$ in 2+1 dimensions. We take $\epsilon^{012} = \sqrt{|g|}$ where $g$ is the determinant of the metric; in flat space, where we will mostly work, $\epsilon^{012} = \epsilon_{012} = +1$. (Similarly, our 1+1 dimensional conventions are $(+-)$ and $\epsilon^{01}>0$.)

We will often neglect the wedge symbol $\wedge$ in writing products of differential forms; for example, if $a$ is a one-form then $a^3 \equiv a \wedge a \wedge a$.

Antisymmetrization of indices is denoted with square brackets so that e.g.
\be T_{[abc]} = \frac{1}{3!} \left(T_{abc} + T_{bca} + T_{cab} - T_{acb} - T_{cba} - T_{bac}\right)\ee
and similarly symmetrization is denoted with round brackets. The wedge product is
\be (X \wedge Y)_{a_1\ldots a_p b_1 \ldots b_q} = \frac{(p+q)!}{p!q!}X_{[a_1\ldots a_p}Y_{b_1 \ldots b_q]} \ee
and the exterior derivative is \be (\rmd X)_{\mu_1\ldots\mu_{p+1}} = (p+1)\partial_{[\mu_1} X_{\mu_2\ldots\mu_{p+1}]}\fstp \ee
The Hodge dual is
\be(\star X)_{a_1\ldots a_{n-p}} = \frac{\sqrt{|g|}}{p!} \epsilon_{a_1\ldots a_{n-p}b_1\ldots b_p} X^{b_1\ldots b_p}\cmma\ee 
so in Lorentzian 3d, $\star \star X = X$.

\paragraph{Gauge Field Conventions}

Gauge fields are given lowercase letters $a_\mu, b_\mu \ldots$ if they are dynamical; $A_\mu, B_\mu, \ldots$ represent non-dynamical fields. A charge one/fundamental field has covariant derivative $D = \partial - ia$; the adjoint is $D = \partial - i[a, \cdot]$. Under a gauge transformation $g$, $a \to g^{-1} a g  + i g^{-1} \rmd g$ whilst for a fundamental field $\phi \to g^{-1} \phi$. (For Abelian $g=e^{-i\chi}$, $\phi\to e^{i\chi}\phi$ and $a \to a + \rmd \chi$.) The corresponding field strength is
\be f = \rmd a - i a\wedge a = \rmd a - i a^2  \fstp \ee


We write $U(1)_{-1/2} + \psi$ for the Abelian theory with a single fermion with the following property: a large positive fermion mass $m \gg 0$ leads to the IR theory containing a free photon ($U(1)_0$) whilst taking $m \ll 0$ leads to an empty theory ($U(1)_1$).

Non-Abelian groups $U(N)$ with a $U(1)$ and $SU(N)$ part have two independent Chern-Simons levels; we write
\be U(N)_{k,k'} = \frac{SU(N)_k \times U(1)_{k'N}}{\mathbb{Z}_N} \ee 
and $U(N)_k \equiv U(N)_{k,k}$. Note that the gauge-invariant theories are $U(N)_{k,k + nN}$ for $k,n\in \mathbb{Z}$.

\paragraph{Spinor Conventions}

The Pauli matrices $\sigma^i$ for $i=1,2,3$ are the standard Hermitian matrices
satisfying $\sigma_i \sigma_j = \delta_{ij} + i \varepsilon_{ijk} \sigma_k$ (where $\varepsilon_{ijk}$ is the standard antisymmetric tensor, with no signs from the signature of spacetime). The $\gamma$ matrices in 2+1 dimensions are can then be taken to be $\gamma^0 = \sigma_2$, $\gamma^1 = i\sigma_1$ and $\gamma^2 = i \sigma_3$:
\be
\gamma^0 = i\begin{pmatrix}0 & -1 \\ 1 & 0\end{pmatrix}, \quad
\gamma^1 = i\begin{pmatrix}0 & 1 \\ 1 & 0\end{pmatrix}, \quad
\gamma^2 = i\begin{pmatrix}1 & 0 \\ 0 & -1\end{pmatrix}
\ee
which satisfy $\{\gamma^\mu, \gamma^\nu\} = 2\eta^{\mu\nu}$. They also satisfy various trace relations:
\be \tr \gamma^\mu \gamma^\nu = 2 \eta^{\mu\nu}, \quad \tr \gamma^\mu \gamma^\nu \gamma^\rho = 2i \epsilon^{\mu\nu\rho}, \quad \tr \gamma^\mu \gamma^\nu \gamma^\rho \gamma^\sigma = 2\eta^{\mu\nu}\eta^{\rho\sigma} - 2\eta^{\mu\rho}\eta^{\nu\sigma} + 2\eta^{\mu\sigma}\eta^{\nu\rho} \ee
A \textit{Dirac spinor} in 2+1 dimensions is a \textit{two}-component, complex object. We have given the matrices in the Majorana basis, in which the matrices are purely imaginary. Majorana spinors are real two-component objects. (There are no Weyl spinors in odd dimensions.)

In 2+1 dimensions, we take the Dirac action to be $S = i \bar{\psi} \gamma^\mu \partial_\mu \psi$. Here, $\bar{\psi} = \psi^\dagger \gamma^0$. In the above basis, charge conjugation acts as $C : \psi \mapsto \psi^\star$; a Majorana fermion is $C$ invariant. Remembering that fermions anti-commute, one can show that not only is the Dirac action invariant under $C$, so is the Dirac mass term $\bar{\psi} \psi$.

Under a parity transformation, taken to invert the first spatial direction only ($P : x_1 \mapsto - x_1$), this action is invariant if we define $P:\psi \mapsto i\gamma^1 \psi$. However, $\bar{\psi} \psi \mapsto - \bar{\psi} \psi$ under this transformation. This operator squares to $P^2 = 1$.\footnote{Note that spatial inversion in the origin, $(x_1,x_2) \mapsto (-x_1,-x_2)$, is actually just a rotation by $\pi$ and in particular does not the orientation of spacetime. Instead, we have taken $P$ to be a reflection in one axis. This is sometimes referred to instead as $R$.}

Under time reversal, which is an anti-unitary symmetry of nature such that $T:x_0 \mapsto -x_0$ and $T:i \mapsto -i$, we define $T:\psi \mapsto i\gamma^0 \psi$. Note that $T:\bar{\psi} \mapsto (i\gamma^0\psi)^\dagger (\gamma^0)^\star = i \psi^\dagger$. It follows that the Dirac action is invariant under $T$. Again, $\bar{\psi} \psi \mapsto - \bar{\psi} \psi$ under this transformation (as was necessary by the CPT theorem). This operator squares to $T^2=(-1)^F$ where $F$ is the number of fermions.

\paragraph{Fermion Conventions}

We adopt the convention that the Lagranigan of a gauged Dirac fermion
\be \lag = i \bar{\psi} \gamma^\mu (\partial_\mu - i a_\mu) \psi \ee 
is implicitly regularized to preserve gauge invariance with a negative-mass Pauli-Villars regulator. We refer to this as
\be U(1)_{-1/2} + \psi \ee
so that it has a time-reversal anomaly of $\lag \to \lag + \frac{1}{4\pi} a \rmd a$ (in flat space).

\mainmatter

\chapter{Introduction}
\labelChapter{introduction}

\begin{introduction}
  We begin by giving some motivation, followed by an outline of the course. Then, we discuss several relatively simple dualities in 2 and 3 dimensions as a warm up for the body of course.
\end{introduction}

\section{Motivation}

\lettrine{Q}{uantum field theories} are, in general, very difficult beasts to work with. Very few explicit calculations are possible, and those that are often given at best asymptotic approximations to physical observables. But physicists are nothing if not persistent, and over the years we have developed many different approaches to understanding QFTs.

Some of these approaches are centered around \textit{kinematics}: the analysis of physical fields, states and operators and the symmetries they enjoy. In this course we will spend some time thinking carefully about both continuous symmetries (global and gauged) and discrete symmetries like charge conjugation and time reversal. Spacetime symmetries are also crucial, of course; the dualities we are most interested in all exhibit conformal invariance. Supersymmetry is perhaps the most constraining of all kinematical considerations; even this will also crop up in this course as a tool in analyzing other, less symmetric systems. 

However, the outstanding problems in quantum field theory usually concern the \textit{dynamics} of poorly understood, strongly-interacting systems of great physical interest. In many cases, this puts exact calculations well out of reach. Without the crutch of supersymmetry, then, what can we possibly hope to say about such a system?

In this course, we provide some tentative answers for a large class of 2+1 dimensional gauge theories by describing several different \textit{dualities}.

\begin{asidebox}[Duality]%
	A \textit{duality} refers to the relationship between two or more theories $A, B, \ldots$ which are equivalent via some surprising dictionary: every object $X$ in the theory $A$ has an dual object $\tilde{X}$ in the dual theory $B$ with identical properties. We write $A \leftrightarrow B$ to assert that the theories are dual; similarly, $X \leftrightarrow \tilde{X}$.
	
	A duality differs from a \textit{symmetry} because generally $A$ and $B$ are completely different theories. If the dual theory is in fact manifestly the same as the original theory (so $A \equiv B$ but the map between $X$ and $\tilde{X}$ is still non-trivial), we refer to $A$ as a \textit{self-dual} theory. This is essentially the statement that theory $A$ has a \textit{hidden symmetry}, often (but not always) a $\mathbb{Z}_2$ involution.
\end{asidebox}

The material we will discuss in this course is the cutting edge of theoretical physics. (Pleasingly, compared to the rest of that edge, it is also relatively easy to grasp.) This has advantages, such as being exciting. It also has disadvantages, like the difficulty of conclusively proving many of the results we want.

\section{Course Outline}

In this introduction, we are going to look at four examples of QFT dualities which we can rigorously establish. This will help in gaining some understanding of what non-supersymmetric dualities look like, and what kind of language might be useful in talking about them.

Then, in \refChapter{particle_vortex}, we will introduce our first IR duality, and one with a very fine pedigree indeed: \textit{particle-vortex} duality. This relates two 3 dimensional bosonic theories, one a gauge theory and the other not. We will follow this up with a discussion of 3d \textit{bosonization} in \refChapter{3d_bosonization}: this is a duality relating an apparently bosonic theory to a fermionic one. In \refChapter{duality_web}, we will then ask how these are related -- and this will set the stage to present the first of many new dualities to be discovered in the last few years. 

The basic ideas established, we will indulge in a large number of different proposals of dualities involving Abelian and then (throughout \refPartOnly{nonabelian}) non-Abelian gauge theories, stopping to see applications to the phase diagram of QCD. Finally, in \refPartOnly{evidence} we discuss briefly how supersymmetric dualities, lattice (and wire) physics and large $N$ calculations lend support to the conjectured dualities we have discussed, before giving an interesting application of dualities to condensed matter physics.

\section{Prototypes}
\labelSection{prototypes}

Before we set about discussing the remarkable zoo of new dualities which we are here to study, however, let us begin by looking at several classic examples of QFT dualities which can be demonstrated exactly. This will help set the stage for a lot of what will follow.

\begin{enumerate}
	\item Firstly, we look at \textsc{T-duality} in 2 dimensions. This simple relation, between two compact scalars of different radii, is a simple and elegant result that is well-known in string theory, but enjoys a much simpler life as a statement about field theory in 2 dimensions.
	\item Then we look at another duality of the 2 dimensional compact scalar: \textsc{2d bosonization}. This relates the scalar to a theory containing a fermion. This is an even a more miraculous result, yet can be demonstrated by direct calculations.
	\item Next up, we look at the \textsc{flux attachment} in 3 dimensions. This again relates fields of apparently different statistics to each other, although in the more limited context of quantum mechanics, or equivalently non-relativistic QFT.
	\item Finally, we look at the \textsc{dual photon} in 3 dimensions. This relates pure gauge theory with yet another compact scalar (albeit one living in 3d).
\end{enumerate}

\subsection{T-duality}\labelSection{tduality}

We are actually going to look at just one particularly simple version of T-duality. Consider the theory
\be \label{compact_scalar_2d}
S = \frac{1}{4\pi} \int \rmd^2x \ \frac{R^2}{2} (\partial \phi)^2
\ee
of a periodic, real scalar $\phi(x) \in [0,2\pi]$. This is actually a conformal field theory; both $R$ and $\phi$ are dimensionless in two dimensions. We refer to $R$ as the \textit{radius} of the scalar $\phi$, thinking of the field $\rho=R\phi$ as a map $\rho : \mathbb{R}\times S^1 \to S^1$ into a circle. Rephrasing this yet again, this is a sigma model whose target space is a circle of radius $R$. (Conformal field theories in 2 dimensions are ubiquitous in string theory, of course; we are using conventions in which $\alpha'=2$.)

The secret which this system conceals is that it is totally equivalent to a sigma model of radius $\tilde{R} = 2/R$. This is the statement of T-duality. (In string theory, with a general string coupling $\alpha'$, this becomes $\tilde{R} = \alpha'/R$.) The way this manifests itself is rather remarkable, and whilst we will not discuss it in detail, it is nice to see it in action.

There is actually a more convenient way to think about the theory: instead of thinking about $\phi$ as the dynamical field, we can think about the vector (or one-form) $b = \rmd \phi$. The action only depends on this one-form, so this is not an unreasonable thing to do.\footnote{We are losing information about the value of $\phi \to \phi + c$, but not the dynamical zero-mode $\phi \to \phi + c(t)$, so we don't really miss anything important.} However, at least for smooth configurations $\phi$, we know that $\rmd (\rmd \phi) = 0$, so this is not a bijective change of variable. There is a simple solution: impose $\rmd b = 0$ using a Lagrange multiplier $\tilde{\phi}$.

We also know that $\phi$ is periodic; this shows up in the fact that $\oint_C \rmd \phi \in 2\pi \mathbb{Z}$ around all cycles $C$. To do this, it suffices to consider $\tilde{\phi}$ which is also $2\pi$ periodic, and write the theory as
%
\be \label{compact_scalar_2d_lagrange}
\tilde{S}
= \frac{1}{4\pi} \int \rmd^2x \ \frac{R^2}{2} b^2 - 2 b \cdot \star \rmd \tilde{\phi} \fstp 
\ee

\begin{exercise}[subtitle=Periodicity of the Dual Scalar]
	Check that this constraint imposes the correctly quantized constraint.
\end{exercise}	

Now all that is left to do is complete the square in \eqref{compact_scalar_2d_lagrange}, integrating out $b$ using its equation of motion $b = (2/R^2)\star \rmd \tilde{\phi}$:
\be \label{compact_scalar_2d_dual}
\tilde{S} = \frac{1}{4\pi} \int \rmd^2x \ \frac{(2/R)^2}{2} (\rmd \tilde{\phi})^2 \fstp
\ee
This is again the action of a compact scalar, but now with the dual radius $\tilde{R} = 2/R$!

Interestingly, we learn that the theory has not only the obvious $U(1)$ symmetry $\phi \to \phi + c$, but an extra, dual symmetry $\tilde{\phi} \to \tilde{\phi} + \tilde{c}$.  Thus the global symmetry consists of $U(1) \times U(1)$. The corresponding Noether currents are
\be j_{\tilde{\phi}}^\mu = \frac{1}{2\pi}\epsilon^{\mu\nu} (\rmd \phi)_\nu \quad \mbox{and} \quad j_\phi^\mu = \frac{R^2}{4\pi} (\rmd \phi)^\mu \ee
and they are effectively interchanged by the duality. Notice that in the $\phi$ description, one of these currents is conserved by virtue of the equations of motion as usual, since $\partial_\mu j_\phi^\mu \propto \nabla^2 \phi$, but the other vanishes automatically, due to what we might call the Bianchi identity $\partial_\mu j_{\tilde{\phi}}^\mu \propto \rmd (\rmd \phi)$ = 0. These roles are also exchanged in the $\tilde{\phi}$ picture.

We can deduce the relationship between the scalars from the equation $b = (2/R^2) \star \rmd \tilde{\phi}$:
\be
\rmd \phi = (2/R^2) \star \rmd \tilde{\phi} \qquad \text{or} \qquad \partial^{\mu} \phi = (2/R^2) \epsilon^{\mu\nu} \partial_\nu \tilde{\phi}
\ee
so that they are essentially harmonic conjugates (on-shell). We can also rewrite this in terms of left-movers and right-movers as
\be
(\partial_t \pm \partial_x) \phi = \pm (2/R^2)(\partial_t \pm \partial_x) \tilde{\phi} \fstp
\ee
In fact, if we write the solution to the $\phi$'s equation of motion in lightcone coordinates we find
\be \phi = \phi_+(t+x) + \phi_-(t-x) \qquad \implies \qquad \tilde{\phi} = \frac{R^2}{2} \left[ \phi_+(t+x) - \phi_-(t-x) \right] \ee
showing that the duality transformation is essentially a relative sign between left-movers $\phi_+$ and right-movers $\phi_-$. (In Euclidean signature, with $\tau = it$ and complex coordinates $z=\tau + ix$, the left-movers are holomorphic and the right-movers anti-holomorphic.)

This factorization into left- and right-movers also explains the extra global symmetry above; we commonly reorganize the $U(1)$ factors into the separately conserved charges associated with $\phi_+$ and $\phi_-$ and write the symmetry group as $U(1)_L \times U(1)_R$.

\def\spinarrowlength{0.4}
\def\spinarrow#1#2{\draw[->,thick,blue] ($#1 +(#2:-\spinarrowlength/2)$) -- ++(#2:\spinarrowlength)}
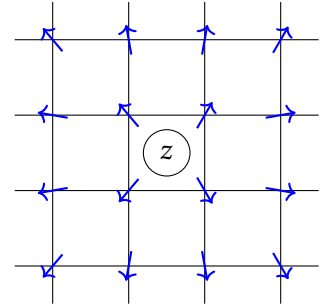
\begin{wrapfigure}{r}{0.4\textwidth}
	\centering
	\begin{tikzpicture}
	\draw [step=1.0] (0.5,0.5) grid (4.5,4.5);
	\spinarrow{(1,4)}{+130}; \spinarrow{(2,4)}{+100}; \spinarrow{(3,4)}{+80}; \spinarrow{(4,4)}{+60};
	\spinarrow{(1,3)}{+170}; \spinarrow{(2,3)}{+130}; \spinarrow{(3,3)}{+60}; \spinarrow{(4,3)}{+10};
	\spinarrow{(1,2)}{-170}; \spinarrow{(2,2)}{-130}; \spinarrow{(3,2)}{-60}; \spinarrow{(4,2)}{-10};
	\spinarrow{(1,1)}{-130}; \spinarrow{(2,1)}{-100}; \spinarrow{(3,1)}{-80}; \spinarrow{(4,1)}{-60};
	\node[draw,circle] at (2.5,2.5) {$z$};
	\end{tikzpicture}
	\caption{A lattice vortex configuration centered at $z$}\labelFigure{lattice_vortex_configuration}
\end{wrapfigure}

It is helpful to be precise about what configurations of the scalar $\phi$ we allow when quantizing this model. Smoothly varying configurations in space are definitely permitted; but the periodicity complicates matters when it comes to (for example) defining a path integral. This is most clearly illustrated by using a lattice regularization. Consider field configurations like that pictured in \refFigureOnly{lattice_vortex_configuration}. At large distances, we see that the field $\phi$ winds once around the target space $S^1$; configurations with this kind of winding are referred as \textit{vortices}. The pictured configuration has vortex number $+1$; in general a configuration carries a charge $w\in \mathbb{Z}$ which is a winding number.

The continuum limit of such vortex configurations is singular at the center $z$; from the lattice point of view these are simply configurations with an action that grows to infinity as the lattice spacing shrinks. Accordingly, we will not include them in the path integral, but we will allow an operation in which we remove a point and introduce a vortex. These can be thought of as \textit{defects}: points omitted from spacetime around which the fields acquire non-trivial boundary conditions.

We also see that a source for a fundamental $\phi$ excitation, $\nabla^2 \phi \sim \delta(x)$, is dual to a source for a vortex of $\tilde{\phi}$, $\rmd (\rmd \tilde{\phi}) \sim \delta(x)$, and vice-versa. Thus T-duality is a \textit{particle-vortex duality} in 2 dimensions. This is often phrased instead as a duality between momentum and winding, which is entirely equivalent.

Indeed, if one is a little more careful, one can show that the operators with well-defined scaling dimension in the original theory are built from $\partial_+\phi_+,\partial_- \phi_-$ and the remaining normal-ordered \textit{vertex operators} $:\exp(i k_+ \phi_+ + i k_- \phi_-):$ for appropriate choices of the momenta $k_+,k_-$. One might naively expect that it was necessary to take $k_+ = k_- = n$ to get a well-defined operator: these are the objects which are classically invariant under adding $2\pi$ to the value of $\phi$.

However, something slightly more general is possible: we are allowed to consider the operator which creates the vortex. We know that $\tilde{\phi}$ excitations are associated with vortices, and the correctly normalized vertex operators must be $V_{n,w}=:\exp(i n \phi + i w \tilde{\phi}):$. Carefully checking the commutation relations, one finds that indeed winding $w$ vortices are created by vertex operators with $k_+ = -k_- = w R^2/2$.
Hence the most general vertex operator has the form
\be \label{compact_boson_vertex_ops}
 V_{n,w} = :\exp(i k_+ \phi_+ + i k_- \phi_-): \qquad k_\pm = n \pm \frac{w R^2}{2} \quad \mbox{for $n,w\in \mathbb{Z}$} \fstp \ee
Then, under the duality, the operators match via $R \to 2/R$, together with $n \leftrightarrow w$ and $\phi_\pm \to \pm (R^2/2) \phi_\pm$.

One last comment: under our duality, $R \to 2/R$. This suggests that there is a special point, $R = \sqrt{2}$, at which the theory has some enhanced structure. This is indeed the case; it is called the $SU(2)$ point because in fact the theory's $U(1)$ symmetries are enhanced to a larger $SU(2)_L \times SU(2)_R$ symmetry. We won't be pursuing this any further here, but this point is certainly of general interest to conformal field theorists and string theorists, and the general message that symmetries are richly enhanced at self-dual points is worth bearing in mind.

\begin{asidebox}[A Simple Anomaly]%
	If the $U(1) \times U(1)$ symmetry sounds a little too good to be true... it is. At most one of them is \textit{non-anomalous}. You cannot consistently gauge both at once. This does not mean to say they are not symmetries of the theory, but it does mean some care should be taken.
	
	There is a simple way to see why this is the case. Suppose you gauge the dual current by adding $A_\mu j^\mu_{\tilde{\phi}} = \frac{1}{2\pi} A \wedge \rmd \phi$ to the action, and integrate by parts:
	\be
	S = \frac{1}{4\pi} \int \rmd^2x \ \frac{R^2}{2} (\partial \phi)^2 + \frac{1}{2\pi} \int \phi \rmd A
	\ee
	This makes clear that the transformation $\phi \to \phi + c$ now leads to an anomalous transformation $S \to S + \int \frac{c}{2\pi} \phi \rmd A$ in the presence of a non-trivial background field $\rmd A$. We cannot now gauge this symmetry.
	
	This situation is referred to as a \textit{mixed 't Hooft anomaly} between the two $U(1)$ symmetries.
\end{asidebox}

\begin{exercise}[subtitle=Optional 2d CFT exercises]
	These exercises are here for 2d CFT enthusiasts or those who enjoy modular forms -- don't worry if the language is unfamiliar, as we won't use these results at all! \cite{yellow} is an excellent reference for hard computations involving partition functions and 2d CFTs in general.
	
	\begin{enumerate}[label=(\alph*)]
	\item
	Show that the conformal dimension of the vertex operator $V_{n,w}$ is $\frac{1}{R^2}(\frac 1 2 k_L^2,\frac 1 2 k_R^2)$.
	\item 
	Compute the partition function of the theory \eqref{compact_scalar_2d} on a torus of modular parameter $\tau$ (a flat torus with $z \sim z + m +n \tau$ for $m,n\in\mathbb{Z}$), showing that it is
	\begin{align} \label{partition_function_2d_scalar}
	Z(\tau, R)
	&:= \tr q^{\textbf{L}_0-\frac{1}{24}} \bar{q}^{\tilde{\textbf{L}}_0-\frac{1}{24}} \qquad q=e^{2\pi i \tau} \\
	& = \frac{1}{|\eta|^2} \sum _{n,w\in\mathbb{Z}} q^{\frac{1}{2}\left(\frac{n}{R}+\frac{w R}{2}\right)^2} \bar{q}^{\frac{1}{2}\left(\frac{n}{R}-\frac{w R}{2}\right)^2}
	\end{align}
	in terms of the Dedekind eta function
	\be \eta(\tau) = q^{\frac{1}{24}} \prod_{n=1}^\infty(1-q^n) \fstp \ee
	This makes the $R \leftrightarrow 2/R$ invariance explicit.
	
	\item
	Note that we have subtracted off $1/24$ for each chiral boson, as appropriate for a pair of CFTs of central charge $c=\frac{1}{2}$ each, in order to obtain expressions which have the appropriate modular properties. If this is unfamiliar, don't worry about it! They just come along for the ride, appearing in the eta function. However, if you fancy a challenge, prove that $Z(\tau, R)$ is a \textit{modular invariant} by proving that it is in invariant under general $SL(2,\mathbb{Z})$ transformations
	\be \tau \to \frac{a \tau + b}{c\tau + d} \qquad \text{where } a,b,c,d \in \mathbb{Z} \text{ and } ad-bc=1 \fstp \ee
	
	[Hints: Firstly, observe that these are generated by $\tau \to \tau + 1$ and $\tau \to -1/z$. You can check these using properties of theta functions and the Poisson resummation formula; see section 10A of \cite{yellow} for some details.]
	
	\item 
	At the self-dual point $R = \sqrt{2}$, the theory has an enhanced symmetry. Find vertex operators with dimension $(1,0)$ at this point, and show that they combine with the $U(1)_L$ current $\partial_z\phi_L$ to generate the algebra of $SU(2)_1$. Show the partition function \eqref{partition_function_2d_scalar} simplifies to
	\be
	Z(\tau, \sqrt{2}) = \left| \frac{1}{\eta} \sum_{m\in\mathbb{Z}} q^{m^2} \right|^2  + \left| \frac{1}{\eta} \sum_{m\in\mathbb{Z}} q^{\left(m + \frac 1 2\right)^2} \right|^2
	\ee
	demonstrating that the theory is a (diagonal) \textit{rational} conformal field theory with just two conformal blocks.
	
	\end{enumerate}
\end{exercise}

\subsection{2d Bosonization}\labelSection{bosonization_2d}

In the previous section, we discussed the compact boson at radius $R$, noting that it had a duality under which $R \to 2/R$. We remarked that this suggests the theory has a special point, $R = \sqrt{2}$, at which the theory gains special extra symmetries. However, this is not the only special point the compact boson possesses.

Let's look again at \eqref{compact_boson_vertex_ops}:
\be
V_{n,w} = :\exp(i k_+ \phi_+ + i k_- \phi_-): \qquad k_\pm = n \pm \frac{w R^2}{2} \quad \mbox{for $n,w\in \mathbb{Z}$} \fstp \relabeleq{compact_boson_vertex_ops} \ee
At the point $R=\sqrt{2}$, the momenta $k_\pm = n \pm w$ have a special structure; all the vertex operators can be expressed in terms of integer powers of the (anti)holomorphic fields $\exp(i \phi_+),\exp(i \phi_-)$ combined with either the identity or $V_{0,1}$. This drastically simplifies the spectrum of the theory.

However, it also makes clear that whenever $R^2$ is rational, a similar simplification will occur.\footnote{Technically, at each such point, the theory is a \textit{rational} CFT. This type of theory has a physical Hilbert space that splits into a finite number of irreducible representations of the chiral symmetry algebra. (An even more special case arises if there are only a finite number of representations -- so-called Verma modules -- of the chiral Virasoro algebra inside the full symmetry algebra. This gives the totally solvable \textit{minimal models}.) Rational CFTs have various nice properties, including that all operator dimensions are rational numbers.} One other especially nice point is at $R=1$ (or equivalently $R=2$). Here, $k_\pm = n \pm \frac{1}{2}w$.

Most of the work is left to an exercise broadly following \cite{tonggt}, but at this point one can show that, for example, the two-point function is
\be \left\langle e^{i\phi_{\pm}\left(x\right)}e^{-i\phi_{\pm}\left(y\right)}\right\rangle = \left(\frac{\epsilon}{\epsilon\pm i\left(x-y\right)}\right)^{1/R^{2}} \ee
where $\epsilon$ is a regularization parameter (see exercise). This can be argued from a standard computation of the dimension of this operator from a CFT perspective, or derived directly from canonical quantization.

\begin{exercise}[subtitle=The Two-Point Function]
	Define $\zeta$ to be the conjugate momentum to $\phi$. Deduce that we can quantize the theory in the Schr\"odinger picture using
	\begin{align}
		\phi &= \frac{\sqrt{4\pi}}{R}\int\frac{\rmd p}{2\pi}\frac{1}{\sqrt{2|p|}}\left(a_{p}e^{ipx}+a_{p}^{\dagger}e^{ipx}\right)e^{-\left|p\right|\epsilon/2} \nn \\
		\zeta &= -i\frac{R}{\sqrt{4\pi}}\int\frac{\rmd p}{2\pi}\sqrt{\frac{\left|p\right|}{2}}\left(a_{p}e^{ipx}-a_{p}^{\dagger}e^{-ipx}\right)e^{-\left|p\right|\epsilon/2} \nn
	 \end{align}
	 if $\left[a_{p},a_{q}^{\dagger}\right]=2\pi\delta\left(p-q\right)$.
	 Note that $\partial_{x}\tilde{\phi} = (R^2/2)\dot{\phi} = 2\pi \zeta$, and hence
	 \be \phi_\pm = \frac{1}{2}\left[\phi\pm\frac{4\pi}{R^{2}}\int_{-\infty}^{x}\zeta \right] \fstp  \ee 
	 Check that
	 \be G_{\pm}	=\left\langle \phi_{\pm}\left(x\right)\phi_{\pm}\left(y\right)\right\rangle -\left\langle \phi_{\pm}\left(0\right)\phi_{\pm}\left(0\right)\right\rangle 
	 =\frac{1}{R^{2}}\log\left(\frac{\epsilon}{\epsilon\pm i\left(x-y\right)}\right) \ee
	 and deduce, using the BCH formulae, that
	 \be \left\langle e^{i\phi_{\pm}\left(x\right)}e^{-i\phi_{\pm}\left(y\right)}\right\rangle =e^{G_{\pm}\left(x,y\right)}=\left(\frac{\epsilon}{\epsilon\pm i\left(x-y\right)}\right)^{1/R^{2}} \fstp \ee
\end{exercise}

But at $R=1$, this is exactly the two-point function of a free fermion! Remarkably, it sounds like we have discovered the result that a compact boson at $R=1$, the theory is a free Dirac fermion with components
\be \label{2d_fermion_ops} \psi_{\pm}=\frac{1}{\sqrt{2\pi\epsilon}}e^{\mp i\phi_{\pm}} \ee 
where our conventions are given by
\be \gamma^0 = \begin{pmatrix}0 & 1 \\ 1 & 0\end{pmatrix}, \quad \gamma^1 = \begin{pmatrix}0 & 1 \\ -1 & 0\end{pmatrix}, \quad \psi = \begin{pmatrix} \psi_+ \\ \psi_- \end{pmatrix} \fstp \ee

It may seem rather alarming that there is such a direct map between bosonic and fermionic operators. After all, these types of particles are distinguished by fundamental statistical properties. However, this is based on fundamentally 3+1 dimensional thinking: we are used to the representation theory of $SO(3)$ and its double cover $\Spin(3)$. But with one spatial dimension, there is no continuous rotation group at all! We cannot hope to distinguish bosons and fermions by smoothly exchanging them; the difference only becomes manifest when we bring two particles to the same point. This pushes such questions into the world of short-distance, non-universal physics, and makes it rather less surprising that the ideas blur together.

Many people are therefore happy to assert that the $R=1$ compact boson is indeed a free fermion. However, this is not really true! The operators \eqref{2d_fermion_ops} which we just identified as corresponding to a free fermion are not allowed operators in the bosonic theory! This is easily seen from \eqref{compact_boson_vertex_ops}; these operators have $k_+ = 1$ and $k_-=0$, which requires $w=1$ but $n=1/2$, which is banned. This subtlety is often overlooked, and is discussed below.

%

\begin{asidebox}[``Emergeability'' and the Sign of the Fermion]%
	In the language of Senthil \cite{senthiltalk}, the theory of a free fermion is not \textit{emergeable} from the theory of the compact boson: the physical Hilbert space of the boson simply does not contain any anticommuting local operators that can possibly be given by the fermion! This seems at odds with the duality we have described. The resolution is in some respects rather mild, though in other ways it is a prime example of the sort of subtleties one has to get right in order to properly understand the sort of dualities we will see throughout this course.
	
	The solution is in some ways fairly intuitive. Morally speaking, the \textit{sign} of the fermions cannot be determined from the duality -- they are effectively square roots of well-defined bosonic operators. This sign ambiguity is what allows them to obey anticommutation relations rather than commutation relations. More precisely, if we attempt to change variables in our path integral, then we find that nothing allows us to predict what sign $\psi$ has at each point, and configurations must be recognized as identical if they differ only by the sign of $\psi$. The resolution is simply that this $\mathbb{Z}_2$ ambiguity can be thought of instead as a gauge redundancy. If we gauge the $\mathbb{Z}_2$ symmetry $\psi \to -\psi$, then technically the fermion must be dressed with a line operator connecting it to a second fermion in order to restore gauge invariance; there is no local fermionic operator, avoiding the problem mentioned above.
	
	In practice this simply means we make two slight modifications to the usual theory of the free fermion. Firstly, the Hilbert space must contain only gauge invariant states -- for us, that is simply those with an even number of fermions. This is a very reasonable restriction. (This corresponds to the restriction $k_+ + k_- \in 2 \mathbb{Z}$.) The second is that we must sum over the value of any $\mathbb{Z}_2$ Wilson lines -- which is a fancy way of saying that we should sum over both periodic and antiperiodic boundary conditions for fermions on compact domains. (This is reflected in the existence of the operator $\exp(i \tilde{\phi}$).) Again, this is in some ways a relatively minor modification of the theory.
\end{asidebox}

This is reflected in the identification of currents that also follows from the above computation:
\begin{align}
	j_{\phi}^\mu = \frac{1}{2\pi} \epsilon^{\mu\nu} (\rmd \phi)_\nu &= \bar{\psi}\gamma^\mu\psi  = j_V^\mu \\
	j_{\tilde{\phi}}^\mu = \frac{1}{4\pi} (\rmd \phi)^\mu &= \frac{1}{2} \bar{\psi}\gamma^\mu\gamma^3\psi = \frac{1}{2} j_A^\mu
\end{align}
The left-hand currents here are integer quantized (corresponding to the $n,w$ quantum numbers), whilst for the case of a true free fermion (with periodic boundary conditions), the right-hand sides should equal $n_+ + n_-$ and $\frac{1}{2}(n_+ - n_-)$ respectively. Clearly these are not consistent.

For completeness, we include a statement of this fixed duality:
\be \mbox{compact scalar at $R=1$} \qquad \longleftrightarrow \qquad \mbox{Dirac fermion} + \mathbb{Z}_2 \mbox{ gauge theory} \fstp \ee 
This ultimately doesn't spoil the simple intuition that a fermion emerges from the left-hand theory, but it is good to be aware of the subtlety. You may wonder whether this can be inverted, to give a free fermion in terms of a gauged scalar. The answer is yes, although it requires a little subtlety to get right: we need to make a theory of a scalar sensitive to a spin structure, which is possible only by introducing a particular topological term for the $\mathbb{Z}_2$ gauge field. See \cite{Karch:2019lnn} and the references therein for a more careful discussion. This actually has a beautiful parallel with the story we will tell.

There are plenty more things which can be said about this system, but we will just outline briefly two of them. Firstly, one can actually work at a general radius by exploiting the bijection of the conserved currents, since
\be (\rmd \phi)^2 = - 4\pi^2 (\bar{\psi}\gamma^\mu \psi)^2 \fstp \ee
Therefore, we can write $R^2 = 1 + g/\pi$ and then we are left with a duality 
\be \frac{R^2}{8\pi} (\rmd \phi)^2 \bigdual i \bar{\psi}\slashed{\partial}\psi - \frac{g}{2} (\bar{\psi}\gamma^\mu \psi)^2 \ee
between a compact boson at a general radius and a ($\mathbb{Z}_2$ gauged) \textit{Thirring model}.
This is a rare example of an \textit{exactly marginal} deformation of a CFT, which simply moves one straight to another conformal theory.

Secondly, the bijection between operators also allows us to extend the above duality by adding a relevant operator. Concretely, it follows from the above identifications that the mass term $\bar{\psi}\psi$ is dual to a potential $\cos \phi$. This leads to the identification of so-called ``Sine-Gordon'' theory with the massive fermion, a famous story going back to Coleman \cite{Coleman:1974bu,Mandelstam:1975hb}.

Some of these details are helpful for understanding higher-dimensional dualities (like the subtleties surrounding statistics, and identification of mass deformations), but others (like the existence of a marginal parameter) are specifically two-dimensional.

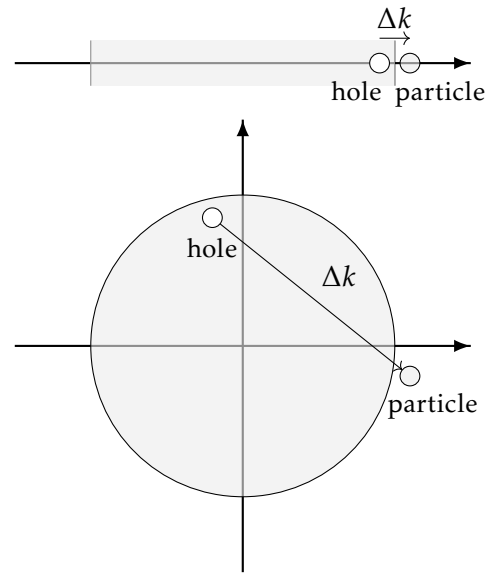
\begin{wrapfigure}{r}{0.4\textwidth}
	\centering
	\begin{tikzpicture}
	\draw[axis] (-3,0) -- (3,0);
	\draw (-2,-0.3) -- (-2,0.3); \draw (2,-0.3) -- (2,0.3);
	\fill [fill=gray!15,fill opacity=0.6] (-2,-0.3) rectangle (2,0.3);
	\coordinate (h) at (1.8,0); \coordinate (p) at (2.2,0);
	\draw[fill=white] (h) circle [radius=0.13]; \draw[fill=gray!15,fill opacity=0.6] (p) circle [radius=0.13];
	\node[anchor=mid] at ([yshift=-0.4cm,xshift=-0.3cm]h) {\small hole};
	\node[anchor=mid] at ([yshift=-0.4cm,xshift=0.4cm]p) {\small particle};
	\draw[->] ([yshift=0.33cm]h) -- node[above] {$\Delta k$} ([yshift=0.33cm]p);
	\end{tikzpicture}
	\begin{tikzpicture}
	\draw[axis] (-3,0) -- (3,0);
	\draw[axis] (0,-3) -- (0,3);
	\draw[fill=gray!15,fill opacity=0.6] circle [radius=2];
	\coordinate (h) at (-0.4,1.7); \coordinate (p) at (2.2,-0.4);
	\draw[fill=white] (h) circle [radius=0.13]; \draw[fill=gray!15,fill opacity=0.6] (p) circle [radius=0.13];
	\node at ([yshift=-0.4cm]h) {\small hole};
	\node at ([yshift=-0.4cm,xshift=0.3cm]p) {\small particle};
	\draw[->,shorten >=0.13cm,shorten <=0.13cm] (h) -- node[above right] {$\Delta k$} (p);
	\end{tikzpicture}
	\caption{Low-energy excitations of Fermi surfaces in one and two spatial dimensions; notice that only in more than one dimension do these low-energy particle-hole pairs generically have large relative momenta $\Delta k$}\labelFigure{fermi_surfaces_2d_3d}
\end{wrapfigure}
There are other physical ways to see that this straightforward type of bosonization belongs in one spatial dimension. One intuitive way to understand the bosonization map $\bar{\psi} \psi \sim \partial \phi$ is to realize that a particle-hole pair of fermions is bosonic in character.

Let's think in terms of a Fermi surface at finite chemical potential, as pictured in \refFigureOnly{fermi_surfaces_2d_3d}. It is easy to see that in one dimension, low-lying particle-hole pairs have low relative momentum (and more relevantly group velocities $\partial E/\partial k$ which are almost identical), so that they propagate coherently. Therefore, they can be thought of loosely as a single particle (and any weak attraction between them will indeed create such a bound state). This is only an intuitive picture, of course, but it helps to illustrate how fermions can combine into bosonic excitations.

In higher dimensions, there are low-energy particle-hole pairs with relative momenta all the way up to $k_F$, and so it is hard to bind particles and holes together in such a naive way. One needs a more sophisticated approach to relate fermions and bosons in higher dimensions; we will see in the next section that in two spatial dimensions bosonization relies crucially on a new mechanism.

\begin{exercise}[subtitle=Another optional 2d CFT exercise]
	2d bosonization suggests another nice exercise for anyone enthusiastic about 2d CFTs. Take the partition function \eqref{partition_function_2d_scalar} and set $R=2$, showing first that it takes the form
	\be \label{partition_function_2d_scalar_fermion}
	Z(\tau, 2) = \frac{1}{2|\eta|^2} \left( \left|\sum_n q^{n^2/2} \right|^2 + \left|\sum_n (-1)^n q^{n^2/2} \right|^2 + \left|\sum_n q^{\frac{1}{2}(n+\frac{1}{2})^2/2} \right|^2 \right)
	\ee
	and then that this can be interpreted as the partition function of a $\mathbb{Z}_2$-gauged fermion:
	\be
	Z_F = \frac{1}{2} \left( \tr_{\text{periodic BCs}} \frac{1 - (-1)^F}{2} q^{\textbf{L}_0 - \frac{1}{24}} \bar{q}^{\tilde{\textbf{L}}_0 - \frac{1}{24}} + \tr_{\text{antiperiodic BCs}}  \frac{1 - (-1)^F}{2} q^{\textbf{L}_0 - \frac{1}{24}} \bar{q}^{\tilde{\textbf{L}}_0 - \frac{1}{24}} \right) \fstp
	\ee
	For the first part, it might be helpful to think about the physics of different $n,w$ in \eqref{partition_function_2d_scalar}. For the last part, notice that the three terms in \eqref{partition_function_2d_scalar_fermion} are theta functions which have useful infinite product representations. ($F$ counts the number of fermions, so the above projects onto sectors with even numbers of fermions.)
\end{exercise}

\begin{asidebox}[Non-Abelian Bosonization]%
	The above story allows us to formulate a bosonic description of a theory of a single complex fermion in 1+1 dimension. This theory has an obvious $U(1)$ symmetry (or more precisely $U(1)_L \times U(1)_R$) which is realized by the dual scalar. But let's follow Witten \cite{Witten-nonab} in asking a sensible question: suppose we take $N$ free complex fermions. If we bosonize them all according to the above recipe, we get a theory of $N$ compact bosons, which seems to have a continuous symmetry group $U(1)_L^N \times U(1)_R^N$. But the fermionic theory has a manifest $U(N)_L \times U(N)_R$ symmetry.
	
	This gives us a concrete example of a hidden symmetry; the theory of $N$ compact bosons has an emergent $U(N)$ structure. However, it also raises the question of whether or not there is a dual theory which makes the $U(N)$ structure manifest. Witten answered this question with a definitive and remarkable "yes": he showed that $N$ complex fermions are equivalent to a bosonic sigma model into $U(N)$ with a very particular topological term. In fact, the overall $U(1)$ factor decouples and gives a single compact boson. The resulting duality goes by the name \textit{non-Abelian bosonization}, and the non-Abelian part of the sigma model is the $SU(N)$ Wess-Zumino-(Novikov-)Witten model (\textit{WZW model}) at level 1, commonly written as $SU(N)_1$. 	
	
	We will not discuss this story in any detail here, since we are ultimately interested in higher dimensional dualities; we will just give the key ingredients of the operator dictionary. In fact, the thinking of the compact boson $\phi$ as lying in the $u(1)$ Lie algebra, we can make some guesses. Firstly, the analogous thing to $e^{i\phi/R} \in U(1)$ is $g_{ij} \in SU(N)$. Accordingly, we expect that this is dual to the mass term:
	\be e^{i \phi/R} g_{ij} \qquad \longleftrightarrow \qquad \psi^\dagger_{+ i} \psi_{j} \ee
	where we have included a separate compact boson to capture the $U(1)$ part. Secondly, the conserved currents lying in the Lie algebra are
	\be \label{nonabelian_2d_op_map} g^{-1} \partial_+ g, \partial_- g g^{-1} \qquad \longleftrightarrow \qquad  \psi^\dagger \gamma_\pm \psi - \mbox{trace} \ee
	where the left and right $SU(N)$ symmetries $g \to h_L g h_R$ are appropriately associated with left- and right-movers.
\end{asidebox}

\subsection{Flux Attachment}\labelSection{flux_attachment}

The type of bosonization discussed in \refSectionOnly{bosonization_2d} seems inherently two-dimensional: as we mentioned above the distinction between fermions and bosons is very weak with only one spatial dimension, since we cannot exchange particles without bringing them through each other.

However, with two spatial dimensions, we are of course free to try to identify the statistics of well-separated particles by orbiting them around each other. This is much more like the familiar story in three spatial dimensions, although now the rotation group $SO(2) \equiv U(1)$ does not have a simply connected double-cover.

It is worth taking a moment to think about this issue. Firstly, consider a single particle. We can certainly define bosons and fermions according to whether a single particle state lies in an integer or half-integer representation of the double-cover $\Spin(2)$. However, there are many \textit{more} ways that particles could in principle transform under rotations. Since there is no compact covering group of $SO(2)$, there is no rotation which must act as the identity on the Hilbert space. Instead, rotations may act as arbitrary unitary operators on the Hilbert space! A state could easily have spin $1/3$, for example, so that the state rotates as $\ket{\psi} \to \exp(i \theta/3) \ket{\psi}$, returning to itself only after a $6\pi$ rotation. This state is neither bosonic nor fermionic; it is an example of an \textit{anyon}.\footnote{If this is a 2+1 dimensional system embedded in a 3+1 dimensional world (with a potential restricting us to a plane, say), then this could not be a true physical state. However, it could be that the fundamental excitations of an effective theory are anyonic. The only restriction would then be that physical states contain combinations of anyons which have (half-)integer spin. This is exactly the situation found in the \textit{Fractional Quantum Hall Effect} (FQHE).}

\begin{figure}[h]
	\centering
	\begin{tikzpicture}
	\node[scale=3] (rot) at (-4,0) {$\circlearrowleft$};
	\node at (-3,0) {$:$};
	\node[below of=rot] {$2\pi$};
	\node[anchor=east] (boson) at (0,1) {$\ket{\,\mathrm{boson}}$};   \node[anchor=west] (boson2) at (1.7,1) {$\ket{\,\mathrm{boson}}$};
	\draw[->] (boson) -- (boson2);
	\node[anchor=east] (fermion) at (0,0) {$\ket{\,\mathrm{fermion}}$}; \node[anchor=west] (fermion2) at (1.7,0) {$-\ket{\,\mathrm{fermion}}$};
	\draw[->] (fermion) -- (fermion2);
	\node[anchor=east] (anyon) at (0,-1) {$\ket{\,\mathrm{anyon}}$};  \node[anchor=west] (anyon2) at (1.7,-1) {$e^{2\pi i s}\ket{\,\mathrm{anyon}}$};
	\draw[->] (anyon) -- (anyon2);
	\end{tikzpicture}
	\caption{The representation theory of the covering group of $SO(2)$ (i.e. the additive group $\mathbb{R}$) allows the $2\pi$ rotation to be represented by an arbitrary phase $e^{2\pi i s}$}\labelFigure{anyon_rotation}
\end{figure}

This is quite remarkable -- decomposing into irreducible representations of the additive covering group $\mathbb{R}$, we see that a general state can have an arbitrary spin $s\in\mathbb{R}$. The fact that this is an Abelian group at least guarantees that we have one-dimensional representations. But things get even more surprising when we start to consider multi-particle states.

Consider a state containing two identical particles. We can act with translation or rotation operators to smoothly exchange them, and then see how the state has transformed. We can impose is that the final state is \textit{equivalent} to the initial state, but not that it is identical. Thus they are related by an arbitrary unitary operator $U$ which need only be a symmetry of the theory. In higher dimensions, one can prove that $U^2 = 1$, but that does not work here. The situation with more particles is even richer: if we have three particles $1,2,3$, then the operators $U_{12}$ and $U_{23}$ which exchange pairs of particles need not commute! Instead, with $n$ particles, the operators $U_{ij}$ form a unitary representation of the braid group $B_n$, as shown in \refFigureOnly{anyon_braiding}. For large $n$, this group is a very complicated non-Abelian group. Particles with such a property are \textit{non-Abelian anyons}.

\begin{figure}[h]
	\centering
	\begin{tikzpicture}
	\filldraw (-1.8,0) circle (3pt); \filldraw (0,0) circle (3pt); \filldraw (1.8,0) circle (3pt);
	\draw[->,semithick,purple]
		(-1.8,-0.2) arc[radius=1.8, start angle=-165, end angle=-15]
		(1.8,0.2) arc[radius=1.8, start angle=15, end angle=165] node[above left] {$U_{13}$};
	\draw[->,semithick,red]
		(-1.6,-0.2) arc[radius=0.8, start angle=-150, end angle=-30]
		(-0.2,0.2) arc[radius=0.8, start angle=30, end angle=150] node[below right]{$U_{12}$};
	\draw[->,semithick,blue]
		(0.2,-0.2) arc[radius=0.8, start angle=-150, end angle=-30]
		(1.6,0.2) arc[radius=0.8, start angle=30, end angle=150] node[below right]{$U_{23}$};
	\node[anchor=west] at (2.5,1.6) {${\color{purple} U_{13}} = {\color{blue} U_{23}} {\color{red}U_{12}} {\color{blue} U_{23}} \quad \neq \quad {\color{red}U_{12}} {\color{blue} U_{23} U_{23}}$};
	\braid[line width=1pt,width=1cm,anchor=north,style floors={1}{fill=blue},style floors={2}{fill=red},style floors={3}{fill=blue},floor command={\fill[fill opacity=0.07] (\floorsx,\floorsy) rectangle (\floorex,\floorey);},style floors={1}{fill=blue},style floors={2}{fill=red},style floors={3}{fill=blue}] at (3.1,1.2) | a_2 | a_1 | a_2;
	\braid[line width=1pt,width=1cm,anchor=north,style floors={1}{fill=blue},style floors={2}{fill=red},style floors={3}{fill=blue},floor command={\fill[fill opacity=0.07] (\floorsx,\floorsy) rectangle (\floorex,\floorey);},style floors={1}{fill=red},style floors={2}{fill=blue},style floors={3}{fill=blue}] at (6.7,1.2) | a_1 | a_2 | a_2;
	\draw[->] (9.2,-2.3) -- node[right] {time}  (9.2,1.2);
	\end{tikzpicture}
	\caption{The action of exchange operators can be an arbitrary unitary matrix; moreover, distinct exchange operators need not commute. On the right, we distinguish between two operator orderings which are not equal in a general non-Abelian representation of the braid group. This can be seen by drawing them as braids which are clearly topologically distinct. Note the first (earliest) operator is on the right in an operator product}\labelFigure{anyon_braiding}
\end{figure}
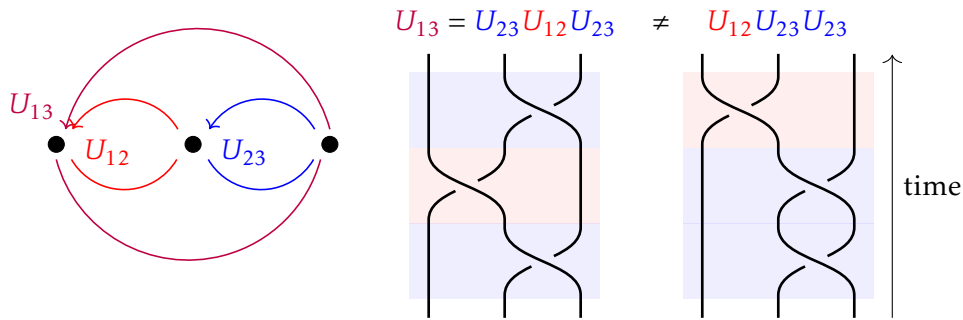

Strikingly, it is possible to have a multi-particle Hilbert space which is larger than the product of single particle Hilbert spaces \cite{simontopquant}; there is extra structure inherent in the entanglement between even widely separated particles. The physics of anyons is very rich indeed.

This is all very interesting, but we haven't found a way to see how bosonization is possible in two spatial dimensions. The trick is to exploit the fact that there can be \textit{multiple} contributions to the phase of a state when we exchange particles. If we can find a way to start with a pair of fermionic particles with a exchange phase of $\pi$, then add something to the theory which generates an extra phase of size $\pi$, we would actually obtain bosonic behaviour. There is even a simple, familiar way to generate phases: the Aharanov-Bohm effect.

Suppose that we start off with a bosonic particle, but then give it both an electric charge $q$ and a magnetic charge $q'$ under a gauge field. (This means it is a \textit{dyon}.) Then the Aharanov-Bohm effect dictates that exchanging two such particles, we will acquire an additional phase $\pi qq'$. Therefore, we postulate a first, very tentative result in 2+1 dimensions:
\be \text{dyonic boson with charge $(1,1)$} \qquad \stackrel{?}{\longleftrightarrow} \qquad \text{fermion} \ee
where both theories are simply quantum mechanics. This process is called \textit{flux attachment}, since we are attaching magnetic flux to a particle.
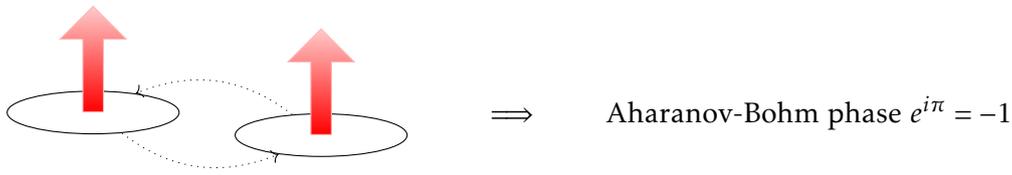
\begin{figure}[h]
	\centering
	\begin{tikzpicture}
	\tikzset{arrowfill/.style={top color=red!20, bottom color=red}}
	\tikzset{arrowstyle/.style={draw=red!30,arrowfill,minimum height=#1,single arrow,single arrow head extend=.3cm,shape border rotate=90,anchor=south}}
	\newcommand{\objwithflux}[2]{
		\node[draw,ellipse,fill=white,inner xsep=0.8cm,inner ysep=0.2cm] (#2) at #1 {};
		\node[arrowstyle=1.4cm] at #1 {};
	}	

	\objwithflux{(0,0)}{p1}
	\objwithflux{(3,-0.3)}{p2}
	\draw[bend right,->,dotted] (p1) to (p2);
	\draw[bend right,->,dotted] (p2) to (p1);
	\node[anchor=west] at (5,0) {$ \implies \qquad $ Aharanov-Bohm phase $e^{i\pi} = -1$};
	\end{tikzpicture}
	\caption{The cartoon of flux attachment: each charge $1$ boson also carries $1$ unit of magnetic flux around with it. (Flux is represented by an arrow, since the magnetic field points out of the plane.) This means that the Aharanov-Bohm effect generates a phase of $2\pi$ when one such particle encircles another, or $\pi$ when two are exchanged, turning the boson into a fermion}\labelFigure{flux_attachment}
\end{figure}

\begin{exercise}[subtitle=Aharanov-Bohm Effect]
	The usual statement of the Aharanov-Bohm effect is that an electric particle of charge $q$ picks up a phase $q \int_C a$ when moving along a contour $C$. Assuming that a particle of magnetic charge $q'$ has total magnetic flux $2\pi q'$, derive carefully from this result that exchanging two dyonic particles with charges $(q,q')$ generates an Aharanov-Bohm phase of $\pi q q'$. There's a factor of 2 you should worry about.
\end{exercise}

Let's formulate this a little more precisely. The quantum mechanics of $n$ charged bosons is 
\be H = \sum_{m=1}^n \frac{1}{2m} (\textbf{p}_i - q \textbf{a}(\textbf{x}_i))^2 + V(\textbf{x}_i) \ee
where $\textbf{a}$ is the gauge field. However, in order to give the bosons a magnetic charge too, we also need to impose
\be \label{flux_attachment_magnetic_field} \frac{1}{2\pi} B(\textbf{x}) = q' \sum_{m=1}^n \delta(\textbf{x} - \textbf{x}_m) \ee
where $B = \nabla \times \textbf{a} = \partial_1 a_2 - \partial_2 a_1$ is the magnetic field strength. We take the bosons to be identical, so the wavefunction must be symmetric under interchange of all pairs of particles.

We can solve \eqref{flux_attachment_magnetic_field} explicitly in the Coulomb gauge $\partial_i a_i=0$ using the Green's function $\nabla^2 \frac{1}{2\pi} \log |\textbf{x} - \textbf{y}| = \delta^{(2)}(\textbf{x}-\textbf{y})$ together with the 2d identity $\epsilon_{ij} \partial_i \epsilon_{jk} \partial_k = - \partial_i \partial_i$:
\be a_i(\textbf{x}) = - q' \epsilon_{ij} \partial_j \sum_{m=1}^n \log |\textbf{x} - \textbf{x}_m| = q' \partial_i \sum_{m=1}^n \arg (\textbf{x} - \textbf{x}_m) \ee
where $\arg$ measures the angle between its argument and the $x_1$-axis. (Recall that $\log z = \log r + i\theta$, so $\theta$ is the harmonic conjugate of $\log r$.)


This seems to suggest that the gauge field is pure gauge, since we have written it as a total derivative. However, because $\arg$ is not single-valued, this is a bit too quick. Suppose we begin with a wavefunction $\Psi(\textbf{x}_1,\ldots,\textbf{x}_n)$ with conventional boundary conditions and then do a gauge transformation with parameter $\lambda(\textbf{x}) = - q' \sum_{m=1}^n \arg(\textbf{x} - \textbf{x}_m)$. Then the new wavefunction, in the $\textbf{a} = 0$ gauge, is
\be \tilde{\Psi}(\textbf{x}_1,\ldots,\textbf{x}_n) = e^{-i  q q' \sum_{m< m'} \arg(\textbf{x}_m - \textbf{x}_{m'})} \Psi(\textbf{x}_1,\ldots,\textbf{x}_n) \ee
where the sum is over all distinct pairs of particles \cite{lerda,jackiwpi}; there is an awkward factor of two one has to get right here. It follows that upon exchanging a pair of particles, $\tilde{\Psi}$ now receives the Aharanov-Bohm contribution $e^{-i \pi q q'}$.

If we take $q = q' = 1$ as proposed, these indeed cancel to give rise to fermion-like boundary conditions!\footnote{Actually, this is still slightly too quick. See Exercise \ref{exercise_flux_attachment_bcs}.} The Hamiltonian acting on $\tilde{\Psi}$ is now simply
\be \tilde{H} = \sum_{m=1}^n \frac{1}{2m} \textbf{p}_i^2 + V(\textbf{x}_i) \fstp \ee
We could clearly also add, for instance, a potential depending on the separation of the particles without spoiling this result.

\begin{exercise}[subtitle=Boundary Conditions on Wavefunctions]
	\label{exercise_flux_attachment_bcs}%
	By considering the wavefunction governing the relative motion of two such particles, show that to regularize the contact interaction when two particles come together, we should modify the boundary conditions to behave as $|\textbf{x}_1-\textbf{x}_2|^{\pm q'/q}$ as particles approach. Show that this translates into 
	\be H_{\mathrm{fixed}} = \sum_{m=1}^n \frac{1}{2m} (\textbf{p}_i - q \textbf{a}(\textbf{x}_i))^2 + V(\textbf{x}_i) + \frac{2\pi q'}{mq} \sum_{i<j} \delta^{(2)} (\textbf{x}_i-\textbf{x}_j) \fstp \ee
\end{exercise}

This establishes our result: bosons with dyonic charge $(1,1)$ are identical to fermions! Moreover, we can even embed this result in a non-relativistic field theory. Consider the action
\begin{align}
	\label{flux_attachment_nonrel_action}
	S &= \int \rmd^3x \ i \phi^\dagger (\partial_0 - iq a_0) \phi - \frac{1}{2m} |(\partial - iqa)\phi|^2 - V(x) |\phi|^2 - \frac{\pi q'}{m q} |\phi|^4 \nn\\
	& \qquad \qquad + \frac{q}{4\pi q'} \epsilon^{\mu\nu\rho} a_\mu \partial_\nu a_\rho \end{align}
for a gauge field $a_\mu$ and a complex field $\phi$. This is an example of a \textit{non-relativistic field theory}. One can take the density of particles to be $|\phi|^2 = \sum \delta^{(2)}(\textbf{x} - \textbf{x}_i)$; this reduces this theory to exactly the quantum mechanics described above if we work in the $a_0 = 0$ gauge. In particular, the $a_0$ equation (Gauss's law) is
\be \frac{1}{2\pi} B = q' |\phi|^2 \ee
implementing the flux attachment which is key to the \textit{statistical transmutation} we have discussed.

The final term in \eqref{flux_attachment_nonrel_action} is something that will become very familiar to us: it is a $U(1)$ \textit{Chern-Simons term} at \textit{level} $k = q/q'$. It is in fact a topological term, independent of the metric, as is clearly seen by writing it in the form
\be S_{\mathrm{CS}} = \frac{k}{4\pi} \int a \rmd a \fstp \ee
Notice that, if we rescale $a$ to set the electric charge to $q=1$, then the magnetic charge and statistical phase are given by $q' = 1/k$ and $\theta = 2\pi/k$ respectively in terms of the level $k$. We will discuss such terms more when we get to \refSection{chern_simons_anomalies}; see also \refAppendix{chern-simons}.


\subsection{The Dual Photon}\labelSection{dual_photon}

Let us now turn to three dimensions, and for the first time consider a gauge theory with $U(1)$ gauge field $a_\mu$. Consider the action
\be
S[a_\mu] = \int \rmd^3 x \ -\frac{1}{4g^2} f_{\mu\nu}f^{\mu\nu}
\ee
of a free photon. The partition function is
\be
Z = \int Da \ \exp(i S[a_\mu])
\ee
where of course one has to treat the gauge-invariance of the path integral correctly.

However, there is another way of looking at this theory. Note that $S$ depends on $a_\mu$ only through the gauge invariant quantity $f_{\mu\nu}$. Therefore, it is possible to replace the path integral over $a_\mu$ with one over $f_{\mu\nu}$. The only thing we have to take care of is the fact that $\rmd f$ = 0. Therefore, let us include a Lagrange multipler $\sigma$ for this constraint. This shows that the above action is equivalent to
\be \label{dual_photon_lagrange}
Z = \int Df D\sigma \ \exp \left[ i\int \rmd^3 x \ -\frac{1}{4g^2} f_{\mu\nu}f^{\mu\nu} + \frac{1}{4\pi} \sigma \epsilon^{\mu\nu\rho} \partial_\mu f_{\nu\rho} \right] \fstp
\ee
But now we can integrate out $f_{\mu\nu}$, which only appears quadratically, by using its equation of motion, $f^{\mu\nu} = - \frac{g^2}{2\pi}\epsilon^{\mu\nu\rho}\partial_\rho\sigma$:
\be
Z = \int D\sigma \ \exp \left[ i\int \rmd^3 x \ \frac{g^2}{8\pi^2} (\partial\sigma)^2 \right]
\ee
Rather remarkably, we discover that in 2+1 dimensional space the free photon $a_\mu$ is totally equivalent to a free scalar $\sigma$! The scalar $\sigma$ is called the \textit{dual photon} for obvious reasons.

Much of the above is more naturally formulated in the language of differential forms, so that, for instance,
\be \label{gaugefield-dualphoton-reln} \rmd a = - \frac{g^2}{2\pi} \star \rmd \sigma \qquad \mbox{or} \qquad \star \rmd a = - \frac{g^2}{2\pi} \rmd \sigma \ee
which makes clear one of the odd properties of this type of duality: Bianchi identities are interchanged with equations of motion. In the photon language, it is obvious that $\rmd \rmd a = 0$ but $\rmd \star \rmd a = 0$ is the standard equation of motion. On the other side of the duality, the former becomes $\rmd \star \rmd \sigma = 0$ which is the equation of motion, whilst the latter becomes $\rmd \rmd \sigma = 0$, which is trivial. In general, in $d$ dimensions, the theory of  a $p$ form and a $d - 2 - p$ form are related in this way. Another classic example of this is in four dimensions, where the $1$ form theory of electromagnetism is dual to another $1$ form theory. This is what is commonly referred to as \textit{electromagnetic duality}, which exchanges $\textbf{B}$ and $\textbf{E}$ fields. In particular, the 2 dimensional T-duality we have already discussed in \refSectionOnly{tduality} is the electromagnetic duality of 0-forms $\phi$ and $\tilde{\phi}$.

The comparison to 2 dimensional T-duality is interesting. It was not obvious that the dual variable $\tilde{\phi}$ in that case should be periodic; this arose from considering the quantization of vortex charge in that theory. It is certainly clear that our dual photon at least has a shift symmetry, $\sigma \to \sigma + c$. We have already learned something remarkable: the humble theory of the free photon in 2+1 dimensions has a secret Abelian symmetry. But to answer questions about its compactness (whether the group is $U(1)$) we need to understand this symmetry much better.

Let us translate this back into the usual language of the one form gauge field $a_\mu$. The conserved current associated to shifts of $\sigma$ (obtained by replacing $\rmd\sigma \to \rmd\sigma - A$ for a background gauge field $A$) is
\be j^\mu = -\frac{g^2}{4\pi^2} \partial^\mu  \sigma \fstp \ee
Using \eqref{gaugefield-dualphoton-reln}, the dual of this current is
\be
j^\mu = \frac{1}{2\pi} \epsilon^{\mu\nu\rho} \partial_\nu a_\rho \quad \text{or} \quad j = \frac{1}{2\pi} \star \rmd a
\ee
which satisfies $\partial_\mu j^\mu \propto \rmd \rmd a = 0$ due to the symmetry of partial derivatives. The corresponding conserved quantity $j^0 = \frac{1}{2\pi}f_{12}$ is magnetic flux, and thus we have an associated $U(1)$ symmetry often referred to as a \textit{magnetic symmetry}.%
\footnote{Notice that the Chern-Simons term of the previous section can be rewritten as $S_{\mathrm{CS}} = k \int \frac{1}{2} a_\mu j^\mu$. Clearly, the Chern-Simons represents a coupling that induces an electric field around magnetic charges: differentiating the action with respect to $a_\mu$ shows that $k j^\mu$ is given a unit electric charge in a theory with this term. In general, the spectrum of Chern-Simons theory consists of dyonic particles whose charges are determined by the Chern-Simons level.} This symmetry can be coupled to a background gauge field with a term $\int \rmd^3 x \, A_\mu j^\mu = \int \frac{1}{2\pi} A \wedge \rmd a$, a so-called \textit{BF term} linking the two gauge fields together.\footnote{The name comes from other contexts where the $A$ is called $B$, and the field strength $\rmd a$ is called $F$. It's a shame we rarely call gauge fields $B$.}

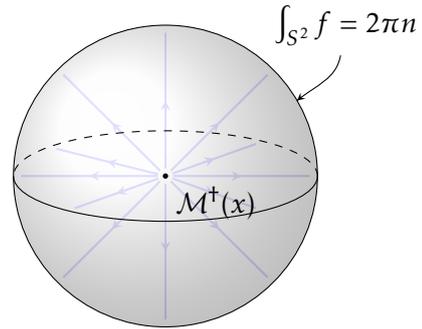
\begin{wrapfigure}{r}{0.4\textwidth}
	\centering
	\begin{tikzpicture}
	\foreach \theta in {0,45,...,315}
	\draw[flux,standoff] (0,0) -- (\theta:2);
	\foreach \theta in {50,140,230,310}
	\draw[flux,standoff] (0,0) -- (\theta:2 and 0.6);
	\shade[ball color = gray!4, opacity = 0.4] (0,0) circle (2cm);
	\draw circle [radius=2cm];
	\draw (-2,0) arc (180:360:2 and 0.6);
	\draw[dashed] (2,0) arc (0:180:2 and 0.6);
	\fill[fill=black] (0,0) circle (1pt) node[below right] {$\monoM^\dagger(x)$};
	\node (formula) [anchor=south west] at (1.3,1.6) {$\int_{S^2} f = 2\pi n$};
	\draw[labelarrow] (formula) to[out=-100,in=40] (30:2);
	\end{tikzpicture}
	\caption{A magnetic monopole is a defect from which there emerges a (quantized) magnetic flux}\labelFigure{monopole_operator}
\end{wrapfigure}

From this point of view, one should ask why this corresponds to a non-trivial symmetry, since it seems the current is \textit{identically} conserved. We normally think of Noether's theorem as relating symmetries to currents which are conserved on-shell, but we do not seem to have used the equations of motion.\footnote{The same questions can be asked in the 2 dimensional T duality. As remarked above, it is a general feature of electromagnetic duality that equations of motion and Bianchi identities are interchanged.} But just as the winding of vortex configurations in 2 dimensions hid a surprise, there is a subtlety here due to the non-trivial topology of the gauge group $U(1)_{\mathrm{gauge}}$.

Suppose we edit the path integral by removing a single point $x$ in spacetime. We must then specify boundary conditions for the gauge field on the $S^2$ surrounding the point $x$. 

The key ingredient is the \textit{monopole operator} $\monoM(x)$. This is a another so-called \textit{defect operator}, defined by editing the path integral to remove the point $x$, and then adding a non-trivial boundary condition for the gauge field on the surface surrounding that point. Acting with the monopole creation operator $\monoM^\dagger(x)$ imposes a flux of $\int_{S^2} f = 2\pi$. Just as with the 2 dimensional vortices, one can compute correlation functions involving $\monoM(x)$ and its dimension in the usual way, by inserting it into the path integral.

In particular, with the above choice of $f$, we can show that this gives the operator charge 1 under $j^\mu$.
Suppose we insert $\monoM^\dagger(x)$ at $x=(t,x_1,x_2)$. Now flatten the sphere so that the surface lies entirely in the spatial $x_1,x_2$ plane, forming two discs $S_+$ and $S_-$ at times $t_+$ and $t_-$ just before and after the insertion of $\monoM^\dagger(x)$. Since $\int_{\partial V} f = \int_{V} \rmd f = 0$, this does not change the value of the surface integral. Therefore,
\be 1 = \frac{1}{2\pi}\int_{S_+ \cup S_-} f = \int_{S_+} \rmd^2x \ \frac{1}{2\pi} f_{12} - \int_{S_-} \rmd^2x \ \frac{1}{2\pi} f_{12} = Q(t_+) - Q(t_-) \ee
showing that inserting this operator has increased the charge $Q = \int \rmd^2x \, j^0$ by 1 unit.

In fact, this is the lowest-charge monopole operator in the theory, and all other monopole operators have a charge which is a multiple of this, as can be seen by a careful mathematical analysis, or a more physical one. (See ``Monopoles from Topology''.)

\begin{asidebox}[Monopoles from Topology]%
The existence of these monopole operators is inextricably linked with the topology of the $U(1)$ gauge group \cite{Nakahara}. Given a gauge group $G$, a gauge field is a connection in a $G$-principal bundle. Now the $G$ bundles over an $\mathbb{R}^3$ spacetime are all trivial, so our usual intuition about gauge theory applies. Yet when we remove a point this is no longer the case. Since $\mathbb{R}^3 \backslash \{0\}$ and $S^2$ are of the same homotopy type, we are really interested in the bundle $P(S^2, G)$. Since $S^2$ is covered by two topologically trivial charts which are glued together along an $S^1$ equator, there can be a non-trivial structure to $P(S^2, G)$ associated with maps from $S^1 \to G$. More precisely, the homotopy classes are classified by $\pi_1(G)$. We have the result
\be \text{gauge group $G$ connected but not simply connected } \ \implies \ \exists \text{ monopole operators} \nn \ee
which gives rise to monopole operators whenever the group $G$ contains a $U(1)$ factor.

In particular, because $\pi_1(U(1)) = \mathbb{Z}$, the theory admits pointlike excitations labelled by a new quantized integer charge, namely which homotopy class the gauge field belongs to. We will check in a moment that gauge connections in the $n$th homotopy class (with first Chern number $n$) satisfy $\int_{S^2} f = 2\pi n$, corresponding to the quantization as claimed above and explained below. The $U(1)_{\mathrm{global}}$ monopole symmetry corresponding to this conserved charge is said to be \textit{topological}, since (a) the monopole operators charged under it only exist because of the topological properties of the gauge group, and (b) the charge is a topological invariant of a bundle.
	
	There is a much more physical way of phrasing the above argument, due to Dirac. We will think about the wavefunction of an electron (or any particle) of charge $1$ in the presence of a point-like magnetic charge such that $\int_{S^2} f = \tilde{q}$. The magnetic field is $\rmd a$, a `pure curl', in empty space, so if $a$ was universally defined, we $\int_{S^2} f = \int_{\partial S^2} a = 0$ since the sphere has no boundary. Therefore $a$ cannot be defined globally; we have to define it on two patches as above, say $a+$ on the upper and $a_-$ on the lower hemisphere. But on the equator, both patches overlap and the electron has two good wavefunctions in two different gauges. They are related by a gauge transformation as $a_+ - a_- = \rmd \chi$ for some $\chi$, which must be well-defined modulo $2\pi$ for our wavefunction to be single-valued. But the equator is the boundary of both the upper and lower hemisphere; hence 
	\be \int_{S^2} f = \int_{\text{upper}} f + \int_{\text{lower}} f = \int_{\text{equator}} a_+ - \int_{\text{equator}} a_- = \int_{\text{equator}} \rmd \chi = 2\pi n \ee
	for some integer $n$, which characterizes the twisting of the bundle. This proves the quantization condition.
	
	This is exactly the phenomenon of \textit{Dirac quantization}: the periodicity of the $U(1)$ gauge field (which is associated with the phenomenon of electric charge quantization) gives rise to the existence of magnetic monopoles (whose quantized charge is fixed in terms of the fundamental electric charge).
\end{asidebox}

So $\sigma$ is another compact scalar: with the above normalization, we have $\sigma \in [0,2\pi)$. Thus we have identified a hidden $U(1)$ symmetry which is a genuine global symmetry of the 2+1 dimensional photon, complete with charged excitations given by monopoles. This type of symmetry will be very important for us in what follows.

In fact, in terms of $\sigma \in [0,2\pi)$, the monopole operator is much less mysterious! Since we know that we want it to be an object carrying unit charge under the $U(1)$ symmetry, and $\sigma \to \sigma + c$ under this symmetry, we can simply look at operators like $\exp(i\tilde{q}\sigma)$. This is clearly a well-defined operator for $\tilde{q}\in\mathbb{Z}$, with the correct quantized charge to be a monopole operator.

This also suggests that the operator is loosely speaking unitary, which makes sense from the point of view of boundary conditions: inserting an antimonopole $\monoM(x)$ right on top of a monopole $\monoM^\dagger(x)$ results in a trivial gauge field boundary condition $\int \rmd a = 0$. (This doesn't stop things being more complicated if we separate these objects and then bring them together, though, particularly in the presence of matter fields, as we will discuss later.) Note that parity inversion changes the sign of the magnetic field and hence should correspond to $\monoM(x) \leftrightarrow \monoM(x)^\dagger$, and thus $\sigma \leftrightarrow -\sigma$. Indeed, from the formula \eqref{gaugefield-dualphoton-reln}, $\sigma$ should be a pseudo-scalar.

Just as with the 2d compact scalar, if you like to think about field theory in terms of the lattice, we get another argument for including on monopoles. The only reason why one might be tempted to exclude monopoles is that they require the excision of a point in spacetime to allow the multivalued nature of the defect to be resolved -- but on the lattice, no such principle can be applied! If you have a small plaquette on a lattice with magnetic flux lines all pointing outwards, that describes a perfectly smooth, finite-energy monopole configuration. Of course, such a field configuration has large derivatives near its core and hence the monopole energy scales with the lattice size, but that is a general phenomenon of all masses on the lattice. One expects that similar reasoning should lead one to include monopoles in any UV regularization.

\part{The Abelian Duality Web}
\labelPart{abelian}
\chapter{Particle-Vortex Duality}
\labelChapter{particle_vortex}

\begin{introduction}
  We describe our first IR duality, analyzing the Wilson-Fisher fixed point in 3 dimensions.
\end{introduction}

\section{IR Dualities}

\lettrine{T}{he examples of dualities} we have discussed so far are remarkable: one can straightforwardly prove in each case that two concrete theories, despite being expressed in very different language, are literally identical. Such statements are tremendously powerful. They are also tremendously rare!

Let us pause a moment to gain some insight into why that should be. One can always rewrite a simple theory -- say, that of a free fermion -- in some other very complicated way, perhaps using awkward non-linear changes of variable. But the resulting theory (setting aside our previous examples) is almost always very artificial and not of much practical use. What do we mean by `artificial'? Typically, the theory will have a long, fiddly Lagrangian with lots of very non-obvious non-linear interaction terms set to very particular values. This lacks \textit{simplicity}, making it hard for us to analyze the new theory, and \textit{universality}, meaning the theory is not interesting for practical applications. But this is not a very mathematical way of talking.

We can do better. There is a lot of machinery developed to understand questions of universality in theoretical physics. The key idea is that of an RG flow.

\begin{asidebox}[Refresher: Renormalization Group Flows]%
Let's briefly establish some useful language. The idea is simple enough: suppose we only care about the low-energy or \textit{infrared (IR)} physics of some theory. We assume the Lagrangian contains many couplings $\lambda_1,\lambda_2,\ldots$ and is defined with a momentum cutoff $\Lambda$. We are interested very low energy scales, so one might imagine taking a limit where we zooming out from the system, rescaling dimensionful quantities like $\Lambda \to \Lambda'$, which increases. But computations with a large $\Lambda$ are hard. So as well as rescaling, we also integrate out modes between $[\Lambda,\Lambda']$, returning the cutoff to $\Lambda$. In this way, we find that zooming out is equivalent to a redefinition of the $\lambda_i$ at a fixed cutoff $\Lambda$.
	
	In general, in an interacting theory, following this \textit{renormalization group (RG) flow} is a very hard problem. However, at least in the weakly interacting regime, a good approximation is simply that $\lambda_i \propto \Lambda^{\Delta_i}$ where $\Delta_i$ is the dimension of $\lambda_i$. Generically, this means that positive dimension or \textit{relevant} terms -- including most significantly the mass term -- grow very large in the IR, and the low-energy physics is usually boring: the theory consists only of very massive particles which we cannot excite. If we throw these massive or \textit{gapped} modes away, we often are left with an empty (gapped) theory.
	
	However, by tuning relevant and possibly \textit{marginal} (i.e. dimensionless) parameters to special values, we might find something more interesting. We often expect it is safe to ignore the remaining \textit{irrelevant} operators, since dimensional analysis suggests they tend to zero. (There can be exceptions to this, with the operators in question known as \textit{dangerously irrelevant operators}.) There are only a finite number of relevant and marginal operators to tune in most sensible theories.
	
	The end-point of the RG flow must be scale-invariant, by definition, as it is a \textit{fixed point} of the dilatations. In fact, it generally has full conformal symmetry. This can happen in two boring ways: the theory could be gapped, or it could be free. However, there is the possibility of landing on a non-trivial \textit{conformal field theory (CFT)}. These interesting theories are the focus of our study.
\end{asidebox}

So how does this idea help us? Well, CFTs are special (and often isolated) points in theory space. This means that we can hope to avoid ending up with unwieldy non-universal Lagrangians. Instead, suppose that theory $A$ can \textit{in principle} be rewritten in the language of theory $B$. Then if we tune to a conformal point in theory $A$, we know that $B$ must also be a CFT at this point. Moreover, if we could assert that there was a unique CFT in theory $B$, we would even be able to deduce that there is a duality between these theories at these precisely defined points. (We can also in principle then move away from the CFT point by identifying corresponding relevant operators on both sides of the duality and turning them on.)

These steps are generally unworkable in practice, since even when we can explicitly rewrite (for example) a lattice theory in new variables, it is virtually impossible to rigorously enumerate CFTs or track RG flows in the continuum limit. But it does give us some hope of finding CFT dualities in superficially unrelated systems. (We will also see later that more generally low-energy physics can display some interesting universal features even in the absence of non-trivial CFTs.)

The duality we are going to investigate in this chapter is along these lines. Consider the theory of a complex scalar $\phi$ in 2+1 dimensions,
\be\label{xymodel}
S_A = \int \rmd^3x \ |\partial_\mu \phi|^2 - \mu |\phi|^2 - \lambda|\phi|^4 + \cdots \fstp
\ee
Here, we impose a $U(1)$ symmetry by insisting the action is invariant under rotations of the phase of $\phi$; this is the global symmetry of \eqref{xymodel}.
In 2+1 dimensions, $\lambda$ has dimension 1; $\mu$ has dimension 2 as always. In principle, we should also be careful with the marginal sextic term, but for simplicity we will focus on the relevant operators. Note that for $\lambda > 0$ the theory is stable without the need for higher-order terms.

This is a famous example of a theory with an interesting, interacting fixed point: the $O(2)$ Wilson-Fisher fixed point, or (the critical point of) the \textit{XY model}. Both of these names refer to the global symmetry. A cartoon of the RG flow for this system is shown in \refFigureOnly{wfpoint}.
\begin{figure}
	\centering
	\begin{tikzpicture}
	\draw[axis] (-6,0)-- node[pos=0.75,below] {$\mu$} ++(12,0);
	\draw[axis] (0,0)-- node[left] {$\lambda$} ++(0,6);
	\draw[flow,tuned] (0,0) to[out=165,in=-70] (-5,3);
	\draw[flow,tuned] (-6,5) to[out=-50,in=110] (-5,3);
	\draw[flow,relevant] (-5,3) to[out=20,in=190] (6,4.4);
	\draw[flow,relevant] (-5,3) to[out=200,in=0] (-6,2.6);
	\draw[flow] (0,0) -- (-6,0);
	\draw[flow] (0,0) -- (6,0);
	\draw[flow] (0,0) to[out=90,in=190] (6,3);
	\filldraw (0,0) circle (3pt) node[below left] {free};
	\filldraw (-5,3) circle (3pt) node[left,xshift=-0.5em] {WF/XY};
	\end{tikzpicture}
	\caption{Cartoon of the RG flows near the O(2) Wilson-Fisher point/XY model, showing only the quadratic coupling $\mu$ and the quartic interaction $\lambda$, both of which are relevant near the free field point in 2+1 dimensions}\labelFigure{wfpoint}
\end{figure}
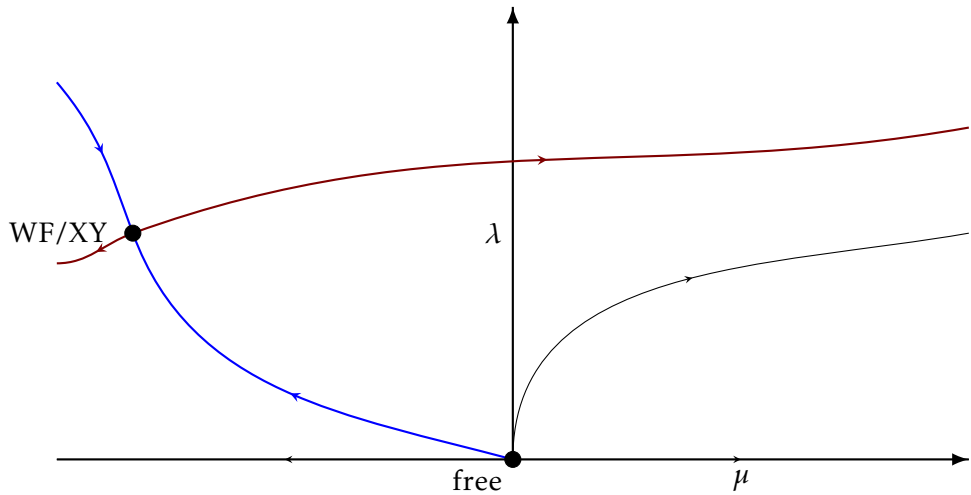

It turns out that this system is dual to the \textit{gauged XY model},
\be\label{gaugedxymodel}
S_B = \int \rmd^3x \ -\frac{1}{4g^2}f_{\mu\nu}f^{\mu\nu} + |D_\mu \tilde{\phi}|^2 - \tilde{\mu} |\tilde{\phi}|^2 - \tilde{\lambda}|\tilde{\phi}|^4 + \cdots
\ee
which we again claim can be tuned to a non-trivial CFT. One can obtain this theory (which actually describes the statistical field theory of 3 dimensional superconductors) from \eqref{xymodel} by gauging the $U(1)$ global symmetry and then once more tuning the coefficients to a critical point. (Note that now $g^2$, which has dimension 1, is also relevant; we will roughly speaking send it to $\infty$, essentially dropping the kinetic term for the photon.) But there is no obvious reason why one would imagine that this would get you back to the same system. For example, in the UV (i.e. the ultraviolet, or at high energies), \eqref{gaugedxymodel} contains a photon, which seems very unlike \eqref{xymodel}.

This duality might be written as
\be
\text{XY model} \qquad \longleftrightarrow \qquad U(1) + \text{XY model}
\ee
although people often use a more schematic notion, as in
\be\label{particle_vortex_schematic}
|\partial\phi|^2 - |\phi|^4 \qquad \longleftrightarrow \qquad -\frac{1}{4g^2} f_{\mu\nu}f^{\mu\nu} + |D\tilde{\phi}|^2 - |\tilde{\phi}|^4
\ee
where we write $|\phi|^4$ to represent that we move away from the free scalar point, but omit the $|\phi|^2$ term to emphasize that this operator is ``tuned away'' at the critical point. One might also send $g^2\to\infty$ (as this coupling grows in the IR) and omit the first term on the right-hand side:
\be\label{particle_vortex_schematic_no_f2}
|\partial\phi|^2 - |\phi|^4 \qquad \longleftrightarrow \qquad |D\tilde{\phi}|^2 - |\tilde{\phi}|^4
\ee
These are essentially mnemonics for the Lagrangian descriptions \eqref{xymodel} and \eqref{gaugedxymodel}. For reasons that will become clear, this is what is usually referred to \textit{particle-vortex duality}.

In order to gain some understanding of how this duality could possibly hold, we will briefly analyze both theory $A$ and and theory $B$. (This duality has a long history, and a lattice version of this was explicitly proven over 40 years ago \cite{Peskin:1977kp,Dasgupta:1981zz}. The continuum limit has been probed with many numerical experiments too. It is definitely correct.)

\section{Theory \texorpdfstring{$A$}{A}}

As already mentioned, the theory \eqref{xymodel} enjoys the $U(1)$ global symmetry $\phi \to e^{i\alpha} \phi$.

Note that, as shown in \refFigureOnly{wfpoint}, there is one relevant deformation around this fixed point. We may as well take that deformation to be $\mu$. Note that as we vary $\mu$, the system enters different phases:
\begin{itemize}
	\item $\mu \gg 0$: The theory is explicitly gapped, and the $U(1)$ is unbroken. The lowest-lying excitation is $\phi$ (with mass $\mu$) which carries unit $U(1)$ charge.
	\item $\mu \ll 0$: Here, $\phi$ acquires a vacuum expectation value (VEV) since $\left<|\phi|^2\right> = v = -\mu/2\lambda$. Therefore, the $U(1)$ symmetry is spontaneously broken, and the theory has a massless Goldstone mode. Explicitly, if we write $\phi = \rho e^{i\sigma}$, then $\sigma$ is massless field, whilst $\rho$ has a large mass.
	
	However, considering static configurations of the form $\phi = \rho(r) e^{i\sigma(\theta)}$ in radial coordinates, we discover that there are also particle-like vortices in which $\sigma$ winds asymptotically. One can show that $\oint_{S^1_{\infty}} \partial_\theta \sigma = 2\pi n$ is quantized, with $n \in \mathbb{Z}$, and that they have a logarithmically divergent energy in infinite space.
	
	Nonetheless, one may consider instead \textit{pairs} of vortices. If we choose the charges $n$ such that there is no winding at infinity, the energy contribution form long distances is finite. Now one can compute the potential energy $V(R)$ of a vortex-anti-vortex pair separated by a distance $R$. One finds that $V(R) \sim \log(vR)$, so that the pairs are logarithmically \textit{confined}.
\end{itemize}

\begin{exercise}[subtitle=Vortices in the XY Model]
	Show that the vortices in \eqref{xymodel} are quantized as stated, and have an energy that scales like $E \sim \log (vL)$ if we regularize the system by integrating out to $r = L$. Show also that the potential energy of the vortex-anti-vortex pair is $V(R) \sim \log (vR)$.
\end{exercise}

Meanwhile, at one intermediate point sits the critical XY model. The location is commonly written as ``$\mu = 0$'' which is not really correct; one really means $\delta\mu = 0$ where $\delta\mu$ is the deviation of $\mu$ from its value at the critical point.

Restricting to just the relevant operators around the XY point, then, the phase diagram reduces to what is shown in \refFigureOnly{xypoint}.

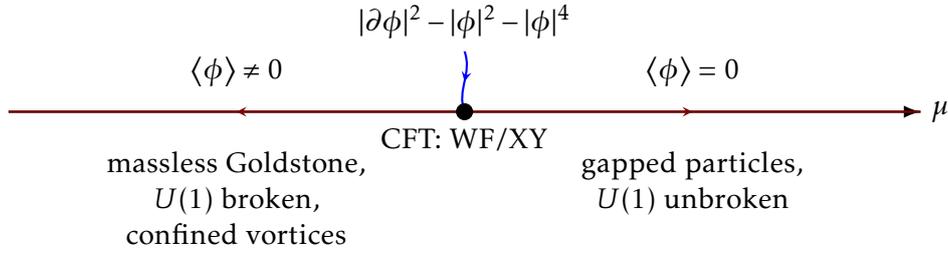
\begin{figure}
	\centering
	\begin{tikzpicture}
	\node at (0,1.2) {$|\partial \phi|^2 - |\phi|^2 - |\phi|^4$};
	\draw[flow,tuned] (0,0.8) to[out=-70,in=110] (0,0);
	\draw[axis] (-6,0) -- ++ (12,0) node[right] {$\mu$};
	\draw[flow,relevant] (0,0) --  node[below,yshift=-1em,align=center] {gapped particles,\\ $U(1)$ unbroken} node[above,yshift=0.5em,align=center] {$\left<\phi\right> = 0$} (6,0);
	\draw[flow,relevant] (0,0) -- node[below,yshift=-1em,align=center] {massless Goldstone,\\ $U(1)$ broken,\\confined vortices} node[above,yshift=0.5em,align=center] {$\left<\phi\right> \neq 0$} (-6,0);
	\filldraw (0,0) circle (3pt) node[below,yshift=-0.2em] {CFT: WF/XY};
	\end{tikzpicture}
	\caption{The phase diagram of Theory $A$ in 2+1 dimensions, showing only the relevant coupling around the critical point. On the right, we ultimately flow to an empty fixed point, whilst on the left-hand side one is left with a massless compact real scalar with a $U(1)$ shift symmetry}\labelFigure{xypoint}
\end{figure}

\section{Theory \texorpdfstring{$B$}{B}}

If theory $B$ is dual to theory $A$, it must also possess a $U(1)$ global symmetry. Since the phase rotations of $\phi$ are now gauged, this is no longer a (faithful) global symmetry of the theory. So what are the global symmetries of theory $B$? It is possible that we have an \textit{emergent symmetry} which is not manifest in the UV description \eqref{gaugedxymodel}; but in fact, that is not the case here. The symmetry is simply of a less familiar kind: it is the magnetic symmetry discussed in \refSectionOnly{dual_photon}. We have now added matter to the mix as well, but a similar argument goes through and tells us that there is indeed a $U(1)$ global symmetry associated with conservation of magnetic charge. It is this magnetic symmetry of theory $B$ which is dual to the global $U(1)$ symmetry of theory $A$.

So what is the phase diagram of this system?

\begin{itemize}
	\item If $\tilde{\mu} \gg 0$, then the $\phi$ excitations are massive and decouple, leaving the $U(1)_{\text{gauge}}$ symmetry unbroken, and hence there is a massless photon. This is therefore called the \textit{Coulomb phase}.
	\item For $\tilde{\mu} \ll 0$, meanwhile, the Higgs mechanism means the theory is gapped and the $U(1)_{\text{gauge}}$ is broken. We call this the \textit{Higgs phase}, for obvious reasons.
\end{itemize}

Can this be related to what we saw for theory $A$? Yes! We need to analyze the global symmetry, and look in a little more detail at the excitations.

\begin{itemize}
	\item Consider the Coulomb phase. We have already seen from \refSection{dual_photon} that a massless photon is equivalent to a compact scalar $\sigma$. Moreover, since that duality looks like $f^{\mu\nu} = -\frac{g^2}{\pi} \epsilon^{\mu\nu\rho} \partial_\rho \sigma$, we find that the current of the global $U(1)$ is $j^\mu = \frac{g^2}{\pi} \partial^\mu \sigma$. This is the current of the $U(1)_\text{global}$ shift symmetry $\sigma \to \sigma + \text{constant}$. Clearly, this $U(1)_\text{global}$ shift symmetry is spontaneously broken in the Coulomb phase. The Goldstone boson is $\sigma$, which is the photon. This exactly matches the behaviour for $\mu \ll 0$ of theory $A$!
	
	We should also ask: what are the theory $B$ duals of the logarithmically confined vortices in theory $A$? Well, look at the $\phi$ excitations. It is a straightforward check that in 2+1 dimensions, the energy of these charged particles due to the electric field lines is again logarithmic. Hence the particles of this theory are the vortices of the other.
	\item In the Higgs phase, meanwhile, $\tilde{\phi}$ has a non-vanishing VEV. This means we have vortices here, but in contrast to the vortices we discussed before, these vortices have a finite energy. One may also check that they are charged under $U(1)_{\text{global}}$. These are dual to the massive $\phi$ excitations of theory $A$. 
	
	One can also deduce that, in the ground state with no vortices, $U(1)_{\text{global}}$ is actually preserved. (Intuitively, the absence of vortices, which are the magnetic charge carriers, this symmetry is preserved.) This now matches with $\mu \gg 0$ in theory $A$.
\end{itemize}

\begin{exercise}[subtitle=Energetics of the Dual Theory]
	Check the logarithmic energies of $\phi$ excitations in the Coulomb phase, and also explain why it is now possible for vortices to have finite energies in the Higgs phase. Prove that such finite energy vortices carry integer charges under $U(1)_{\text{global}}$.
\end{exercise}

This shows that the phase diagram of theory $B$, depicted in \refFigureOnly{gaugedxypoint}, is identical to that of theory $A$ with the identification $\mu \leftrightarrow - \tilde{\mu}$.\footnote{Again, this statement is meant very loosely. We should at the very least write $\delta\mu \leftrightarrow - \delta\tilde{\mu}$; but we still don't know what numerical factors relate these quantities, or what values of $\mu,\tilde{\mu}$ we are perturbing around.} We can also see that
\begin{align}
\phi \text{ excitations} \quad &\longleftrightarrow \quad \tilde{\phi} \text{ vortices} \nn\\
\phi \text{ vortices} \quad &\longleftrightarrow \quad \tilde{\phi} \text{ excitations} \nn
\end{align}
earning this duality the name particle-vortex duality.

This fits in nicely with
\begin{align}
\left<\phi\right> = 0 \quad &\longleftrightarrow \quad \left<\tilde{\phi}\right> \neq 0 \nn\\
\left<\phi\right> \neq 0 \quad &\longleftrightarrow \quad \left<\tilde{\phi}\right> = 0 \nn
\end{align}
which can be seen as follows. Consider, for example, the $\left<\phi\right> \neq 0$ phase in which $\phi$ excitations have condensed. The duality tells us that this is equivalent to $\tilde{\phi}$ vortices having condensed. Since vortices are points at which $\tilde{\phi} = 0$, it makes sense that the dual phase has $\left<\tilde{\phi}\right> = 0$.

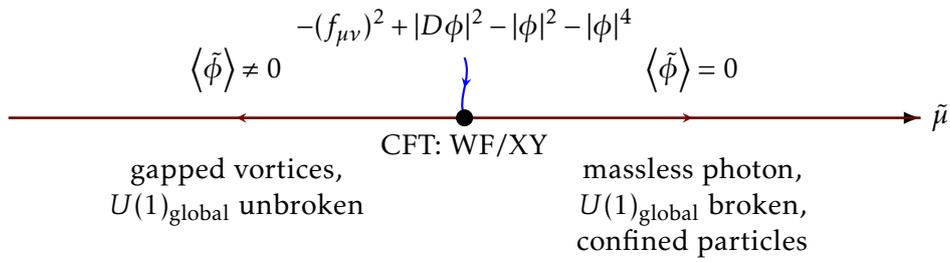
\begin{figure}
	\centering
	\begin{tikzpicture}
	\node at (0,1.2) {$ -(f_{\mu\nu})^2 + |D \phi|^2 - |\phi|^2 - |\phi|^4$};
	\draw[flow,tuned] (0,0.8) to[out=-70,in=110] (0,0);
	\draw[axis] (-6,0) -- ++ (12,0) node[right] {$\tilde{\mu}$};
	\draw[flow,relevant] (0,0) -- node[below,yshift=-1em,align=center] {massless photon,\\ $U(1)_\text{global}$ broken,\\confined particles} node[above,yshift=0.5em,align=center] {$\left<\tilde{\phi}\right> = 0$} (6,0);
	\draw[flow,relevant] (0,0) -- node[below,yshift=-1em,align=center] {gapped vortices,\\ $U(1)_\text{global}$ unbroken} node[above,yshift=0.5em,align=center] {$\left<\tilde{\phi}\right> \neq 0$} (-6,0);
	\filldraw (0,0) circle (3pt) node[below,yshift=-0.2em] {CFT: WF/XY};
	\end{tikzpicture}
	\caption{The phase diagram of Theory $B$, showing only the relevant coupling around the critical point}\labelFigure{gaugedxypoint}
\end{figure}

\section{The Critical Point}

Our real interest, however, lies not in analyzing the complicated details of the two separate phases, but in the critical point describing the (supposed) second-order transition between them. Our central claim is that the CFT sitting at this critical point is actually the same on both sides.

We should emphasize that many phase transitions in nature are first-order. In such a situation, a theory has two disconnected ``vacuum'' states, and as we vary parameters in the system, which one dominates changes. In such a situation, we do not expect to find an interesting conformal theory describing the transition at the special value of the parameter where they exchange stability. The much more interesting case is where the transition is second-order, corresponding to a genuine scale-invariant theory emerging at low energies, containing massless modes. We are claiming that this is the situation in both the ungauged scalar and Abelian-Higgs theories -- and, moreover, that these conformal field theories are identical.

\begin{asidebox}[What is a CFT?]%
	Apart from free (or empty!) theories, it is often hard to get much intuition about conformal field theories. One reason for this is that we tend to arrive at them as strongly coupled fixed points in some Lagrangian field theory; but this is a tremendously roundabout way to characterize them. Another is that scale invariance means that CFTs do not have states with particles in any familiar sense. It is helpful to at least bear in mind an intrinsic definition of a CFT that does not make reference to some complicated field theory embedding.
	
	The right way to present a CFT is to look at a list of all of the (local) operators ${O_i}$ in the theory.\footnote{This should perhaps say all of the local \textit{primary} operators, which basically means ignoring total derivatives.} If you are told the dimension and spin of every operator, then conformal symmetry fixes the two-point function $\left<O_i O_j\right>$ up to normalization.
	
	However, each $\left<O_i O_j O_k\right> \propto c_{ijk}$ is only determined up to a further constant, which we cannot eliminate by rescaling. Therefore in the \textit{CFT data} we include dimensions, spin representations, and three-point functions: $(\Delta_i, s_i, c_{ijk})$. It turns out that this is enough to completely solve the theory: all higher correlation functions can be expanded in terms of two- and three-point functions! In particular, therefore, this suffices to define a CFT.
	
	The claim of the dualities we are looking at, then, is that two different theories flow to a CFT which can be given in terms of the same CFT data. In particular, there should be a one-to-one map of operators at the fixed point. The CFT data also specifies the critical exponents of the theory in the dimensions of low-lying operators.
	
	As a further aside which we will return to in \refSection{conclusion_bootstrap}, we should mention the \textit{bootstrap program} \cite{Poland}. It turns out that the above data cannot be specified totally freely. There are huge numbers of consistency conditions; this follows from the associativity of the correlator. This means that by analyzing these, one can derive constraints on the spectra of arbitrary CFTs. This program is computationally tricky, but very interesting. It seems to suggest that CFTs may be rare, rigid, isolated objects -- not just in the phase diagram of a particular field theory, but in total generality!
\end{asidebox}

This means that, in theory, we should be able to identify a correspondence between operators in theory $A$ and $B$, and show that the critical exponents of the phase transition in both theories coincide. The first is something we can outline easily enough for the most important operators; the second is feasible only numerically. In fact, theory $A$ can be studied relatively easily in the $\epsilon$ expansion, but theory $B$ is much harder to analyze.

Nonetheless, as we emphasized, there is a lattice ``proof'' of this and overwhelming evidence that it is correct.

We will finish by presenting an (approximate) operator map for this duality in \refTableOnly{xy_opmap}. Whilst thinking about the Ward identity suggests the current is protected, we emphasize there will be corrections to the map between other operators.

\begin{table}[h]
	\centering
	\begin{tabular}{Cc Cc}
		Theory $A$ & Theory $B$ \\ \hline 
		$j^\mu = i(\phi^\dagger \partial^\mu \phi - \partial^\mu\phi^\dagger \phi)$ & $j^\mu = \frac{1}{2\pi} \epsilon^{\mu\nu\rho}\partial_\nu a_\rho$ \\
		$|\phi|^2$ & $-|\tilde{\phi}|^2$ \\
		$\phi(x)$ & monopole operator $\monoM(x)$ \\
	\end{tabular}
	\caption{Operator map for the duality of the XY model and the gauged XY model}\labelTable{xy_opmap}
\end{table}

We can add sources for all the operator in \refTableOnly{xy_opmap} into the Lagrangian. For example, we can couple both theory to a background (non-dynamical) $U(1)$ gauge field as follows:
\begin{subequations}
	\begin{align}
	S_{\mathrm{XY}}[\phi; A_\mu] &= \int \rmd^3x \ |(\partial_\mu - iA_\mu) \phi|^2 - \mu |\phi|^2 - \lambda|\phi|^4 + \cdots \\
	S_{\text{gauged XY}}[\tilde{\phi},a_\mu;A_\mu] &= S_{\mathrm{XY}}[\tilde{\phi};a_\mu] - \int \rmd^3 x \ \frac{1}{4g^2} f_{\mu\nu}f^{\mu\nu} + \frac{1}{2\pi} \epsilon^{\mu\nu\rho} A_\mu \partial_\nu a_\rho + \cdots
	\end{align}
\end{subequations}
where we emphasize $A_\mu$ is again simply a source one can use to probe the theory.

Here, we use the common convention that dynamical gauge fields like $a_\mu$ are given lowercase letters, whilst non-dynamical gauge fields like $A_\mu$ are given uppercase letters. The duality asserts that 
\be
Z_A[A_\mu] = \int D\phi \ e^{i S_{\mathrm{XY}} [\phi; A_\mu]} \overset{!}{=} \int D\tilde{\phi} Da \ e^{i S_{\text{gauged XY}} [\tilde{\phi}, a_\mu; A_\mu]} = Z_B[A_\mu]
\ee
where both theories are tuned to the critical point.

We should emphasize that it is impractical to use either Lagrangian description of this CFT. Both descriptions are \textit{strongly coupled}. Note that this is a property of the description, and not necessarily of the CFT. In the next section we will see our first example of a CFT which is strongly coupled in one description, but which is actually a free theory.

\chapter{3d Bosonization}
\labelChapter{3d_bosonization}

\begin{introduction}
  We describe our first 2+1 dimensional relativistic bosonization duality. This involves understanding the physics of fermions in three dimensions, and analyzing some subtle questions about gauge invariance.
\end{introduction}

\section{The Duality}

\lettrine{W}{ay back in} \refSection{flux_attachment}, we discussed a very simple notion of flux attachment, in which a non-relativistic boson coupled to a Chern-Simons theory at level $1$ turned out to be identical to a fermion. With the notion of an IR duality now established, it is very natural to speculate about the possibility that there might be a much more exciting version of this duality describing two dual conformal field theories.

The form this takes is hinted at by the work we did back there; in particular, in Exercise \ref{exercise_flux_attachment_bcs}, we observed the importance of including a $|\phi|^4$ term in the bosonic theory in getting the short-distance behaviour of the theory correct. Inspired by this, we claim the following duality holds:
\be
U(1)_1 + \text{XY model}  \qquad \longleftrightarrow \qquad \text{free Dirac fermion}
\ee
where both sides are tuned to a CFT. (In particular, the fermion is massless.)
In the notation of \eqref{particle_vortex_schematic},
\be\label{3d_bosonization_schematic}
\frac{1}{4\pi} a\rmd a + |D\phi|^2 - |\phi|^4  \qquad \longleftrightarrow \qquad i \bar{\psi} \gamma^\mu \partial_\mu \psi
\ee
where we conventionally drop the Maxwell term on the left-hand side since its coefficient becomes small in the infrared.\footnote{As we will discuss below, the theory does not have a massless photon any more, so one might argue that the Maxwell term is less crucial in understanding the dynamics of this theory anyway.} This is one of several dualities we will encounter going by the name \textit{3d bosonization}. Whilst the basic idea goes back to \cite{Polyakov:1988md} and many others, many of the details have been filled in relatively recently \cite{Metlitski:2015eka,Wang:2015qmt}.

This is actually a remarkable statement; it's a far cry from the simple non-relativistic quantum mechanics of \refSectionOnly{flux_attachment}. The theory on the left is a fully fledged strongly interacting conformal field theory, and a priori doing any computations with it would be totally intractable. However, the claim of this duality is that the theory on the left-hand side has been tuned to a point where it is a totally free CFT, containing only a massless Dirac fermion.

We should mention that this is generally believed to be true (and we will see various reasons why, from basic evidence now to evidence from e.g. supersymmetry breaking later on), but the closest we have to a proof is probably the ``wire construction'' \cite{2016PhRvL.117a6802M} in which one direction is discretized, and each separate line is individually bosonized.

\subsection{The Bosonic Theory}

Firstly, we should see what global symmetries we can spot based on the UV description of \eqref{3d_bosonization_schematic}. The situation is more or less as with the gauged XY model of \refChapterOnly{particle_vortex}. The $U(1)$ symmetry of the complex scalar $\phi$ is gauged, so the only possibility for a non-trivial global symmetry is a monopole symmetry. You may be wondering whether the theory with a Chern-Simons term still supports a monopole symmetry, since the manipulations that led us to the construction of the dual photon relied on us being able to express the action in terms of $\rmd a$. It turns out there isn't anything to worry about; see \refAppendix{chern-simons}.

We can play the same game with this bosonic theory as we did with the theory in \refChapterOnly{particle_vortex}, deforming the theory with the $|\phi|^2$ operator with a large coefficient $\mu$. It will be useful to couple the theory to a background field $A_\mu$ for the monopole symmetry; schematically,
\be \lag = \frac{1}{4\pi} a\rmd a + |D_a\phi|^2 - |\phi|^4 + \frac{1}{2\pi} A \rmd a \ee
where we have added the BF term discussed in \refChapterOnly{particle_vortex}. We have also used the notation $D_a\phi \equiv (\rmd - i a) \phi$ to concisely denote the covariant derivative is for a field with charge 1 under the dynamical field $a$.

\paragraph{$\mu \ll 0$ Phase}

Subject to a large negative mass squared term, $\phi$ develops a vacuum expectation value. As before, the condensed $\phi$ field Higgses the gauge field $a$, essentially setting it to $0$. At low energies, there is nothing left.

The result is that the partition function in this phase is essentially $Z[A] = 1$, independent of the background $A$.

\paragraph{$\mu \gg 0$ Phase}

If we make $\phi$ very massive, we can integrate it out. This essentially just means dropping it from the theory, leaving only a pure Chern-Simons theory called $U(1)_1$.  But this theory, it turns out, is also trivial. We will postpone discussion of the precise sense in which this is trivial to \refAppendix{chern-simons}. The key result is that a Chern-Simons term actually gives a (gauge-invariant) mass to the photon proportional to the Chern-Simons level and the gauge coupling. Since the theory has only a massive excitation in its spectrum, we can drop this at low energies.\footnote{Actually, we are being a little quick here, since in general Chern-Simons theories can possess non-trivial structure at low energy even though there are no light \textit{local} excitations. We will discuss these issues briefly in \refSectionOnly{abelian_technicalities}, and extensively in \refSectionOnly{tqfts}. However, $U(1)_1$ is almost as trivial as it gets!}

In fact, we can integrate out the field $a$, which appears quadratically, as follows:
\be \lag = \frac{1}{4\pi} a \rmd a + \frac{1}{2\pi} A \rmd a = \frac{1}{4\pi} (a+A) \rmd (a+A) - \frac{1}{4\pi} A \rmd A \fstp \ee
The partition function in this phase is actually different from what we had above. We find $\tilde{Z}[A] = \exp(-\frac{i}{4\pi}A \rmd A)$ now has a so-called \textit{contact term} for the background field $A$.

\subsection{The Fermionic Theory}

Meanwhile, the fermionic theory is free, so it is rendered entirely massive if we deform it using a fermion mass term. Naively, we can take
\be \tilde{\lag}_{\text{naive}} = i \bar{\psi} \gamma^\mu (\partial_\mu - i A_\mu) \psi - \tilde{\mu} \bar{\psi} \psi \fstp \ee 
Now obviously, the fact that both phases are gapped matches the bosonic theory. However, it seems the contact term cannot ever appear.

It turns out that there is a key fact we have overlooked about fermionic theories in three dimensions \cite{Redlich:1983dv}. These issues will be the subject of our next section. Let us just quickly preview a rough outline of the solution; we will clarify various subtleties later on.

Firstly, when we integrate out a Dirac fermion, we actually generate a contact term for the gauge field. This shift,
\be \tilde{\lag} \to \tilde{\lag} + \frac{\sign \tilde{\mu}}{2} \times \frac{1}{4\pi} A \rmd A \cmma \ee
depends on the sign of the mass deformation, and is a Chern-Simons term of level $\frac{1}{2}$. Understanding what precisely we mean by this incorrectly-quantized Chern-Simons term will occupy us below. It will be useful, however, to notice one particular aspect of this which makes sense: the dependence upon the sign of $\tilde{\mu}$.

Consider the $\mathbb{Z}_2$ symmetry of time reversal, $T : t \to -t$, or equivalently parity $P : x_1 \mapsto -x_1$.\footnote{Since CPT is a good symmetry of interesting physical theories, and charge conjugation is easy to understand, one can essentially swap ``time reversal'' for ``parity'' in everything we do.}
The free massless fermion Lagrangian $i \bar{\psi} \gamma^\mu (\partial_\mu - i A_\mu) \psi$ is symmetric under $T$ and $P$ in any number of dimensions. However, the mass term $\bar{\psi}\psi$ is odd under both time reversal and parity in odd numbers of dimensions.

\begin{exercise}[subtitle=Discrete Fermion Symmetries]
	Check the discrete transformation properties of 2+1d fermions described on page~\pageref{conventions}.
\end{exercise}

Consequently, if we time-reverse our whole theory, the sign of the mass term changes. This fits in perfectly with the fact that the Chern-Simons term $A \rmd A$ is clearly odd under both $T$ and $P$.

Secondly, nothing has told us we can't include a contact term in a full statement of the duality:
\be \tilde{\lag} = i \bar{\psi} \gamma^\mu (\partial_\mu - i A_\mu) \psi - \tilde{\mu} \bar{\psi} \psi - \frac{1}{2} \frac{1}{4\pi} A \rmd A \fstp \ee 
Again, this incorrectly-quantized Chern-Simons term needs explaining.

If we were to assume these modifications were sensible, we see that
\be \mu \qquad \longleftrightarrow \qquad - \tilde{\mu} \ee
matches up the phases with the bosonic theory. This is promising -- so let's try and understand all these slightly strange contact terms more carefully.

\section{Chern-Simons Terms \& Anomalies}\labelSection{chern_simons_anomalies}

In this section, we will start off by doing a concrete computation to learn what is left behind when we integrate out a fermion coupled to a non-dynamical gauge field. Then, we will try and gain a deeper understanding of the result by introducing the concept of an anomaly.

\subsection{The Effective Action of a Fermion}\labelSection{effective_action_abelian_fermion}

We want to know what happens when we integrate out a Dirac fermion. Consider the path integral
\be Z[A;m] = \int D\psi D\bar{\psi} \ \exp \left[ i \int \rmd^3x \ i \bar{\psi} \gamma^\mu (\partial_\mu - i A_\mu) \psi - m\bar{\psi}\psi \right] = \det \left[ i\gamma^\mu (\partial_\mu - i A_\mu) - m \right] \ee
corresponding to a Dirac fermion of mass $m$ coupled to a background $U(1)$ gauge field $A$. We want to know the \textit{effective action} for $A$ which remains upon integrating out $\psi$. We obtain this by defining $Z[A;m] = \exp(i S_\text{eff}[A;m])$:
\be S_\text{eff}[A;m] = -i\log Z[A;m] = -i\tr \log \left[ i \slashed{\partial} + \slashed{A} - m \right] \fstp \ee
This can now be computed order-by-order in $A$, using the Taylor expansion of $\log$, as
\be S_\text{eff}[A;m] = -i \tr \log \left[ i \slashed{\partial} - m \right]  -i\tr \left[ \frac{1}{i \slashed{\partial} - m} \slashed{A} \right] -\frac{i}{2}\tr \left[ \frac{1}{i \slashed{\partial} - m} \slashed{A} \frac{1}{i \slashed{\partial} - m} \slashed{A} \right]  + \cdots \fstp \ee
The term we care about is the one quadratic in $A$. (The leading term can be normalized away; the second term is a tadpole diagram which must vanish anyway.)

Graphically, this is represented by the Feynman diagram in \refFigureOnly{renormalization_cs_term}. The corresponding loop integral is 
\be \Gamma^{\mu\nu}(p;m) = \tr \int \frac{\rmd^3 \ell}{(2\pi)^3} \gamma^\mu \frac{\slashed{\ell} + m}{\ell^2 - m^2 + i \epsilon}  \gamma^\nu \frac{\slashed{\ell} - \slashed{p} + m}{(\ell-p)^2 - m^2 + i \epsilon} \ee
and it contributes to the quadratic effective action the term
\be S_\text{eff}[A;m] = \cdots  -\frac{i}{2} \int \frac{\rmd^3 p}{(2\pi)^3} A_\mu(-p) \Gamma^{\mu\nu}(p;m) A_\nu(p) + \cdots  \fstp \ee
Taking the trace over the spinor indices, 
\be \Gamma^{\mu\nu}(p;m) = \int \frac{\rmd^3 \ell}{(2\pi)^3}\frac{2m^2 \eta^{\mu\nu} + 2mi \epsilon^{\mu\rho\nu} p_\rho + 4\ell^\mu \ell^\nu - 2\ell^\mu p^\nu - 2\ell^\nu p^\mu - 2\eta^{\mu\nu} \ell \cdot (\ell - p)}{(\ell^2 - m^2 + i \epsilon)((\ell-p)^2 - m^2 + i \epsilon)} \ee
where the most interesting term is the one containing the antisymmetric tensor $\epsilon^{\mu\rho\nu}$, arising from the trace of three $\gamma$ matrices. No such term occurs in four dimensions.

\begin{figure}
	\centering
	\begin{tikzpicture}
	\draw[feynman photon] (0,0) node[xshift=-1em] {$A_\mu$} -- (2,0);
	\draw[feynman momentum] (0.7,0.3) -- node[yshift=0.6em] {$p$} (1.3,0.3);
	\draw[feynman fermion] (5,0) arc (0:-180:1.5cm);
	\draw[feynman fermion] (2,0) arc (-180:-360:1.5cm);
	\draw[feynman photon] (5,0) -- (7,0) node[xshift=+1em] {$A_\nu$};
	\draw[feynman momentum] (5.7,0.3) -- node[yshift=0.6em] {$p$} (6.3,0.3);
	\draw[feynman momentum] (3,2) -- node[yshift=0.6em] {$\ell$} (4,2);
	\end{tikzpicture}
	\caption{The renormalization of the photon two-point function due to a fermion loop}\labelFigure{renormalization_cs_term}
\end{figure}
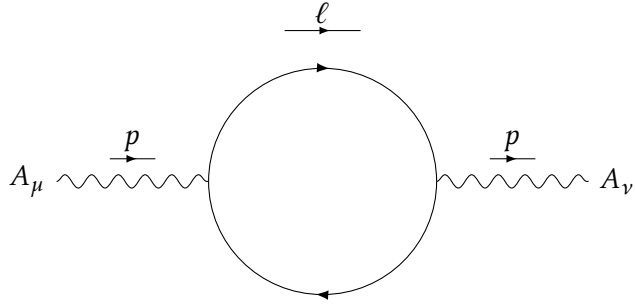

Let's focus on this parity-violating term, since only this term can contribute a Chern-Simons term. (The other terms can be regularized in the usual manner.) We see that
\begin{align*}
S_\text{quad, parity-odd}[A;m] &= - \frac{i}{2} \int \frac{\rmd^3 p}{(2\pi)^3} i \epsilon^{\mu\rho\nu} A_\mu(-p) p_\rho A_\nu(p) \int \frac{\rmd^3 \ell}{(2\pi)^3}\frac{2m}{(\ell^2 - m^2 + i \epsilon)((\ell-p)^2 - m^2 + i \epsilon)} \\
&= - \frac{i}{2} \int \rmd^3 x \ \epsilon^{\mu\nu\rho} A_\mu \partial_\nu A_\rho \int \frac{\rmd^3 \ell}{(2\pi)^3}\frac{2m}{(\ell^2 - m^2 + i \epsilon)((\ell-p)^2 - m^2 + i \epsilon)} 
\end{align*}
To compute this integral, we perform a Wick rotation to evaluate it in Euclidean signature:
\begin{align}
S_\text{quad, parity-odd}[A;m] &= 2\pi S_\mathrm{CS}[A] \int \frac{\rmd^3 \ell}{(2\pi)^3}\frac{2m}{(\ell^2 + m^2)((\ell-p)^2 + m^2)} \nn \\
&= S_\mathrm{CS}[A] \frac{m}{|p|} \arcsin \left( \frac{|p|}{\sqrt{p^2+4m^2}} \right) \nn \\
&= \frac{1}{2} S_\mathrm{CS}[A] \frac{m}{|m|} + O\left(\frac{p^2}{m^2} \right)
\end{align}
where we are neglecting terms which are small for a large mass $m$. (This is a standard effective field theory attitude: we expect to describe physics well only at energies below the characteristic scale of the fields we integrate out. Here, this means we look at momenta satisfying $|p| \ll m$.)

This is exactly the result we were touting above: if one integrates out a heavy fermion of mass $m$, one generates a Chern-Simons term at the level $\frac{1}{2} \sign m$:
\be
\int D\psi D\bar{\psi} \ \exp \left[ i \int \rmd^3x \ i \bar{\psi} \gamma^\mu (\partial_\mu - i A_\mu) \psi - m\bar{\psi}\psi \right] = \exp \left[ i \int \frac{\sign m}{2} \frac{1}{4\pi} A \rmd A \right]
\ee
This immediately raises all sorts of questions!

\subsection{Anomalies and the Fermion Determinant}

In general, when we discuss a quantum field theory, we need to be careful that what we define is mathematically sensible. Assuming that we are working in an action-based formalism, the non-trivial issue we must address is whether the path integral is well-defined. Famously, most quantum field theories require some sort of \textit{regularization} before we can compute anything. In renormalizable field theories, this leads directly to a calculable, well-defined procedure to extract physical quantities which amounts to a prescription for computing divergent integrals.

However, sometimes this process can unearth some surprises. Suppose that the original (classical) theory had a symmetry group $G$. It may be that there is no regularization which preserves the action of $G$. If this is the case, then we say the symmetry is \textit{anomalous}. This can be fatal, if $G$ was supposed to be a gauge group, since then the entire theory is ill-defined. Otherwise, this is typically a very interesting feature of a theory. One can ask, for instance, exactly how the partition function $Z$ transforms under the action of $G$.

Since $G$ is a symmetry of the action $S$, and $Z = \int D \phi \exp(iS[\phi])$, we can deduce that anomalies can be thought of as arising because the path integral measure $D\phi \to A D\phi$ transforms non-trivially under $G$. Writing $A = \exp(i \delta S_\text{anom})$ for some $\delta S_\text{anom}$, we can instead think of the anomalous transformation as a transformation $S \to S + S_\text{anom}$. Moreover, locality tells us that we should be able to express $S_\text{anom}$ as a spacetime integral of a local function of the fields in the theory, or an anomalous Lagrangian $\lag_{\text{anom}}$.

It turns out that the theory of fermions in 3 dimensions is an excellent example of an theory with an anomaly. The relevant symmetry can be taken to be the $\mathbb{Z}_2$ symmetry of time reversal, $T : t \to -t$. As we mentioned above, a good classical symmetry of the free massless fermion Lagrangian $i \bar{\psi} \gamma^\mu (\partial_\mu - i A_\mu) \psi$ in any number of dimensions. However, the mass term $\bar{\psi}\psi$ is odd under time reversal in spacetimes of odd dimension.

One reason to be suspicious of this symmetry is revealed by thinking about a standard way to regularize matter theories called \textit{Pauli–Villars (PV) regularization}, which simply amounts to adding to the theory auxiliary copies of every field with mass $M$, and then take $|M| \to \infty$ to decouple it. By artificially giving a duplicate fermion the opposite statistics to the physical one (so that its propagator has the opposite sign to the physical field), we can cancel UV divergences well enough for our purposes. In fact, it is very easy to understand what this means from the point of view of the path integral: we can define the PV-regularized effective action of a fermion of mass $m$ via
\be S^{\text{PV}}_{\text{eff}}[A;m] = S_{\text{eff}}[A;m] - \lim_{M\to \pm \infty} S_{\text{eff}}[A;M] \fstp \ee

However, we have a choice of sign in taking $M \to \pm\infty$, and time reversal maps $M \to -M$. It is certainly possible that physical quantities could depend upon this sign, and there could be an anomaly associated with this symmetry.\footnote{Different regularizations reveal that some analogous choice is always necessary, but Pauli-Villars offers a particularly clear illustration. If one instead uses dimensional regularization, for example, spacetime symmetries like time reversal become somewhat subtle.} And in fact, given the computation we have already completed, we find that for any non-zero mass $m \neq 0$, we have a well-behaved effective action for the gauge field
\be S^{\text{PV}}_{\text{eff}}[A;m] = \frac{\sign m + \sign M}{2} S_\mathrm{CS}[A] \ee
which manifestly depends upon the sign of $M$! We have two choices of regularization, and they lead to effective Chern-Simons levels as follows:
\be
M \to \infty: \begin{cases} k_\mathrm{eff} = 1 & m > 0 \\ k_\mathrm{eff} = 0 & m < 0 \end{cases}
\hspace{4em}
M \to -\infty: \begin{cases} k_\mathrm{eff} = 0 & m > 0 \\ k_\mathrm{eff} = -1 & m < 0 \end{cases}
\ee
In the massless case $m=0$, we cannot integrate out the fermion as it is important at arbitrarily low energies, but we still have the option of using either regularization.

Notice that using PV regularization, which fully respects gauge invariance, has eliminated the possibility of ending up with fractional Chern-Simons levels, which would represent a fatal \textit{gauge anomaly}. But it has left us with a clear \textit{time reversal anomaly} and \textit{parity anomaly}. Defining $Z_\pm[A]$ to be the partition function of the massless fermion with the regularization $M \to \pm\infty$, we see that the two differently regularized theories $Z_\pm$ have different phases and are clearly non-equivalent, since we end up with different effective Chern-Simons levels for the background gauge field under mass deformations. Essentially, $(Z_+, Z_-)$ fill out a non-trivial representation of the $\mathbb{Z}_2$ symmetry group.

However, it is inconvenient to have to constantly specify which of the two regularization we are using. Accordingly, we will adopt the convention that we \textit{always} use the $Z_-$ regularization. Heuristically, this means including a bare Chern-Simons term at level $-\frac{1}{2}$:
\be i \bar{\psi} \gamma^\mu(\partial_{\mu} - i A_\mu) \psi \quad \equiv \quad 
``i \bar{\psi} \gamma^\mu(\partial_{\mu} - i A_\mu) \psi - \frac{1}{2}\frac{1}{4\pi} A \rmd A'' \fstp \ee
However, formally, you can think of the theory as being defined by the choice of a negative-mass Pauli-Villars regulator. (See \refSectionOnly{fermion_det_parity_anom} for a little more discussion on this issue.) We will never explicitly write such a level $1/2$ Chern-Simons term.

We can now rephrase our understanding of the anomaly in the language of an anomalous transformation, since $Z_+/Z_- = \exp(i\delta S_\text{anom})$ where $\delta S_\text{anom} = S_\text{CS}[A]$.  The action of time-reversal upon the massless theory is to add a Chern-Simons term at level $1$! Notice that this is an involution, since
\begin{align}
i \bar{\psi} \gamma^\mu (\partial_\mu - i A_\mu) \psi &\stackrel{T}{\longrightarrow} i \bar{\psi} \gamma^\mu (\partial_\mu - i A_\mu) \psi + \underbrace{ \frac{1}{4\pi} A \rmd A }_\text{anomaly} \nn \\
&\stackrel{T}{\longrightarrow} i \bar{\psi} \gamma^\mu (\partial_\mu - i A_\mu) \psi - \frac{1}{4\pi} A \rmd A + \underbrace{ \frac{1}{4\pi} A \rmd A }_\text{anomaly} = i \bar{\psi} \gamma^\mu (\partial_\mu - i A_\mu) \psi
\label{fermion_anomaly}
\end{align}
after being applied twice.\footnote{Technically, the $T$ operator squares to $T^2 = (-1)^F$ where $F$ counts fermion number. This does not affect the Lagrangian, which necessarily contains an even number of fermions.} Here, all fermions are understood to be regularized using a negative mass PV regulator.

\subsection{Quantization in a Monopole Background}\labelSection{quantization-monopole-bg}

There is another way to uncover the lack of gauge invariance in the above theory, and that is to ask about what happens if we choose $A$ to be a monopole background. To answer this question, we need to understand the quantization of fields in such a background. This will prove useful later on, too. 

Our approach is to first understand the spectrum of the conserved angular momentum operators in such a background for a \textit{scalar} particle \cite{Wu:1976ge}. We will do this in Euclidean signature for simplicity.

Suppose that we have a monopole of charge $\tilde{q} \in \mathbb{Z}$ in 3 spatial dimensions, and we look at the quantum mechanics of a spinless particle of electric charge $q$ in its background. Recall, from our earlier discussion in \refSection{flux_attachment}, we expect that the dyonic combination of these two objects has spin $s = \frac{1}{2}q \tilde{q}$. This number $s$ will be important. (Often, $q$ is used for what we call $s$.)

Consider the quantum mechanical operator
\be \textbf{L} = \textbf{r} \times \left( \textbf{p} - 2s \textbf{A}^{(1)} \right) - s \textbf{e}_r \ee
which one may verify has the property that it generates gauge-invariant rotations:
\be [L_i, v_j] = \epsilon_{ijk} v_k \qquad \text{for } \textbf{v} = \textbf{x}, \textbf{p} - 2s\textbf{A}^{(1)} \fstp \ee
Here, $\textbf{A}^{(1)}$ is taken to be the gauge field of a rotationally symmetric charge 1 monopole, given by spherical coordinates by
\be
A^{(1)} = \begin{cases}
	\frac{1}{2}(1 - \cos \theta) \rmd \varphi & \mbox{in northern hemisphere} \\
	\frac{1}{2}(1 + \cos \theta) \rmd \varphi & \mbox{in southern hemisphere}
\end{cases}
\ee
which is a perfectly regular bundle. The corresponding magnetic field is $\mathbf{B}^{(1)} = \frac{1}{2r^2} \mathbf{e}_r$ with flux $2\pi$ through the sphere. The transition function for our theory is $\exp(2is \varphi)$, in the sense that the gauge transformation linking the two patches is $-i \rmd \log \exp(2is \varphi) = 2s \cos \theta \rmd \varphi$. This is single-valued for half-integer values $s$, as we expect.

The extra term in $\textbf{L}$ is needed classically to make sure this is a conserved quantity; from the point of view of single particle physics in the background of the monopole, we would expect
\be \frac{\rmd}{\rmd t} \left(\textbf{r} \times \dot{\textbf{r}}\right) = \textbf{r} \times m \ddot{\textbf{r}} = \textbf{r} \times (q \dot{\textbf{r}} \times \tilde{q}\textbf{B}^{(1)} ) = s \frac{\rmd}{\rmd t} \textbf{e}_r \ee
which goes back to an 1896 result of Poincar\'e.
However, the physical origin of the angular momentum is perhaps more illuminating:
\begin{exercise}[subtitle=Origins of the Extra Angular Momentum]
	Show that the extra angular momentum is precisely the angular momentum of the electomagnetic field
	\be \textbf{L}_{\mathrm{em}} = \int \textbf{r} \times (\textbf{E} \times \textbf{B}) \ee
	where $\textbf{E}$ is due to the inserted charge $q$ particle, whilst $\textbf{B}$ is the charge $\tilde{q}$ monopole background.
\end{exercise}

Now $[\textbf{L}, r^2] = [\textbf{L}, L^2] = 0$, as $L^2,r^2$ are scalars, so as usual we may choose to investigate wavefunctions which simultaneously diagonalize $r^2, L^2, L_3$. Thus a complete basis of wavefunctions exists of the form $R(r) \Theta(\theta) \Phi(\varphi)$. Moreover, neglecting the radial dependence, these must fall into the usual representations of the $so(3)$ generated by $L_i$, with
\be L^2 Y(\theta,\varphi) = l(l+1) Y(\theta,\varphi) \qquad \mbox{and} \qquad L_3 Y(\theta,\varphi) = m Y(\theta,\varphi) \ee
for some $l \in \{0,\frac{1}{2},1,\ldots\}$ and $m \in \{-l,-l+1,\ldots,l-1,l\}$.

In the usual case of the scalar field, each representation with integer $l=0,1,2,\ldots$ shows up precisely once, and we define $Y_{l,m}$ to be the unique spherical harmonic of the given angular momenta. Something similar happens in the monopole background, except that now a different set of $l$ are possible. We label the possible functions as the so-called \textit{monopole harmonics}\footnote{These are essentially identical to the \textit{spin-weighted spherical harmonics} studied elsewhere \cite{Dray:1984gy}.}
\be \label{monopoleharmonics} Y_{s,l,m}(\theta, \varphi) \quad \mbox{for} \quad l = |s|, |s|+1, |s|+2, \ldots \quad \mbox{and} \quad m=-l,\ldots,l \ee
where again, each term occurs exactly once.

This follows from two key observations. Firstly, notice that the $L_3 = (-i \partial/\partial \varphi \mp s)$ equation enforces
\be Y_{s,l,m} \propto \begin{cases}
	e^{i(m+s)\varphi} & \mbox{northern hemisphere} \\
	e^{i(m-s)\varphi} & \mbox{southern hemisphere}
\end{cases}
\ee
and so $m \pm s$ are integers. Secondly, one may compute that
\be 
L^2 = \left( \textbf{r} \times \left(\textbf{p} - 2s \textbf{A}^{(1)}\right) \right)^2 + s^2
\ee
and so $l(l+1) \ge s^2$. One can then explicitly show that there are unique solutions for every allowed value of $l,m$ as in \eqref{monopoleharmonics}.

Therefore, a scalar particle in the background of a monopole naturally lives in the representations
\be V_{s} = \spinrep{s} \oplus \spinrep{s+1} \oplus \spinrep{s+2}  \oplus \cdots \ee
with the lowest-energy bound state being in the spin $s$ representation, as we expected. The higher states appear as higher spin excitations of that bound state. (Technically, we should check the $r$ dependence obeys suitable boundary conditions too, but we neglect this subtlety here.)

Intuitively, the lower bound on the spin arises because of the extra contribution $\textbf{L} = \cdots - s\hat{\textbf{r}}$ to the angular momentum. (Indeed, this is also where the lower bound arises in the calculation.) This can be thought of as due to an extra contribution to the angular momentum about the monopole (along the direction of displacement from the monopole) received by any particle interacting with its gauge field.

Now what does this tells us about the unit charge fermion in a unit charge monopole background? Well, the fermion has a bare spin of $1/2$. An admittedly slightly slick bit of reasoning suggests that the quantized system has a Hilbert space
\be \spinrep{\frac{1}{2}} \times V_{1/2} = \spinrep{0} + \cdots \ee
that contains a singlet. This conclusion turns out to be sound. Moreover, this state is actually a zero-energy mode of the Dirac equation in this background.

An intuitive reason for this is to remember that the Dirac equation, restricted to the 2d sphere, splits up into left and right chiralities. The usual spectrum has particles with helicity ($J_z$ spin) at the point $(0,0,1)$ equal to $+\frac{1}{2}$ for the positive chirality particle, and $-\frac{1}{2}$ for the negative chirality one. Therefore, there are modes of total angular momentum $l=\frac{1}{2},\frac{3}{2},\ldots$ for both chiralities, and they are all mixed by the Dirac operator and generically have non-zero mass terms. But the presence of the monopole breaks the parity symmetries relating the chiralities, subtracting $\frac{1}{2}$ from the helicities of all modes. Now the spectrum of the positive chirality modes, with helicity $0$, consists of $l=0,1,2,\ldots$ states whilst the negative chirality modes, with helicity $1$, may only consist of modes $l=1,2,3,\ldots$. Consequently, the Dirac operator -- which maps positive chirality modes onto negative chirality ones -- must annihilate the $l=0$ mode of the positive chirality spectrum! It simply cannot map them to anything else.

So what is the issue? Well, now consider the full field theory problem of a fermionic field $\psi$ in a monopole background. The vacuum state is taken to be $\ket{0}$. But there is another state, $\ket{0'} = \psi^\dagger_0 \ket{0}$, containing a single excitation living in the singlet $\spinrep{0}$. These are degenerate states.

Now clearly, the electric charge of these two states differs by $1$. On the other hand, suppose time reversal was a good symmetry, along with charge conjugation. Then the two relations
\be \ket{0'} = \psi^\dagger_0 \ket{0} \qquad \mbox{and} \qquad \ket{0} = \psi_0 \ket{0'}  \ee
are exchanged under $CT$ symmetry, so the states must have physical charges of the same magnitude. (Normally, in a half-integral spin representation, this would instead be a parity transformation interchanging the positive and negative spins.) This forces us to ascribe them both charges $\pm \frac{1}{2}$; but this violates gauge invariance.

Alternatively, we can keep gauge invariance, but the price we must pay is losing complete time reversal invariance. We effectively privilege one of $\ket{0}$ over $\ket{0'}$, declaring it to have charge $0$. Then time reversal \textit{creates} a charge by mapping us to $\ket{0'}$. This can be seen by looking at the form of the time reversal anomaly: a Chern-Simons term $+\frac{1}{4\pi} A \rmd A$ appears, which says that the gauge field $A$ experiences one extra unit of charge at the location of any charge $\rmd A = 2\pi \delta(x - x_{\text{monopole}})$. This extra unit of charge is the fermion zero mode $\psi_0^\dagger$.

\begin{asidebox}[Index Theorems]%
	A particularly elegant class of mathematical theorem underlies the result that there is a single fermion zero mode in the background of a magnetic monopole. So-called \textit{index theorems} govern questions about zero-mode counting, and generally tell us the number of zero modes of some operators on a manifold $M$ are related in some exact way to topological invariants of the background, plus very particular contributions from the boundary if $\partial M \neq \emptyset$.
	
	A particularly simple version of these theorems is to consider the 2 spatial dimension problem of the Dirac operator $i\slashed{D}$ on the sphere $S^2$, with the round metric. The \textit{index} is 
	\be\operatorname{ind} (i\slashed{D}) = \#(\text{positive chirality zero modes}) - \#(\text{negative chirality zero modes}) \ee
	which counts the difference between positive chirality and negative chirality zero modes. It turns out that this is given (for a charge $q=1$ spinor field) precisely by the magnetic charge,
	\be\operatorname{ind} (i\slashed{D}) = \frac{1}{2\pi}\int \rmd a = \tilde{q} \ee
	and in fact the lowest, spin $s-\frac{1}{2}$ representation, which has dimension $2s = \tilde{q}$, always consists of zero modes of the Dirac operator. This also follows from the pairing reasoning implemented above for the case of the singlet arising at $s=\frac{1}{2}$.
	
	Combined with a check that there are no negative chirality modes (which can be checked to follow from a positivity criterion), the index is actually enough to perform the counting of zero modes exactly. But in general, the robust quantity is not the number of zero modes but the index. This follows from a generalized version of the above pairing argument: one can argue that one can create and destroy zero modes by smoothly altering mass terms in the theory, but only from pairs of positive and negative chirality modes.
\end{asidebox}

\section{Operator Matching and Spin}

In presenting the proposed duality, we have already identified the two conserved $U(1)$ symmetries with each other. Accordingly, their currents must also match. This tells us that
\be \label{bosonization-current-matching} \frac{1}{2\pi} \left(\star \rmd a \right)^\mu \qquad \longleftrightarrow \qquad \bar{\psi} \gamma^\mu \psi \fstp \ee
This is a protected matching, in the sense that based solely on the UV theory we can write down precisely what the correspondence is. But these are very special operator in both theories. What about more general operators?

One of the striking differences between the relativistic flux attachment we have described in this section and our previous (much simpler) quantum mechanical version from \refSectionOnly{flux_attachment} is that we now have to match a theory based on the spinor representation of the Lorentz group with a theory based on scalar particles. This a priori sounds rather unlikely. The resolution, however, can be naturally understood by looking at what the operator matching must be and recalling the discussion of \refSectionOnly{quantization-monopole-bg}.

Let us focus on understanding the dual of the free fermion operator $\psi$. The first observation we make is that this object carries charge $1$ under the global $U(1)$ conserved charge of the theory. This means that its dual must have the same charge under the $U(1)$ monopole symmetry of the bosonic theory. Hence in particular, we must include a monopole operator $\monoM$ of charge 1.

But the dual operator must also be gauge invariant. This is a problem, since as we discussed previously the presence of the Chern-Simons term $\frac{1}{4\pi}a\rmd a$ means that the monopole operator carries a gauge charge of size $1$. But this is easy enough to address; the field $\phi$ has charge $1$ too. Hence the natural guess is that
\be \label{bosonization-correspondence} \phi^\dagger \monoM \qquad \longleftrightarrow \qquad \psi \fstp \ee

This now puts us in precisely the situation of \refSectionOnly{quantization-monopole-bg}! The gauge-invariant operators in the bosonic theory are built from of a unit charge scalar field quantized in the background of a unit charge monopole, and we have already argued that such configurations carry half-integer spins! Intuitively, the dynamics of the strongly-coupled CFT can project us down onto the lightest $\spinrep{\frac{1}{2}}$ representation in the combined $\phi$-$\monoM$ system as we flow into the IR, leaving only this spin $1/2$ operator surviving. The higher modes are presumably gapped.

This means that the notation $\phi^\dagger \monoM$ is slightly misleading. In the background of $\monoM$, there are \textit{two} possible $\phi$ modes which one can turn on, and they transform into each other under rotations, filling out a spinor representation. However, since we have suppressed the spinor indices of $\psi$, it is consistent to suppress those of $\phi^\dagger \monoM$.

Meanwhile, we have already argued that
\be \label{bosonization-correspondence-squared} |\phi|^2 \qquad \longleftrightarrow \qquad - \bar{\psi}\psi \ee
in analyzing the phase diagram of the system. How does this square\footnote{No pun intended.} with the result of \eqref{bosonization-correspondence}? Well, heuristically, the argument of \refSectionOnly{dual_photon} points out that indeed $\monoM \monoM^\dagger$ could be trivial, and so
\be \phi^\dagger\phi \approx \left(\phi^\dagger \monoM\right) \left(\phi^\dagger \monoM\right)^\dagger \longleftrightarrow \psi \psi^\dagger \approx - \psi^\dagger\psi \ee
up to subtleties with normal-ordering and so forth.

But note that, unlike the relation \eqref{bosonization-current-matching}, these identities can be arbitrarily corrected by operators of matching charges anyway. At strong coupling, it is generally impossible to make precise statements about issues like numerical coefficients of operators, which in general are regularization-dependent anyway. (There are exceptions to this, as we will see in the context of supersymmetric dualities in \refChapter{susybreaking}.) We should not place too much faith in this kind of reasoning.
 
\section{The Duality, Summarized}\labelSection{bosonization-summarized}

We should summarize what we have learned, and give a clear a statement of our claim. We assert that
\be\label{bosonization-with-bg-fields}
\boxed{
	|D_a\phi|^2 - |\phi|^4 + \frac{1}{2\pi} A \rmd a + \frac{1}{4\pi} a \rmd a \qquad \longleftrightarrow \qquad i \bar{\psi} \gamma^\mu (\partial_{\mu} - i A_\mu) \psi
}
\ee
where the fermion determinant is understood to be regularized with a ``Chern-Simons term at level $-\frac{1}{2}$''. The left-hand side is at the Wilson-Fisher-type fixed point arising from tuning the relevant coupling $|\phi|^2$ in the presence of a $|\phi|^4$ term.

The underlying operator correspondence is repeated below.
\begin{align}
\phi^\dagger \monoM \qquad &\longleftrightarrow \qquad \psi
\relabeleq{bosonization-correspondence} \\
|\phi|^2 \qquad &\longleftrightarrow \qquad - \bar{\psi}\psi \relabeleq{bosonization-correspondence-squared}
\end{align}

The two phases reached by deforming the fixed point Lagrangians as $\lag - \mu |\phi|^2 \equiv \tilde{\lag} + \mu |\psi|^2$ are described by the contact terms
\be
\lag_{\mathrm{eff}} = 
\begin{cases}
 	0 & \mu > 0 \\
	- \frac{1}{4\pi} A \rmd A & \mu < 0
\end{cases} 
\ee
as suggested by either classical manipulations on the scalar side, or by considering fermion determinants on the right-hand side.

This is the key claim of this section: there is a second-order phase transition between these two phases which is mediated by a free fermion, or equivalently a bosonic Chern-Simons-matter system.

\section{Aside: Technicalities and More Anomalies}\labelSection{abelian_technicalities}

We've been glossing over a lot of the finer points of these dualities in favour of trying to focus on the big picture. However, from a mathematical point of view there are several things to discuss.

\subsection{A Brief Note on TQFTs}\labelSection{brief_tqfts}

Pure Chern-Simons theory, with the Lagrangian taken to be exactly
\be \lag = \frac{k}{4\pi} a \rmd a \cmma \ee
is a \textit{Topological Quantum Field Theory}, in that the action is independent of the metric. None of the resulting physics can depend on things like the separation of points. This is much stronger than even the constraints on CFTs. Any physical quantity can only depend upon topological properties of the observables and the spacetime!

We will discuss these theories in a little more detail in \refSectionOnly{tqfts} and \refAppendixOnly{chern-simons}, including issues like gauge invariance requiring integer $k$; for now, we will briefly outline the key properties of the pure Chern-Simons theory $U(1)_k$. (Adding a small Maxwell term proportional to $1/g^2$ introduces states of a large mass $\sim k g^2$, as discussed in \refAppendixOnly{chern-simons}. We don't care about these.)

Firstly, observe that $a_0$ is a Lagrange multiplier, and its equation of motion is
\be \rmd a = 0 \fstp \ee
Hence on a topologically trivial manifold (like flat space), and in the absence of any charged insertions, there is only one state in the Hilbert space. The partition function is simply a phase. There are still operators -- Wilson lines -- which can be inserted, and the algebra of those operators is all that remains of the structure of $U(1)_k$. Essentially, the Aharonov-Bohm effect makes the Wilson lines into anyonic operators.

Meanwhile, on topologically non-trivial manifolds, there is actually a very rich story to tell \cite{aspectscs,Witten:1988hf}.

\subsubsection*{Topological Degeneracy and Anyons}

Let's work in the gauge $a_0 = 0$. Now the Lagrangian is
\be \lag = \frac{k}{4\pi} \left( \dot{a}_1 a_2 - \dot{a}_2 a_1 \right) \ee
which makes clear that the two fields $a_1,a_2$ are canonically conjugate to each other:
\be [a_i(\textbf{x}),a_j(\textbf{x}')] = \frac{2\pi i}{k} \epsilon_{ij} \delta^{(2)}(\textbf{x}-\textbf{x}') \fstp \ee

Let's suppose the theory is on a spatial torus, so the spacetime is $\mathbb{R} \times T^2$. We will write $x \in [0,L_x)$ and $y \in [0,L_y)$ for the torus coordinates. Now we know the $a_0$ always fixes $\rmd a$, so that there are no local excitations in the theory -- they are all pure gauge. It can only be large fluctuations of some kind that can give rise to physically distinct states.

What other gauge-invariant quantities are there in this theory? We can consider integrals like $\int_C a$ for some curve $C$, but these still transform by total derivatives which do not generally vanish. We can do slightly better by enforcing that $C$ is closed, and computing things like
\be \theta_x = \oint a_x \rmd x \quad \mbox{and} \quad \theta_y = \oint a_y \rmd y \ee
although since $\rmd a = 0$, these are the only non-equivalent quantities available. Now these are also not necessarily invariant, due to the existence of large gauge transformations under which
\be a_x \to a_x + \partial_x \left(\frac{2\pi n x}{L_x} + \frac{2\pi m y}{L_y}\right) \quad \mbox{and} \quad a_y \to a_y + \partial_y \left(\frac{2\pi n x}{L_x} + \frac{2\pi m y}{L_y}\right) \ee
which means that
\be \theta_x \sim \theta_x + 2\pi \quad \mbox{and} \quad \theta_y \sim \theta_y + 2\pi \ee
are periodic. This leads to a standard conclusion: the gauge-invariant observables are the \textit{Wilson lines}
\be W_x = e^{i\theta_x} \quad \mbox{and} \quad W_y = e^{i\theta_y} \ee
and that is it!

Now we find that the quantum operators obey $[\theta_x, \theta_y] = 2\pi i/k$, and hence
\be W_x W_y = e^{-2\pi i/k} W_y W_x \ee
so that in particular $(W_x)^k,(W_y)^k$ commute with everything in the theory. The Hilbert space is the smallest representation of this algebra, and without loss of generality we define states 
$\left| l \right>$ for $l=0,\ldots, k-1$ upon which the Wilson lines act as
\be W_y \ket{l} = e^{-2\pi il/k} \ket{l} \quad \mbox{and} \quad W_x \ket{l} = \ket{l+1 \operatorname{mod} k} \fstp \ee

The striking result is that there are $k$ degenerate states on the torus for $U(1)_k$. Putting the theory on a higher-genus Riemann surface increases this degeneracy further: a genus $g$ surface has $k^g$ states. This is a robust result -- no small deformation of the theory can remove this degeneracy.

There is a nice story that goes with this. Firstly, one can understand the operation
\be W_x W_y W_x^{-1} W_y^{-1} = e^{-2\pi i/k} \ee
as corresponding to moving two anyonic particles around each other. This works as follows:  suppose we add a heavy test particle (with electric charge 1) and an anti-particle (with electric charge -1) to the torus at a point. Now move the particle one way around the torus, then the other, then back around the first loop, and finally back around the second loop. If you play with this a little, you should be able to convince yourself that this is equivalent to moving the particle around the anti-particle. (Mathematically, the world lines have linking number $1$ \cite{tongqhe}.)

Then this result expresses the Aharanov-Bohm effect for anyonic particles. Concretely, electric charge $1$ particles carry a magnetic charge of magnitude $1/k$, and therefore we expect a phase of $2\pi/k$ when we move one around the other. This is exactly what we get!

Secondly, this sort of description also gives a nice interpretation to the topological degeneracy: we can reach a new state by pair-creating an anyon and an anti-anyon, moving the anyon around a loop of the torus, and then annihilating the anyons. The non-triviality of the anyonic statistics guarantees that one must be able to end up in a physically distinct state.


\subsubsection*{The Level $1$ Theory}

As the discussion above suggests, the theory $U(1)_1$ is essentially trivial. There is only ever one state in the Hilbert space, and the partition function is only ever a phase; there is no dynamics to be observed.

However, the presence of a $U(1)_1$ theory is not quite entirely trivial. Even defining it requires a spin structure (or coupling to a \spinc field as discussed below). The reason is it possesses a single line operator (the above Wilson line, placed anywhere) that one can show has spin $1/2$. The theory also comes with a so-called \textit{framing anomaly}. This can be expressed in terms of a gravitational Chern-Simons term, as discussed in Appendix B of \cite{ssww}. Intuitively, whilst the $U(1)_1$ partition function is a phase, that phase necessarily changes when we mess with the manifold.

Nonetheless, from a dynamical point of view, $U(1)_1$ is irrelevant since it does not alter the Hilbert space of the theory. We refer to it as "almost trivial".

\subsubsection*{A Time Reversal Invariant TQFT}

One consequence of the triviality of $U(1)_{\pm 1}$ is that 
\be U(1)_2 \bigdual U(1)_{-2} \ee 
up to to subtlety that we should add $U(1)_{-1}$ on the left and $U(1)_{+1}$ on the right.\footnote{Technically, there is a time reversal anomaly associated to the gravitational part of these theories. We won't discuss this.}

Intuitively, one might reason that time reversal acts roughly like complex conjugation on Wilson lines in the spectrum, but when the Wilson line obeys $W_x^2 = 1$, it is real. This is a bit of a fairytale, but it suggests why there might be something special about $U(1)_k$ when $k=2$.

Concretely, start with the $U(1)_2$ side of the theory, letting
\be \lag = \frac{2}{4\pi} b \rmd b + \frac{1}{2\pi} b \rmd B - \frac{1}{4\pi} c \rmd c \ee
which indeed also contains a decoupled $U(1)_{-1}$ sector.
Now let $b = b'+c'-B$ and $c=c'+2b'-B$. Then
\be \lag = -\frac{2}{4\pi} b' \rmd b' + \frac{1}{2\pi} b' \rmd B - \frac{1}{4\pi} B \rmd B + \frac{1}{4\pi} c \rmd c \ee
takes the form of a $U(1)_{-2}$ theory, combined with a decoupled $U(1)_1$ sector. (Note that we have coupled the non-trivial theory to a background field, revealing there is a slightly non-trivial background term differentiating $U(1)_{\pm 2}$.)

This is discussed further in Section 3.1 of \cite{hsinseiberg}. (It forms part of a series of non-Abelian time reversal invariant cases, namely those we will later refer to as $U(N)_{N,2N}$.)

\subsubsection*{Aside: $\mathbb{Z}_N$ Gauge Theory}

One other quirky theory to be aware of -- though we won't work with it -- is $\mathbb{Z}_N$ theory. At level $k$,
\be
\lag = \frac{k}{4\pi} a \rmd a + \frac{N}{2\pi} a \rmd b + \frac{1}{2\pi} a \rmd A
\ee
which at first looks trivial, since the $b$ equation is classically $\rmd a = 0$. But in fact, this "BF theory" defines $(\mathbb{Z}_N)_k$ gauge theory. This theory only has line operators, but is still non-trivial. It arises from Higgsing $U(1)_k$ with a scalar of charge $N$, as can be seen by dualizing $b$.

\begin{exercise}
	What non-trivial gauge-invariant observables exist in this theory? Argue that there is a new type of $\mathbb{Z}_N$ symmetry which acts on these non-local operators.
\end{exercise}	

We will not pursue this story here, but this is called a \textit{1-form} symmetry. In general, $q$-form symmetries (or generalized symmetries) act on $q$-dimensional objects. Conventional symmetries, which we would in this language call $0$-form symmetries, act on particles, but higher form symmetries like this act only on extended objects, in the case above lines or strings. Even the familiar free Maxwell theory in 4d possesses two such symmetries, one associated with the electric Wilson lines, and one with the magnetic 't Hooft lines, and one can even show that there is a mixed 't Hooft anomaly between them. (This precisely parallels the situation of the compact boson in 2d, where there were two $U(1)$ symmetries associated with the vector and axial currents but only one may be gauged. Moreover, dimensional reduction of 1-form symmetries gives rise to 0-form symmetries, and one can directly connect these properties of Maxwell theory in 4d with 3d and 2d theories.) See \cite{Gaiotto:2014kfa} for an accessible introduction to these ideas.

\subsection{Fermion Determinants and the Parity Anomaly}\labelSection{fermion_det_parity_anom}

It is worth understanding the parity anomaly of fermions in odd dimensions properly; some useful references include \cite{fermionpathintegrals,anomaliesodddimensions}. Here, we will just outline a couple of issues that crop up.

Firstly, let's think about the unregularized path integral of the massless theory in Euclidean signature, that is
\be \det \underbrace{ \left[i\gamma^\mu (\partial_\mu - i A_\mu) \right]}_{\mathcal{D}} \fstp \ee
The operator $\mathcal{D}$ is Hermitian, and therefore has real eigenvalues; loosely speaking,
\be \det \mathcal{D} = \prod_a \lambda_a \fstp \ee
Now $|\det \mathcal{D}|$ is well-defined (in particular, unique) in any regularization scheme; for instance, $|\det \mathcal{D}|^2$ can be computed by regularizing a theory of two fermions using two Pauli-Villars fields of opposite mass. (The regulators are exchanged under $T$, so their combination is $T$ invariant.) However, the overall sign of $\prod_a \lambda_a$ is impossible to uniquely determine, since it in general consists of a product of infinitely many positive and negative numbers. This leaves us with a choice.

One might now hope that we can just arbitrarily choose $Z$ to be positive for some reference configuration $A_0$, and then continuously define it for other $A$ by changing sign every time an eigenvalue $\lambda_a$ passes through $0$. But this would uniquely define a real partition function $Z$ without violating $T$ symmetry (recalling that $T$ contains a complex conjugation), which must be impossible. In fact, in the above, we saw that this would violate gauge invariance under large gauge transformations (since it would give an incorrectly quantized Chern-Simons level). This is indeed the case. Suppose we follow the above procedure and define $A_s = (1-s)A_0 + s A$, tracking how many eigenvalues pass through zero along the way. Now take $A$ to differ from $A_0$ by a large gauge transformation. Then the gauge-invariance of the Dirac equation guarantees that the set of eigenvalues $\{\lambda_a\}$ (counted with multiplicity) is the same for both $A$ and $A_0$. But it is entirely possible that an odd number of eigenvalues have passed through zero along the way -- this is called \textit{spectral flow}. (See \refFigureOnly{spectralflow}. Clearly, since the eigenvalues at the left and right must coincide, this is only possible for infinite sets of eigenvalues.) Consequently, we can encounter violation of gauge-invariance in the overall sign of the determinant. In this formulation, the theory has a gauge anomaly and is sick.

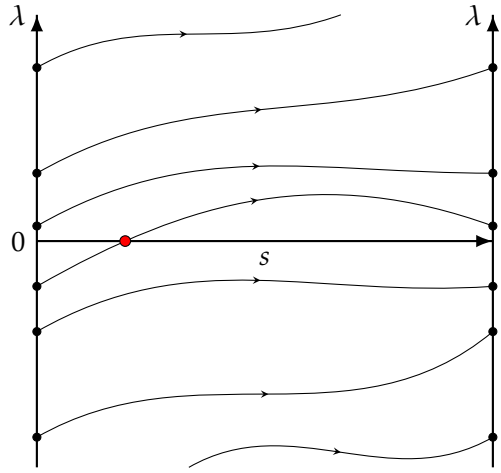
\begin{figure}
	\centering
	\begin{tikzpicture}
	\draw[axis,name path=P1] (0,3) node[left] {$0$} -- node[pos=0.5,below] {$s$} ++(6,0);
	\draw[axis] (0,0) -- ++(0,6) node[left] {$\lambda$};
	\draw[axis] (6,0) -- ++(0,6) node[left] {$\lambda$};
	\draw[flow] (2,0) to[out=30,in=-150] (6,0.4);
	\draw[flow] (0,0.4) to[out=30,in=-140] (6,1.8);
	\draw[flow] (0,1.8) to[out=30,in=184] (6,2.4);
	\draw[flow,name path=P2] (0,2.4) to[out=30,in=160] (6,3.2);
	\draw[flow] (0,3.2) to[out=30,in=-180] (6,3.9);
	\draw[flow] (0,3.9) to[out=30,in=-160] (6,5.3);
	\draw[flow] (0,5.3) to[out=30,in=-160] (4,6);
	
	\path [name intersections={of=P1 and P2, by={E}}];
	\filldraw[fill=red] (E) circle (2pt);
	
	\filldraw (0,0.4) circle (1.5pt); \filldraw (6,0.4) circle (1.5pt);
	\filldraw (0,1.8) circle (1.5pt); \filldraw (6,1.8) circle (1.5pt);
	\filldraw (0,2.4) circle (1.5pt); \filldraw (6,2.4) circle (1.5pt);
	\filldraw (0,3.2) circle (1.5pt); \filldraw (6,3.2) circle (1.5pt);
	\filldraw (0,3.9) circle (1.5pt); \filldraw (6,3.9) circle (1.5pt);
	\filldraw (0,5.3) circle (1.5pt); \filldraw (6,5.3) circle (1.5pt);

	\end{tikzpicture}
	\caption{Spectral flow of eigenvalues of Dirac operator under a large gauge transformation, illustrating one sign change and hence a violation of gauge invariance}\labelFigure{spectralflow}
\end{figure}

Suppose now we used our simple PV prescription with mass $M$ to regularize the determinant. Then we obtain a regularized expression
\be \det\!' \mathcal{D} \propto \prod_a \frac{\lambda_a}{\lambda_a - i M} \ee
for the determinant, where the factor of $i$ arises in the Wick rotation to Euclidean signature. Now taking $M\to-\infty$, 
\be \det\!' \mathcal{D} = |\det \mathcal{D}| \exp \left( -\frac{i \pi}{2} \sum_a \sign \lambda_a \right) = |\det \mathcal{D}| \exp \left( -\frac{i \pi \eta}{2} \right) \ee
where $\eta$ is the Atiyah-Patodi-Singer (APS) $\eta$-invariant, which is defined with a regularization such as
\be \eta = \lim_{s \to 0} \eta(s) = \lim_{s \to 0} \sum_a \sign \lambda_a |\lambda_a|^{-s} \ee
for example. (This actually corresponds to $\zeta$-function regularization; one defines it by analytic continuation from sufficiently large $\operatorname{Re} s$.) It is a measure of the spectral asymmetry of the Dirac operator.

All that remains is to relate $\eta$ somehow to the Chern-Simons action. In flat space, the result is that
\be \pi \eta = \int \frac{1}{4\pi} A \rmd A \pmod {2\pi} \ee
though this is actually modified in curved space; more generally,
\be \pi \eta = \int \frac{1}{4\pi} A \rmd A + 2 \mathrm{CS}_\mathrm{grav} \pmod {2\pi} \ee
where the gravitational Chern-Simons term (which looks like a Chern-Simons term for the spin connection $\omega_\mu$) can naturally be defined by taking a curved spacetime to be the boundary of a 4-manifold (cf. the Chern-Simons case as in \refAppendixOnly{chern-simons})
\be \mathrm{CS}_\mathrm{grav} = \frac{1}{192\pi} \int_X \tr R \wedge R \ee
and in general is well-defined modulo $2\pi$ either in the APS combination appearing in $\pi \eta$, or when multiplied by $16$. (On spin manifolds, it is well-defined modulo $2\pi$ on its own.) We will not dress all the dualities we discuss with the correct gravitational Chern-Simons term, however. See \cite{hsinseiberg} for a more careful analysis of these issues.

The key point is that now, having come up with a gauge-invariant regularization, the theory no longer has time-reversal or parity invariance; this is most straightforwardly understood by observing that the partition function is no longer real. We have a time-reversal anomaly, and it is given by $2\times \frac{\pi}{2} \eta = \pi \eta = S_\mathrm{CS}$.

We will discuss this for fermions in general representations of more general gauge groups in \refSectionOnly{generalize_representation}.

\subsection{Spin Structures and \texorpdfstring{\Spinc}{Spin c} Fields} \labelSection{spinc}

For the more mathematically minded, there is a large objection to be raised to the contents of both this section and the next. We will try and briefly explain both the concern and its resolution in this section.

\subsubsection*{Fermions}

Let's assume we have in our possession an orientable, Riemannian 3-manifold $\mathcal{M}$. (We cannot define Chern-Simons theory on non-orientable manifolds. We will also mainly use the language of Riemannian manifolds rather than Lorentzian ones for simplicity.) Imagine first trying to define a vector field $A^\mu$ on this manifold. In order to do that, one has to have a notion of what vector space $A(x)$ lives in at each point $x \in \mathcal{M}$. The answer is that $A$ is a \textit{section} of a \textit{vector bundle} of $SO(3)$. This means we divide the manifold up into open patches $U_\alpha$, and at each point $x$ in each patch we have some $A^{(\alpha)}(x) \in \mathbb{R}^3$. Now consider a point $x \in U_\alpha\cap U_\beta$ in the overlap of two patches, so that we have two values of the field, $A^{(\alpha)}(x)$ and $A^{(\beta)}(x)$. There is then a compatibility criterion: we must have $A^{(\alpha)}(x) = g(x) A^{(\beta)}(x)$ for some $g(x) \in SO(3)$.\footnote{There is also a triple overlap condition, which is actually important for these bundles.} The choice of $g(x)$ is specified by  the bundle.

One particularly nice way to think of the choice of $SO(3)$ bundle is simply as a choice of (oriented) basis of the tangent space $\mathbb{R}^3$ at each point of $\mathcal{M}$; the bundle then once more specifies the rotations relating different patches. This is the so-called \textit{frame bundle}. A simple argument shows that in fact the space of all oriented bases is isomorphic to $SO(3)$.\footnote{This makes the frame bundle is a \textit{principal $SO(3)$ bundle}. The tangent bundle in which $A$ lives is then an \textit{associated vector bundle} for the vector representation of $SO(3)$. Specifying one is equivalent to specifying the other.} One simply fixes a reference basis $B_0$, and then notes that given another basis $B$ there certainly exists a rotation $g$ with $B=gB_0$; and then that this is obviously unique. This defines an isomorphism $g \leftrightarrow B$.

But now consider the theory of a free fermion (either real or complex for now),
\be S = \int_{\mathcal{M}} \rmd^3 x i \bar{\psi} \gamma^\mu \partial_{\mu} \psi \fstp \ee
Recall that the spinor field $\psi$ transforms not under $SO(3)$ but its double cover, $\Spin(3) \cong SU(2)$. A very similar argument to the above tells us that we need to construct a spinor bundle, which we can do by considering a frame bundle for $SU(2)$ instead. But now we have various questions, like ``is there such a bundle?'' and ``is it uniquely determined by the vector bundle?'' for example.

Let's assume we have an oriented manifold, with its $SO(3)$ frame bundle. Then there is a notion called a \textit{spin structure}, which is simply an $SU(2)$ frame bundle which is compatible with the $SO(3)$ one.\footnote{What do we mean by "compatible"? Recall that there is a double-cover map $\rho : SU(2) \to SO(3)$. We require that the $SU(2)$ transition functions $g$ of our new bundle are such that $\rho(g)$ are the transition functions of the old bundle.} It turns out that:
\begin{itemize}
	\item There need not be a spin structure for a given vector bundle in general dimensions; however, in three dimensions, orientable manifolds do always have at least one spin structure. (In general, one requires that something called the second Stiefel-Whitney class of the vector bundle vanishes.)
	\item When there are spin structures, they are in one-to-one correspondence with the homology group $H^1(M,\mathbb{Z}^2)$ (though there is no canonical bijection between these sets). For instance, the circle has 2 spin structures whilst a Riemann surface of genus $g$ has $2^{2g}$ spin structures.
\end{itemize}
Intuitively, choices of spin structure correspond to choices of boundary conditions for a spinor field $\psi$. For every loop in the manifold, we can either impose that $\psi \to \psi$ when we encircle the loop (periodic or Ramond boundary conditions) or $\psi \to -\psi$ (antiperiodic or Neveu-Schwarz boundary conditions).

Clearly, in order to even define what a typical fermionic theory means, we must make a choice of spin structure. This is a bit troubling, since we claimed that a free fermion is totally equivalent to a bosonic theory $U(1)_1 + \text{boson}$, which seems like it does not need a choice of spin structure!

\subsubsection*{Chern-Simons Theories}

So let's look more closely at the Chern-Simons theories. Clearly, the term
\be \frac{1}{4\pi} \int a \rmd a \ee 
is not manifestly gauge invariant. In fact, in \refAppendixOnly{chern-simons}, it turns out that gauge invariance of the action is only guaranteed in the presence of a spin structure! Otherwise, it is well-defined only modulo shifts of $\pi$. This means that the path integral without a spin structure is well defined only for even Chern-Simons levels.

One way this has reared its head is in \refSectionOnly{quantization-monopole-bg}, where we saw that the monopole operator in $U(1)_1$ necessarily transforms in a spin $1/2$ representation.

\begin{asidebox}[Spin in Abelian Chern-Simons Theory]%
	Another analysis which supports this conclusion arises from computing the spin of certain states in the pure field theory. Consider level $k$ Chern-Simons theory. Then a magnetic flux with magnetic charge $1$ can be accompanied by Wilson lines or bosons of total charge $k$ to give a physical state.
	
	This can be done by a careful quantization, or a careful regularization of the classical theory. The conserved angular momentum operator in the quantum mechanical theory is
	\be L = - \frac{k}{4\pi} \int x^i \epsilon^{ij} (a_j f + f a_j) \fstp \ee
	Suppose we have a background charge density $\rho(x)$, due to the insertion of bosons or Wilson lines. Then one can solve Gauss's law as an operator statement. The resulting states are eigenstates of $L$, and
	\be L \ket{\rho} = \boxed{\frac{Q^2}{2k}} \ket{\rho} \ee
	where $Q = \int \rmd^2x \, \rho$ is the total charge of the state \cite{aspectscs}. This anomalous quadratic growth of the spin (which can be thought of as a one-loop effect in a $1/k$ expansion) is typical of anyons.
	
	Hence the total angular momentum of our configuration is $J = k^2/2k = k/2$, matching the above analysis of the monopole background. (Note that $Q^2$ is the quadratic Casimir of the charge $k$ representation of $U(1)$. This generalizes to other gauge groups and representations too.)
\end{asidebox}

Another physical (but still subtle) illustration of this is given in the appendix of \cite{ssww}. The subtlety of the action is highlighted by the fact that it is hard to evaluate in the presence of both magnetic flux/Dirac strings and Wilson lines. Suppose $\mathcal{M} = T^2 \times S^1$, and consider a configuration with flux $\int_{T^2} \rmd b = 2\pi$ through the torus. The Chern-Simons term means that this has one unit of electric charge. We can make this gauge-invariant by also inserting a Wilson line $\exp(i\int_{S^1} a)$ into the path integral. It turns out that regularizing this line insertion requires a choice of spin structure, and \cite{ssww} works through the details which shows that you get an action of $0,\pi$ for two different spin structures.

In each case, it is clear that the presence of both electric and magnetic charge is what makes the argument fly. Ultimately, this is very little more than the Aharanov-Bohm effect one last time!

\subsubsection*{The \texorpdfstring{\Spinc}{Spin c} Solution}

One last thing we should mention is that it is in fact possible to generalize our notion of gauge field in such a way that the $U(1)_1$ Chern-Simons theory is gauge invariant without a choice of spin structure. One slightly tweaks the quantization condition of the gauge field, allowing
\be \frac{1}{2\pi} \int_C \rmd A = \frac{1}{2} \int_C w_2  \pmod{\mathbb{Z}} \ee 
where $C \subset \mathcal{M}$ is an oriented two-cycle and $w_2$ is something called the second Stiefel-Whitney class (which measures the obstruction to choosing two linearly independent vector fields on the manifold). The technical details don't matter hugely, but the conclusion does: it gives a unique value to the path integral for \textit{arbitrary} integer Chern-Simons level. This all works on any manifold with a \textit{\spinc structure}, which includes everything in three dimensions.

Assuming you're happy with the procedure for evaluating the Chern-Simons action as a boundary term of a 4-manifold $X$ with $\partial X = \mathcal{M}$, you can think of this as follows. There are different choices for $X$, corresponding to different spin structures on $\mathcal{M}$, giving either $S_\mathrm{CS} \equiv 0 \pmod{2\pi}$ or $S_\mathrm{CS} \equiv \pi \pmod{2\pi}$ for the non-\spinc theory. But a $\spinc$ connection on $\mathcal{M}$ specifies that these different choices of filling manifold $X$ actually have different boundary conditions on a \spinc field. This shifts $S_\mathrm{CS}$ for half of the fillings, leaving a consistent choice for its value for arbitrary $X$.

More intuitively, you can think of this as (roughly) \textit{gauging the choice of spin structure}. This can be seen by the fact we get to choose the spin structure of the filling $X$ freely when evaluating the path integral: it's a \textit{gauge choice}. We will see this again as we now discuss spinors.

For the duality to work out on \spinc manifolds, we'd better have a prescription for evaluating the path integral of Dirac fermions without a spin structure too! It turns out this does work. Recall that the choice of spin structure is essentially the choice of boundary conditions for fermions, $\psi \to \pm \psi$ around each loop of the manifold. But now, \textbf{if} we give $\psi$ charge 1 under a gauge field so that $\psi \to e^{i\alpha}\psi$, \textbf{and} we allow large gauge transformations such that $\int \rmd \alpha = \pi$ around circles, \textbf{then} both choices of boundary condition are equivalent up to this large gauge transformation. But allowing a large gauge transformation carrying winding $\pi/2\pi = 1/2$ is intimately related with allowing half-integer fluxes $\frac{1}{2\pi} \int_C \rmd A \in \frac{1}{2}\mathbb{Z}$. This is essentially the reason for the tweaked quantization condition of \spinc fields.

The correct statement is that, without a spin structure, we are unable to define the operator $\slashed{D}$ acting on a \textit{neutral} fermion -- but given a \spinc structure, we \textit{can} define $\slashed{D}$ acting a charge 1 complex fermion. In fact, we can make sense precisely of complex fermions carrying odd charge under a \spinc gauge field.

This fixes the duality to work in a completely spin-structure-ambivalent way.

\chapter{The Duality Web}
\labelChapter{duality_web}

\begin{introduction}
  We discuss how to obtain new dualities from existing ones, deriving a so-called web of dualities from our initial `seed' duality. This includes something familiar in the form of particle-vortex bosonization, as well as some novel dualities.
\end{introduction}

\lettrine{I}{n two almost} simultaneous papers in June 2016, \cite{karchtong} and \cite{ssww}, it was observed that there is a deep interrelation between the bosonization duality of \refChapterOnly{3d_bosonization} and particle-vortex duality, as presented in \refChapterOnly{particle_vortex}. In fact, it turned out that using just a couple of simple operations, it is possible to derive all sorts of different dualities from a single seed duality. For us, the seed will be 3d bosonziation.
\be
|D_a\phi|^2 - |\phi|^4 + \frac{1}{2\pi} A \rmd a + \frac{1}{4\pi} a \rmd a \qquad \longleftrightarrow \qquad i \bar{\psi} \gamma^\mu (\partial_{\mu} - i A_\mu) \psi
\relabeleq{bosonization-with-bg-fields}
\ee

\section{Reversing Bosonization}

Our first aim in this chapter is to explain how to derive particle-vortex duality from 3d bosonization. We will see that there are two key ingredients: gauging global symmetries (making background fields dynamical by integrating over them in the partition function), and time-reversal.

We are aiming for
\be
|\partial\phi|^2 - |\phi|^4 \qquad \longleftrightarrow \qquad |D\tilde{\phi}|^2 - |\tilde{\phi}|^4 \relabeleq{particle_vortex_schematic_no_f2}
\ee
which looks distinctly different from \eqref{bosonization-with-bg-fields} since the only scalar field there was gauged and had a Chern-Simons term. But this is actually not so hard to address.

Suppose that we made the field $A$ in \eqref{bosonization-with-bg-fields} dynamical; then the equation of motion of $A$ simply sets $a=0$. This immediately leaves us with a theory like that on the left-hand side of \eqref{particle_vortex_schematic_no_f2}. Moreover, we've actually just derived our first new duality! We write the schematic Lagrangians as
\be\label{reverse-bosonization-lag}
\boxed{
	|\partial\phi|^2 - |\phi|^4 \qquad \longleftrightarrow \qquad i \bar{\psi}\gamma^\mu(\partial_{\mu} - i a_\mu) \psi
}
\ee
where the lower-case $a$ is now understood to be dynamical on the right-hand side. In our other kind of notation for dualities, we might write
\be
\text{WF scalar} \qquad \longleftrightarrow \qquad U(1)_{-1/2} + \text{fermion}
\ee
where the heuristic half-integral level is written explicitly to remind us of the regularization we are using. (Note in \eqref{reverse-bosonization-lag} we do not write such a term.)

This is another type of bosonization relation, and can be thought of as attaching flux to a fermion in order to turn it into a boson (instead of attaching flux to bosons to make them fermions). We know that the lowest mode of the fermion in a monopole background will be a spacetime scalar, so this makes sense; based on \eqref{bosonization-correspondence} we would propose
\be \label{reverse-bosonization-correspondence} \phi^\dagger \qquad \longleftrightarrow \qquad \monoM \psi \fstp \ee

The global symmetries still match, of course, with both theories still possessing a $U(1)$ symmetry. Now the bosonic theory has a simple $\phi \to e^{i\alpha} \phi$ rotation, whereas the fermionic theory has a monopole symmetry. We could alternatively derive this by adding a term $\frac{1}{2\pi} A \rmd B$ to \eqref{bosonization-with-bg-fields} before making $A$ dynamical. On the right-hand side, this clearly becomes a standard monopole coupling. On the left-hand side, meanwhile, we see that the equation of motion of $A$ sets $a = -B$. Hence in fact we deduce that
\be\label{reverse-bosonization-lag-bg}
\boxed{
	|(\partial_\mu + i B_\mu)\phi|^2 - |\phi|^4 + \frac{1}{4\pi} B \rmd B \qquad \longleftrightarrow \qquad i \bar{\psi}\gamma^\mu(\partial_{\mu} - i a_\mu) \psi + \frac{1}{2\pi} B \rmd a 
}
\ee
where, slightly surprisingly, we have given the field $\phi$ a negative charge under the global $U(1)$. This actually matches up perfectly with the situation in \eqref{reverse-bosonization-correspondence}.

This sign does not really matter: the theory of the boson has an obvious charge-conjugation symmetry we can use to simply swap $\phi \leftrightarrow \phi^\dagger$, effectively redefining the theory to have the opposite charge under $B$.

What about the phase diagram of these theories? We know they must match, but it would be nice to know what kind of transition we are capturing. Let us begin on the left-hand side. Here, if we deform using the relevant operator $\lag \to \lag - \mu|\phi|^2$, then we find:
\begin{itemize}
	\item $\mu \gg 0$ leads to an empty theory with only the contact term
	\be \lag_{\mathrm{eff}} = \frac{1}{4\pi} B \rmd B \cmma \ee
	whilst
	\item $\mu \ll 0$ leads to a theory with a Goldstone boson associated to $\alpha$ in $\phi \to e^{i\alpha}\phi$, where the shift symmetry is coupled to $B$. $B$ also still has its contact term:
	\be \lag_{\mathrm{eff}} = (\partial_{\mu} \alpha - B_{\mu})^2 + \frac{1}{4\pi} B \rmd B \fstp \ee
	(If $B$ were dynamical, the Higgs mechanism would kill it, after it ate the $\alpha$ mode.)
\end{itemize}
Meanwhile, on the right-hand side, $\lag \to \lag - \tilde{\mu}|\psi|^2$ leads to either
\begin{itemize}
	\item $\tilde{\mu} \gg 0$ leads to a free photon $a$ coupled via a BF term to $B$,
	\be \lag_{\mathrm{eff}} = \frac{1}{2\pi} a \rmd B \cmma\ee
	or
	\item $\tilde{\mu} \ll 0$ leads to a theory with Lagrangian
	\be \lag_{\mathrm{eff}} = -\frac{1}{4\pi} a \rmd a + \frac{1}{2\pi} B \rmd a \equiv \frac{1}{4\pi} B \rmd B \cmma \ee
	 where we have solved the $a$ equation of motion directly as we have done previously.
\end{itemize}
Clearly, $\mu \gg 0$ and $\tilde{\mu} \ll 0$ match perfectly.

The situation with $\mu \ll 0$ and $\tilde{\mu} \gg 0$ perhaps seems a little more subtle. The first thing to appreciate is that the Goldstone boson of the bosonic theory can be realized as a dual photon to a dynamical field $b$. Rewritten in terms of this variable, the Lagrangian becomes simply $\frac{1}{2\pi} b \rmd B + \frac{1}{4\pi} B \rmd B$. This almost matches the situation in the fermionic theory, except for the contact term. But this is not a problem at all. The equation of motion of $b$ actually imposes a constraint on the background field $B$: it sets $B$ to be pure gauge. This guarantees that we can simply ignore the contact term, since it will never contribute for any value of $B$.

Hence
\be \label{reverse-bosonization-correspondence-sq} |\phi|^2 \qquad \longleftrightarrow \qquad - \bar{\psi} \psi \ee
completes our analysis of this duality.

Writing all this in terms of partition functions, we have learned that
\be
\int Da D\phi \ e^{ iS_{\mathrm{BF}}[A;a] + i S_{\mathrm{CS}}[a] + iS_{\mathrm{scalar}}[\phi;a] }
= \int D\psi \ e^{ iS_{\mathrm{fermion}}[\psi;A] }
\ee
implies
\be
\int DA Da D\phi \ e^{ iS_{\mathrm{BF}}[A;B] + iS_{\mathrm{BF}}[A;a] + i S_{\mathrm{CS}}[a] + iS_{\mathrm{scalar}}[\phi;a] }
= \int DA D\psi \ e^{ iS_{\mathrm{BF}}[A;B] + iS_{\mathrm{fermion}}[\psi;A] }
\ee
and then we simplified the left-hand side (relabelling $A \to a$ on the right-hand side) to give
\be
\int D\phi \ e^{ iS_{\mathrm{CS}}[B] + iS_{\mathrm{scalar}}[\phi;-B] }
= \int Da D\psi \ e^{ iS_{\mathrm{BF}}[a;B] + iS_{\mathrm{fermion}}[\psi;a] }
\ee
which was our result. Charge-conjugating the scalar allows the similar conclusion
\be
\int D\phi \ e^{ iS_{\mathrm{CS}}[B] + iS_{\mathrm{scalar}}[\phi;B] }
= \int DA D\psi \ e^{ iS_{\mathrm{BF}}[A;B] + iS_{\mathrm{fermion}}[\psi;A] } \fstp
\ee

\section{The Trick: Particle-Vortex Duality from Bosonization}

This is progress, but it was not quite what we were aiming for. We still do not have a gauged Wilson-Fisher theory without a Chern-Simons level. But looking at \eqref{reverse-bosonization-lag-bg}, we can easily find one.

If we were to play the same game as before, and make $B$ dynamical, we would end up with a Wilson-Fisher theory coupled to $U(1)_1$. But there is nothing to stop us simply subtracting off the background contact term $S_{\mathrm{CS}}[B]$ before we promote $B$ to a dynamical field -- provided we do so on both sides of the duality!

This means gauging the duality
\be |D_B\phi|^2 - |\phi|^4 \qquad \longleftrightarrow \qquad i \bar{\psi} \slashed{D}_a \psi + \frac{1}{2\pi} B \rmd a - \frac{1}{4\pi} B \rmd B \label{wf-dual-gauged-fermion} \ee
(where for convenience we have also charge-conjugated the boson) by promoting $B \to b$. We find
\begin{align}
|D_b\phi|^2 - |\phi|^4 \qquad \longleftrightarrow &\qquad i \bar{\psi} \slashed{D}_a \psi + \frac{1}{2\pi} b \rmd a - \frac{1}{4\pi} b \rmd b \nn \\
\equiv& \qquad i \bar{\psi}\slashed{D}_a \psi + \frac{1}{4\pi} a \rmd a \label{pv-from-bos-intermediate}
\end{align}
by solving for $b$ using its equation of motion. This is intriguing -- but we would like to be able to further dualize this right-hand theory back to an ungauged scalar.

This sounds plausible; the right-hand theory describes fermions with flux attached. Yet it has an extra Chern-Simons term $S_\mathrm{CS}[a]$ relative to \eqref{reverse-bosonization-lag-bg}; this is the theory we would call $U(1)_{+1/2} + \text{fermion}$ instead of $U(1)_{-1/2} + \text{fermion}$. But these are related -- by time-reversal invariance!

Indeed, our trick is to apply time-reversal to \eqref{reverse-bosonization-lag-bg}. The anomalous transformation of the fermionic theory leads to the result
\be\label{reverse-bosonization-lag-bg-time-reversed}
\boxed{
	|D_B\phi|^2 - |\phi|^4 - \frac{1}{4\pi} B \rmd B \qquad \longleftrightarrow \qquad i \bar{\psi}\slashed{D}_a \psi + \frac{1}{4\pi} a \rmd a - \frac{1}{2\pi} B \rmd a 
}
\ee
which we can directly apply to \eqref{pv-from-bos-intermediate} (sans background terms for now) to find exactly the result we wanted:
\be
|D\phi|^2 - |\phi|^4 \qquad \longleftrightarrow \qquad |\partial\phi|^2 - |\phi|^4
\ee
in agreement with \eqref{particle_vortex_schematic_no_f2}.

\begin{exercise}[subtitle=Particle-Vortex Duality from Bosonization]
	Check that the background terms work out as they should. Confirm that the particle-vortex operator matching follows from that of the bosonization dualities.
\end{exercise}

Thus, particle-vortex duality is a direct logical consequence of 3d bosonization:
\be
\mbox{3d bosonization} \qquad \implies \qquad \mbox{particle-vortex duality} \fstp
\ee
There does not seem to be a natural way to reverse the implication here, however; it is logically possible for particle-vortex duality to hold, but 3d bosonization to fail.

Nonetheless, we can think of it as a rather non-trivial piece of evidence in favour of 3d bosonization that we can derive another (more familiar) duality from it. We could certainly have got nonsense out of these manipulations, a priori, but instead we have landed on our feet.

Since particle-vortex duality follows from applying the bosonization duality twice, combined with time-reversal, one might be tempted to write
\be
\mbox{particle-vortex duality} \quad = \quad  |\mbox{3d bosonization}|^2
\ee
where complex conjugation represents time-reversal.

\begin{asidebox}[3d Dualities from $SL(2,\mathbb{Z})$ in 4d]%
	Actually, the operations which we are implementing to move between dualities can be related to the action of the famous $SL(2,\mathbb{Z})$ symmetry group which acts on 3+1 Maxwell fields. This is part of an older story to do with understanding boundary conditions for 3+1d gauge fields \cite{Gaiotto:2008ak,Witten:2003ya}. A full discussion of this can be found in \cite{ssww}, but it is nice to understand the general idea. 
	
	The first ingredient is to recall that in 3+1d, the electromagnetic action
	\be S = \int \rmd^4x \ \left( - \frac{1}{4g^2} F_{\mu\nu}F^{\mu\nu} + \frac{\theta}{32\pi^2} \epsilon^{\mu\nu\rho\sigma}F_{\mu\nu}F_{\rho\sigma} \right) \ee
	allows for a $\theta$ term, and that one can define the complex coupling parameter
	\be \tau = \frac{\theta}{2\pi} + \frac{2\pi i}{e^2} \ee
	which lives in the upper half-plane and naturally parametrizes the theory. Electromagnetic duality of the by now familiar type exchanges $F$ for its dual $\tilde{F}$ (or $A$ for $\tilde{A}$) and, as you might like to check, maps
	\be \tau \mapsto S(\tau) = -\frac{1}{\tau} \fstp \ee
	Another transformation which leaves the theory invariant (at least on a spin manifold, a subtlety we will not explore) is $\theta \to \theta + 2\pi$ or
	\be \tau \mapsto T(\tau) = \tau + 1 \ee
	which combines with $S$ to generate a group of transformations $SL(2,\mathbb{Z})$. This is the S-duality group of the 3+1d theory. (Note that $-1 \in SL(2,\mathbb{Z})$ acts trivially on $\tau$ but not on the theory; one may verify it acts as charge conjugation, $A \to -A$.)
	
	One may now prove that, in the presence of a boundary, this group of dualities now acts on the restriction of the gauge group to the boundary in a non-trivial way. In fact, the generator $S$ acts by adding a BF to a new dynamical field. Meanwhile, $T$ simply adds a Chern-Simons term of size $-1$ to the action. These are precisely the operations we have been working with.
	
	Now one proposes that one can couple the 3+1d theory defined on a half-space to (say) a Wilson-Fisher boson on its 2+1d boundary whilst \textit{preserving S-duality} at the point $\tau=i$. (This is not supposed to be obvious!) This can be thought of a description of a topological insulator \cite{ssww}. If one has this, then one can deform away from the S-dual point to get one weakly-coupled and one strongly-coupled 3+1d theory. S-dualizing the strongly coupled theory leaves two arbitrarily weakly-coupled theories which are now dual. Freezing the weakly-coupled fields as background fields now gives bosonic particle-vortex duality. One can construct a similar story for other dualities.
	
	(There is also nice way of thinking about these dualities as different choices of basis for the electromagnetic charge lattice of the same theory; $S,T$ act on this basis. In this picture, for instance, the Aharanov-Bohm effect tells us that some choices will be fermions and some will be bosons. Theories with hidden T-reversal symmetry and simply those for which an asymmetric basis of the lattice is chosen.)
\end{asidebox}

\begin{exercise}[subtitle=Gauge Fields on Both Sides]
	\label{ex_hidden_so3}%
	Show that, assuming that there is a fixed point for these theories in the IR, we should find a duality
	\be U(1)_{2} + \text{WF boson} \bigdual U(1)_{-3/2} + \text{fermion} \label{ex_hidden_so3_eq} \ee
	i.e.
	\be |D_a \phi|^2 - |\phi|^4 + \frac{2}{4\pi} a \rmd a \bigdual i \bar{\psi} \slashed{D}_b \psi - \frac{1}{4\pi} b \rmd b \fstp \ee
	Argue that this fixed point has a global $O(2)$ symmetry. (We will return to this example in \refSection{so3_symmetry}.)
\end{exercise}	

\section{A New Duality: Fermionic Particle-Vortex Duality}\labelSection{fermionic-pv}

We seem to have a new favourite game to play! Let's make a list of the tricks we have thought up so far.
\begin{itemize}
	\item Gauging global symmetries by promoting background fields to dynamical ones.
	\item Adding background contact terms to both sides of a duality.
	\item Time-reversing a duality.
\end{itemize}
We have not nearly exhausted all the possible theories we can derive using this kind of approach!

One obvious possibility we have not yet come across is a \textit{purely fermionic} duality. We can engineer this by playing similar sorts of games to the above. Let's start with 
\be
	|D_{-B}\phi|^2 - |\phi|^4 + \frac{1}{4\pi} B \rmd B \qquad \longleftrightarrow \qquad i \bar{\psi}\slashed{D}_a \psi + \frac{1}{2\pi} B \rmd a 
\relabeleq{reverse-bosonization-lag-bg}
\ee
and look at the left-hand side. As we discussed above, if we simply promote $B \to b$ to a dynamical field, then on the right-hand side we simply set $a = 0$ and obtain the $U(1)_1 + \text{boson} \leftrightarrow \text{free fermion}$ duality again. This doesn't tell us anything new.

But if we use our ability to add background terms, then we can make some interesting. Specifically, suppose that we add $-\frac{2}{4\pi} B \rmd B - \frac{1}{2\pi} B \rmd C$ to both sides of \eqref{reverse-bosonization-lag-bg} before making $B \to -b$ dynamical, with a change of sign for convenience. Then we obtain instead
\be
|D_b\phi|^2 - |\phi|^4  -\frac{1}{4\pi} b \rmd b + \frac{1}{2\pi} b \rmd C \qquad \longleftrightarrow \qquad i \bar{\psi}\slashed{D}_a \psi + \frac{1}{2\pi} b \rmd a - \frac{2}{4\pi} b \rmd b + \frac{1}{2\pi} b \rmd C \nn
\ee
which is a bit of a mouthful. Nonetheless, the left-hand side is now $U(1)_{-1} + \text{boson}$ which we can relate to $U(1)_1 + \text{boson}$ using time-reversal.

The time-reversal of the seed duality \eqref{bosonization-with-bg-fields}, using \eqref{fermion_anomaly} and also charge-conjugating the fermion, is
\be
|D_a\phi|^2 - |\phi|^4 - \frac{1}{4\pi} a \rmd a -\frac{1}{2\pi} A \rmd a \qquad \longleftrightarrow \qquad i \bar{\tilde{\psi}} \slashed{D}_{-A} \tilde{\psi} + \frac{1}{4\pi} A \rmd A
\ee
and applying this to the above duality we find
\be
\boxed{
	i \bar{\psi}\slashed{D}_a \psi + \frac{1}{2\pi} b \rmd a - \frac{2}{4\pi} b \rmd b + \frac{1}{2\pi} b \rmd C
	\qquad \longleftrightarrow \qquad
	i \bar{\tilde{\psi}} \slashed{D}_C \tilde{\psi} + \frac{1}{4\pi} C \rmd C
} \label{fermion-fermion}
\ee
which is our first fermion-fermion duality.

If we are willing to work with incorrectly quantized Chern-Simons terms, we may as well subtract off $-\frac{1}{8\pi} C \rmd C$:
\be
	i \bar{\psi}\slashed{D}_a \psi + \frac{1}{2\pi} b \rmd a - \frac{2}{4\pi} b \rmd b + \frac{1}{2\pi} b \rmd C - \frac{1}{2} \frac{1}{4\pi} C \rmd C
	\qquad \longleftrightarrow \qquad
	i \bar{\tilde{\psi}} \slashed{D}_C \tilde{\psi} + \frac{1}{2} \frac{1}{4\pi} C \rmd C
	\label{fermion-fermion-t-invariant}
\ee
where the right-hand theory is actually time-reversal invariant.

\begin{exercise}[subtitle=Hidden Time Reversal Invariance]
	Check that the left-hand theory is also T-invariant, writing down the action of T on all the fields. Show that this transformation squares to the identity (on-shell).
\end{exercise}

As pointed out in \cite{ssww}, this makes contact with various proposals \cite{Son:2015xqa,Wang:2015qmt,Metlitski:2015eka,Wang:2016gqj} of a fermionic particle-vortex duality that is believed to exist for reasons originating in the quantum Hall effect (discussed briefly in \refChapterOnly{condmat}). In particular, if one were to naively integrate out $b$ using its equation of motion -- setting $b = (a+C)/2$ and therefore violating the charge quantization -- one would conclude
\be
`` i \bar{\psi}\slashed{D}_a \psi + \frac{1}{2} \frac{1}{4\pi} a \rmd a + \frac{1}{2} \frac{1}{2\pi} a \rmd C 
\qquad \longleftrightarrow \qquad
i \bar{\tilde{\psi}} \slashed{D}_C \tilde{\psi} + \frac{1}{2} \frac{1}{4\pi} C \rmd C ''
\ee
with various incorrectly quantized terms on the left-hand side. These are the sort of issues that affected the previous formulations of this duality, but it does convey the sense of a duality along the lines of ``$U(1)_0 + \text{fermion} \leftrightarrow \text{free fermion}$''.

Another interesting formulation of \eqref{fermion-fermion-t-invariant} comes from setting $C = 2 c$ to be an even multiple of a correctly quantized dynamical gauge field. If we do this, including also a negative BF coupling of $c$ to a new background field $A$, then \eqref{fermion-fermion-t-invariant} becomes
\be
i \bar{\psi}\slashed{D}_a \psi + \frac{1}{2\pi} b \rmd a - \frac{2}{4\pi} b \rmd b + \frac{2}{2\pi} b \rmd c - \frac{2}{4\pi} c \rmd c - \frac{1}{2\pi} A \rmd c
\qquad \longleftrightarrow \qquad
i \bar{\tilde{\psi}} \slashed{D}_{2c} \tilde{\psi} + \frac{2}{4\pi} c \rmd c - \frac{1}{2\pi} A \rmd c
\nn
\ee
or, shifting $c \to c + b$, we find 
\be
i \bar{\psi}\slashed{D}_a \psi + \frac{1}{2\pi} b \rmd (a - A) - \frac{2}{4\pi} c \rmd c - \frac{1}{2\pi} A \rmd c
\qquad \longleftrightarrow \qquad
i \bar{\tilde{\psi}} \slashed{D}_{2c} \tilde{\psi} + \frac{2}{4\pi} c \rmd c - \frac{1}{2\pi} A \rmd c
\nn
\ee
so that the $b$ equation of motion enforces $a = A$. Renaming the dynamical gauge fields, we conclude
\be
\boxed{i \bar{\psi}\slashed{D}_A \psi - \frac{2}{4\pi} c \rmd c - \frac{1}{2\pi} A \rmd c
	\qquad \longleftrightarrow \qquad
	i \bar{\tilde{\psi}} \slashed{D}_{2c} \tilde{\psi} + \frac{2}{4\pi} c \rmd c - \frac{1}{2\pi} A \rmd c }
\ee
which is another form of this fermionic particle-vortex duality \cite{Cordova:2017kue}. Note that the explicit Chern-Simons term on the right-hand side cancels the implicit $-\frac{1}{2} \frac{1}{4\pi} (2c) \rmd (2c)$ term built into the regularization of the charge 2 fermion. We can therefore write it as
\be
	\mbox{free fermion} + \mbox{decoupled } U(1)_{-2} \qquad \longleftrightarrow \qquad U(1)_0 + \mbox{charge 2 fermion}
\ee
which is now a much more precise statement, very close to the earlier proposal of \cite{Son:2015xqa}. (The latter discussion missed the decoupled topological sector on the left-hand side, which is important to correctly reproduce more subtle properties of the right-hand theory.)

\section{More Matter and Self-Dual Theories}

We can actually add another trick to the list at the start of \refSection{fermionic-pv}: multiplying different partition functions together! This simply means adding two non-interacting theories; of course, we can subsequently couple them together by making use of their background fields.

\subsection{Self-Dual QED with Two Fermions}\labelSection{self-dual-qed}

It has been proposed \cite{2016PhRvB..94u4415C,PhysRevB.92.220416} that 2+1 dimensional quantum electrodynamics is in fact \textit{self-dual}. This exercise will guide you through deriving this result, following \cite{hsinseiberg,karchtong}.

\begin{exercise}[subtitle=Self-Dual QED from Fermion Particle-Vortex Duality]
		Making use of \eqref{fermion-fermion} twice, demonstrate that
		\be i \bar{\psi}^1\slashed{D}_{A+X} \psi_1 + i \bar{\psi}^2\slashed{D}_{A-X} \psi_2 + \frac{1}{4\pi} (A+X) \rmd (A+X) \ee
		is dual to
		\begin{align}& i \bar{\chi}^1\slashed{D}_{a} \chi_1 + i \bar{\chi}^2\slashed{D}_{a'} \chi_2
		+ \frac{1}{2\pi} a \rmd b
		- \frac{2}{4\pi} b \rmd b
		+ \frac{1}{4\pi} a' \rmd a'
		+ \frac{2}{4\pi} b' \rmd b'
		- \frac{1}{2\pi} a' \rmd b' \nn\\& \qquad
		+ \frac{1}{2\pi} (b-b') \rmd A
		+ \frac{1}{2\pi} (b+b') \rmd X \fstp
		\end{align}
		
		Find appropriate background terms to add and subtract before gauging $A$ to deduce that
		\begin{align}
		\lag_{\text{s.d. QED}} = i \bar{\psi}^1\slashed{D}_{a+X} \psi_1 + i \bar{\psi}^2\slashed{D}_{a-X} \psi_2 + \frac{1}{4\pi} a \rmd a + \frac{1}{2\pi} a \rmd Y - \frac{1}{4\pi} Y \rmd Y \qquad
		& \\
		\longleftrightarrow\quad
		\tilde{\lag}_{\text{s.d. QED}} = i \bar{\chi}^1\slashed{D}_{b-Y} \chi_1 + i \bar{\chi}^2\slashed{D}_{b+Y} \chi_2 + \frac{1}{4\pi} b \rmd b + \frac{1}{2\pi} b \rmd X - \frac{1}{4\pi} X \rmd X
		& \nn
		\end{align}
		and deduce that $U(1)_0 + 2 \text{ fermions}$ is self-dual, with the currents exchanged as $X \leftrightarrow Y$. (We haven't proved that this theory has a non-trivial CFT as a fixed point, but this argument suggests if there is, one should be able to reach it in two different ways.)
\end{exercise}

This result is intriguing, because the first theory has a manifest $SU(2)$ global symmetry rotating $\psi_i$, and the latter theory has a manifest $SU(2)$ global symmetry rotating $\chi_{i'}$ -- but these are not the same $SU(2)$ group!\footnote{This discussion follows those in \cite{hsinseiberg,Benini:2017dus}.}

Let's look at the left-hand theory. The manifest global symmetry on the left-hand side is $SU(2) \times U(1)$, where $\psi$ is in the fundamental of the $SU(2)$, whilst the other $U(1)$ is a monopole symmetry coupled to $Y$. If we couple the $SU(2)$ symmetry to a background $SU(2)$ gauge field $x = x^I \sigma^I$, then we see that $x^3$ couples to $\psi$ in the same way as $X$. In fact, $X$ couples to the $U(1) \subset SU(2)$. (The $\sigma^3$ generates the Cartan subalgebra.)

\begin{exercise}[subtitle=The Exact Global Symmetry]
	Show that including charge conjugation (specifically one under which $\Psi_i \to \epsilon_{ij}\bar{\Psi}^j$) enhances the symmetry slightly to $SU(2) \times O(2)$. On the other hand, argue that all gauge-invariant operators either have even charge under both $X$ and $Y$, or odd charge under both. The manifest global symmetry which acts faithfully is therefore
	\be (SU(2) \times O(2))/\mathbb{Z}_2 \fstp \ee 
\end{exercise}

Thus we can write the apparent global symmetry of the left-hand side as $(SU(2)^X \times O(2)^Y)/\mathbb{Z}_2$. Similarly, of course, the right-hand side has an apparent symmetry $(SU(2)^Y \times O(2)^X)/\mathbb{Z}_2$. This is a classic example of \textit{symmetry enhancement} in a self-dual theory\footnote{Recall, for instance, that a 2d compact boson at the self-dual radius has a hidden $SU(2)$ symmetry.}: if one assumes that the above theory flows to a non-trivial CFT in the IR, the \textit{full} quantum symmetry of that CFT should actually include the larger group
\be (SU(2)^X \times SU(2)^Y)/\mathbb{Z}_2 \cong SO(4) \ee
which is not manifest in either of the classical Lagrangians. Yet there it is! In each framework, one of the $SU(2)$ symmetries is an \textit{emergent} IR symmetry.

Actually, the duality tells us about an extra $\mathbb{Z}_2^{\text{dual}}$ factor which is a symmetry of the theory, under which $X$ and $Y$ charge is exchanged. One can check that
\be SO(4) \rtimes \mathbb{Z}_2^{\text{dual}} \cong O(4) \ee
is therefore the full symmetry group.

We should emphasize that emergent symmetries like these $SU(2)$ factors are a property of the low-energy modes only; the gapped modes in each theory need not have this symmetry. The idea is that as we flow to the IR, we are left with only a subset of the full UV theory possessing a larger symmetry group than the UV theory.

\begin{exercise}[subtitle=Larger Charges]
	Suppose we modify the above theory by giving one of the fermions charge $q$. Surprisingly, this model retains self-duality \cite{2016PhRvB..94u4415C,Benini:2017dus}. Prove that the usual fermion-fermion dualities imply that
	\be i \bar{\psi}^1 \slashed{D}_a \psi_1 + i \bar{\psi}^2 \slashed{D}_{qa} \psi_2 + \frac{q^2+1}{2} \frac{1}{4\pi} a \rmd a \ee
	is indeed self-dual, and give the correct coupling to background fields.
\end{exercise}
 
\subsection{Self-Dual QED with Two Scalars}

Similarly, we can work with
\be
\lag_{\text{scalar}} = |D_{a}\phi_1|^2 - |\phi_1|^4 + |D_{a-C}\phi_2|^2 - |\phi_2|^4 + \frac{1}{2\pi} a \rmd A \fstp
\ee
Notice that this has very little symmetry in the UV compared to the above fermionic theories; only something like $U(1)\times U(1) \times \mathbb{Z}_2$ is manifest acting upon the fields (where the $\mathbb{Z}_2$ factor we include exchanges $\phi_1 \leftrightarrow \phi_2$). However, it is dual to
\begin{align}
\lag_{\text{scalar}} \smallerdual & i\bar{\psi}^1 \slashed{D}_{b} \psi_1
+ \frac{1}{2\pi} b \rmd a - \frac{1}{4\pi} a \rmd a \\
& \quad + i\bar{\psi}^2 \slashed{D}_{b'} \psi_2
+ \frac{1}{4\pi} b' \rmd b' - \frac{1}{2\pi} b' \rmd (a-C) + \frac{1}{4\pi} (a-C) \rmd (a-C)
+ \frac{1}{2\pi} a \rmd A & \nn
\end{align}
by \eqref{wf-dual-gauged-fermion}. Integrating out $a$ imposes $b'=b + (A-C)$, and so this simplifies to
\be
\lag_{\text{scalar}} \smallerdual i\bar{\psi}^1 \slashed{D}_{b} \psi_1 + i\bar{\psi}^2 \slashed{D}_{b+(A-C)} \psi_2
+ \frac{1}{4\pi} (b+A) \rmd (b+A) \fstp
\ee

If we write $A=Y-X$ and $C=X+Y$ and then shift variables using $b = a + X$, we reduce to
\be
\lag_{\text{scalar}} \smallerdual  i\bar{\psi}^1 \slashed{D}_{a+X} \psi_1 + i\bar{\psi}^2 \slashed{D}_{a-X} \psi_2
+ \frac{1}{4\pi} (a+Y) \rmd (a+Y) \fstp
\ee
which is (remarkably!) exactly the same theory as we discussed in \refSectionOnly{self-dual-qed}, differing only by a background term for $Y$.

Thus taking this background term into account, we conclude that
\be
\lag_{\text{EP}} = |D_{a}\phi_1|^2 - |\phi_1|^4 + |D_{a-(X+Y)}\phi_2|^2 - |\phi_2|^4 + \frac{1}{2\pi} a \rmd (Y-X) - \frac{2}{4\pi} Y \rmd Y
\ee
flows to an identical fixed point to that discussed in the previous section. But that theory has the property that it is invariant under $X \leftrightarrow Y$, and therefore so does this one, revealing a duality between $\lag_{\text{EP}}$ and
\be
\tilde{\lag}_{\text{EP}} = |D_{a}\tilde{\phi}_1|^2 - |\tilde{\phi}_1|^4 + |D_{a-(X+Y)}\tilde{\phi}_2|^2 - |\tilde{\phi}_2|^4 - \frac{1}{2\pi} a \rmd (Y-X) - \frac{2}{4\pi} X \rmd X \fstp
\ee
The subscript "EP" is used because we sometimes use the name \textit{easy-plane} theory to describe the theory of two scalars subject to the asymmetric $|\phi_1|^4+|\phi_2|^4$ potential.\footnote{The model with an explicit $SU(2)$ symmetry is also supposedly self-dual, but with a different fermionic dual: namely one at the so-called Gross-Neveu fixed point with $|\psi|^4$ interactions. This is claimed to possess an even larger $SO(5)$ symmetry at the fixed point \cite{Nahum:2015vka}.}

We have rather remarkably stumbled upon a duality between two self-dual theories. Labelling them with their manifest UV symmetries, we get the following set of dualities:
\be
\underbrace{\tilde{\lag}_{\text{s.d. QED}}}_\text{$G^{UV} = \frac{SO(2) \times SU(2)}{\mathbb{Z}_2}$}
\smallerdual \underbrace{\lag_{\text{s.d. QED}}}_\text{$G^{UV} = \frac{SU(2) \times SO(2)}{\mathbb{Z}_2}$}
\smallerdual \underbrace{\lag_{\text{EP}}}_\text{$G^{UV} = U(1)^2 \times \mathbb{Z}_2$}
\smallerdual \underbrace{\tilde{\lag}_{\text{EP}}}_\text{$G^{UV} = U(1)^2 \times \mathbb{Z}_2$}
\nn\ee
In the easy-plane theories, the $\mathbb{Z}_2$ factor either exchanges $\phi_1 \leftrightarrow \phi_2$ or $\tilde{\phi}_1 \leftrightarrow \tilde{\phi}_2$. Meanwhile, the $\mathbb{Z}_2$ quotient reflects the fact that the objects again carry even charge under $X+Y$. Again, we conclude that despite the small symmetry group visible in the UV of the easy-plane theories, there is an emergent $O(4)$ symmetry in their IR. Again, we must stress there is no proof here that there is actually a CFT with these properties which we can reach by tuning the above UV Lagrangians. Yet the above reasoning does suggest that there ought to be some interesting structure in these theories.

\begin{asidebox}[Walking on the Lattice and the Bootstrap]%
	\label{walking-lattice}%
	The question of whether a genuine, unitary CFT exists with these properties -- and whether this can be obtained by tuning the above Lagrangians appropriately -- is a fascinating and subtle one. Such questions have often proven somewhat controversial.
	
	There are at least two tools one might reasonably turn to in an attempt to address these issues. Perhaps the most obvious thing we could try is a direct numerical simulation. In particular, a lot of effort goes into trying to find these CFTs in lattice simulations, and even estimating anomalous dimensions of various operators. Hints of these sorts of enhanced symmetries have indeed been seen in lattice experiments \cite{PhysRevX.7.031052}, although it is notoriously difficult to obtain conclusive results about these strongly interacting field theories, and simulations typically extend only to lattices with on the order of 20 sites along each axis. Essentially, one tries to identify the location of a phase transition, checking to see if it is continuous or first-order; then, assuming one finds a continuous phase-transition, one uses the scaling behaviour of correlation functions near that point to extract anomalous dimensions. This has met with some success, giving indications of this type of enhanced symmetry: miraculous agreement between the apparent dimensions of naively unrelated operators.
	
	The other approach is something we alluded to in the introduction: the \textit{conformal bootstrap}. This proceeds by using the fundamental properties any unitary CFT must obey, and some input data such as the symmetry group of a theory, and the existence of a relevant coupling in some representation of that symmetry group. One can then attempt to analytically or (more often) numerically investigate the equations imposed by conformal symmetry to see if the input data is consistent. This is in principle a mathematically rigorous approach to the problem, though one must be careful about the assumptions about the spectrum which are made. As reviewed in some detail in \cite{Poland} (see the references therein for more detail), it seems like it is mathematically impossible for a fixed point consistent with the properties posited above to exist.
	
	So what is the explanation for the lattice's apparent support of the naive hypothesis that there is a fixed point of enhanced symmetry? There are two key ideas which form an interesting alternative picture that embraces all the known facts, again discussed briefly in \cite{Poland}, and presented in more detail in e.g. \cite{Gorbenko:2018ncu}. The first is the concept of \textit{walking}. We have discussed the idea of RG \textit{flows} a lot, but we have said almost nothing about the \textit{speed} of those flows. It is entirely possible that certain RG flows actually `walk' quite slowly. If this happens, one can \textit{naturally} find a \textit{weakly} first-order transition which is very hard to detect on the lattice, giving rise to the appearance of a second-order fixed-point.
	
	The second idea offers some explanation for why this walking behaviour might arise. The idea is that, much as in particle physics one finds resonances corresponding to unstable states at complex values of the energy, perhaps there \textit{is} a conformal field theory with enhanced symmetry -- it is just non-unitary. Just as resonances can have a very long lifetime if they sit close to the real axis, so a complex CFT with operators possessing dimensions with small imaginary parts can cause the RG flow to look (over a wide range of e.g. lattice spacings) like it is approaching a CFT, even though ultimately it `misses', because the real flow is constrained to the space of unitary field theories.
	
	In this way, even if our argument for symmetry enhancement does not lead to the discovery of a genuine duality of unitary CFTs, it may be telling us (a) there is something interesting to look for in the larger space of non-unitary CFTs; and (b) that we therefore expect some unusual, yet still fairly universal behaviour at fairly low energies!
\end{asidebox}

\subsection{Quiver Theories}\labelSection{abelian_quiver}

Our last generalization in this chapter is to allow for many gauge groups, following \cite{Karch:2016aux}. The idea is simple enough. Let's start with $N$ free fermions:
\be \sum_{i=1}^N |D_{a_i}|^2 - |\phi_i|^4 + \frac{1}{2\pi} A \rmd a+i + \frac{1}{4\pi}a_i \rmd a_i \bigdual i \sum_{i=1}^N \bar{\psi}^i \slashed{D}_{A_i} \psi_i \ee
We have included $N$ background gauge fields to play with. We are going to gauge $r$ linear combinations of them, say $\tilde{a}_a$ (for $a=1,\ldots,r$) in such a way that we give the $i$th fermion charge $R^a_i$. This would mean taking $A_i = R^a_i \tilde{a}_a$. We will assume that $R^a_i$ has the maximal rank of $r$, so that the only gauge-invariant products of fermions are products of $\bar{\psi}^i \psi_i$.

However, it is helpful to keep track of the global currents too. There are $N-r$ ungauged linear combinations of the $A_i$ to handle, labelled $\alpha = 1,\ldots,N-r$. Then we can define the matrix $S_i^\alpha$ with rank $N-r$ such that
\be \sum_{i=1}^N R_i^a S_i^\alpha = 0 \qquad \text{for }a=1,\ldots,r \text{ and } p=1,\ldots,N-r \fstp \ee 
Then the gauge fields can be decomposed as
\be A_i = R^a_i \tilde{a}_a + S^\alpha_i C_\alpha \ee
where $C_\alpha$ are the remaining ungauged combinations of the $A_i$. We should also introduce $r$ new background gauge fields $\tilde{C}_a$ coupling to the topological currents.

Therefore, the fermionic theory is
\be \textbf{Fermionic Theory: } U(1)^r + N \text{ fermions of charge } R^a_i \ee
with no explicit Chern-Simons terms. The Lagrangian is
\be \lag_{\text{fermion}} = i \bar{\psi}^i \slashed{D}_{R^a_i \tilde{a}_a + S^\alpha_i C_\alpha} \psi_i - \frac{\kappa^{ab}}{2\pi} \tilde{C}_a \rmd \tilde{a}_b  \ee
for
\be \kappa^{ab} = \sum_i R^a_i R^b_i \fstp \ee
Notice that the naive Chern-Simons levels of $U(1)^r$ are also described by this so-called \textit{K-matrix}, in that
\be \sum_{i=1}^N -\frac{1}{2}\frac{1}{4\pi} (R^a_i \tilde{a}_a + S^\alpha_i C_\alpha)^2 = - \frac{1}{2}\frac{\kappa^{ab}}{4\pi} \tilde{a}_a \rmd \tilde{a}_b - \frac{1}{2}\frac{\kappa^{\prime \alpha\beta}}{4\pi} C_\alpha \rmd C_\beta \ee
where similarly
\be \kappa^{\prime \alpha\beta} = \sum_i S^\alpha_i S^\beta_i \fstp \ee

Now the dual description, which is simplified by writing $a_i = R^a_i a_a + S_i^\alpha a_\alpha$, is
\be
\lag_{\text{boson}} = |D_{R^a_a a_a+S_i^\alpha a_\alpha} \phi_i|^2
+ \frac{\kappa^{ab}}{4\pi} a_a \rmd a_b
+ \frac{\kappa^{\prime \alpha\beta}}{4\pi} a_\alpha \rmd a_\alpha
+ \frac{\kappa^{ab}}{2\pi} (a_a-\tilde{C}_a) \rmd \tilde{a}_b
+ \frac{\kappa^{\prime \alpha\beta}}{2\pi} a_\alpha \rmd C_\beta
\label{bosonic_quiver_dual_unintegrated} \ee
which is again quite a mouthful.

However, if we set aside questions of correct flux quantization for now, then the $a_a$ and $\tilde{a}_a$ equations of motion can be solved and substituted back into the action. This gives
\be
\lag_{\text{boson}} \approx |D_{R^a_a C_a+S_i^\alpha a_\alpha} \phi_i|^2
+ \frac{\kappa^{\prime \alpha\beta}}{4\pi} a_\alpha \rmd a_\alpha
+ \frac{\kappa^{ab}}{2\pi} \tilde{C}_a \rmd \tilde{C}_b
+ \frac{\kappa^{\prime \alpha\beta}}{2\pi} a_\alpha \rmd C_\beta \fstp 
\ee 
Therefore, up to questions of flux quantization, we can conclude that
\be \textbf{Bosonic Theory: } \approx U(1)^{N-r} + N \text{ WF scalars of charge } S^\alpha_i \text{ and Chern-Simons levels } \kappa^{\alpha\beta} \ee
is the dual to the above fermionic theory.

\subsubsection*{Example: QED}

A simple example of this is to consider $N$ fermions all coupled to $U(1)$ gauge theory.
\be \textbf{Fermionic Theory: } U(1)_{-N/2} + N \text{ fermions} \ee
This theory has $r=1$ and $R^1_i = 1$.

We can take the dual to be of the form described in the previous section, with
\be
S^1_i = \begin{pmatrix}+1 \\ -1 \\ 0 \\ \vdots \\ 0 \\ 0 \end{pmatrix}, \ 
S^2_i = \begin{pmatrix}0 \\ +1 \\ -1 \\ \vdots \\ 0 \\ 0 \end{pmatrix}, \ \cdots, \ 
S^{N-1}_i = \begin{pmatrix}0 \\ 0 \\ 0 \\ \vdots \\ +1 \\ -1 \end{pmatrix}
\ee
and then
\be
\kappa^{\alpha\beta} = \begin{pmatrix}
	2 	& -1 	& 			& 			& \\
	-1 	& 2 	& -1 		& 			& \\
	& -1 	& \ddots 	& \ddots 	& \\
	& 		& \ddots 	& 2 		& -1 \\
	& 		&  			& -1 		& 2
\end{pmatrix}
\ee
is the K-matrix. The dual is therefore
\be \textbf{Bosonic Theory: } U(1)^{N-1} + N \text{ WF bosons of charge } S^\alpha_i \text{ and Chern-Simons matrix } \kappa^{\alpha\beta} \label{bosonic-qed-quiver-theory}\ee

This is a special type of theory, known as a \textit{quiver}. The matter content and gauge group can be communicated by a so-called quiver diagram using a simple prescription. One draws a circle for each gauge group factor, often labelling it with the rank of the gauge group; for us, we will label with $n$ nodes corresponding to the gauge group $U(n)$. One can also include square boxes to indicate flavour groups. Then, one draws a line between certain pairs of nodes, whether square or circular. Each line represents (for now) a Wilson-Fisher scalar transforming in the bifundamental representation of the two attached nodes. The above theory comes out like \refFigureOnly{bosonic-qed-quiver}.\footnote{There is an alternative, circular quiver which one can use to describe the same theory more symmetrically. To obtain this representation, rather than integrating out the dynamical field under which no matter was charged, we return to \eqref{bosonic_quiver_dual_unintegrated}. There are $N$ gauge nodes, but under the overall $U(1)$ factor all matter is neutral. This is also the $U(1)$ which is killed by the gauging process.}
\begin{center}
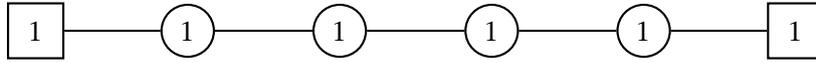

	\centering
	\begin{tikzpicture}[every node/.style={circle,draw},thick]
	\node[square](k0) at (-2,0){$1$};
	\node(k1) at (0,0){$1$};
	\node(k2) at (2,0){$1$};
	\node(k3) at (4,0){$1$};
	\node(k4) at (6,0){$1$};
	\node[square](k5) at (8,0){$1$};
	\draw(k0.east)--(k1.west);
	\draw(k1.east)--(k2.west);
	\draw(k2.east)--(k3.west);
	\draw(k3.east)--(k4.west);
	\draw(k4.east)--(k5.west);
	\end{tikzpicture} 
	\captionof{figure}{The quiver of \eqref{bosonic-qed-quiver-theory} for $N=5$. There is one line per boson, with 4 dynamical gauge fields and 2 flavour nodes at each end. These flavour nodes are actually redundant, since all operators have the same charge under the $U(1)$ factors associated with the two endpoints }\labelFigure{bosonic-qed-quiver}
\end{center}

Such quiver theories are widely studied, especially in the context of supersymmetric field theory dualities. In fact, the duality presented here is non-supersymmetric version of the original supersymmetric \textit{mirror symmetry} \cite{Intriligator:1996ex}. We will talk more about mirror symmetry in \refChapter{susybreaking}.

\begin{exercise}[subtitle=Operator Matching for the Linear Quiver]
	Derive the operator correspondence that underlies this duality.
\end{exercise}

One other reason to be interested in quivers is that they can "deconstruct" higher-dimensional theories. Imagine starting in 3+1 dimensions, and discretizing one dimension into $N$ copies of 2+1 dimensional theories. Having a separate gauge theory living along each of the $N$ nodes now looks rather like one of the quivers above. This suggests one might be able to derive 3+1d dualities from 2+1d ones, as suggested for S-duality in \cite{Aitken:2018joz}. This is a tantalizing direction for future work, though it is hard to see how to control the physics properly.

\part{Non-Abelian Dualities}
\labelPart{nonabelian}
\chapter{Level-Rank Duality}
\labelChapter{level_rank}

\begin{introduction}
  In order to move beyond the world of Abelian dualities, we must introduce a famous set of exact dualities of pure Chern-Simons theory: the so-called level-rank dualities.
\end{introduction}

\section{Non-Abelian Chern-Simons and Topological Field Theories}\labelSection{tqfts}

\lettrine[nindent=-2pt]{T}{he interesting part} of the dualities we have seen so far has been the dynamics of the matter fields, either bosons or fermions. However, in moving beyond the Abelian case to study non-Abelian gauge theory, it turns out there is something to say even about the pure Chern-Simons gauge theories.

In this section, we will improve our understanding of Chern-Simons field theory by introducing the Chern-Simons term for $SU(N)_k$,
\be
\lag_{SU(N)_k} = \frac{k}{4\pi} \tr \left( a \wedge \rmd a - \frac{2i}{3} a \wedge a \wedge a \right)
\ee
and studying the contents of this theory. We will leave to \refAppendix{chern-simons} questions like how one shows this is gauge invariant. (The time reversal of $SU(N)_k$ is $SU(N)_{-k}$, of course; parity acts in the same way. We can take $k\ge 0$ if we want.)

\subsection{Some Preliminaries}

Firstly, we should remind ourself about how non-Abelian gauge theory $SU(N)_0$ behaves in the absence of a Chern-Simons term. The main thing to remember is that this is actually a strongly interacting theory, entirely governed by the Maxwell term we will often be too lazy to even write,
\be \lag_{\mathrm{Maxwell}} = \frac{1}{4g^2} \tr f_{\mu\nu}f^{\mu\nu} \fstp \ee 
The general expectation is that this kind of theory has a mass gap of order $g^2$, with all finite-energy states being \textit{glueballs} which are heavy colour singlets. Such a theory always \textit{confines} the charge of the gauge group: coupling matter to this theory forces the matter to form colour-singlet states like mesons or baryons. (We will return to this briefly in \refSectionOnly{generalize_quiver}.) The intuition is that the field lines linking electric charges do not spread out, but are forced together like flux lines in a superconductor, forming a string. The energy cost then scales linearly with the length of this string, preventing charged particles moving far apart.

If we are at energies much below the scale set by $g$, then we would be totally oblivious to the existence of the gauge field -- the IR limit is trivial. This is to be contrasted with TQFTs, where the low-energy physics is sensitive to the existence of the gauge group, as we will discuss shortly.

Now the non-Abelian \textit{pure} Chern-Simons theory $SU(N)_k$ is a topological field theory, and has no propagating modes of any mass at all. As we will discuss shortly, the only physical observables of this model are a finite number of topological quantities. There is no mass scale at all: it is a CFT. 

There is an interesting question, however, about the Maxwell-Chern-Simons theory $SU(N)_k$ with both a Maxwell term and a Chern-Simons term. This is a gapped theory, but there are two competing effects involved. There is the mass gap of confinement, arising only from the attractive gluon-gluon interactions and the energetics of glueball states; however, the quadratic Chern-Simons term also provides a tree-level mass term of order $m_{\mathrm{top}} \sim |k|g^2$. This theory therefore also has a \textit{topological mass}. These effects are in competition. In fact, the theory ultimately is believed not to confine -- the theory is in the topological phase at low energies.

Finally, a useful bit of notation. There are two independent levels for a $U(N)$ group, since it has both a $U(1)$ part and an $SU(N)$ part; we write
\be U(N)_{k,k'} = \frac{SU(N)_k \times U(1)_{k'N}}{\mathbb{Z}_N} \ee 
and $U(N)_k \equiv U(N)_{k,k}$. Note that the gauge-invariant theories are $U(N)_{k,k + nN}$ for $k,n\in \mathbb{Z}$. This follows from writing
\be \lag = \frac{k}{4\pi} \tr \left( a \rmd a -\frac{2i}{3} a^3 \right) + \frac{n}{4\pi} \tr a \rmd \tr a \fstp \ee
(The parity of $k+n$ determines whether this is a spin theory or not.)

\subsection{Topological Degeneracy and A Trivial Theory}

Let's pick up where we left off in \refSectionOnly{brief_tqfts}, looking now at non-Abelian theories.

Pure Chern-Simons theory, with the Lagrangian taken to be exactly
\be \lag = \frac{k}{4\pi} \tr \left( a \rmd a -\frac{2i}{3} a^3 \right) \cmma \ee
is still a TQFT, and we still have an equation of motion $f = 0$ which eliminates local degrees of freedom.

The local operators of the Abelian theory were the Wilson lines $W_m = \exp(im\oint a)$ around each non-contractible loop. We observed that $W_m \sim W_{m+k}$ were equivalent, so that there were $k^g$ states on a genus $g$ surface.

What is the analogous story in the non-Abelian case? The answer is most clearly understood by thinking of the Abelian $W_m$ as corresponding to the Wilson line in the charge $m$ representation of $U(1)$, and then we find that certain representations should be dropped -- those of charge of $k$ or more. In the non-Abelian case, we can compute the Wilson lines
\be \tr_R \exp(i \oint a^A t_R^A) \ee
for $t^A$ the generators of a general representation $R$. One again finds that there are interrelations between these Wilson lines which mean that we do not get contributions from arbitrarily large representations.

One finds that, on the torus, $SU(N)_k$ has a degeneracy of $(N+k-1)/k!/(N-1)!$.

\begin{asidebox}[Some Underlying Representation Theory]%
	The operators in $U(N)_k$ Chern-Simons theory carry an $SU(N)$ representation, and as such can be labelled by a Young diagram. A single box $\yng(1)$, for instance, is something in the fundamental representation. A row of $\ell$ boxes represents the $\ell$th symmetric representation, whilst a column represents a purely antisymmetric representation. A general representation of $SU(N)$ consists of up to $N-1$ rows of $\ell_1 \ge \ell_2 \ge \cdots \ell_{N-1} \ge \ell_N = 0$ boxes:
	\be \yng(7,3,3,2) \ee
	Note that a column of $N$ boxes would represent an $N$-fold antisymmetrization, which in $SU(N)$ leaves only a trivial singlet.
	In $SU(N)_k$, the width of the diagram is also restricted to $\ell \le k$, giving the \textit{integrable representations}. So-called \textit{fusion rules} generalize the familiar process of computing tensor products of representations when we multiply objects together: only a subset of the possible product representations occur. There are $n = (k+N-1)!/k!/(N-1)!$ such representations, with height at most $N-1$ and width at most $k$.
	
	Transporting each species of anyon around a torus gives rise to the above degeneracy.
\end{asidebox}

However, on more complicated surfaces, there is usually not such a simple formula. (See e.g. \cite{Blau:1993tv} for the \textit{Verlinde formula} giving the general result.) But for $k=1$, we find the $SU(N)_1$ theory has exactly $N^g$ states, just like $U(1)_N$. (This is not a coincidence, as we shall discuss below when we look at level-rank duality.)

In fact, something interesting happens when we look at the theory $U(N)_1$. This is almost ``trivial'' in the same sense as $U(1)_1$. (One can even show that $N$ copies of $U(1)_1$ are equivalent to $U(N)_1$.) This may seem a little surprising as $U(N)_1 = SU(N)_1 \times U(1)_N / \mathbb{Z}_N$ and both $SU(N)_1$ and $U(1)_N$ have $N$ states on a torus. But this simplistic analysis ignores the $\mathbb{Z}_N$ quotient, and ultimately a careful analysis shows that we only have a single operator in the theory.

\begin{asidebox}[Boundary Theories]%
	The above discussion of topological field theories is centered around the 3d perspective, but it is clear that the bulk of the 3d spacetime is irrelevant. Instead, all of the interesting structure is associated to non-contractible loops -- and in fact, also boundaries.
	
	If one has a theory on a manifold with boundary, then gauge invariance of the bulk theory requires a particular, non-trivial boundary theory transforming under the gauge group restricted to the boundary. This is known as the Wess-Zumino(-Novikov)-Witten (WZW) model \cite{Witten-nonab}. For $U(1)$, it consists simply of a chiral boson. For $SU(N)_k$, it is a non-trivial 1+1 dimensional conformal field theory. The conformal dimensions of operators in this theory give determine the physics of Chern-Simons theory. including quantities such as the spins of Wilson lines. (This is natural if we think of a Wilson line as cutting a hole through a manifold.)
\end{asidebox}

\subsection{Aside: Knots}

Recall that TQFTs necessarily compute topological invariants. We won't discuss this in detail, but the classic paper \cite{Witten:1988hf} of Witten shows that non-Abelian Chern-Simons theory on general manifolds computes particular polynomials known as knot invariants.

This is a beautiful story: the partition function is a topological invariant of the manifold, and the expectation value of a product of Wilson lines is given in terms of knot invariants of the configuration of Wilson lines.

More specifically, in three dimensions, lines can have a \textit{linking number} measuring how they weave through one another. Each Wilson line also has an ambiguous self-linking which is fixed by a so-called \textit{framing} of the line. (Mathematically, this is a choice of a normal vector along the line. Pictorially, this is like making the line into a ribbon.) The expectation of a product of Wilson lines can be expressed entirely in terms of these quantities (together with representation theoretic factors), with the ambiguity of self-linking being naturally related to the spin of the Wilson line.

It is worth reading \cite{Witten:1988hf} to understand this in detail.

\section{Level-Rank Duality}

The above TQFTs are certainly interesting, and there is a lot one can say about them. In this section, we are going to explore one of their most fascinating aspects, which is that they obey \textit{level-rank duality}:
\be
U(N)_k \bigdual SU(k)_{-N} \label{level_rank_basic}
\ee
This result is quite subtle, but it is possible to get some intuition about why it ought to hold.

Ultimately, the origins of this identity lie in 1+1 dimensional conformal field theory. the fact that $Nk$ free fermions form a representation of the $SU(Nk)_1$ algebra, whilst also faithfully representing $SU(N)_k$ and $SU(k)_N$. (We could include an overall $U(1)_{Nk}$ on both sides too.) This describes an embedding
\be SU(N)_k \times SU(k)_N \subset SU(Nk)_1 \fstp \ee
One can then define something called a GKO coset theory
\be SU(k)_N \bigdual \frac{SU(Nk)_1}{SU(N)_k} \ee
in 1+1 dimensions, which ultimately leads to an identification of the Chern-Simons theories
\be SU(k)_N \bigdual \frac{SU(Nk)_{1} \times SU(N)_{-k}}{\mathbb{Z}_N} \fstp \ee
By remembering $U(Nk)_1$ is trivial, and writing $SU(Nk)_1 = U(Nk)_1 \times U(1)$ where the second $U(1)$ kills the first but then instead dropping the trivial $U(Nk)_1$ factor, one finds
\be SU(k)_N \bigdual \frac{U(1)_{-Nk} \times SU(N)_{-k}}{\mathbb{Z}_N} = U(N)_{-k} \fstp \ee

However, whilst this does point the way to a rigorous proof of the result, it is beyond the scope of this course to truly understand the above manipulations.

\subsubsection*{An Intuitive Sketch in 2+1d}

This section comes with a massive health warning for mathematicians (which is to say it is tantamount to nonsense) but may make the proposal seem less outrageous. Consider the theory $U(Nk)_1$, taking the gauge field to be $a \in u(Nk)$. As we discussed above, this is essentially trivial. However, it turns out it is in some sense a parent theory from which we will can extract the physics of both $U(N)_k$ and $SU(k)_{-N}$. Intuitively, we think of elements of $u(Nk)$ as made up of $k \times k = k^2$ elements of $u(N)$:
\be
a \ =  \  
\underbrace{
\begin{pmatrix}
	\underbrace{
	\begin{pmatrix}
		\cdot & \cdot \\ \cdot & \cdot
	\end{pmatrix}
	}_N
    &
    \begin{pmatrix}
    	\cdot & \cdot \\ \cdot & \cdot
    \end{pmatrix}
	&
	\begin{pmatrix}
		\cdot & \cdot \\ \cdot & \cdot
	\end{pmatrix}
	\\
	\begin{pmatrix}
		\cdot & \cdot \\ \cdot & \cdot
	\end{pmatrix}
	&
	\begin{pmatrix}
		\cdot & \cdot \\ \cdot & \cdot
	\end{pmatrix}
	&
	\begin{pmatrix}
		\cdot & \cdot \\ \cdot & \cdot
	\end{pmatrix}
	\\
	\begin{pmatrix}
		\cdot & \cdot \\ \cdot & \cdot
	\end{pmatrix}
	&
	\begin{pmatrix}
		\cdot & \cdot \\ \cdot & \cdot
	\end{pmatrix}
	&
	\begin{pmatrix}
		\cdot & \cdot \\ \cdot & \cdot
	\end{pmatrix}
\end{pmatrix}
}_k
\  \in \  u(Nk)
\ee
Suppose that we could enforce that the $SU(k)$ part of the structure was proportional to the identity, so that $a$ was actually of the form
\be
a =  
\underbrace{
	\begin{pmatrix}
		\hat{a} & 0 & 0 \\
		0 & \hat{a} & 0 \\
		0 & 0 & \hat{a}
	\end{pmatrix}
}_k
\ee
for some $\hat{a} \in u(N)$. For these configurations, we find that the Lagrangian
\be
\lag_{U(Nk)_1} = \frac{1}{4\pi} \int_M \tr_{u(Nk)} \left( a \wedge \rmd a - \frac{2i}{3} a \wedge a \wedge a \right) = \frac{k}{4\pi} \int_M \tr_{u(N)} \left( \hat{a} \wedge \rmd \hat{a} - \frac{2i}{3} \hat{a} \wedge \hat{a} \wedge \hat{a} \right) 
\ee
reduces to that for $U(N)_k$ Chern-Simons theory. Thus our goal is to implement this constraint. The problem is that it is clearly not a gauge-invariant constraint! We are privileging a particular $SU(k) \subset U(Nk)$ when imposing our constraint. 

This is a problem we could have seen coming. For pure Chern-Simons theory, the equation of motion sets the field strength to zero and hence the gauge field is always pure gauge. There are no local gauge-invariant objects in the bulk of this theory! The constraint we want to impose is really a constraint on non-local aspects of the theory (that the $SU(k)$ structure of the bundle is trivial) and possibly on boundary dynamics (if there are any boundaries).

Let us solve the equations of motion by locally representing the $U(Nk)$ gauge field as a derivative of a group-valued field, $a = i g^{-1} \rmd g$. We can write e.g. $\lag_{U(Nk)_1}[g]$ for the Chern-Simons action of $a$. What we want is to insist that there is a globally-defined gauge transformation which maps $g$ to something with no $SU(k)$ structure at the boundary or around loops (i.e. something proportional to the $SU(k)$ identity).

With this in mind, and glossing over lots of subtleties, let's take $h \in SU(k)$ to act in the obvious way upon $U(Nk)$ elements, and note that (essentially because of the Polyakov-Wiegmann property, cf. \eqref{cs_direct_variation} in \refAppendix{chern-simons}) that 
\begin{align*}
 \lag_{U(Nk)_1}[g] + \lag_{SU(k)_{-N}}[h] 
 &= \lag_{U(Nk)_1}[g] + \lag_{U(kN)_{1}}[h^{-1}] \\
 &= \lag_{U(Nk)_1}[g h^{-1}] - \frac{1}{4\pi} \rmd \tr \left[ (i h^{-1} \rmd h) \wedge (i g^{-1} \rmd g) \right]
\end{align*}
One can then change variables to $\tilde{g} = g h^{-1}$ and argue that the effect of $h$ is precisely to constrain $\tilde{g}$ to have trivial $SU(k)$ structure.

Hence thinking of the right-hand side as being the correctly constrained version of the $U(Nk)_1$ Chern-Simons theory, we see it is equivalent to the trivial theory $U(Nk)_1$ evaluated on general configurations \textit{plus} a new $SU(k)_{-N}$ theory!

Now clearly the above is only a very vague sketch of a much more sophisticated argument, but it hopefully does make the result at least more plausible. Either way, it is a rigorously true statement that the theory of $U(N)_k$ differs from $SU(k)_{-N}$ only by trivial terms (in the sense of $U(M)_1$ theories and so forth). We can indeed write
\be
U(N)_k \bigdual SU(k)_{-N}
\ee
where it is understood that we suppress a trivial sector.

\begin{asidebox}[Representation Theory of Level-Rank Duality]%
	You might be curious how such a bizarre correspondence works at the level of a bijection between operators. It is actually a remarkably elegant story \cite{yellow}, which we will just briefly allude to here. 
	
	Recall from the above that in $U(N)_k$, there are $n = (k+N-1)!/k!/(N-1)!$ integrable representations -- those whose Young diagrams have height at most $N-1$ and width at most $k$.
	
	Meanwhile, in $SU(k)_{N}$ (and so in $SU(k)_{-N}$), we have diagrams of height at most $k$ and width at most $N-1$. There are $n' = (N+k-1)!/N!/(k-1)!$ such representations. It seems like there is a slight discrepancy between this and the dual theory, with $n/N = n'/k$. Similarly, transposing all Young diagram by switching row and column lengths (or equivalently reflecting in the diagonal) does not quite give a bijection, since we have a discrepancy of 1 in the width and height limits. However, it turns out that we do not expect a bijection between representations, but between the \textit{orbits} of representations under what is called the \textit{outer automorphism group} of the gauge group.
	
	The outer automorphism group of $SU(N)$ is $\mathbb{Z}_N$, the generator of which acts on the Young tableau of $SU(N)_k$ by adding a row of $k$ boxes to the top of the diagram. Any columns of $N$ boxes on the left of the diagram may then be removed. You can check that doing this $N$ times gets us back to where we started. There is now a bijection between the $n/N$ orbits of the $SU(N)$ theory and the $n'/k$ orbits of the $SU(k)$ theory: we transpose and then add an appropriate number of $k$-box rows!
	
	For example, $SU(4)_2$ and $SU(2)_4$ match as follows (where we write $\bullet$ for the singlet representation):
	\def\yngsep{\ , \ \ }
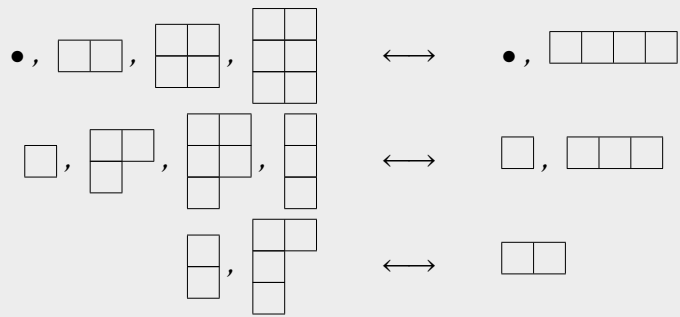
	\begin{align*}
		\Yvcentermath1
		\bullet\yngsep\yng(2)\yngsep\yng(2,2)\yngsep\yng(2,2,2) &\bigdual \bullet\yngsep\yng(4) \\
		\Yvcentermath1
		\yng(1)\yngsep\yng(2,1)\yngsep\yng(2,2,1)\yngsep\yng(1,1,1) &\bigdual \yng(1)\yngsep\yng(3) \\
		\Yvcentermath1
		\yng(1,1)\yngsep\yng(2,1,1) &\bigdual \yng(2)
	\end{align*}
	\undef\yngsep
	
	Why is it natural that we should have to worry about these extra rows of boxes? Well, the simplest monopole operator of $U(N)_k$ lives in precisely the $k$th symmetric representation of $SU(N)$, and the higher ones have Young diagrams consisting of  $m$ rows of $k$ boxes. Hence we can essentially translate the identification of representations up to outer automorphisms into identification of Wilson lines up to monopole operators.
\end{asidebox}

For now, we will just mention that there are generalizations of what we have discussed to other gauge groups other than the unitary ones (namely, the orthogonal and symplectic groups). We will see those briefly in \refSection{generalize_gauge_group}.

\chapter{The Master Duality}
\labelChapter{master_duality}

\begin{introduction}
  Having established level-rank duality for various topological quantum field theories, we can now look at critical theories representing phase transitions between these TQFTs. It turns out that bosonization goes hand-in-hand with level-rank duality. We will first explore the first class of such dualities proposed by Aharony, then look at a more recent generalization.
\end{introduction}

\section{Aharony's Dualities}\labelSection{aharony}

\lettrine{E}{quipped with level-rank duality,} we are now ready to couple matter to our non-Abelian gauge theories. Let's leap straight in and state three dualities proposed by Aharony in 2016 \cite{Aharony:2015mjs}:
\begin{align}
SU(N)_k + N_f \text{ WF scalars} &\bigdual U(k)_{-N+N_f/2} + N_f \text{ fermions} \label{aharony_sun} \\
U(N)_k + N_f \text{ WF scalars} &\bigdual SU(k)_{-N+N_f/2} + N_f \text{ fermions} \label{aharony_un} \\
U(N)_{k,k+N} + N_f \text{ WF scalars} &\bigdual U(k)_{-N+N_f/2,-N-k+N_f/2} + N_f \text{ fermions} \label{aharony_other}
\end{align}
which are each believed to hold for $N_f$ flavours of fundamental bosons/fermions where
\be N_f \le N \fstp \ee
We will see a reason for this flavour bound later, in \refSectionOnly{rg_flows_dualities}. Note that the special case $N=k=N_f=1$ of \eqref{aharony_un} refers to the case of Abelian 3d bosonization we studied in \refChapterOnly{3d_bosonization}, since $SU(1)$ is a trivial gauge group.

Both of these theories must be tuned to a critical point with a manifest flavour symmetry $SU(N_f)$ that rotates the matter fields. In particular, the scalars $\phi_i$ for $i=1,\ldots,N_f$ are subject to a potential
\be V(|\phi|^2) = \mu \phi^{\dagger i} \phi_i + \lambda \phi^{\dagger i} \phi_i \phi^{\dagger j} \phi_j  + \cdots \ee
which is $SU(N_f)$ invariant.\footnote{The $SU(N)$ gauge indices are suppressed here. In fact there are generally \textit{two} possible quartic terms with different $SU(N)$ index contractions \cite{hsinseiberg}; if $M^i_j = \phi^{\dagger i}_\alpha \phi_j^\alpha$, then the two options are $\tr M^2$ and $(\tr M)^2$. In these lectures we will generally avoid thinking about the details of potentials, but in \refSectionOnly{aharony_technicalities} we will see we want $\tr M^2$ to dominate the potential, so we propose this is a lower dimension operator.} The coefficient $\mu$ must of course be tuned. This is what we have in mind when we write schematic Lagrangians for the above dualities. For example, \eqref{aharony_un} can be written as
\begin{align}
|D_a \phi|^2 - |\phi|^4 + \frac{k}{4\pi} \tr \left[ a \rmd a - \frac{2i}{3} a^3 \right]
\bigdual
i \bar{\psi} \slashed{D}_b \psi + \frac{-N+N_f}{4\pi} \tr \left[ b \rmd b - \frac{2i}{3} b^3 \right] 
\end{align}
where $a \in u(N)$ and $b \in su(k)$.

Note that most of these dualities are more in the spirit of the duality of a Wilson-Fisher scalar $\leftrightarrow$ gauged fermion of \eqref{reverse-bosonization-lag} than of $U(1)_1 + \text{WF scalar}$ $\leftrightarrow$ fermion of \eqref{3d_bosonization_schematic}, in that in almost every case both sides are non-trivial interacting field theories. (The exception is $k=1$ in \eqref{aharony_un}, where we have $N_f$ free fermions on the right-hand side.) This means the best picture to have in mind is that there are two distinct UV field theories which can be tuned flow to the same IR fixed point.
In fact, we make a slightly stronger claim, which is that as we vary the coefficient of the symmetric mass terms $|\phi|^2$ and $\bar{\psi}\psi$, we find a \textit{unique} fixed point. Hence we postulate that the phase diagram of both theories looks like \refFigureOnly{aharony_un_uv_flow}.

\begin{center}
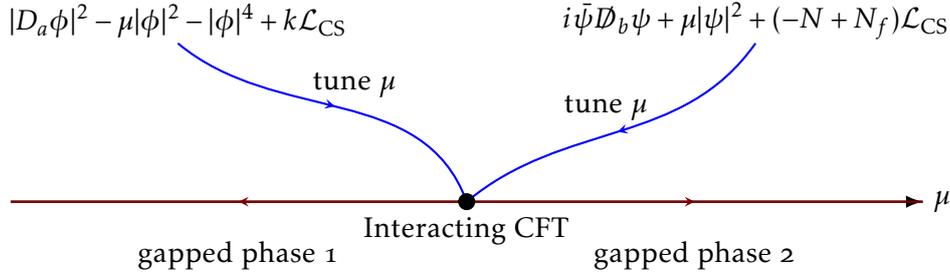

	\centering
	\begin{tikzpicture}
	\node at (-3.8,2.4) {$|D_a \phi|^2 - \mu|\phi|^2 - |\phi|^4 + k \lag_{\mathrm{CS}}$};
	\draw[flow,tuned] (-3.8,2.1) to[out=-40,in=110] node[above,xshift=0.4em] {tune $\mu$} (0,0);
	\node at (3.8,2.4) {$i \bar{\psi} \slashed{D}_b \psi + \mu|\psi|^2 + (-N+N_f) \lag_{\mathrm{CS}}$};
	\draw[flow,tuned] (3.8,2.1) to[out=-125,in=40] node[above,xshift=-0.5em] {tune $\mu$} (0,0);
	\draw[axis] (-6,0) -- ++ (12,0) node[right] {$\mu$};
	\draw[flow,relevant] (0,0) --  node[below,yshift=-1em,align=center] {gapped phase 2} (6,0);
	\draw[flow,relevant] (0,0) -- node[below,yshift=-1em,align=center] {gapped phase 1} (-6,0);
	\filldraw (0,0) circle (3pt) node[below,yshift=-0.2em] {Interacting CFT};
	\end{tikzpicture}
	\captionof{figure}{The meaning of the duality \eqref{aharony_un}. We will analyze the gapped phases in \refSectionOnly{rg_flows_dualities}. Note that the $SU(N_f) \times U(1)$ symmetry is manifest all the way along the flow, and indeed throughout this diagram}\labelFigure{aharony_un_uv_flow}
\end{center}

Since both sides have a global $SU(N_f)$ symmetry, we can couple them to a background gauge field. In fact, both sides also have a $U(1)$ global symmetry: this is the monopole symmetry of $\phi$ in the $U(N)_k$ theory, and the overall phase of the fermion in the $SU(k)$ theory. These combine to form a global $U(N_f)$ symmetry, and so we introduce a background gauge field $A \in u(N_f)$. The correct expression turns out to be
\begin{align} \label{aharony_un_bg}
|D_{a + A} \phi|^2 - |\phi|^4 + \frac{k}{4\pi} \tr \left[ a \rmd a - \frac{2i}{3} a^3 \right]
\bigdual
i \bar{\psi} \slashed{D}_{b + A} \psi + \frac{-N+N_f}{4\pi} \tr \left[ b \rmd b - \frac{2i}{3} b^3 \right] \\ + \frac{k}{4\pi} \tr \left[ A \rmd A - \frac{2i}{3} A^3 \right] \quad \nn
\end{align}
and we will see how this could be guessed in \refSectionOnly{rg_flows_dualities}. The slightly surprising thing about this is that $A$ does not appear to couple to the monopole symmetry, but instead is attached to the gauged $U(1)$ symmetry of $\phi$. This can be addressed, and this is the first of part of Exercise \ref{relating_aharony}.

Moreover, once we have a handle on this $U(1)$ global symmetry, the door is opened to gauging it. Remarkably, all of Aharony's dualities equivalent to each other because they can be related by gauging a global $U(1)$ symmetry, as discussed in Exercise \ref{relating_aharony}.

\begin{exercise}[subtitle=Equivalence of Aharony's Dualities]
	\label{relating_aharony}%
	\begin{enumerate}[label=(\alph*)]
		\item
		By absorbing the $U(1)$ part of $A$ into $b$, show that in fact $A$ does couple to the monopole symmetry of the right-hand side.
		\item 
		Assuming one of Aharony's dualities \eqref{aharony_sun}-\eqref{aharony_other}, derive a broader class of dualities \cite{Radicevic:2016wqn}
		\be U(N)_{k,k+nN} + N_f \text{ WF scalars} \bigdual U(1)_n \times U(k)_{-N+N_f/2} + N_f \text{ fermions} \label{aharony_alln} \ee
		that hold for any integer $n \in \mathbb{Z}$. (You should find that the fermions are charged under $U(k)$ and the extra $U(1)_n$ factor couples via a BF term to the $U(1) \subset U(k)$.)
		\item
		Show that Aharony's dualities are special cases of \eqref{aharony_alln} for various values of $n$, and derive another special case \cite{hsinseiberg}
		\be U(N)_{k,k-N} + N_f \text{ WF scalars} \bigdual U(k)_{-N+N_f/2,-N+k+N_f/2} + N_f \text{ fermions} \fstp \ee
		\item
		If you are feeling particularly enthusiastic, feel free to write a generalization with a general $U(1)$ level for the fermion theory.
		\item
		How does time reversal interact with \eqref{aharony_un_bg}?
	\end{enumerate}
\end{exercise}

\subsection{Fermions and Non-Abelian Gauge Fields}

We find ourselves working with a fermion coupled to a non-Abelian gauge field, say
\be \lag_{\mathrm{bare}} = i\bar{\psi} \gamma^\mu(\partial_\mu - i A_\mu) - \mu \bar{\psi} \psi \ee
where we have also included a mass term. This sort of coupling is subject to the same kind of subtleties as the Abelian case. Indeed, the computation of section \refSectionOnly{effective_action_abelian_fermion} goes through more or less unchanged for the quadratic terms, generating a bare effective Lagrangian of
\be \lag_{\mathrm{eff}} = \frac{\sign \mu}{2} \frac{1}{4\pi} \tr A \rmd A \ee
at quadratic order when a fermion is integrated out. But this term is not invariant under non-Abelian gauge transformations, even if we regularize the theory. But that is fine, since in the language of Feynman diagrams, there is now also a contribution to the cubic term as depicted in \refFigureOnly{renormalization_cs_term_nonab}.

\begin{figure}[h]
	\centering
	\begin{tikzpicture}
	\draw[feynman fermion] (0,-2) -- (-1,-3.73);
	\draw[feynman fermion] (-1,-3.73) -- (1, -3.73);
	\draw[feynman fermion] (1,-3.73) -- (0, -2);
	\draw[feynman photon] (0,-1) node[xshift=-1em] {} -- (0,-2);
	\draw[feynman photon] (-1,-3.73) -- (-1.86,-4.23) node[xshift=+1em] {};
	\draw[feynman photon] (+1,-3.73) -- (1.86,-4.23) node[xshift=+1em] {};
	\end{tikzpicture}
	\caption{The renormalization of the photon three-point interaction due to a fermion loop}\labelFigure{renormalization_cs_term_nonab}
\end{figure}
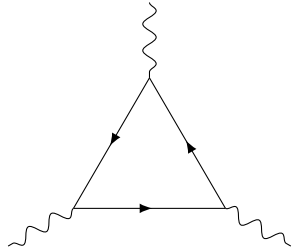

This generates precisely the cubic term needed to complete the Lagranigan to the level $\frac{1}{2}$ Chern-Simons term
\be \lag_{\mathrm{eff}} = \frac{\sign \mu}{2} \frac{1}{4\pi} \tr \left[ A \rmd A - \frac{2i}{3} A \wedge A \wedge A \right] \fstp \ee
That means we can use exactly the same approach of adding a Pauli-Villars regulator to preserve gauge invariance, so that in our conventions 
\be
\lag = i \bar{\psi} \slashed{D}_A \psi - \mu \bar{\psi}\psi \qquad \longrightarrow \qquad
\lag_{\mathrm{eff}} = \begin{cases}
	0 & \mu > 0 \\
	- \frac{1}{4\pi} \tr \left[ A \rmd A - \frac{2i}{3} A^3 \right] & \mu < 0 \\
\end{cases} \fstp
\ee

\subsection{RG Flows Between Dualities}\labelSection{rg_flows_dualities}

Each of the above dualities comes with an operator correspondence
\be
\phi^{\dagger i} \phi_i \bigdual -\bar{\psi}^{i} \psi_i \label{nonab_mass_correspondence}
\ee
which holds for each $i$, with no summation. There is an obvious game we can play: deform both sides of the duality with this mass term and see where we land in the infrared.

Let's focus on the duality \eqref{aharony_un}
\be U(N)_k + N_f \text{ WF scalars} \bigdual SU(k)_{-N+N_f/2} + N_f \text{ fermions} \nn\ee
in the form of \eqref{aharony_un_bg}
\be 
|D_{a + A} \phi|^2 - |\phi|^4 + \frac{k}{4\pi} \tr \left[ a \rmd a - \frac{2i}{3} a^3 \right]
 \ \longleftrightarrow \ 
i \bar{\psi} \slashed{D}_{b + A} \psi + \frac{-N+N_f}{4\pi} \tr \left[ b \rmd b - \frac{2i}{3} b^3 \right] + \frac{k}{4\pi} \tr \left[ A \rmd A - \frac{2i}{3} A^3 \right] \quad \nn
\ee
and investigate what happens when we add mass terms.

Firstly, suppose we turn on the $SU(N_f)$ symmetric mass term
\be \delta\lag = - \mu |\phi|^2 = - \mu \sum_{i=1}^{N_f} \phi^{\dagger i}\phi_i \bigdual \delta\tilde{\lag} = \mu |\psi|^2 = \mu \sum_{i=1}^{N_f} \psi^{\dagger i}\psi_i \fstp \ee
Let us look at each sign of $\mu$ in turn:
\begin{itemize}
	\item $\mu \gg 0$: In this case, we simply decouple all of the bosonic matter modes, leaving an effective Lagrangian
	 \be \lag_{\mathrm{eff}} = \frac{k}{4\pi} \tr \left[ a \rmd a - \frac{2i}{3} a^3 \right] \ee
	 describing the TQFT $U(N)_k$. Meanwhile, the fermions receive a negative mass, shifting both the dynamical and background Chern-Simons theories to leave
	 \be \tilde{\lag}_{\mathrm{eff}} = \frac{-N}{4\pi} \tr \left[ b \rmd b - \frac{2i}{3} b^3 \right] \ee
	 which describes pure $SU(k)_{-N}$, again with no background terms. But we already know these are dual by level-rank duality:
	 \be
	 U(N)_k \bigdual SU(k)_{-N} \relabeleq{level_rank_basic}
	 \ee
	 
	 \item $\mu \ll 0$: Here, the physics is a little different. We expect that the bosons acquire a vacuum expectation value which generically breaks the gauge group
	 \be U(N) \to U(N-N_f) \fstp \ee
	 In fact, something interesting happens when we remember the background field $A$. Firstly, observe that the nature of the Higgs mechanism is such that the gauge fields acquire a mass term
	 \be |\partial_\mu \phi - i A \phi - i \phi a^T|^2 = \cdots + \tr|A \left<\phi\right> + \left<\phi\right> a^T|^2 + \cdots \ee
	 at low energies. Let us assume we can fix a gauge in which the boson's VEV takes the form
	 \be \left<\phi\right> = \labeldims{\begin{pmatrix} v & 0 & 0 & 0 & 0 \\ 0 & v & 0 & 0 & 0 \\ 0 & 0 & v & 0 & 0 \end{pmatrix}}{N}{N_f} \fstp  \ee
	 Then the above mass term forces $a$ to take the values
	 \be \arraycolsep=0.8em\def\arraystretch{1.4}
	 a = \left(\begin{array}{c|c}
	 	-A^T & 0 \\\hline
	 	0 & c
	 \end{array}\right) \ee
    where $c$ is the unconstrained, dynamical $U(N - N_f)$ field which remains after Higgsing.
    Evaluating the Chern-Simons term for $a$ now generates a Chern-Simons term for the background field $A$, leaving
    \be \lag_{\mathrm{eff}} = \frac{k}{4\pi} \tr \left[ c \rmd c - \frac{2i}{3} c^3 \right] + \frac{k}{4\pi} \tr \left[ A \rmd A - \frac{2i}{3} A^3 \right] \fstp \ee 
    The story on the fermionic side is happily much simpler! We simply scrub out the fermions in \eqref{aharony_un_bg}, leaving
    \be \tilde{\lag}_{\mathrm{eff}} = \frac{-N+N_f}{4\pi} \tr \left[ b \rmd b - \frac{2i}{3} b^3 \right] + \frac{k}{4\pi} \tr \left[ A \rmd A - \frac{2i}{3} A^3 \right] \ee
    which is dual to $\lag_{\mathrm{eff}}$ by level-rank duality in the form
    \be U(N-N_f)_k \bigdual SU(k)_{-N+N_f} \fstp \ee
\end{itemize}
So far so good.

Now let us instead add only a mass term $-\mu |\phi_{N_f}|^2 \leftrightarrow + \mu |\psi_{N_f}|^2$ to both Lagrangians. Note that this explicitly breaks the global symmetry $U(N_f) \to U(N_f-1) \times U(1)$. It is useful to write $\tilde{A}$ and $A_{N_f}$ for the corresponding parts of $A$.

\begin{itemize}
\item $\mu \gg 0$: Clearly, if we turn on a positive mass for the boson, then the only effect on the bosonic side is to decouple $\phi_{N_f}$. This leaves us with $U(N)_k + (N_f-1) \text{ WF scalars}$. On the dual side, we shift the level of the $b$ Chern-Simons term by $1$, landing on $SU(k)_{-N + (N_f-1)/2} + (N_f-1)\text{ fermions}$. This actually reproduces the duality \eqref{aharony_un} but with parameters $(N,k,N_f) \to (N,k,N_f-1)$!

We can also ask what happens to the background terms. On the bosonic side, it is clear that the variable $A_{N_f}$ decouples. The fermionic side receives a correction to the $A_{N_f}$ contact term; since the fermion $\psi_{N_f}$ is an $SU(k)$ fundamental field, there are effectively $k$ fermions, and the $A_{N_f}$ Chern-Simons term is shifted by $-k$, leaving nothing behind, matching the bosonic theory. This again nicely matches \eqref{aharony_un_bg}.

\item $\mu \ll 0$: Alternatively, if we turn on a negative mass squared for the boson, then we expect that it will partially Higgs the gauge group $U(N)$, breaking it down to $U(N-1)_k + (N_f-1) \text{ WF scalars}$.\footnote{Once we have broken the $U(N_f)$ symmetry, it gets even harder to control the potentials generated in this theory. We will briefly mention the potential in \refSectionOnly{aharony_technicalities}.} The fermion meanwhile receives a positive mass and so we can simply remove it from the Lagrangian without adding any contact terms. It is easy to verify that this leaves us with $SU(k)_{-(N-1)+(N_f-1)/2} + (N_f-1) \text{ fermions}$. This is precisely the duality \eqref{aharony_un} with parameters $(N,k,N_f) \to (N-1,k,N_f-1)$.

Again, the background terms can also be dealt with. We will not work through the details.

\end{itemize}

\begin{wrapfigure}{r}{0.4\textwidth}
	\centering
	\begin{tikzpicture}
		\node at (0,0) {$(N,k,N_f)$};
		\node at (-1.8,-1.6) {$(N-1,k,N_f-1)$};
		\node at (1.8,-1.6) {$(N,k,N_f-1)$};
		\draw[standoff long,->] (0,0) -- node[left,yshift=0.1cm] {$\mu \ll 0$} (-1.8,-1.6);
		\draw[standoff long,->] (0,0) -- node[right,yshift=0.1cm] {$\mu \gg 0$} (1.8,-1.6);
		
	\end{tikzpicture}
	\caption{Flows of the duality \eqref{aharony_un} upon integrating out a single flavour of matter.}\labelFigure{aharony_un_flows}
\end{wrapfigure}
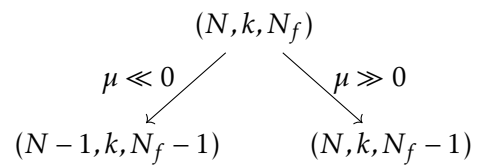
These flows are depicted in \refFigureOnly{aharony_un_flows}. This analysis applies equally well to the other dualities in this family too, since they are all equivalent. The fact that these flows agree forms an obvious consistency check of the dualities. In particular, even if we integrate out all of the matter fields one at a time, we still land on the level-rank duality
\be
U(N)_k \bigdual SU(k)_{-N} \fstp \relabeleq{level_rank_basic}
\ee
This is an encouraging sign for our proposed dualities, though it is very far from constituting a real proof of them.

We promised to give some motivation for the flavour bound $N_f \le N$ in this section. Suppose instead that $N_f > N$; then we can Higgs $N$ of the scalars and we are left with the case $(0, k, N_f - N)$ which describes $N_f - N$ ungauged Wilson-Fisher scalars on one side and $SU(k)_{-(N_f-N)/2}$ coupled to $N_f - N$ fermions in the dual theory. This seems implausible; the fermionic theory seems to have a phase with the non-trivial TQFT $SU(k)_{-N_f+N}$, but no such phase is visible in the dual theory. We infer that indeed,
\be N_f \le N \fstp \ee 

\subsection{Operator Matching}

We have already mentioned how the mass terms map:
\be
\phi^{\dagger i} \phi_i \bigdual -\bar{\psi}^{i} \psi_i \relabeleq{nonab_mass_correspondence}
\ee
We also coupled both theories to background fields, which implicitly defines the current correspondence
\be i (\phi^\dagger D^\mu \phi - \text{c.c.}) \bigdual i \bar{\psi} \gamma^\mu \psi \fstp \ee

This leaves us to understand how to match various interesting gauge-invariant operators like the fermionic baryon
\be
\tilde{\calO} = \epsilon^{\alpha_1 \cdots \alpha_k} \psi_{i_1 \alpha_1} \cdots \psi_{i_k \alpha_k}
\ee
across the duality. Note that this lies in the $k$th symmetric representation of the flavour group $SU(N_f)$. We know the dual to this must be a monopole operator of the lowest possible charge, from the matching of the $U(1)$ currents. As mentioned in \refChapterOnly{level_rank}, the simplest monopole operators in $U(N)_k$ actually transform in the $k$th symmetric representation of $U(N)$. To make them gauge-invariant, therefore, they must be contracted with another appropriate operator in the conjugate representation. We can easily build such an object from $k$ bosons. Neglecting gauge indices, this looks like
\be
\calO = \monoM \phi_{i_1} \cdots \phi_{i_k}
\ee
which again is also in the $k$th symmetric representation of $SU(N_f)$. This is exactly how the matching works; one can even compute approximations to the dimensions of these operators in the limit of large $N$ \cite{Aharony:2015mjs,Radicevic:2015yla} and show that they match.
\be  \monoM \phi_{i_1} \cdots \phi_{i_k} \bigdual \epsilon^{\alpha_1 \cdots \alpha_k} \psi_{i_1 \alpha_1} \cdots \psi_{i_k \alpha_k} \ee
This matching can be extended to versions of these operators containing derivatives too.


\subsection{An Example of Enhanced Symmetry}\labelSection{so3_symmetry}

Back in Exercise \ref{ex_hidden_so3}, we discussed the duality
\be U(1)_{2} + \text{WF boson} \bigdual U(1)_{-3/2} + \text{fermion} \fstp \relabeleq{ex_hidden_so3_eq} \ee
This had a global symmetry of $O(2)$, including complex conjugation.

However, the dualities we have discussed in this section tell us that there are more things we can add to this duality! In fact, one can easily see that
\be SU(2)_{1} + \text{WF boson} \qquad \mbox{and} \qquad SU(2)_{-1/2} + \text{fermion} \ee
are \textit{also} dual to the above theories. This is especially interesting because these theories actually have a larger symmetry group \cite{Aharony:2016jvv,Benini:2017dus}! Let's see why that is, using an argument which is actually relatively well-known in the case of the Standard Model. (Look up ``custodial symmetry''.)

Consider an object $\phi_\alpha$ which carries a fundamental $SU(2)$ index $\alpha$, where the $SU(2)$ is a gauge field. Now there is a peculiarity of the fundamental representation of $SU(2)$, due to the fact it is \textit{pseudoreal}. This is a fancy way of saying that the generators $t^a$ obey $t^{a \dagger} = -V^{-1}t^a V$ for some $V$. For our case, $V=\sigma_2$. Therefore, both
\be \phi = \begin{pmatrix} \phi_1 \\ \phi_2 \end{pmatrix} \qquad \mbox{and} \qquad \hat{\Phi} = i \sigma_2 \Phi = \begin{pmatrix} \phi_2^\star \\ -\phi_1^\star \end{pmatrix} \ee
transform in the same way under left-multiplication by an $SU(2)$ matrix. This is easy to check, so feel free to do so!

Now one can construct
\be \Phi = \begin{pmatrix} \phi_1 & \phi_2^\star \\ \phi_2 & -\phi_1^\star \end{pmatrix} \ee
which transforms under left-multiplication by $SU(2)$, and then the potential is expressed in terms of objects like $\tr \Phi^\dagger \Phi = |\phi|^2$.

But something more is true: we can now consider \textit{right}-multiplication by an entirely independent $SU(2)_R$ matrix! Taking account of the fact that the $-1\in SU(2)_R$ is actually a gauge transformation, we obtain a $SU(2)/\mathbb{Z}_2 \cong SO(3)$ global symmetry.\footnote{Another, arguably more straightforward way to derive this result is to expand $\phi$ into four real degrees of freedom, and find that $o(4) \cong su(2) \oplus su(2)$ contains two $su(2)$ factors, only one of which was gauged.}


We conclude that we suspect all four of these theories flow to an IR fixed point exhibiting $SO(3)$ symmetry, although only the non-Abelian field theories make this symmetry obvious:
\begin{center}
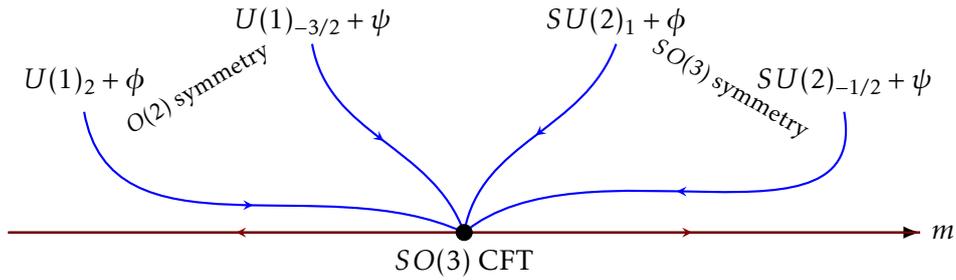

	\centering
	\begin{tikzpicture}
	\node at (-5,2) {$U(1)_{2} + \phi$};
	\draw[flow,tuned] (-5,1.6) to[out=-80,in=155] (0,0);
	\node at (-2,2.8) {$U(1)_{-3/2} + \psi$};
	\draw[flow,tuned] (-2,2.5) to[out=-80,in=110] (0,0);
	
	\node[rotate=30,font=\smaller] at (-3.5,1.9) {$O(2)$ symmetry};
	
	\node at (2,2.8) {$SU(2)_{1} + \phi$};
	\draw[flow,tuned] (2,2.5) to[out=-110,in=80] (0,0);
	\node at (5,2) {$SU(2)_{-1/2} + \psi$};
	\draw[flow,tuned] (5,1.6) to[out=-80,in=40] (0,0);
	
	\node[rotate=-30,font=\smaller] at (3.5,1.9) {$SO(3)$ symmetry};
	
	\draw[axis] (-6,0) -- ++ (12,0) node[right] {$m$};
	
	\draw[flow,relevant] (0,0) -- (6,0);
	\draw[flow,relevant] (0,0) -- (-6,0);
	\filldraw (0,0) circle (3pt) node[below,yshift=-0.2em] {$SO(3)$ CFT};
	\end{tikzpicture}
	\captionof{figure}{Four theories flowing to the same $SO(3)$-symmetric critical point, with only two of them possessing this symmetry in the UV}\labelFigure{so3_emergent_flow}
\end{center}

\subsection{Some Technicalities}\labelSection{aharony_technicalities}

Firstly, we have not been careful about including potential almost-trivial factors in our dualities. It turns out that it is quite simple to fix this; focussing on our favourite duality \eqref{aharony_un}, we can simply add a factor to the $U(N)$ side of the duality as follows:
\be U(N)_k + N_f \text{ WF scalars} + U(N k)_{-1} \bigdual SU(k)_{-N+N_f/2} + N_f \text{ fermions} \fstp \nn\ee
This new factor, an almost trivial spin-TQFT, does not couple to the matter in any way. It also encodes a gravitational Chern-Simons term $2k N \mathrm{CS}_g$.

Secondly, we can be a little more specific about the global symmetry which acts on this theory \cite{Benini:2017dus}. We have focussed on $U(N_f)$, which is not wrong, but it does miss two issues which we should emphasize here. For one thing, there is also a charge conjugation symmetry, which is an additional $\mathbb{Z}_2$ factor, commonly written as $\mathbb{Z}_2^{C}$. This combines with $U(N_f)$ to give $U(N_f) \rtimes \mathbb{Z}_2^{C}$ rather than $U(N_f) \times \mathbb{Z}_2^{C}$, since charge conjugation also maps $U(N_f)$ representations to their conjugates too.

Also, the fact that we only look at gauge singlets means that we don't actually get all representations of $U(N_f)$ cropping up. For instance, on the fermion side, operators must have a fermion number which is a multiple of $k$ in order to form an $SU(k)$ singlet. Therefore, there is a $\mathbb{Z}_k \subset U(1) \subset U(N_f)$ factor which does not act on any physical states, and hence $U(N_f)$ does not act faithfully. A similar argument applies to the monopole operators on the left-hand side. In total, therefore, the manifest faithful global symmetry of these theories is given by
\be G = \frac{U(N_f)}{\mathbb{Z}_k} \rtimes \mathbb{Z}_2^{C} \fstp \ee
It is of course possible that the CFT which emerges at low energies has even more symmetry, with global symmetry $G^{\mathrm{IR}} \supset G$, or even that there is more manifest extra symmetry we didn't notice. We won't worry about this for now (although in \refSectionOnly{qcd_flows}, we discuss an extra global $SU(2)$ symmetry present in $SU(2)$ gauge theories).

Note that when the gauge group is $SU(2)$, the analysis is complicated by the fact that we can write more invariants \cite{Benini:2017aed,Benini:2017dus} and thereby construct various different potentials preserving different global symmetries. We also won't worry about this in great detail.

We will, however, note that the proposal for the way that the Higgs mechanism works did make some implicit assumptions about the nature of the potential. Firstly, we assumed that the only terms we needed to worry about were the quadratic deformation we added, and a quartic term. Note that sextic terms like $|\phi|^6$ are classically marginal in the UV theory, for example, so you might worry about such terms. We will however make the natural assumption that the quartic terms (which are classically relevant in the UV theory) dominate. But even then, in terms of the gauge-invariant object $M^i_j = \phi^{\dagger i}_\alpha \phi_j^\alpha$, there are two flavour-symmetric operators at quartic order $\tr M^2$ and $(\tr M)^2$. Of course, generically these will have different dimensions and we expect one of them to dominate the physics a low energies. Now if the potential is assumed to be purely the former operator before we deform it, so
\be V = \lambda \tr M^2 + \mu \tr M = \lambda \tr  \left[ M^{i}_{j} + \frac{\mu}{2\lambda} \delta^{i}_{j} \right]^2 + \text{const.} \fstp \ee
then for $\mu \ll 0$, the Higgs mechanism indeed forces $\phi$ to condense as proposed in \refSectionOnly{rg_flows_dualities}. We assume that the operator $(\tr M)^2$ is simply irrelevant at this fixed point, and plays no role in the physics.

For some motivation, one can look to the limit of large $N,k$ (with $N/k$ fixed), where there is a natural distinction between \textit{single-trace operators} like $\tr M^2$ and \textit{multi-trace operators} like $(\tr M)^2$. In particular, single-trace operators generally have lower dimensions at large $N$, since multi-trace operators receive extra anomalous dimensions from every extra trace beyond the first. In fact, the large $N$ limit predicts $(\tr M)^2$ to be irrelevant, which is a good sign!

\section{QCD in Three Dimensions}\labelSection{qcd3}

The way we have discussed these dualities so far makes them seem like something of a curiosity; most of them are ``strong-strong'' dualities, relating two theories we do not really understand to each other. There are exceptions, of course, like the $k=1$ case of \eqref{aharony_un} which states that
\be U(N)_1 + N_f \text{ WF scalars} \bigdual N_f \text{ free fermions} \ee
which expresses the more intriguing fact that an apparently strongly coupled theory is in fact free in the IR.

But there are other ways to use proposed dualities to make more conjectures about the behaviour of interesting theories. In this section, we will discuss the proposal of Komargodski and Seiberg for the structure of QCD in three dimensions \cite{Komargodski:2017keh}.

Concretely, let's consider the theory
\be SU(N)_{k} + N_f \text{ fundamental fermions} \ee 
for general values of $k,N,N_f$. Importantly, $k + \frac{1}{2}N_f \in \mathbb{Z}$ since we have not explicitly written the level as $k-N_f/2$. A classical analysis suggests the only relevant operators are the various quadratic operators $\bar{\psi}^i \psi_j$. These are the operators we will play with. In fact, we will chiefly work with the flavour symmetric mass $m$, so
\be \lag = i\bar{\psi} \slashed{D}_a \psi - m \bar{\psi}\psi + \frac{k+N_f/2}{4\pi} \left[ a \rmd a - \frac{2i}{3} a^3 \right] \fstp \ee
Here, we see that it is indeed necessary that $k+N_f/2 \in \mathbb{Z}$ for gauge invariance. We might as well take $k \ge 0$, since time reversal simply takes $k \to -k$ and switches the sign of the mass term. The massless theory with $k=0$ has time-reversal symmetry.

As usual, we can tell that there must be \textit{some} kind of transition as we vary $m$, since
\be SU(N)_{k} + N_f \ \psi \to 
\begin{cases}
	SU(N)_{k+N_f/2} & \text{if } m \gg 0 \\
	SU(N)_{k-N_f/2} & \text{if } m \ll 0
\end{cases}
\ee
leads to two distinct gapped phases (both TQFTs, except for when $k = \pm N_f/2$ and we obtain the trivial confining theory $SU(N)_0$ on one side). The details of what happens for intermediate values of $m$ are unclear, however. Notice that there is in fact another scale in the theory, namely the gauge coupling $g^2$. In principle, when $|m| \sim g^2$, the dynamics could be very different to when $|m| \gg g^2$.

We will stick with the proposal of \refSectionOnly{aharony} for $N_f \le 2k$:
\begin{center}
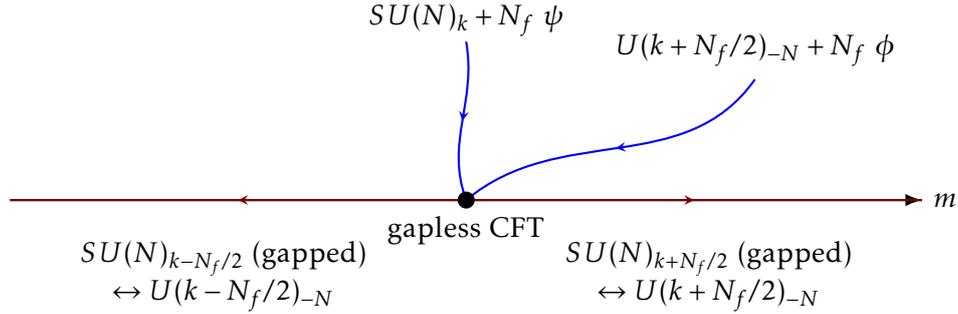

	\centering
	\begin{tikzpicture}
	\node at (3.8,2) {$U(k+N_f/2)_{-N} + N_f \ \phi$};
	\draw[flow,tuned] (3.8,1.6) to[out=-125,in=40] (0,0);
	\node at (0,2.4) {$SU(N)_{k} + N_f \ \psi$};
	\draw[flow,tuned] (0,2.1) to[out=-80,in=110] (0,0);
	\draw[axis] (-6,0) -- ++ (12,0) node[right] {$m$};
	\draw[flow,relevant] (0,0) --  node[below,yshift=-1em,xshift=0.5em,align=center] {$SU(N)_{k+N_f/2}$ (gapped) \\ $\leftrightarrow U(k+N_f/2)_{-N}$ } (6,0);
	\draw[flow,relevant] (0,0) -- node[below,yshift=-1em,xshift=-0.5em,align=center] {$SU(N)_{k-N_f/2}$ (gapped) \\ $\leftrightarrow U(k-N_f/2)_{-N}$ } (-6,0);
	\filldraw (0,0) circle (3pt) node[below,yshift=-0.2em] {gapless CFT};
	\end{tikzpicture}
	\captionof{figure}{The behaviour of QCD for $N_f \le 2k$ (identical to \refFigureOnly{aharony_un_uv_flow}, though we have switched $N$ and $k$ relative to the previous section)}\labelFigure{qcd_small_nf}
\end{center}
Our perspective now is that this theory is gapped except at one point, at which there is a second-order transition. That gapless point has a dual description as a theory of $N_f$ gauged scalars. Notice that the $U(N_f)$ symmetry is unbroken everywhere in the diagram.

This cannot be exactly what happens for $N_f > 2k$, since we have $k-N_f/2 < 0$, and $U(k-N_f/2)$ does not exist! Let's take a moment to ask exactly what happens to the bosonic theory
\be U(k+N_f/2)_{-N} + N_f \ \phi \label{qcd_bosonic_1} \ee
in this situation.

\subsection{The Grassmannian in the Bosonic Theory}

The fate of
\be U(M) + N_f \ \phi \qquad \text{with } M < N_f \label{bosonic_over_higgsed} \ee
is that the VEV looks something like
\be \left<\phi\right> = \phi_0 := \labeldims{\begin{pmatrix} v & 0 & 0 \\ 0 & v & 0 \\ 0 & 0 & v \\ 0 & 0 & 0 \\ 0 & 0 & 0 \end{pmatrix}}{M}{N_f} \ee
which actually spontaneously breaks the $U(N_f)$ global symmetry. In fact, any $U(N_f)$ rotation of $\left<\phi\right>$ would do equally well. Taking $|m| \to \infty$, we conclude that the low-energy physics consists of a map into the space of all such VEVs, since they all minimize the potential. Let's parametrize this by writing $\phi(x) = g(x)\phi_0$ where $g \in U(N_f)$. Clearly, there is a subgroup $U(M) \times U(N_f-M) \subset U(N_f)$ which acts trivially on $\phi_0$. Therefore, the low-energy physics of \eqref{bosonic_over_higgsed} is a sigma model with target space
\be \mathrm{Gr}(M,N_f) = \frac{U(N_f)}{U(M) \times U(N_f - M)} \cmma \ee
a space referred to as a \textit{Grassmannian}. From the above, it can be thought of as the set of all $M$-dimensional linear subspaces of an $N_f$ dimensional vector space.

In our case, the theory \eqref{qcd_bosonic_1} in the Higgs phase is described by a sigma model into
\be \mathcal{M}(N_f,k) = \frac{U(N_f)}{U(N_f/2 + k) \times U(N_f/2 - k)}  \fstp \label{grassmannian} \ee
So far we have not given the Chern-Simons level $N$ in \eqref{qcd_bosonic_1} any role to play in this theory, which seems wrong. Indeed, it turns out that there is a topological term which can be added to our $\mathcal{M}(N_f,k)$ sigma model.

At a mechanical level, it is simple enough to get some sense of why this topological term should arise. Notice that the bosonic kinetic terms become
\be |(\rmd g \phi_0 - i \phi_0 a)|^2 \ee 
and so the $a$ equation of motion, neglecting the kinetic terms for the gauge field by setting $g^2 = \infty$, is something like
\be \frac{N}{2\pi} \rmd a = i ( \phi_0^\dagger\rmd g \phi_0 - i v^2 a - \mathrm{c.c.}) \ee
which relates the gauge field to the gradient of the dynamical sigma model field. Therefore, the Chern-Simons term $-N \lag_{\mathrm{CS}}[a]$ must reduce to some non-trivial function of $g$. Actually evaluating this term is a little harder; in \refAppendixOnly{chern-simons}, however, we will see that Chern-Simons terms can be expressed entirely in terms of $\rmd a$ by embedding our 3d spacetime in a 4d one. It follows that is a natural 4d expression for the topological term, which we refer to as a \textit{Wess-Zumino-Witten term}. We will not discuss its details any further than this, leaving that to the literature \cite{Komargodski:2017keh,Freed:2006mx}. We will simply refer to this interesting theory as
\be \mathcal{M}(N_f,k)_N \ee 
in what follows, denoting the presence of the topological term with coefficient $N$.

The analysis above was essentially semi-classical; since we can take the mass-squared to be arbitrarily negative in this bosonic theory, we can understand the coset theory fairly well. At intermediate values of the mass, the size of the target space $\mathcal{M}(N_f,k)_N$ which is essentially $v^2$ can be small. Then quantum effects are important in getting a handle on the physics. But there are some properties of the Higgs phase which are fairly robust. The most important thing is that the broken global symmetry guarantees the presence of gapless modes in the spectrum: the Nambu-Goldstone bosons.

\subsection{The Grassmannian in the Fermionic Theory}

Now it is entirely possible that this has nothing to do with the dynamics of $SU(N)_{k} + N_f \ \psi$, but intriguingly there are hints that it does -- for at least some values of $N_f > N$. We claim that as $m$ is lowered from infinity down to values of order $g^2$, there is a useful description in terms of
\be U\left(k+N_f/2 \right)_{-N} + N_f \ \phi \ee
and that in particular, both theories flow to the same CFT for a special tuned value of $m$. Below, we will argue that we expect this to hold to $N < N_f < N_\star(N,k)$ for some unknown $N_\star$.

From the above, this requires that if we deform by lowering the fermion mass $m$ further, we actually enter a phase described by the sigma model $\mathcal{M}(N_f,k)_N$. Let us take this at face value. Since we are assuming that there is still a duality between the relevant operators of both theories, which we take to be the quadratic operators $|\phi_i|^2$ and $|\psi_i|^2$, it is natural to suggest the condensation of the scalar corresponds to the condensation of the quark bilinear, with for instance
\be \bar{\psi}^i\psi_j = \diag (\underbrace{x, x, \ldots, x\vphantom{y}}_{N_f/2+k}, \underbrace{y, y, \ldots, y}_{N_f/2-k}) \qquad \mbox{for } x \neq y \fstp \ee
Obviously this has the same phenomenology as the bosonic theory: symmetry breaking and accompanying gapless bosonic excitations.

In fact, this is an old proposal when $N_f$ is even and $k=0$, and sometimes goes by the name of ``chiral symmetry breaking'' \cite{Vafa:1984xg,Vafa:1984xh,Vafa:1983tf}. This language comes from four dimensional conventions, where we can think of an even numbers of flavours of fermion as $N_f/2 = N_4$ four-component Dirac fermions $\chi_I$. The full symmetry group of $N_4$ massless 4-component fermions is $U(2N_4)$, since each Dirac fermion has an internal ``chiral'' symmetry. There is then the possibility of generating $\bar{\chi}^I\chi_I$ mass terms for these Dirac fermions, which would represent the onset of symmetry breaking of these chiral rotations.\footnote{We emphasize that at $k=0$, the theory has time-reversal invariance, and these 4-component mass terms preserve a $\mathbb{Z}_2$ which combines time-reversal invariance with a flavour rotation, which can be seen by writing them in terms of two 2-component mass terms $\bar{\chi}\chi = \bar{\psi}^1 \psi_1 - \bar{\psi}^2 \psi_2$ and noting that time-reversal combined with $\psi_1 \leftrightarrow \psi_2$ is a symmetry of this term. In four dimensions, this is the $\mathbb{Z}_2$ which we would take to \textit{define} time reversal acting on a theory with a Dirac spinor.}

In the late '80s, an expansion in large $N_f$ for the case $k=0$ \cite{Appelquist:1989tc,Appelquist:1988sr} was used to argue that for $N_f < N_\star$, for some threshold $N_\star \approx 128(N^2-1)/3\pi^2 N \approx 4.3N$, precisely such a breaking pattern does occur. In \refSectionOnly{qcd_flows} we will see how this implies it should also happen for certain other values of $(k,N_f)$.

One way to think about this new phase is in terms of \textit{confinement}. Intuitively, by confinement, we refer to the idea that we never see particles carrying gauge charge physically separated -- there is a growing potential if we break a singlet state up into its component pieces and move them apart.\footnote{In some circumstances, like fermions with a $\Spin(N)$ gauge group where we can test whether the one-form $\mathbb{Z}_2$ global symmetry is broken, there are rigorous order parameters which test for confinement, but that is not the case for this theory. It is famously difficult to define confinement rigorously.} As we discussed in \refChapterOnly{level_rank}, the pure Chern-Simons phases $SU(N)_{k_\mathrm{eff}}$ are to be thought of as non-confining -- inserting heavy quarks has relatively little energy cost beyond the mass of the quark itself. However, in our new phase, it is suspected \cite{Komargodski:2017keh} that we can think of the condensate as a diagnostic of very strong attractive interactions with the gluons that lead to confinement.


But unlike with the bosonic theory, larger and larger mass deformations do \textit{not} simply lead to this sigma model becoming more and more weakly coupled. For $m \ll -g^2$, the low-energy dynamics has to be $SU(N)_{k-N_f/2}$. The idea is that the sigma model is never parametrically large as we vary $m$, so that it is a ``purely quantum'' phase \cite{Komargodski:2017keh}. It follows that we have to include \textit{another} phase transition between $\mathcal{M}(N_f,k)_N$ and the phase $SU(N)_{k-N_f/2}$ which reigns supreme at $m \ll 0$. Remarkably, there is another theory which captures such a phase transition:

\begin{exercise}[subtitle=The Other Bosonic Dual]
	Show that there is a new bosonic theory with precisely these two phases, namely
	\be U(N_f/2 - k)_N + N_f \ \tilde{\phi} \fstp \ee
\end{exercise}

This leads to a rather intriguing proposal for the QCD phase diagram:
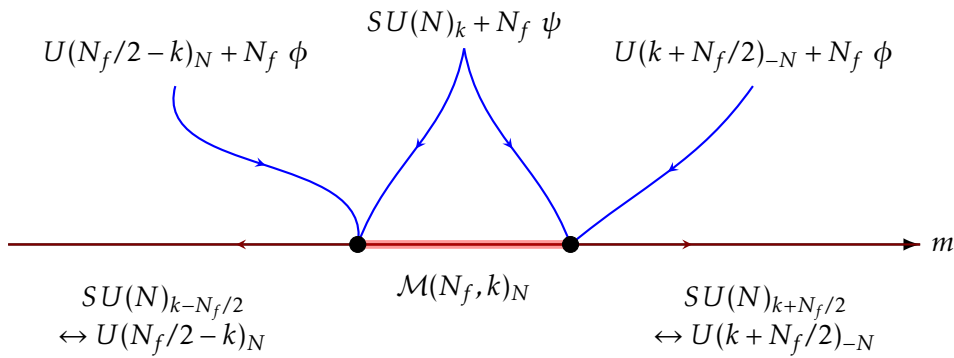
\begin{center}
	\centering
	\begin{tikzpicture}
	\node at (3.8,2.5) {$U(k+N_f/2)_{-N} + N_f \ \phi$};
	\draw[flow,tuned] (3.8,2.1) to[out=-125,in=40] (1.4,0);
	\node at (-3.8,2.5) {$U(N_f/2-k)_{N} + N_f \ \phi$};
	\draw[flow,tuned] (-3.8,2.1) to[out=-105,in=80] (-1.4,0);
	\node at (0,2.9) {$SU(N)_{k} + N_f \ \psi$};
	\draw[flow,tuned] (0,2.6) to[out=-80,in=110] (1.4,0);
	\draw[flow,tuned] (0,2.6) to[out=-100,in=70] (-1.4,0);
	\draw[axis] (-6,0) -- ++ (12,0) node[right] {$m$};
	\draw[flow,relevant] (0,0) --  node[below,yshift=-1em,xshift=2.5em,align=center] {$SU(N)_{k+N_f/2}$ \\ $\leftrightarrow U(k+N_f/2)_{-N}$ } (6,0);
	\draw[flow,relevant] (0,0) -- node[below,yshift=-1em,xshift=-2.5em,align=center] {$SU(N)_{k-N_f/2}$ \\ $\leftrightarrow U(N_f/2-k)_{N}$ } (-6,0);
	\draw[phase line] (-1.4,0) -- node[below,yshift=-0.5em,opacity=1] {$\mathcal{M}(N_f,k)_N$} (1.4,0);
	\filldraw (1.4,0) circle (3pt);
	\filldraw (-1.4,0) circle (3pt);
	\end{tikzpicture}
	\captionof{figure}{The behaviour of QCD for some values of $N_f > 2k$ ($N_f < N_\star$), displaying an intermediate range of symmetry breaking, and two proposed dual descriptions valid only near the marked points}\labelFigure{qcd_large_nf}
\end{center}

It is important to appreciate that the dual bosonic descriptions are each only useful in some region of this phase diagram. As we dial $m$ away from the right-hand marked point into the Grassmannian phase, the fermionic RG flow eventually leads us to phases which cannot be reached as deformations of the right-hand bosonic theory.

For sufficiently large $N_f$, however, our conjecture must fail. This is because of what is known about the large flavour limit of CFTs. The approach of \cite{Appelquist:1989tc,Appelquist:1988sr} shows that when $N_f$ is very large, no flavour symmetry breaking mass terms can be dynamically generated in the IR, and so the theory is conformal at one point, and gapped away from that point. We do not know a simple scalar dual for the CFT.

\subsection{Flows Between QCD Theories}\labelSection{qcd_flows}

We can of course play the by now very familiar game of integrating out individual flavours of fermion to flow to new dualities. You are asked to check this as an exercise:

\begin{exercise}[subtitle=Bounds from Flows]
	Argue that if the Grassmannian phase $\mathcal{M}(N_f,k)_N$ occurs at $(N,N_f,k)$, then
	\be (N,N_f-1,k\pm 1/2) \mbox{ has a phase } \mathcal{M}(N_f-1,k\pm 1/2)_N \fstp \ee
	Let $N_\star(N,k)$ be such that the Grassmann phase occurs precisely for $2k < N_f < N_\star$. We can think of $N_\star$ as a differentiable function of $k$ if we just smoothly link its value between adjacent values of $k$. Show that we must be able to take
	\be \left| \frac{\partial N_\star}{\partial k} \right| \le 2 \ee
	and deduce that if there is any $SU(N)$ theory with a Grassmannian phase, then there must be a $SU(N)_0$ theory with such a phase.
\end{exercise}


In total, therefore, we predict that the various $N_f$ and $k$ should look something like \refFigureOnly{qcd_phase_boundaries}. We emphasize that we have not really proven that $N_\star(N,k) > 0$ for any $N,k$. This means we cannot guarantee the region with a Grassmannian phase exists, but it is striking that there is a nice, consistent picture based around this idea.

\begin{figure}
	\centering
	\begin{tikzpicture}
	\draw[axis] (-6,0)-- ++(12,0) node[below] {$k$};
	\draw[axis] (0,0)--  ++(0,6) node[left] {$N_f$};
	
	\draw (0,0) -- (4,4.8);
	\draw[dashed,dash pattern=on 4pt off 4pt,dash phase=4pt] (4,4.8) -- (5,6) node[right] {$N_f = 2k$};
	\draw (0,0) -- (-4,4.8);
	\draw[dashed,dash pattern=on 4pt off 4pt,dash phase=4pt] (-4,4.8) -- (-5,6) node[left] {$N_f = 2|k|$};
	\draw[dotted] (0,3.5) to[out=0,in=160] (4,4.8) node[above left] {$N_\star$};
	\draw[dotted] (0,3.5) to[out=180,in=20] (-4,4.8);
	
	\node[fill=white,fill opacity=0.8,text opacity=1] at (0,2.3) {has broken phase};
	\node[fill=white,fill opacity=0.8,text opacity=1] at (0,4.4) {unique CFT};
	\node[align=center] at (4,2) {unique CFT,\\ scalar dual};
	\node[align=center] at (-4,2) {unique CFT,\\ scalar dual};
	\end{tikzpicture}
	\caption{This diagram shows the postulated behaviour of $SU(N)_k$ QCD with $N_f$ fermions as we vary $(k,N_f)$ for fixed $N > 1$. We assume $N_\star > 0$ so that theories with the interesting intermediate phase exist}\labelFigure{qcd_phase_boundaries}
\end{figure}
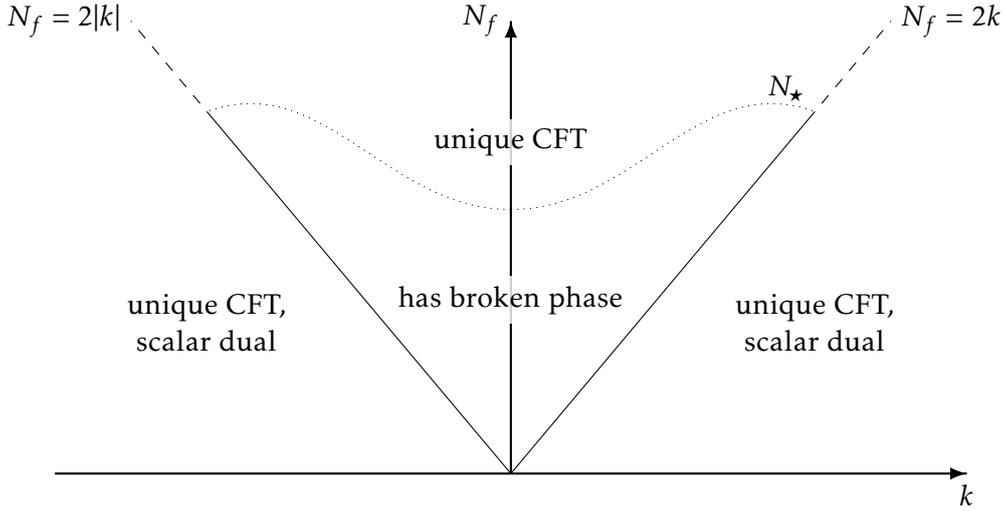

\begin{asidebox}[Skyrmions]%
	There is plenty more to say about these theories. One interesting point is that the effective sigma model $\mathcal{M}(N_f,k)_N$ supports solitonic configurations. These go by the name of \textit{Skyrmions}. The presence of the Wess-Zumino-Witten term is crucial for understanding the nature of these excitations, since they are bosonic for even $N$ and fermionic for odd $N$.
	
	This is interesting -- the Skyrmions are gauge-invariant solitons whose statistics are given by $(-1)^N$. This might remind you of the baryons of the fermionic field theory. One can indeed show that the Skyrmion excitations are dual to the baryons of the fermionic field theory, by noting that they are the monopole operators of the bosonic field theory, then arguing that these flow to the Skyrmion excitations of the sigma model. (The mesons, of course, are precisely what have condensed to give the fermion bilinear.)
\end{asidebox}

\section{The Master Duality}\labelSection{master_duality}

Although very general, it turns out that Aharony's are part of a still broader class of proposed dualities. One of the most obvious things one might wish is to include both fermions and bosons on the \textit{same side} of the duality. Late in 2017, Jensen \cite{Jensen:2017bjo} and Benini \cite{Benini:2017aed} both proposed that there is indeed a very similar class of dualities of this kind. (This was partially anticipated back in \cite{Jain:2013gza}.) They claim that
\be SU(N)_{k-N_f/2} + N_s \text{ scalars} + N_f \text{ fermions} \longleftrightarrow U(k)_{-N+N_s/2} + N_f \text{ scalars} + N_s \text{ fermions} \label{master_duality} \nn\ee
for fundamental Wilson-Fisher scalars and fermions subject to
\be N_s \le N \quad \mbox{and} \quad N_f \le k \quad \mbox{and} \quad (N_s, N_f) \neq (N, k) \fstp \ee
We call this the ``master duality'' \cite{Jensen:2017bjo}, since it makes for an exciting name, and also because we can derive not only all the previous dualities but many more by taking this as a seed.

Note that there is a global symmetry $U(N_f) \times U(N_s)$ which acts \textit{separately} on the two different types of matter. We will discuss the background terms needed to make this duality work in a moment.

This new duality also comes with an operator map, of course. Let us begin by listing the operators which preserve the global symmetry and which might be needed to understand how we tune the theory.
\begin{itemize}
	\item At quadratic order, the $SU(N)$ theory has two mass terms preserving the global symmetry:
	\be \boxed{|\phi|^2 \qquad \mbox{and} \qquad \bar{\psi} \psi}  \ee
	We expect to have to tune these away at the critical point, but that they correspond to relevant deformations of the IR theory. (For each of $i=1,\ldots,N_s$ and $a=1,\ldots,N_f$, the other mass terms $|\phi_i|^2$ and $|\psi_a|^2$ are also important relevant deformations, of course, but they will break the global symmetry.)
	\item At quartic order, we again have multiple possibilities for how we contract our various indices, as discussed in \refSectionOnly{aharony_technicalities}. Suppressing gauge indices, there are four operators
	\be
	\phi^{\dagger i}\phi_{j} \ \phi^{\dagger j}\phi_{i} \,, \quad
	\phi^{\dagger i}\phi_{i} \ \phi^{\dagger j}\phi_{j} \,, \quad
	\phi^{\dagger i}\phi_{i} \ \bar{\psi}^{a}\psi_{a} \,, \quad
	\phi^{\dagger i}\psi_{a} \ \bar{\psi}^{a} \phi_{i}
	\nn\ee
	which we expect to need to think about near our IR fixed point. The fermionic ones
	\be \psi^{\dagger a}\psi_{a} \ \bar{\psi}^{b} \psi_{b} \,, \quad
	\psi^{\dagger a}\psi_{b} \ \bar{\psi}^{b} \psi_{a} \nn\ee
	are classically irrelevant, so we ignore them. Of the purely bosonic operators, we expect that the story of \refSectionOnly{aharony_technicalities} goes through again so that only the first, single-trace operator survives. Of the remaining (classically marginal) terms which mix bosons and fermions, the first is multi-trace but the second is single-trace. Dropping the multi-trace operators, then, we see that we anticipate seeing only
	\be \boxed{\phi^{\dagger i}\phi_{j} \ \phi^{\dagger j}\phi_{i} \qquad \mbox{and} \qquad \phi^{\dagger i}\psi_{a} \ \bar{\psi}^{a} \phi_{i}} \ee 
	in the IR. We will take both to be present in the potential at low energies. In particular, we will take the mixed term
	\be \lag_{\mathrm{mix}} = - \lambda_{\mathrm{mix}} (\phi^{\dagger i} \psi_a)(\bar{\psi}^a \phi_i) \ee
	to appear with a \textit{negative} coefficient $\lambda_{\mathrm{mix}} < 0$ in the IR potential.
	\item Finally, at sextic order, the only operators which are not irrelevant in the UV are the classically marginal operators \be \sim |\phi|^6 \nn\ee
	with various index contractions. Since we do not tune away the quartic operators, we expect that we need not worry about these, or indeed higher-order operators.
\end{itemize}

This means that we expect the only relevant deformations at the IR fixed point to ultimately be the mass deformations. A very similar story goes through on the $U(N)$ side, and the mapping between the IR-relevant mass operators across the duality is straightforward to guess from what we have seen before. \refSectionOnly{master_phase_diagram} contains the details of how this works.

Before we start analyzing the phases of the system, the last thing we want to do is describe how the background flavour terms work out. Here are the schematic Lagrangians of the system:
\be
\lag = |D_{a+A+D} \phi|^2 + i \bar{\psi} \slashed{D}_{a-A+C} - |\phi|^4 - |\phi|^2|\psi|^2 + k \lag_{CS}^{SU(N)} [a] + \frac{1}{2\pi} c \rmd (B - \tr_{SU(N)} a)
\ee
and
\begin{align}
\tilde{\lag}
=& |D_{b-A+C} \phi|^2 + i \bar{\psi} \slashed{D}_{b+A+D} - |\phi|^4 + |\phi|^2|\psi|^2 \nn\\
 & - (N-N_s) \lag_{CS}^{SU(k)} [b] + \frac{1}{2\pi} \left(\tr_{SU(k)} b\right) \rmd (B + N_s A) + N_s k \lag_{CS}[A]  + k \lag_{CS}^{SU(N_s)}[D] 
\end{align}
where $a,b,c$ are dynamical $U(N)$, $U(k)$ and $U(1)$ fields. We also have $A,B$ as background $U(1)$ fields, and $C,D$ as background $SU(N_f)$ and $SU(N_s)$ fields. It's quite a mouthful!

\subsection{Phase Diagram of the Flavour-Symmetric Theory}\labelSection{master_phase_diagram}

Let's start by thinking about what happens as we vary the flavour symmetric mass terms, which are given by
\be \delta \lag = - \mu_1 |\psi|^2 - \mu_2 |\phi|^2 \bigdual \delta \tilde{\lag} = - \mu_1 |\Phi|^2 + \mu_2 |\Psi|^2 \fstp \ee 
The analysis is somewhat different according to whether or not $N_s < N$ or $N_s = N$. We will do the analysis for the first case, and mostly leave the second as an exercise.

We should emphasize that we do not have much control near the proposed CFT. For example, there is no quantum number that prevents $|\phi|^2$ and $|\psi|^2$ from mixing with each other. This means that we certainly cannot assert with great confidence what happens for small perturbations $\mu_1,\mu_2$. It is still sensible to perform a classical analysis which we expect to be valid for large $|\mu_i|$; we will follow \cite{Benini:2017aed} in marking the portion of the phase diagram which we feel less confident about.

\subsubsection*{The Phases of $SU(N)_{k-N_f/2} + N_s \phi + N_f \psi$}

Let us begin by giving a large positive mass $\mu_1 \gg 0$ for the scalars of the left-hand theory. This means simply decoupling them, leaving us with
\be SU(N)_{k-N_f/2} + N_f \ \psi \ee 
whose time reversal we have already studied in \refSectionOnly{aharony}. We already understand this theory, so let us move on.

For $\mu_2 \gg 0$, we expect the scalars to condense, and take on a VEV
\be \left<\phi\right> = \labeldims{\begin{pmatrix} v & 0 & 0 & 0 & 0 \\ 0 & v & 0 & 0 & 0 \\ 0 & 0 & v & 0 & 0 \end{pmatrix}}{N}{N_s} \ee
that reduces us to a gauge group $SU(N-N_s)_{k-N_f/2}$. Now it is important that we work with $N_s < N$.

However, something interesting now happens to the $N_f$ fermions which transformed in the fundamental of $SU(N)$. Firstly, they split up into $N_s N_f$ ``singlet fermions'' $\chi_{Ia}$ -- which do not transform under the surviving gauge group -- and $N_f$ fermions $\eta_a$ which are in the fundamental of the gauge group. Secondly, the presence of $\lag_{\mathrm{mix}} = - \lambda_{\mathrm{mix}} (\phi^{\dagger i} \psi_a)(\bar{\psi}^a \phi_i)$ in the Lagrangian now generates a negative mass for the singlet fermions,
\be \lag_{\mathrm{mix}} \approx \lambda_{\mathrm{mix}} v \bar{\chi} \chi \fstp \ee
Recall that $\lambda_{\mathrm{mix}} < 0$, so this represents a positive mass.

This doesn't affect things much when $\mu_2 \gg 0$; we obtain a TQFT $SU(N-N_s)_{k-N_f}$. However, as we decrease $\mu_2$ from $0 \to -\infty$, we expect that at some point (heuristically $\mu_2 = \lambda_{\mathrm{mix}} v$) the fermions $\chi$ become massless. Since these fermions are not charged under a dynamical gauge group, this transition does not affect the IR TQFT, which is always $SU(N-N_s)_{k}$.

This means the phase diagram is predicted to take the form of \refFigureOnly{master_sun_phase}.
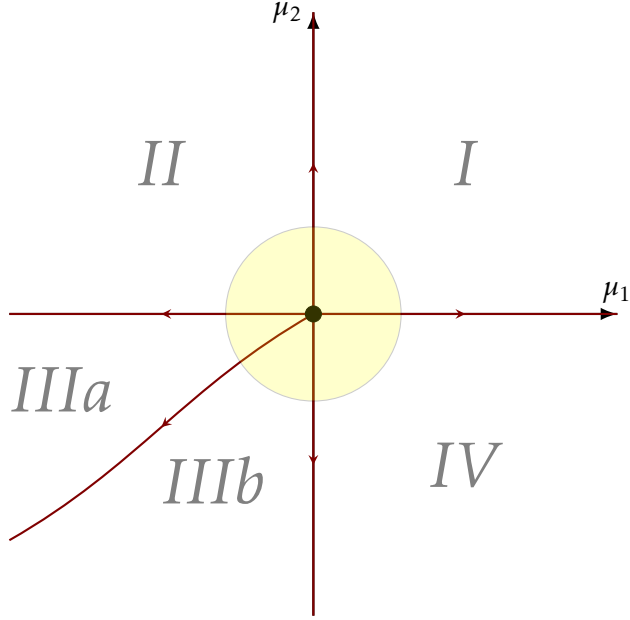
\begin{figure}
	\centering
	\begin{tikzpicture}
	\draw[axis] (-4,0)-- ++(8,0) node[above] {$\mu_1$};
	\draw[axis] (0,-4)-- ++(0,8) node[left] {$\mu_2$};
	\draw[flow,relevant] (0,0) to (4,0);
	\draw[flow,relevant] (0,0) to (-4,0);
	\draw[flow,relevant] (0,0) to (0,4);
	\draw[flow,relevant] (0,0) to (0,-4);
	\draw[flow,relevant] (0,0) to[out=-150,in=30] (-4,-3);
	\node[region label] at (2,2) {I};
	\node[region label] at (-2,2) {II};
	\node[region label] at (2,-2) {IV};
	\node[region label] at (-1.3,-2.2) {IIIb};
	\node[region label] at (-3.3,-1) {IIIa};
	\filldraw (0,0) circle (3pt); 
	\filldraw[opacity=0.2,fill=yellow] (0,0) circle (3em);
	\end{tikzpicture}
	\caption{Cartoon of the phase structure of $SU(N)_{k-N_f/2} + N_s \ \phi + N_f \ \psi$ with mass terms $-\mu_1 |\psi|^2 - \mu_2 |\phi|^2$, for $N < N_s$. The marked region represents where we are less certain of the details}\labelFigure{master_sun_phase}
\end{figure}
In summary, the TQFTs in the labelled regions are as follows.
\begin{wideitemize}
	\item[I] $SU(N)_k$
	\item[II] $SU(N)_{k-N_f}$
	\item[IIIa, IIIb] $SU(N-N_s)_{k-N_f}$
	\item[IV] $SU(N-N_s)_k$
\end{wideitemize}
The critical theories we propose describe the transitions between these phases are given by the following theories:
\begin{wideitemize}
	\item[I-II] $SU(N)_{k-N_f/2} + N_f \ \psi$
	\item[II-IIIa] $SU(N)_{k-N_f} + N_s \ \phi$
	\item[IIIa-IIIb] $SU(N-N_s)_{k-N_f} + N_f N_s \text{ singlet } \chi$
	\item[IIIb-IV] $SU(N-N_s)_{k-N_f/2} + N_f \ \eta$
	\item[IV-I] $SU(N)_{k} + N_s \ \phi$
\end{wideitemize}

\begin{exercise}[subtitle=Too Many Scalars]
	We have not yet handled the case of $N_s = N$. Argue that here, there are at most 4 distinct regions, one of which is not a TQFT. Similarly, show that there are at most four transitions, and explain what they are.
	
	What do you think happens for $N_s > N$?
\end{exercise}

\subsubsection*{The Phases of $U(k)_{-N+N_s/2} + N_f \ \Phi + N_s \ \Psi$}

There is a very similar story here, with the phase diagram taking the form of \refFigureOnly{master_un_phase}.
\begin{figure}
	\centering
	\begin{tikzpicture}
	\draw[axis] (-4,0)-- ++(8,0) node[above] {$\mu_1$};
	\draw[axis] (0,-4)-- ++(0,8) node[left] {$\mu_2$};
	\draw[flow,relevant] (0,0) to (4,0);
	\draw[flow,relevant] (0,0) to (-4,0);
	\draw[flow,relevant] (0,0) to (0,4);
	\draw[flow,relevant] (0,0) to (0,-4);
	\draw[flow,relevant] (0,0) to[out=-150,in=30] (-4,-3);
	\node[region label] at (2,2) {I'};
	\node[region label] at (-2,2) {II'};
	\node[region label] at (2,-2) {IV'};
	\node[region label] at (-1.3,-2.2) {IIIb'};
	\node[region label] at (-3.3,-1) {IIIa'};
	\filldraw (0,0) circle (3pt); 
	\filldraw[opacity=0.2,fill=yellow] (0,0) circle (3em);
	\end{tikzpicture}
	\caption{Cartoon of the phase structure of $U(k)_{N_s/2-N} + N_f \ \Phi + N_s \ \Psi$ with mass terms $-\mu_1 |\Phi|^2 + \mu_2 |\Psi|^2$ for $N < N_s$. The marked region represents where we are less certain of the details}\labelFigure{master_un_phase}
\end{figure}
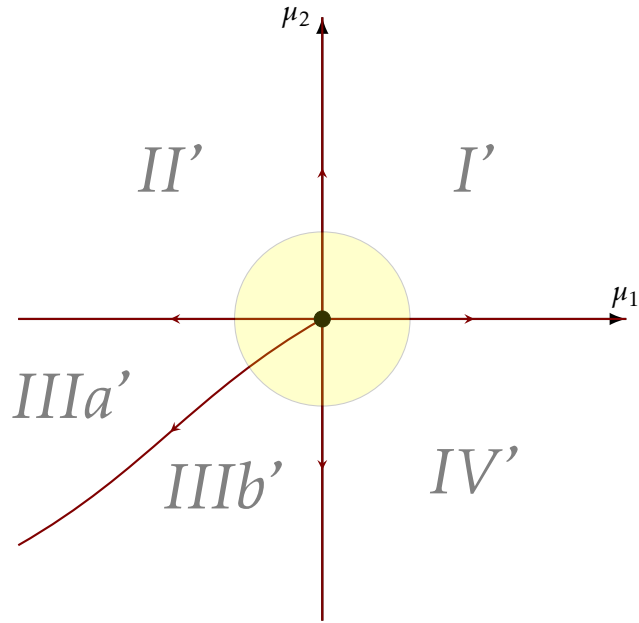

It should not be too hard to convince yourself we have the following phases for $N<N_s$:
\begin{wideitemize}
	\item[I'] $U(k)_{-N}$
	\item[II'] $U(k-N_f)_{-N}$
	\item[IIIa', IIIb'] $U(k-N_f)_{N_s-N}$
	\item[IV'] $U(k)_{N_s-N}$
\end{wideitemize}
The critical theories are as follows:
\begin{wideitemize}
	\item[I'-II'] $U(k)_{-N} + N_f \ \Phi$
	\item[II'-IIIa'] $U(k-N_f)_{N_s/2-N} + N_s \ \Psi$
	\item[IIIa'-IIIb'] $U(k-N_f)_{N_s-N} + N_f N_s \text{ singlet } \tilde{\chi}$
	\item[IIIb'-IV'] $U(k)_{N_s-N} + N_f \ \Phi$
	\item[IV'-I'] $U(k)_{N_s/2-N} + N_s \ \Psi$
\end{wideitemize}
By Aharony's dualities, these match across the duality!

\begin{exercise}[subtitle=Bits and Bobs]
	\begin{enumerate}[label=(\alph*)]
		\item Check that the Aharony dualities are reproduced with the correct background terms.
		\item Explain the flavour bound by considering first the case $N < N_s$ and $k > N_f$, and then $N = N_s$ and $k \ge N_f$.
	\end{enumerate}
\end{exercise}

\subsection{Technicalities Again}

The master duality \eqref{master_duality} can again be dressed with an almost trivial spin TQFT as follows:
\be SU(N)_{k-N_f/2} + N_s \ \phi + N_f \ \psi \longleftrightarrow U(k)_{-N+N_s/2} + N_f \ \Phi + N_s \ \Psi + U(k(N-N_s))_1 \ee

Additionally, the global symmetries can be dressed with charge conjugation and quotiented by $\mathbb{Z}_N$ to give the faithful group
\be G = \frac{U(N_f) \times U(N_s)}{\mathbb{Z}_N} \rtimes \mathbb{Z}_2^{C} \fstp \ee
We have not taken care with exceptional symmetries, like that of $SU(2)$ gauge theory.

\section{Aside: Further Generalizations}

The dualities we have seen here turn out to still be only the tip of the iceberg. There are more dualities which can be deduced from the master duality (some rigorously, some subject to additional assumptions). There are some which are independent. In this section, we will briefly outline the road towards three kinds of generalizations:

\begin{itemize}
	\item In \refSection{generalize_gauge_group}, we will briefly discuss versions of the master duality appropriate for orthogonal and symplectic groups. By and large, the story is very similar.
	\item Then, in \refSection{generalize_representation}, we will turn to the question of what happens when there is matter in other representations.
	\item Finally, in \refSection{generalize_quiver}, we will look at what happens when we consider matter transforming under product gauge groups.
\end{itemize}

\subsection{Other Gauge Groups}\labelSection{generalize_gauge_group}

We have so far worked entirely with unitary and special unitary groups, based essentially around a level-rank duality of TQFTs that looks like
\be SU(N)_k \bigdual U(k)_{-N} \fstp \ee
However, there are other level-rank dualities out there.

\subsubsection{Orthogonal Groups}

Firstly, let's look at $SO(N)$ gauge theories. Consider a gauge field $A \in so(N)$, and work in the vector representation. Then there is a Chern-Simons term which can be added to the action, 
\be k \lag_\mathrm{CS}[A] =  \ \frac{k}{8\pi} \tr \left[ A \rmd A - \frac{2i}{3}A^3 \right] \qquad \mbox{for } k \in \mathbb{Z} \cmma \ee
whose gauge invariance follows from essentially the same argument given in \refAppendixOnly{chern-simons}.

It turns out that essentially the same correspondence exists as for unitary gauge fields \cite{Aharony:2016jvv}:
\be SO(N)_k \bigdual SO(k)_{-N} \ee
This suggests, for example, that $SO(N)_1$ is trivial, since $SO(1)$ certainly is.
\footnote{
	This indeed is true, up to similar caveats to those discussed for unitary groups: $SO(N)_1$ is a trivial spin-TQFT, with two transparent lines of spin $\{0,\frac{1}{2}\}$ and a framing anomaly of $c=\frac{N}{2}$ \cite{Seiberg:2016rsg}. (In particular, $SO(1)_1$ can actually be taken to be the spin Ising CFT.) A more precise version of the level-rank duality is
	\be SO(N)_k \times SO(0)_1 \smallerdual SO(k)_{-N} \times SO(kN)_1 \ee 
	Here, $SO(0)_1$ is defined by the simple presence of two transparent lines, but the absence of any framing anomaly. It serves simply to make the left-hand side a spin TQFT. $SO(kN)_1$ gives the right-hand side a framing anomaly and makes it also a spin TQFT.
}

It is believed that these can be dressed with matter in much the same way as the unitary ones can be. We propose
\be
 SO(N)_{k} + N_f \text{ real Wilson-Fisher scalars } \phi \smallerdual SO(k)_{-N+N_f/2} + N_f \text{ real fermions } \psi
\ee
is another family of valid IR dualities, for some $(k,N,N_f)$. All the matter fields are in the vector representation of the gauge group.\footnote{The left-hand theory should strictly be supplemented with $SO(0)_1$ and the right-hand theory with $SO(k(N-N_f))_1$.} The half-integer quantization of the fermionic theory's Chern-Simons term reflects a very similar story to the unitary case; a Majorana fermion $\psi$ must be regularized in a gauge-invariant way (loosely speaking, with a negative half-integer Chern-Simons term), which introduces a time-reversal anomaly equal to a unit Chern-Simons term.

The basic global symmetry of each theory is $O(N_f)$. As usual, there are various subtleties to get right here \cite{Aharony:2013kma}. Firstly, because the gauge group is $SO(N)$, there is an ungauged $\mathbb{Z}_2 \subset O(N)$ which acts as a global symmetry on the matter. Call this $\mathbb{Z}_2$ symmetry $C$. (If we have $i=1,\ldots,N_f$ as a flavour index and $a=1,\ldots,N$ as a colour index, this transformation can be taken to be $C:\phi_{i1} \to -\phi_{i1}$ and $C:\phi_{ia}\to\phi_{ia}$ for $a>1$.) Note that one can gauge this $\mathbb{Z}_2$ factor to generate an $O(N)$ gauge theory.

Secondly, there is a monopole symmetry associated to $SO(N)$ gauge theories. Recalling our earlier discussions, monopoles are associated with the fundamental group $\pi_1(SO(N))$. For $N=2$, we have $\pi_1(SO(2)) = \pi_1(U(1)) = \mathbb{Z}$ since we can have arbitrary winding around the circle. This is nothing new -- there is a monopole symmetry $U(1)$ associated to a $U(1)$ gauge group. However, for $N>2$, this is no longer the case, and instead, $\pi_1(SO(N)) = \mathbb{Z}_2$. This means that in $SO(N)$ for $N\ge3$, we do not have operators with arbitrary monopole charge, but only $\pm 1$. This gives rise to a global monopole symmetry $\mathbb{Z}_2$.

The global symmetry is therefore $O(N_f) \times \mathbb{Z}_2 \times \mathbb{Z}_2$. We will not address questions of how much of this acts faithfully.

The flavour bound restricting the range of validity of these orthogonal dualities is thought to be a little stranger than the simple $N_f \le N$ bound of the unitary case. Instead,
\be N_f \le \begin{cases} N-2 & k=1 \\ N-1 & k=2 \\ N & k \ge 3 \end{cases} \ee
As before, these bounds emerge naturally from considering the RG flow in the presence of mass deformations.

One can flow from $(N,k,N_f)$ to either $(N,k,N_f-1)$ or $(N-1,k,N_f-1)$ by adding a mass term for a single flavour. The phenomenology is the same; the scalar may Higgs the gauge group, and the fermion can shift the Chern-Simons term. If we can integrate out $N_f > N$ fermions, the non-trivial TQFT which remains in the fermionic theory cannot be dual to the trivial gapped theory which we find on the bosonic side. This is certainly never possible.

Meanwhile, if $N_f=N$ and $k=2$, then we can flow down to the theory for $(0,2,0)$, where we find $SO(0)_2 \leftrightarrow SO(2)_0 \equiv U(1)$. This is impossible: the right-hand theory contains a gapless photon. Therefore, this theory is excluded too. Finally, if $N_f = N-1$ and $k=1$, then we can reach $(2,1,1)$ which would require a duality
\be SO(2)_1 + \text{Wilson-Fisher scalar} \stackrel{?!}{\bigdual} \text{Majorana fermion} \fstp\ee
This is surely also wrong! The left-hand theory is equivalent to $U(1)_1 + \text{complex WF scalar}$ which we know should actually be dual to a \textit{Dirac} fermion, not a \textit{Majorana} fermion. Indeed, the global symmetries do not seem to work out.

Otherwise, we ultimately reach consistent level-rank duals and we do not find matter dualities we know to be inconsistent. We hypothesize that the other dualities do in fact hold. There is plenty to say about these dualities, but we will leave this alone for now, save for a couple of exercises inspired by \cite{Aharony:2016jvv} which you might like to try out.

\begin{exercise}[subtitle=Two Orthogonal Exercises]
	\begin{enumerate}[label=(\alph*)]
		\item Consider the case $N=k=2$ and $N_f=1$. Show that this described the same $U(1)_2 + \phi \leftrightarrow U(1)_{-3/2} + \psi$ fixed point discussed at length in \refSectionOnly{so3_symmetry}. By considering the global symmetries, argue that the duality is realized in a different way by the orthogonal duality to the $U(1)$ duality. Finally, show that this is consistent with the enhanced IR symmetry of these theories.
		\item Derive a duality
		\be SU(N)_1 + N_f \text{ complex scalars} \bigdual SO(N)_2 + N_f \text{ real scalars} \ee
		and state what range of $N_f$ this should hold for. What goes wrong for $N_f = N$?
	\end{enumerate}
\end{exercise}

\subsubsection{Symplectic Groups}

There is also a story to be told for the symplectic groups defined by
\be USp(2N) = SU(2N) \cap Sp(2N,\mathbb{C}) \fstp \ee
The fundamental level-rank pair here is
\be USp(2N)_k \bigdual USp(2k)_{-N} \ee
which is again very similar. Since symplectic groups are likely less familiar, it is worth mentioning there are some accidental isomorphisms for small ranks:
\be USp(2)_k \cong SU(2)_k \qquad \mbox{and} \qquad USp(4)/\mathbb{Z}_2 \cong SO(5) \fstp \ee
A further useful fact is that these groups are simply connected, and hence do not support monopole operators.

Now we can also add matter to these level-rank duals. We will add matter in the fundamental $\textbf{2N}$ representation of $USp(2N)$. This is another example of a pseudoreal (or \textit{quaternionic}) representation, which will be useful in a moment. For now, let us state the duality:%
\footnote{
	The missing trivial TQFTs can be added to give the pure level-rank duality
	\be USp(2N)_k \times SO(0)_1 \bigdual USp(2k)_{-N} \times SO(4kN)_1 \ee
	or
	\be USp(2N)_k \times SO(0)_1 + N_f \phi \bigdual USp(2k)_{-N} \times SO(4k(N-N_f))_1 + N_f \psi \fstp \ee
}
\be
 USp(2N)_k + N_f \text{ scalars with $\phi^4$ } \phi \smallerdual USp(2k)_{-N+N_f/2} + N_f \text{ fermions } \psi
\ee
This we claim has the more familiar bound of $N_f \ge N$.

Now using pseudoreality, one can actually write the $2N$ complex components of each $\textbf{2N}$ representation as $4N$ complex scalars subject to a sort of reality condition. Taking $a=1,\ldots,2N$ and $i=1,\ldots,2N_f$, we can impose
\be \varphi_{ai}\Omega^{ab}\tilde{\Omega}^{ij} = \varphi^\dagger_{bj} \ee
where $\Omega,\tilde{\Omega}$ are invariant symplectic tensors of $USp(2N)$ and $USp(2N_f)$ respectively. Taking account of the fact that the center $\mathbb{Z}_2 = USp(2N) \cap USp(2N_f)$ of both groups is gauged, this makes clear that there is in fact a $USp(2N_f)/\mathbb{Z}_2$ global symmetry in this theory. There is no monopole symmetry of course, and charge conjugation is swallowed up in the global symmetry by the pseudoreality condition. Note that there are potentials which do not respect the full global symmetry, so in general we need to observe the full $USp(2N_f)$.

Notice that $N=k=N_f=1$ reproduces the dualities of \refSectionOnly{so3_symmetry} yet again. One can also derive various other interesting results using the $USp(2) \cong SU(2)$ isomorphism, such as
\be USp(2k)_{-1/2} + \psi \bigdual U(k)_{-3/2} + \tilde{\psi} \ee
which implies an emergent $SU(2)$ symmetry in the right-hand theory; and similarly for $N_f \le N$ we have
\be USp(2N)_{1} + N_f \phi \bigdual U(N)_{2} + N_f \tilde{\phi} \ee
which implies the $U(N_f) \rtimes \mathbb{Z}_2$ symmetry of the right-hand side is enhanced to $USp(2N_f)$.

Another more intriguing possibility arises from considering a possibly $SO(5)$ symmetric theory, but we will not discuss this here.


\subsection{Other Representations and Adjoint QCD}\labelSection{generalize_representation}

Through what we have discussed, we have almost always been looking at matter transforming in the fundamental representation of whatever group we had. (The exceptions are the $U(1)$ fields of charge greater than 1.) But there is plenty to say about theories with matter in other representations.

The most interesting example which has been studied in the literature is the case of \textit{adjoint QCD} \cite{Gomis:2017ixy}:
\be SU(N)_k + \text{ adjoint Majorana fermion } \lambda \fstp \ee
Notice that because the adjoint representation of $SU(N)$ is real, we have a choice as to whether we couple a real or complex matter field to the gauge field. We will consider the case of a real fermion.

\subsubsection{Fermions in General Representations}

We need to know how fermions in the adjoint representation shift the Chern-Simons level. We might as well quote the general result for a representation $R$ of a group $G$, generalizing the discussion of \refSectionOnly{fermion_det_parity_anom} \cite{Atiyah:1975jf,Aharony:2016jvv}. The fermion determinant is
\be \det\!' \mathcal{D} =
\begin{cases}
	|\det \mathcal{D}| \exp \left( -\frac{i \pi \eta}{2} \right) & R \text{ is a complex representation} \\
	|\det \mathcal{D}| \exp \left( -\frac{i \pi \eta}{4} \right) & R \text{ is a real representation}
\end{cases}
\ee
where the APS index theorem now tells us that
\be \pi \eta = \int 2x_R \lag_{\mathrm{CS}} + 2 \dim R \mathrm{CS}_\mathrm{grav} \pmod {2\pi} \ee
and we need to explain what the \textit{Dynkin index} $x_R$ is. The concisest definition is given by computing the trace $\tr_R [t^A_R t^B_R] = 2 x_R \delta^{AB}$ where $t^A_R$ are the generators of the gauge group in the representation $R$. They are normalized so that $\tr [t^A t^B] = \delta^{AB}$ in the representation used to compute $\lag_{\mathrm{CS}} = \frac{1}{4\pi} \tr [ A \rmd A - \frac{2i}{3}A^3 ]$. There is a relation between this and the quadratic Casimir $C_2(R)$, given by
\be x_R = \frac{\dim R \cdot C_2(R)}{2 \dim G} \fstp \ee
For a few examples with the conventional normalizations in each case:
\begin{subequations}
\begin{align}
 SU(N): & \qquad  x_{\text{charge }q} = \frac{q}{2} \\
 SU(N): & \qquad  x_{\text{fundamental}} = \frac{1}{2}  && x_\text{adjoint} = N \\
 SO(N): & \qquad  x_{\text{vector}} = 1  && x_\text{adjoint} = N-2 \\
 USp(2N): & \qquad  x_{\text{fundamental}} = \frac{1}{2}  && x_\text{adjoint} = N+1
\end{align}
\end{subequations}
(The index $x_\text{adjoint} \equiv h$ is known as the dual Coxeter number of $G$.)

Integrating out a single fermion of mass $m$ therefore shifts the Chern-Simons level by $\sign(m) x_R$ for complex representations or $\frac{\sign(m)}{2}x_R$ for real representations.

\subsubsection{Adjoint QCD}

Returning to adjoint QCD, we have a real fermion in the adjoint of $SU(N)$, and hence integrating it out therefore shifts
\be SU(N)_k + \lambda \to \begin{cases} SU(N)_{k+\frac{N}{2}} \\ SU(N)_{k-\frac{N}{2}} \end{cases}\ee
which means that $k$ must be an integer for even $N$ and a half-integer for odd $N$. We will take $k \ge 0$ without loss of generality.

This also tells us what the limiting TQFTs are for large values of the fermion mass. But, as in \refSectionOnly{qcd3}, the middle of the phase diagram is in principle more complicated. What happens as we vary the mass? Let's follow the analysis of \cite{Gomis:2017ixy}.

The first point of interest is that the matter content of adjoint QCD is precisely the content of $\mathcal{N} = 1$ pure supersymmetric gauge theory, with the fermion $\lambda$ being the gaugino in the vector multiplet. This means that there is at least one special point in the phase diagram: the supersymmetric point!

Neglecting auxiliary fields, the supersymmetric Chern-Simons Lagrangian is
\begin{align}
k\lag_{\mathrm{CS}}^{\mathcal{N}=1} = \frac{k}{4\pi} \tr\left(a \rmd a - \frac{2i}{3} + \bar{\chi}\chi\right) 
\lag_V = \frac{1}{g^2} \tr\left(-\frac{1}{4} f^2 + i\bar{\chi}\slashed{D}\chi\right)
\end{align}
so writing $\chi=g\lambda$, we conclude that the supersymmetric fermion mass sits at $m = m_\text{SUSY}\sim-kg^2$. Let's define 
\be m = m_\text{SUSY} + \mu\ee
to put the SUSY point at $\mu = 0$.

Notice that for large $k$, the mass of the dynamical fields are very large. We can therefore trust a semi-classical analysis: we integrate out the fermion with a large negative mass $m \ll -g^2$, leaving the theory $SU(N)_{k-N/2}$. In fact, it is believed that the theory behaves in exactly this way for all $k \ge N/2$ \cite{Witten:1999ds}, with supersymmetry being unbroken at low energies. We propose that the phase diagram looks like \refFigureOnly{adjoint_qcd_1}.
\begin{center}
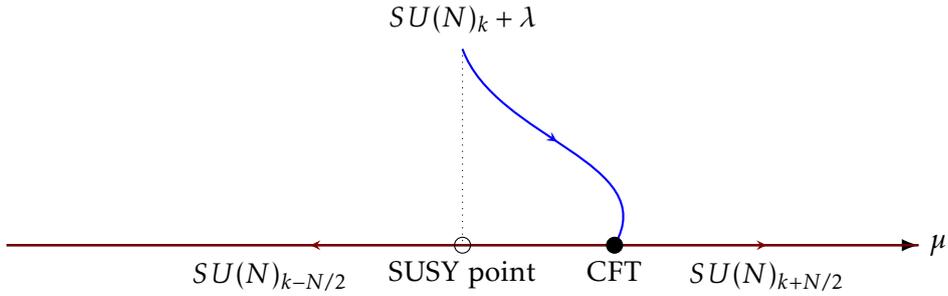

	\centering
	\begin{tikzpicture}
	\node at (0,3) {$SU(N)_{k} + \lambda$};
	\draw[flow,tuned] (0,2.6) to[out=-70,in=60] (2,0);
	\draw[dotted] (0,2.6) -- (0,0);
	
	\draw[axis] (-6,0) -- ++ (12,0) node[right] {$\mu$};
	
	\draw[flow,relevant] (2,0) -- node[below,yshift=-0.2em] {$SU(N)_{k+N/2}$} (6,0);
	\draw[flow,relevant] (2,0) -- node[below,yshift=-0.2em,xshift=-1.4em] {$SU(N)_{k-N/2}$} (-6,0);
	\filldraw (2,0) circle (3pt) node[below,yshift=-0.2em] {CFT};
	\draw (0,0) circle (3pt) node[below,yshift=-0.2em] {SUSY point};
	\end{tikzpicture}
	\captionof{figure}{The proposed phase diagram of adjoint QCD for $k \ge N/2$, the range for which SUSY is does not spontaneously break. There is no phase transition at the SUSY point; it is simply a gapped theory whose massive excitations are organized into SUSY multiplets. Assuming the other transition is second-order, there is a CFT at some value of $\mu \sim kg^2$}\labelFigure{adjoint_qcd_1}
\end{center}

Let us now consider $k < N/2$. In this range, it is believe that at low energies, SUSY is broken \cite{Witten:1999ds}. At the supersymmetric point, this means there must be a massless Majorana fermion mode which transforms under the broken symmetry. We call this the \textit{Goldstino}. The minimal proposal encapsulating this possibility is illustrated in \refFigureOnly{adjoint_qcd_2}. However, we suggest this happens only at one value of $k$, specifically $k=N/2-1$.
\begin{center}
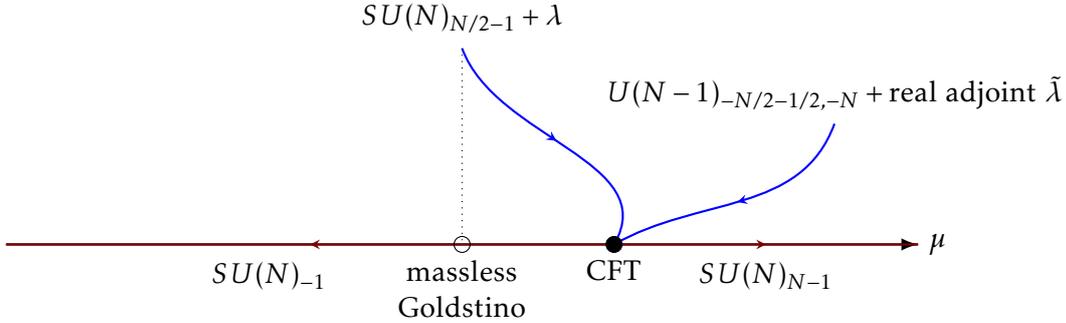

	\centering
	\begin{tikzpicture}
	\node at (0,3) {$SU(N)_{N/2-1} + \lambda$};
	\draw[flow,tuned] (0,2.6) to[out=-70,in=60] (2,0);
	\draw[dotted] (0,2.6) -- (0,0);
	
	\node at (4.9,2) {$U(N-1)_{-N/2-1/2,-N} + \text{real adjoint } \tilde{\lambda}$};
	\draw[flow,tuned] (4.9,1.6) to[out=-110,in=30] (2,0);
	
	\draw[axis] (-6,0) -- ++ (12,0) node[right] {$\mu$};
	
	\draw[flow,relevant] (2,0) -- node[below,yshift=-0.2em] {$SU(N)_{N-1}$} (6,0);
	\draw[flow,relevant] (2,0) -- node[below,yshift=-0.2em,xshift=-1.4em] {$SU(N)_{-1}$} (-6,0);
	\filldraw (2,0) circle (3pt) node[below,yshift=-0.2em] {CFT};
	\draw (0,0) circle (3pt) node[below,yshift=-0.2em,align=center] {massless \\ Goldstino};
	\end{tikzpicture}
	\captionof{figure}{The proposed phase diagram of adjoint QCD for $k = N/2-1$. There is a massless Goldstino for spontaneously broken SUSY at $\mu=0$. Again, assuming the other transition is second-order, there is a CFT at some value of $\mu \sim kg^2$. We also give a proposed dual of this theory}\labelFigure{adjoint_qcd_2}
\end{center}
Assuming that there is a CFT sitting at the transition point at $\mu \sim k g^2$, there is a plausible dual of this point:
\be \text{CFT transition} \bigdual U(N-1)_{-N/2-1/2,-N} + \text{real adjoint } \tilde{\lambda} \fstp  \ee
We will come back to this in a moment.

Why should this not hold for all $k$? Well, let us look at theories with even $N$ so that $k=0$ is permitted. This theory is time-reversal invariant, and at the special point $\mu = m = 0$, we have both time-reversal symmetry (which is preserved in the IR, because of a result known as the Vafa-Witten theorem which states that ``in parity-conserving vectorlike theories such as QCD, parity conservation is not spontaneously broken'' \cite{Vafa:1984xg}) and supersymmetry (which should be spontaneously broken \cite{Witten:1999ds}).

Now suppose there was only a single transition in this theory. Then by symmetry, the transition sits at the special point $\mu=m=0$. It mediates between $SU(N)_{-N/2}$ and $SU(N)_{+N/2}$. Since these are distinct except for the case $N=2$ (and then we are in the $k=N/2-1$ case from above again), there would have to be a complicated interacting field theory sitting at the transition. No such theory is known.

Instead, it is suggested that the picture of \refFigureOnly{adjoint_qcd_3} could be correct: we have two transitions, and an intermediate quantum phase, much like that of normal QCD discussed in \refSectionOnly{qcd3}. This time, it is not gapless (except at the supersymmetric point where a Goldstino briefly rears its head). Instead, it is described by a new gapped TQFT. It turns out that there are natural theories which could describe the transitions between that intermediate phase and the two asymptotic phases. These are shown in the picture as duals flowing to the two (tentative) CFTs at those transitions.

\begin{center}
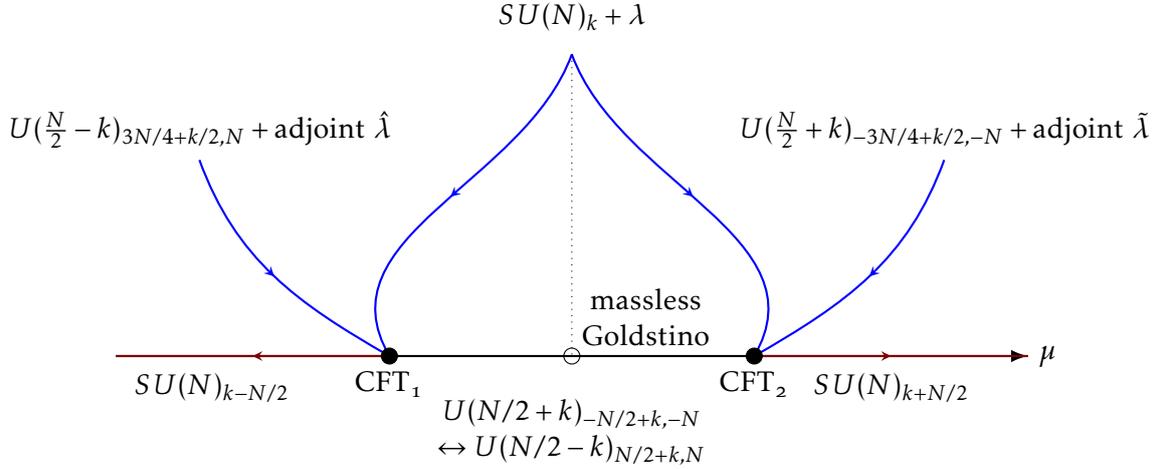

	\centering
	\begin{tikzpicture}
	\node at (0,4.5) {$SU(N)_{k} + \lambda$};
	\draw[flow,tuned] (0,4) to[out=-70,in=60] (2.4,0);
	\draw[flow,tuned] (0,4) to[out=-110,in=120] (-2.4,0);
	\draw[dotted] (0,4) -- (0,0);
	
	\node at (4.9,3) {$U(\frac{N}{2}+k)_{-3N/4+k/2,-N} + \text{adjoint } \tilde{\lambda}$};
	\draw[flow,tuned] (4.9,2.6) to[out=-110,in=30] (2.4,0);
	\node at (-4.9,3) {$U(\frac{N}{2}-k)_{3N/4+k/2,N} + \text{adjoint } \hat{\lambda}$};
	\draw[flow,tuned] (-4.9,2.6) to[out=-70,in=150] (-2.4,0);
	
	\draw[axis] (-6,0) -- ++ (12,0) node[right] {$\mu$};
	
	\draw[flow,relevant] (2.4,0) -- node[below,yshift=-0.2em] {$SU(N)_{k+N/2}$} (6,0);
	\draw[flow,relevant] (-2.4,0) -- node[below,yshift=-0.2em,xshift=-1.4em] {$SU(N)_{k-N/2}$} (-6,0);
	
	\filldraw (2.4,0) circle (3pt) node[below,yshift=-0.2em] {CFT\textsubscript{2}};
	\filldraw (-2.4,0) circle (3pt) node[below,yshift=-0.2em] {CFT\textsubscript{1}};
	\draw (0,0) circle (3pt) node[above right,align=center] {massless \\ Goldstino};
	\node[align=center] at (0,-1) {$U(N/2+k)_{-N/2+k,-N}$ \\ $ \leftrightarrow U(N/2-k)_{N/2+k,N}$};
	\end{tikzpicture}
	\captionof{figure}{The proposed phase diagram of adjoint QCD for $k < N/2-1$. There is a massless Goldstino for spontaneously broken SUSY at $\mu=0$. There are assumed to be two second-order transitions either side of this, each with a distinct dual theory. The phase between these is described by a new gapped TQFT (apart from at the SUSY point)}\labelFigure{adjoint_qcd_3}
\end{center}
This actually makes for a fairly easy-to-understand phase diagram: just understanding the semi-classical physics of the two dual theories matches the TQFTs and so forth perfectly. All of the symmetries work out straightforwardly because of level-rank duality.

The key parts of this claim are that there are two different fixed points we can tune to by looking at larger or smaller $\mu$, and that they can also be reached by tuning one of two different dual theories:
\begin{align}
SU(N)_{k} + \text{ adjoint } \lambda &\stackrel{\text{larger }\mu}{\bigdual} U\left(\frac{N}{2}+k \right)_{-3N/4+k/2,-N} + \text{adjoint } \tilde{\lambda}_+ \\
SU(N)_{k} + \text{ adjoint }\lambda &\stackrel{\text{smaller }\mu}{\bigdual} U\left(\frac{N}{2}-k \right)_{3N/4+k/2,N} + \text{adjoint } \tilde{\lambda}_-
\end{align}
Note that the $U(1)$ factors on the right decouple completely from their respective theories. Hence the interacting part of each dual theory is given by
\be SU(N_\pm)_{\mp k_\pm} + \text{adjoint fermion }\lambda_\pm \qquad \mbox{where } N_\pm = N/2 \pm k \mbox{ and } k_\pm = N_\pm/2 + (N/2 \mp k) \fstp \ee
Hence, assuming $k < N/2 - 1$, $k_\pm \ge N_\pm/2$. This means that each dual theory lies in the regime of \refFigureOnly{adjoint_qcd_1}, near the single CFT point. (Of course, one should also include the decoupled $U(1)$ sector.) The duality is however not valid far from that point. We do not generally expect the other points of interest to be accessible in under RG flow from this point.

Note that at $k=0$, the diagram must have time reversal symmetry. To see this, one needs the non-trivial result that actually 
\be U(N/2)_{-N/2,-N} \bigdual U(N/2)_{N/2,N} \ee
is time-reversal invariant \cite{hsinseiberg}.

\begin{exercise}[subtitle=The $SU(2)_0$ Case]
	Argue that no phase transition is needed in the case of $SU(2)_0$ plus an adjoint real fermion (although there is still a massless Goldstino at $m=\mu=0$). This suggests that the interacting theory flows in the IR to a free theory.
	\be SU(2)_0 + \text{adjoint } \lambda \qquad \to \qquad U(1)_2 \text{ TQFT } + \text{ neutral Majorana fermion } \tilde{\lambda} \ee
\end{exercise}

\subsection{Quiver Theories}\labelSection{generalize_quiver}

Finally, just as in \refSectionOnly{abelian_quiver}, it is possible to construct \textit{quiver theories} with product gauge groups. However, armed with the technology of the master duality of \refSectionOnly{master_duality}, we can derive some more interesting quivers! We shall follow the approach of \cite{Jensen:2017dso,Aitken:2018cvh}.

\subsubsection{Two Node Quivers}

The basic technique we will use is the familiar approach of gauging global symmetries. Let's start with a simple example, using only Aharony's duality
\be U(N)_{k} + N_f \text{ } \phi \bigdual SU(k)_{-N+N_f/2} + N_f \text{ } \psi \ee
and remembering that there is an explicit level $k$ Chern-Simons term for the $U(N_f)$ global symmetry on the right-hand side. Suppose that we add  $k_2$ to the Chern-Simons term for this group on both sides before gauging it. Then if we call $(N_1,k_1)=(N,k)$ and $(N_2,k_2)=(N_f,k_2)$, we find that we gain a theory
\be U(N_1)_{k_1} \times U(N_2)_{k_2} + \text{ bifund. } \phi \smallerdual SU(k_1)_{-N_1+N_2/2} \times U(N_2)_{k_2 + k_1/2} + \text{ bifund. } \psi \label{simplest_quiver} \ee
where as usual the scalars have $|\phi|^4$ interactions. There is now only one species of matter on both sides, but it transforms in the fundamental of both gauge groups (i.e. the bifundamental). We stress that this only holds for $N_2 \le N_1$. We will also assume all variables are positive.

As a quiver diagram, we can draw this as \refFigureOnly{simplest_quiver}. We think of this duality as \textit{dualizing the first node of the quiver}. Starting from the left-hand theory, we see that the effect is essentially level-rank duality at that node; combined with a change of the statistics of its charged matter; and finally a shift of the Chern-Simons level at the connected nodes.
\begin{figure}
	\centering
	\begin{tikzpicture}[every node/.style={quiver gauge},thick]
	\node(k1) at (0,0) {$U(N_1)$};
	\node[quiver level] at (k1.south east) {$k_1$};
	\node(k2) at (3,0) {$U(N_2)$};
	\node[quiver level] at (k2.south east) {$k_2$};
	\draw(k1.east)--(k2.west);
	
	\draw[<->] (4.5,0) -- (5.5,0);
	
	\node(kk1) at (7,0) {$SU(k_1)$};
	\node[quiver level] at (kk1.south east) {$-N_1+\frac{N_2}{2}$};
	\node(kk2) at (10,0) {$U(N_2)$};
	\node[quiver level] at (kk2.south east) {$k_2+\frac{k_1}{2}$};
	\draw[dashed] (kk1.east)--(kk2.west);
	
	\draw[<->] (8.5,-1.5) -- (8.5, -2.5);
	
	\node(kkk1) at (7,-3.5) {$SU(k_1)$};
	\node[quiver level] at (kkk1.south east) {$-N_1+N_2$};
	\node(kkk2) at (10,-3.5) {$SU(k_1+k_2)$};
	\node[quiver level] at (kkk2.south east) {$-N_2$};
	\draw (kkk1.east)--(kkk2.west);

	\end{tikzpicture} 
	\caption{The dual quivers of \eqref{simplest_quiver} and \eqref{simplest_quiver_more}. We adopt the convention that a straight line denotes a Wilson-Fisher boson, whilst a dashed line refers to a fermion. We think of the relationships as dualizing the first node (requiring $N_2 \le N_1$) and then the second node (requiring $k_1 \le N_2$) of the quiver}\labelFigure{simplest_quiver}
\end{figure}
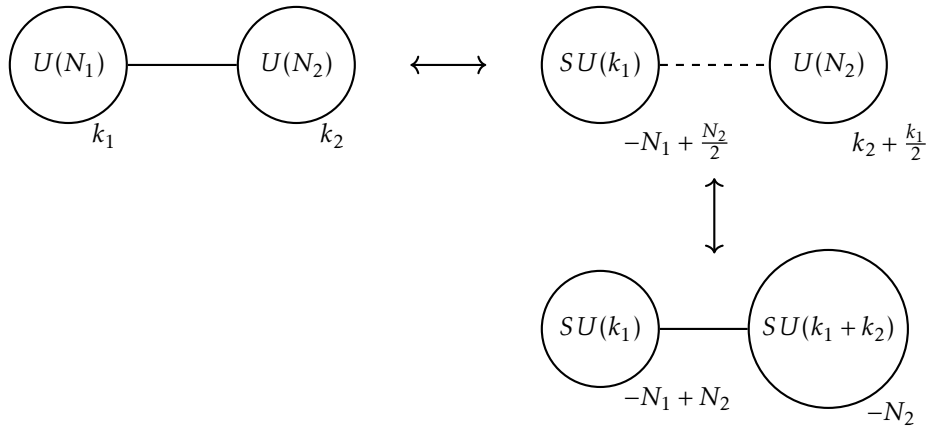

We could also now dualize the second node of the quiver using a similar approach. This results in a third dual theory,
\begin{align}
U(N_1)_{k_1} \times U(N_2)_{k_2} + \text{ bifund. } \phi
&\smallerdual SU(k_1)_{-N_1+N_2/2} \times U(N_2)_{k_2 + k_1/2} + \text{ bifund. } \psi \nn\\
&\smallerdual SU(k_1)_{-N_1+N_2} \times SU(k_1+k_2)_{-N_2} + \text{ bifund. } \tilde{\phi}
\label{simplest_quiver_more}
\end{align}
provided that not only $N_2 \le N_1$ but also $k_1 \le N_2$.

Note that if we also impose $N_1=N_2=N$, then the first theory can be dualized at the second node instead. In this case, we obtain the top and left-hand edges of \refFigureOnly{simplest_quiver_special_case}. The other edges can be obtained by further dualizations of course, one requiring that $k_1 \le N$ and the other that $k_2 \le N$. However, since we can follow the dualities around the loop either way, it must be that \textit{both} dualities actually hold when \textit{either} $k_1 \le N$ \textit{or} $k_2 \le N$.

\begin{figure}
	\centering
	\begin{tikzpicture}[every node/.style={quiver gauge},thick]
	\node(k1) at (0,0) {$U(N)$};
	\node[quiver level] at (k1.south east) {$k_1$};
	\node(k2) at (3,0) {$U(N)$};
	\node[quiver level] at (k2.south east) {$k_2$};
	\draw(k1.east)--(k2.west);
	
	\draw[<->] (4.5,0) -- (5.5,0);
	
	\node(kk1) at (7,0) {$SU(k_1)$};
	\node[quiver level] at (kk1.south east) {$-\frac{N}{2}$};
	\node(kk2) at (10,0) {$U(N)$};
	\node[quiver level] at (kk2.south east) {$k_2+\frac{k_1}{2}$};
	\draw[dashed] (kk1.east)--(kk2.west);
	
	\draw[<->] (8.5,-1.5) -- (8.5, -2.5);
	
	\node(kkk1) at (7,-3.5) {$SU(k_1)$};
	\node[quiver level] at (kkk1.south east) {$0$};
	\node(kkk2) at (10,-3.5) {$SU(k_1+k_2)$};
	\node[quiver level] at (kkk2.south east) {$-N$};
	\draw (kkk1.east)--(kkk2.west);
	
	\draw[<->] (1.5,-1.5) -- (1.5, -2.5);
	
	\node(kkkk1) at (0,-3.5) {$U(N)$};
	\node[quiver level] at (kkkk1.south east) {$k_1+\frac{k_2}{2}$};
	\node(kkkk2) at (3,-3.5) {$SU(k_2)$};
	\node[quiver level] at (kkkk2.south east) {$-\frac{N}{2}$};
	\draw[dashed] (kkkk1.east)--(kkkk2.west);
	
	\draw[<->] (4.5,-3.5) -- (5.5, -3.5);
	
	\end{tikzpicture} 
	\caption{This illustrates the special case of \refFigureOnly{simplest_quiver} when $N_1=N_2=N$. The top and left-hand dualities hold for all $(k_1,k_2,N)$ but the right-hand duality requires $k_1 \le N$ and the left-hand duality requires $k_2 \le N$}\labelFigure{simplest_quiver_special_case}
\end{figure}
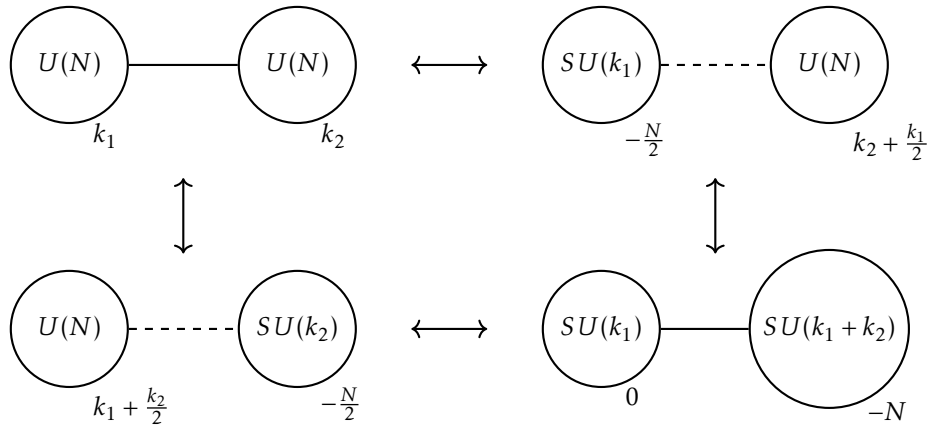

As a further note, if we assume that the theory of $SU(k_1)_0$ confines so that all matter fields form singlets, there is a natural further ansatz for the IR behaviour of the bottom-right theory in \refFigureOnly{simplest_quiver_special_case}. Writing $\phi_{a\alpha}$ where $a=1,\ldots,k_1$ and $\alpha=1,\ldots,k_1+k_2$, then it is plausible that the only light matter field is the simplest singlet operator, $T_\alpha^\beta = \phi_{a\alpha} \bar{\phi}^{a\beta}$. This is certainly only an educated guess, but it makes for an interesting conclusion:
\be  SU(k_1)_{0} \times SU(k_1+k_2)_{-N} + \text{ bifund. } \phi \qquad \to \qquad SU(k_1+k_2)_{-N} + \text{ adjoint } \tilde{\phi} \fstp \ee

%

\subsubsection{Many Nodes}

Even just with Aharony's dualities, we can do more:

\begin{exercise}[subtitle=Many Node Dualities from Aharony]
	\begin{enumerate}[label=(\alph*)]
		\item Firstly, establish the duality
		
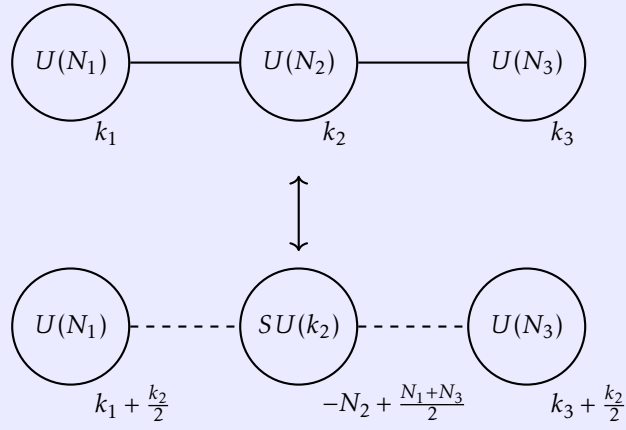
\begin{center}
			\centering
			\begin{tikzpicture}[every node/.style={quiver gauge},thick]
			\node(k1) at (0,0) {$U(N_1)$};
			\node[quiver level] at (k1.south east) {$k_1$};
			\node(k2) at (3,0) {$U(N_2)$};
			\node[quiver level] at (k2.south east) {$k_2$};
			\node(k3) at (6,0) {$U(N_3)$};
			\node[quiver level] at (k3.south east) {$k_3$};
			\draw(k1.east)--(k2.west);
			\draw(k2.east)--(k3.west);
	
			\draw[<->] (3,-1.5) -- (3, -2.5);
		
			\node(kk1) at (0,-3.5) {$U(N_1)$};
			\node[quiver level] at (kk1.south east) {$k_1+\frac{k_2}{2}$};
			\node(kk2) at (3,-3.5) {$SU(k_2)$};
			\node[quiver level] at (kk2.south east) {$-N_2+\frac{N_1+N_3}{2}$};
			\node(kk3) at (6,-3.5) {$U(N_3)$};
			\node[quiver level] at (kk3.south east) {$k_3+\frac{k_2}{2}$};
			\draw[dashed] (kk1.east)--(kk2.west);
			\draw[dashed] (kk2.east)--(kk3.west);
			\end{tikzpicture}
		\end{center}
		by dualizing the central node. You can assume $N_2 \ge N_1 + N_3$. Explain what assumption must be made about the symmetry breaking patterns when the scalars get a mass.
	\item Secondly, establish this duality, valid for $2N_2 \le N_1$:
	
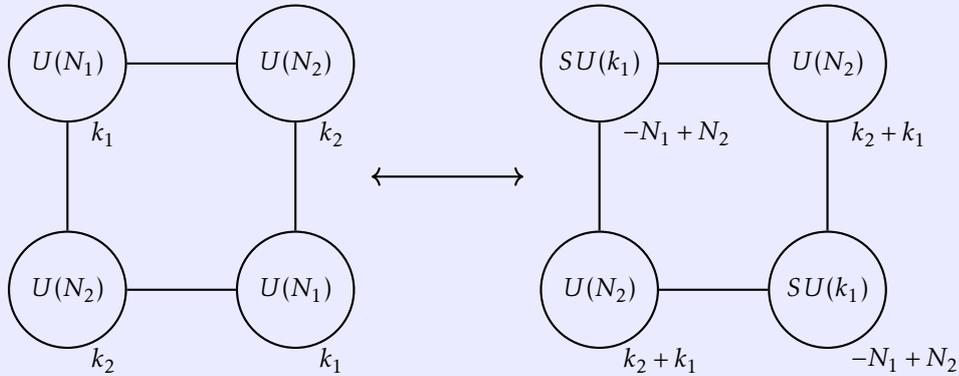
\begin{center}
		\centering
		\begin{tikzpicture}[every node/.style={quiver gauge},thick]
		\node(k1) at (0,0) {$U(N_1)$};
		\node[quiver level] at (k1.south east) {$k_1$};
		\node(k2) at (3,0) {$U(N_2)$};
		\node[quiver level] at (k2.south east) {$k_2$};
		\node(k3) at (3,-3) {$U(N_1)$};
		\node[quiver level] at (k3.south east) {$k_1$};
		\node(k4) at (0,-3) {$U(N_2)$};
		\node[quiver level] at (k4.south east) {$k_2$};
		\draw(k1.east)--(k2.west);
		\draw(k2.south)--(k3.north);
		\draw(k3.west)--(k4.east);
		\draw(k4.north)--(k1.south);
		
		\draw[<->] (4,-1.5) -- (6, -1.5);
		
		\node(kk1) at (7,0) {$SU(k_1)$};
		\node[quiver level] at (kk1.south east) {$-N_1+N_2$};
		\node(kk2) at (10,0) {$U(N_2)$};
		\node[quiver level] at (kk2.south east) {$k_2+k_1$};
		\node(kk3) at (10,-3) {$SU(k_1)$};
		\node[quiver level] at (kk3.south east) {$-N_1+N_2$};
		\node(kk4) at (7,-3) {$U(N_2)$};
		\node[quiver level] at (kk4.south east) {$k_2+k_1$};
		\draw(kk1.east)--(kk2.west);
		\draw(kk2.south)--(kk3.north);
		\draw(kk3.west)--(kk4.east);
		\draw(kk4.north)--(kk1.south);
		\end{tikzpicture}
	\end{center}
	\end{enumerate}
\end{exercise}	

However, the master duality of \refSectionOnly{master_duality} gives us the power to do even more elaborate things, because we can dualize theories with both scalars and fermions, and with more flavours of matter in total. In particular, we can start with a linear bosonic quiver, then dualize the nodes one at a time from left to right. Since every edge is attached to two nodes, it effectively gets dualized twice, and we end up with a new linear bosonic quiver; you are asked to work through this as an exercise. A similar trick works for fermions.

\begin{exercise}[subtitle=Quivers from Master Dualities]
	Starting from a quiver
	\be SU(N_1)_{-k_1} \times SU(N_2)_{-k_2} \times SU(N_3)_{-k_3} \times SU(N_4)_{-k_4} \ee
	with purely bosonic matter, and applying Aharony's duality on the first node, derive a theory whose gauge group is
	\be U(k_1)_{N_1-N_2/2} \times SU(N_2)_{-k_2-k_1/2} \times SU(N_3)_{-k_3} \times SU(N_4)_{-k_4} \fstp \ee
	Without worrying about the details of the way the interactions must be tuned, now apply the master duality to the second node to obtain the gauge group
	\be U(k_1)_{N_1-N_2} \times U(k_1+k_2)_{N_2-N_3/2} \times SU(N_3)_{-k_3-k_1/2-k_2/2} \times SU(N_4)_{-k_4} \ee	
	and to the third to get
	\be U(k_1)_{N_1-N_2} \times U(k_1+k_2)_{N_2-N_3} \times U(k_1+k_2+k_3)_{N_3-N_4/2} \times SU(N_4)_{-k_4+k_1/2+k_2/2+k_3/2} \fstp \ee
	Finally, apply Aharony's dualities one last time to derive a purely bosonic theory
	\be U(k_1)_{N_1-N_2} \times U(k_1+k_2)_{N_2-N_3} \times U(k_1+k_2+k_3)_{N_3-N_4} \times U(k_1+k_2+k_3+k_4)_{N_4} \fstp \ee
	You should find this is valid for
	\be k_i \ge 0 \quad \mbox{and} \qquad N_1 \ge N_2 \ge N_3 \ge N_4 \fstp \ee
	How do the global symmetries match?
	
	State how the duality of the first and last theories generalizes to higher numbers of nodes. Now set $N_1 = N_2 = \cdots = N_n = N$ and $k_i = 1$ for all $i=1,\ldots,n$, and write down the resulting duality. You should find a theory with many massless Goldstone bosons under various mass deformations. Identify a term which could be added to the potential to break the symmetry protecting these modes. What is the dual of this term, and how does the elimination of the massless modes work work in the dual theory? Finally, you should find some confining nodes. Come up with a simple guess for the low-energy description of the theory.
\end{exercise}

These results have interesting applications to $SU(N)$ QCD in 3+1 dimensions \cite{Aitken:2018cvh}, since this theory is believed to support domain walls on which theories like $SU(N)_k$ live \cite{Gaiotto:2017tne}. In particular, one sees that there is the possibility of an interesting phase transition between $k$ separate theories $\left[SU(N)_1\right]^k$ living on widely separated domain walls, and a single theory $SU(N)_k$ when the walls coincide.  This can be described by an $\left[ SU(N)_1 \right]^k$ theory with bifundamental matter, and dualized using the above results.

\part{Evidence and Interrelations}
\labelPart{evidence}
\chapter{Supersymmetry Breaking}
\labelChapter{susybreaking}

\begin{introduction}
  We briefly review how supersymmetric dualities are related to the non-supersymmetric cases studied above.
\end{introduction}

\section{Mirror Symmetry}

\lettrine{S}{upersymmetric dualities} have a rather better pedigree than non-supersymmetric ones -- many such dualities have been subjected to an impressive array of analytic tests, including exact computations of partition functions, operator dimensions, and more. We are interested in \textit{mirror symmetry}, a term which refers to a huge collection of dualities in different dimensions that in some sense generalize the supersymmetric version of T-duality. They can generally be derived from string theoretic constructions, giving us some reason to believe them even before starting on the amassed evidence that supports them.

The idea of this section is to take one of these well-known supersymmetric dualities in 2+1 dimensions and find a way to deform them, breaking supersymmetry, to obtain our non-supersymmetric bosonization duality \cite{Kachru:2016aon}. Our derivation shows that, with fairly mild assumptions, the SUSY dualities imply the non-SUSY ones.

\subsection{\texorpdfstring{$\mathcal{N}=2$}{N=2} Supersymmetry in 2+1 Dimensions}

In 2+1 dimensions, $\mathcal{N}=2$ supersymmetry implies the theory has 2 Majorana supercharges, for a total of 4 real supercharges. Relative to the more familiar 3+1 dimensional world, this means that it is like $\mathcal{N}=1$ supersymmetry in 3+1 dimensions; in fact, one can dimensionally reduce $(\mathcal{N},d)=(1,4) \to (\mathcal{N},d)=(2,3)$.

The $(\mathcal{N},d)=(1,4)$ chiral multiplet reduces to
\be \mathcal{N} = 2 \text{ chiral multiplet: } \text{complex scalar } \phi\text{, Dirac fermion } \psi \ee
whilst the vector multiplet reduces to
\be \mathcal{N} = 2 \text{ vector multiplet: } \text{vector field } a_\mu \text{, Dirac fermion } \lambda \text{, real scalar } \sigma \fstp \ee
The vector multiplet also contains an auxiliary (non-propagating) real scalar $D$.

The chiral multiplet comes with a conserved $U(1)$ charge given by phase rotations, which can be gauged using a $U(1)$ vector multiplet. The whole theory also comes with a $U(1)_R$ symmetry which associated to the relative phase between the fermions and the scalars. The $U(1)$ vector multiplet also comes with a monopole symmetry as usual.

As well as the usual kinetic terms, the supersymmetric theory supports Chern-Simons and more generally BF terms between vector muliplets. The trick of coupling background fields to symmetries generalizes to SUSY theories too -- except now there is a whole background superfield which we can couple to $U(1)_J$.

\subsection{The Simplest \texorpdfstring{$\mathcal{N}=2$}{N=2} Mirror Pair}

The theory we will study is the simplest $\mathcal{N}=2$ mirror pair \cite{Tong:2000ky}, which can be itself be obtained by partially breaking supersymmetry in the simplest $\mathcal{N}=4$ theory \cite{Intriligator:1996ex}. It is given by
\be
\text{free chiral multiplet} \bigdual U(1) \text{ vector multiplet } + \text{ chiral multiplet }
\ee
which looks remarkably like both the bosonization dualities and the particle-vortex dualities discussed in \refPartOnly{abelian}. We will see that it both of these emerge naturally when we break the SUSY of this $\mathcal{N}=2$ pair. This is an IR duality: far below the scale of the gauge interactions in the right-hand theory, the dynamics is that of the free, left-hand theory.

Let us define the free theory of the chiral multiplet $(\phi,\psi)$. This theory enjoys both a $U(1)_J$ symmetry and a $U(1)_R$ symmetry, with charges
\begin{center} \centering
\begin{tabular}{c|cc}
		& $U(1)_J$ 	& $U(1)_R$ \\ \hline
$\phi$ 	& $1$		& $1$ \\
$\psi$ 	& $1$		& $0$
\end{tabular}
\end{center}
for the matter fields. We can couple the $U(1)_R$ symmetry to a background field $A_R$, and the $U(1)_J$ symmetry to a background vector multiplet with vector field $A_J$, real scalar $\sigma_J$ and auxiliary scalar $D_J$. Defining
\be m_\phi^2 = \sigma_J^2 + D_J \quad \mbox{and} \quad m_\psi = \sigma_J \cmma \ee 
we find
\be
\lag_{\mathrm{free}} = |D_{A_J+A_R}\phi|^2 - m_\phi^2 |\phi|^2 + i \bar{\psi} \slashed{D}_{A_J} \psi - m_\psi \bar{\psi}\psi  \fstp
\ee 

The dual, gauged theory has a chiral multiplet $(\tilde{\phi},\tilde{\psi})$ and also a vector multiplet $(a_\mu,\lambda,\sigma,D)$. (Here, $\lambda$ is the gaugino.) The charged objects in this theory are as follows:
\begin{center} \centering
\begin{tabular}{c|ccc}
					& $U(1)_J$ 		& $U(1)_R$ 	& $U(1)_a$ \\ \hline
	$\tilde{\phi}$ 	& $0$			& $0$		& $-1$ \\
	$\tilde{\psi}$ 	& $0$			& $-1$		& $-1$ \\
	$\sigma$ 		& $0$			& $0$		& $0$  \\
	$\lambda$ 		& $0$			& $-1$		& $0$  \\
	$e^{i\rho}$ 	& $1$			& $0$		& $0$
\end{tabular}
\end{center}
where $\rho$ is the dual photon to $a$.

These are described by the following Lagrangian:
\begin{align}
	\lag_{\mathrm{gauge}} =
	&\frac{1}{g^2} \left( -\frac{1}{4}f_{\mu\nu}f^{\mu\nu} + \frac{1}{2} (\partial \sigma)^2 + i \bar{\lambda} \slashed{D}_{-A_R}\lambda + \frac{1}{2}D^2 \right) \nn\\
	&+ |D_{-a}\tilde{\phi}|^2 + i \bar{\tilde{\psi}} \slashed{D}_{-a-A_R} \tilde{\psi} - (\sigma^2 - D) |\tilde{\phi}|^2 + \sigma \bar{\tilde{\psi}}\tilde{\psi} + \bar{u}\bar{\lambda}\tilde{\psi} + \bar{u}\bar{\tilde{\psi}}\lambda \nn\\
	&+ \frac{1}{4\pi} \left( a \rmd a + 2 D \sigma + \bar{\lambda} \lambda \right) \nn\\
	&- \frac{1}{2\pi} \left( a \rmd A_J + D_J \sigma + \sigma_J D \right) + \frac{1}{2\pi} a \rmd A_R + \frac{1}{4\pi} A_R \rmd A_R
\end{align}
This is an interacting theory, but the claim is that at low energies, the dynamics is that of the free theory $\lag_{\mathrm{free}}$.

(Note that our conventions differ from \cite{Kachru:2016aon} since we use an implicit Pauli-Villars regulator. We are using a supersymmetric version of the regulator we used previously.)

\section{Breaking Supersymmetry}

So how do we see non-supersymmetric dualities emerging from this setup? We have already given the game away with our notation: we will simply use the various scalars in the background vector fields to give a large mass to the fields in our theory, allowing us to identify the low-energy TQFTs and infer what the gauge transitions are as we vary these masses \cite{Kachru:2016aon}.

One thing that is missing is the $|\phi|^4$ interactions; without these, even the free scalar theory is unstable under negative mass deformations. A neat trick is to add
\be \lag_D = \frac{1}{2h^2} (D_J - \tilde{D}_J)^2 \ee
to both Lagrangians and then integrate over $D_J$. This generates a scalar potential
\be V(\phi) = (\sigma_J^2 + \tilde{D}_J)|\phi|^2 + \frac{h^2}{2} |\phi|^4 \ee
in the free theory, for instance. We now take
\be m_\phi^2 = \sigma_J^2 + \tilde{D}_J \quad \mbox{and} \quad m_\psi = \sigma_J \fstp \ee 

With this done, we find the phase diagram looks like \refFigureOnly{susy_breaking_free}.
\begin{figure}
	\centering
	\begin{tikzpicture}
	\draw[axis] (-5,0)-- ++(10,0) node[below] {$-\sigma_J|\sigma_J|$};
	\draw[axis] (0,-5)--  ++(0,10) node[left] {$\tilde{D}_J$};
	
	\draw[phase line] (0,0) -- node[left,xshift=0.0em,opacity=1,rotate=90,anchor=south] {free fermion} (0,5);
	\draw[phase line] (0,0) -- node[left,xshift=-0.0em,opacity=1,rotate=90,anchor=south] {free fermion} (0,-5);
	\draw[phase line] (0,0) -- node[left,yshift=0.0em,opacity=1,rotate=31,anchor=south] {Wilson-Fisher} (-5,-3);
	\draw[phase line] (0,0) -- node[left,yshift=0.0em,opacity=1,rotate=-31,anchor=south] {Wilson-Fisher} (5,-3);
	
	\node[region label] (I) at (3,3) {I}; \node[region sublabel] at (I.south) {$m_\phi^2 > 0$, $m_\psi < 0$};
	\node[region label] (II) at (-3,3) {II}; \node[region sublabel] at (II.south) {$m_\phi^2 > 0$, $m_\psi > 0$};
	\node[region label] (III) at (-2,-4) {III}; \node[region sublabel] at (III.south) {$m_\phi^2 < 0$, $m_\psi > 0$};
	\node[region label] (IV) at (2,-4) {IV}; \node[region sublabel] at (IV.south) {$m_\phi^2 < 0$, $m_\psi < 0$};
	\end{tikzpicture}
	\caption{The phase diagram of the free theory}\labelFigure{susy_breaking_free}
\end{figure}
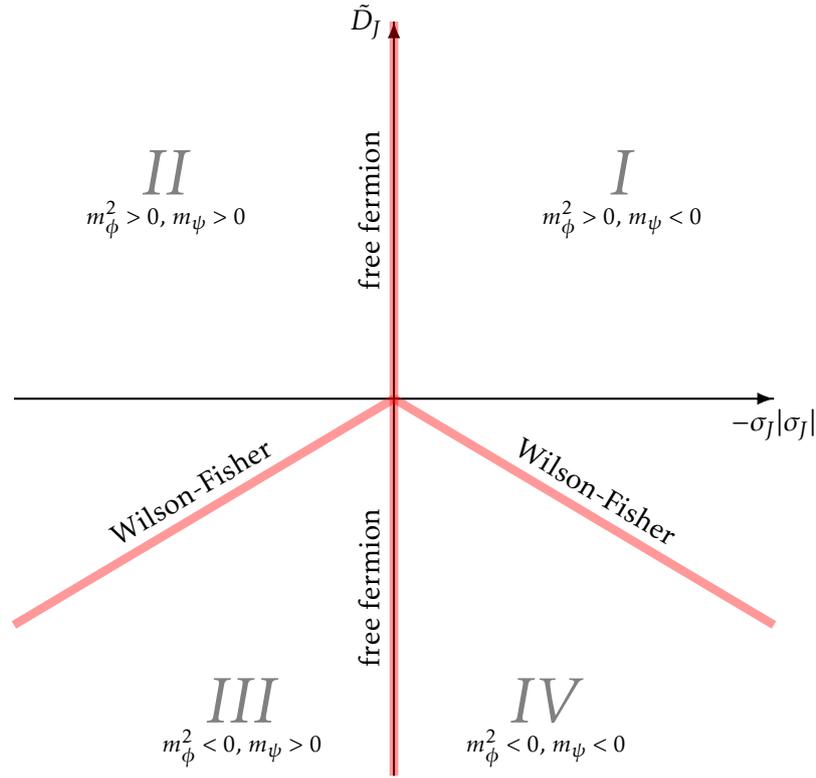

Along particular lines in this diagram, either the scalar field or the fermion become massless. These are to be thought of as critical theories governing various phase transitions arising as we vary $\tilde{D}_J,\sigma_J$.

\begin{exercise}[subtitle=Annotating the Phases]
	In the phases \textit{I}-\textit{IV}, what is the low-energy physics? What background terms are there? Write down a description of each critical theory including such terms.
\end{exercise}

The dual theory is a little more interesting. There are various extra dynamical fields here, and they all interact with each other non-trivially.

\begin{exercise}[subtitle=The Interacting Theory]
	For various regimes of $\tilde{D}_J,\sigma_J$, compute the coefficients in the various scalar potentials including $m^2_{\tilde{\phi}}$, and deduce some approximations to the vacuum expectation values of the scalars. Notice that when $\tilde{\phi}$ condenses, there are terms in the action which cause the chiral fermion $\tilde{\psi}$ and the gaugino $\lambda$ to mix. Diagonalize this mass matrix, identifying the fermion masses $m_+ > m_-$. Otherwise, compute the masses $m_\lambda$ and $m_{\tilde{\psi}}$.
	
	You should find the structure of \refFigureOnly{susy_breaking_gauge}. Check that the phases \textit{I}-\textit{IV} match.
\end{exercise}

\begin{figure}
	\centering
	\begin{tikzpicture}
	\draw[axis] (-5,0)-- ++(10,0) node[below] {$-\sigma_J|\sigma_J|$};
	\draw[axis] (0,-5)--  ++(0,10) node[left] {$\tilde{D}_J$};
	
	\draw[phase line] (0,0) -- node[left,xshift=0.0em,opacity=1,rotate=90,anchor=south] {scalar QED} (0,5);
	\draw[phase line] (0,0) -- node[left,xshift=-0.0em,opacity=1,rotate=90,anchor=south] {free fermion} (0,-5);
	\draw[phase line] (0,0) -- node[left,yshift=0.0em,opacity=1,rotate=31,anchor=south] {scalar QED} (-5,-3);
	\draw[phase line] (0,0) -- node[left,yshift=0.0em,opacity=1,rotate=-31,anchor=south] {fermionic QED} (5,-3);
	
	\node[region label] (I) at (3,3) {I}; \node[region sublabel] at (I.south) {$m_{\tilde{\phi}}^2 > 0$, $m_{\tilde{\psi}} > 0$, $m_\lambda < 0$};
	\node[region label] (II) at (-3,3) {II}; \node[region sublabel] at (II.south) {$m_{\tilde{\phi}}^2 < 0$, $m_{+} > 0$, $m_- < 0$};
	\node[region label] (III) at (-2,-4) {III}; \node[region sublabel] at (III.south) {$m_{\tilde{\phi}}^2 > 0$, $m_{\tilde{\psi}} < 0$, $m_\lambda > 0$};
	\node[region label] (IV) at (2,-4) {IV}; \node[region sublabel] at (IV.south) {$m_{\tilde{\phi}}^2 > 0$, $m_{\tilde{\psi}} < 0$, $m_\lambda < 0$};
	\end{tikzpicture}
	\caption{The phase diagram of the gauge theory}\labelFigure{susy_breaking_gauge}
\end{figure}
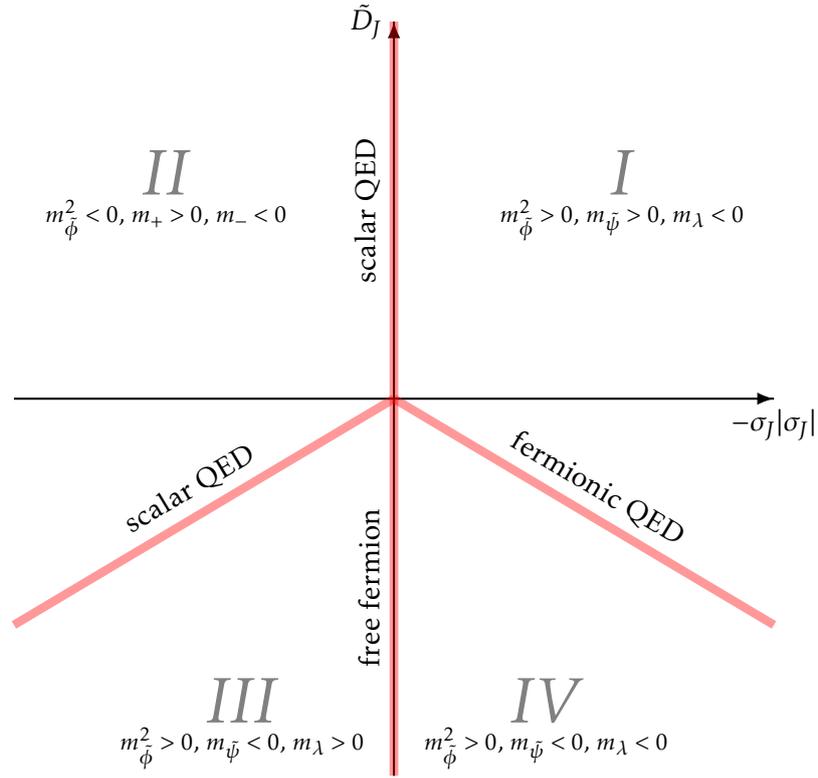

We can therefore deduce that four transitions should be described by identical CFTs:

\begin{exercise}[subtitle=The Dualities]
	Show that this implies the following dualities, and identify them:
	\begin{subequations}\begin{align}
		i \bar{\psi} \slashed{D}_{A_J} \psi
		&\bigdual |D_{-a}\tilde{\phi}|^2 - |\tilde{\phi}|^4 + \frac{1}{4\pi} a \rmd a - \frac{1}{2\pi} A_J \rmd a\\
		|D_{A_J+A_R}\phi|^2 - |\phi|^4
		&\bigdual |D_{-a}\tilde{\phi}|^2 - |\tilde{\phi}|^4 - \frac{1}{2\pi} \left( A_J + A_R \right) \rmd a\\
		i \bar{\psi} \slashed{D}_{A_J} \psi - \frac{1}{2\pi} b \rmd (A_J + A_R)
		&\bigdual i \bar{\lambda} \slashed{D}_{-A_R} \lambda - \frac{1}{2\pi} a \rmd (A_J + A_R) \\
		|D_{A_J+A_R}\phi|^2 - |\phi|^4 - \frac{1}{4\pi} A_J \rmd A_J
		&\bigdual i \bar{\tilde{\psi}} \slashed{D}_{-a-A_R}\tilde{\psi} + \frac{1}{4\pi} a \rmd a + \frac{1}{2\pi} a \rmd A_J
	\end{align}\end{subequations}
\end{exercise}

There is an analogous story in 2 dimensions \cite{KarchTongTurner1}, which is the dimensional reduction of the story we have told here.
\chapter{More Evidence}
\labelChapter{more_evidence}

\begin{introduction}
  We very briefly outline how various discretizations and large $N$ limits of dualities can be fruitful sources of evidence in support of them.
\end{introduction}

\section{Lattice Physics and the Wire Construction}

\lettrine{H}{istorically, many dualities} were understood in terms of dualizations of lattice theories. This is the case here too; for example, it is possibly to explicitly rewrite the lattice partition function of a complex scalar to look like that of a gauged complex scalar. This is central to the history of particle-vortex duality \cite{Peskin:1977kp,Dasgupta:1981zz}.

Of course, this does not \textit{prove} anything about the continuum limit of those lattice theories. We retreat to non-rigorous arguments about what we expect to see emerging in the continuum limit, supported by numerical evidence arising from taking larger and larger lattices. As we emphasized on page~\pageref{walking-lattice}), numerical evidence from the lattice is not always a reliable guide!

Analyzing these questions in detail is beyond our scope. We will briefly discuss some discretized constructions and reserve further judgement.

\subsection{Some Typical Lattice Constructions}

Let us briefly outline how a typical lattice argument might look. We will look at Peskin's original argument \cite{Peskin:1977kp} for particle-vortex duality.

We begin with the usual lattice presentation of the XY model, which consists of compact variables $\theta_n \in[0,2\pi)$ at every lattice site $n \in \mathbb{Z}^3$, together with a free energy
\be
	F = -\frac{1}{T} \sum_n \sum_{\mu} \cos (\theta_{n + \mu} - \theta_n)
\ee
where $n + \mu$ runs over all neighbours of the lattice site $n$. If this is unfamiliar, the general idea is that the mean-field approximation of $\phi = \exp(i\theta)$ gives rise to a complex scalar. Tuning $T$ is then thought of a proxy for tuning the usual $|\phi|^2$ mass term: clearly as $T \to \infty$, the theory should become disordered, whilst as $T \to 0$, we expect the free energy to prefer configurations where $\theta$ is a constant in space, and hence $\phi$ acquires a vacuum expectation value.

Now one can rewrite the partition function as
\begin{align}
	Z = \exp(-F)
	&= \int \rmd\theta \prod_{n,\mu} \exp\left(\frac{1}{T} \cos(\theta_{n + \mu} - \theta_n) \right) \\
	&\propto \int \rmd\theta \prod_{n,\mu} \sum_{m_{n,\mu}} \exp\left(-\frac{1}{2T} (\theta_{n + \mu} - \theta_n - 2\pi m_{n,\mu})^2 \right) \\
	&\propto \int \rmd\theta \rmd b_{n,\mu} \prod_{n,\mu} \sum_{m_{n,\mu}} \exp\left(-\frac{T}{2} b_{n,\mu}^2 + i b_{n,\mu}(\theta_{n + \mu} - \theta_n - 2\pi m_{n,\mu}) \right)
\end{align}
where $m_{n,\mu}$ are some auxiliary integers and $b_{n,\mu}$ are new continuous variables. Together, $m,\theta$ now exhibit a gauge redundancy, and by fixing the gauge we may take $\theta_n \in (-\infty,\infty)$ instead, provided we fix, say, $0 = \sum_\mu m_{n+\mu,\mu} - m_{n,\mu}$. We must introduce a new periodic field $\tilde{\theta}_n$ to impose this integer constraint.

Now the integral over $\theta_n$ is a Lagrange multiplier imposing that $0 = \sum_\mu b_{n-\mu,\mu} - m_{n,\mu}$. This is analogous to $\nabla \cdot \mathbf{b} = 0$ and has pure curl solutions $b_{n\mu} = \sum_{\nu,\sigma} \epsilon_{\mu\nu\sigma} (a_{n-\sigma,\sigma} - a_{n-\nu-\sigma, \sigma})$; here, $a$ has a gauge symmetry.

Altogether, we end up with a presentation in terms of the gauge field $a$ and the new compact variable $\tilde{\theta}$. This is believed to correspond to the dual $U(1) + \phi$ theory with appropriate tuning.

\subsection{The Wire Construction}

One particularly nice discretization argument for 3d dualities relies on using 2d dualities -- intuitively, much of the hard work has already been done, so we should take advantage of it! The idea is to construct a 3 dimensional system by taking many long, parallel wires. By coupling these wires together in a cunning way, the claim is that one can build three-dimensional dualities \cite{WireCon}.

The idea is simply to define a Hamiltonian which is a sum of decoupled 2d field theories, then add a hopping term allowing particles to move between the wires. (This is in the spirit of our earlier discussions of quiver theories.) Suppose we start with fermions. Then we can bosonize the description on each wire, obtaining a sequence of coupled bosons. There is now a simple prescription to construct some dual bosonic fields. These turn out to still have local hopping terms, but what used to be the normal kinetic terms become highly non-local. This can be remedied by introducing an appropriate new variable into the path integral to induce those long-range interactions: it will be no surprise that this take the form of a gauge field. Some simple formal manipulations reveal that there is an entirely local expression for the path integral of this new theory of a gauged boson: hence 3d bosonization indeed seems to follow fairly naturally from 2d bosonization, at least in an appropriate discretization.

Of course, we still have to take the continuum limit, and the dramatic lack of isotropy might be a cause for concern. As promised, we will content ourselves with the knowledge that at least the discrete version of bosonization does make sense.

\section{Large \texorpdfstring{$N$}{N} Physics}

Changing direction somewhat, we should emphasize that one of the many hints which led Aharony to his set of dualities was that their large $N$ limit was relatively well-established. The large $N$ limit is a common crutch we use in situations where we apparently lack any small parameter in which to expand, an obvious problem in our study of 3d gauge theory.

Happily, one can compute many things in the large $N$ limit (at least to leading order) relatively simply. We will not do any of the detailed calculations (and see \cite{Aharony:2015mjs} for extensive references), but we will outline one of the main observations about the simplest non-Abelian dualities,
\be
SU(N)_k + 1 \text{ scalar} \bigdual U(k)_{-N+1/2} + 1 \text{ fermion} \nn
\ee

In particular, we will ask about the minimal dimension of a baryon on the left-hand-side. In the large $N$ limit, it suffices to work classically, in a nearly-free approximation. This means multiplying together fundamental fields $\phi_a$ and their spatial derivatives $\partial_{\mu_1}\cdots\partial_{\mu_n}\phi$, seeking the lightest operator we can write down containing $N$ scalar field insertions $\phi$ contracted with $\epsilon^{a_1 \cdots a_N}$. Derivatives increase the dimension of an operator, and so to find the lightest operator we need to minimize the number of derivatives. However, we should remember that the classical equation of motion $\partial^2 \phi = 0$ effectively eliminates certain operators. This counting problem can be solved directly; there are $2j+1$ operators with $j$ derivatives. Hence we can build $j^2$ objects with fewer than $j$ derivatives each, and using a total of $\sum_{i=0}^{j-1} i(2i+1) \sim 2 j^3/3$ derivatives. Taking $N \sim j^2$ then gives a dimension $\Delta \sim 2 N^{3/2}/3$.

Now consider the right-hand side of the duality. Baryons are not gauge invariant here, but we can build gauge-invariant operators from monopoles. Monopoles are normally labelled by their magnetic (or GNO) charges under the $U(1)^k$ maximal torus of the $U(k)$ group. The simplest monopole has charges $(1,0,\ldots,0)$. The Chern-Simons term means this is not gauge-invariant; even from the $U(1)$ part we can see it has charge $N$. In terms of Young diagrams, as we discussed in \refChapterOnly{level_rank}, it transforms as a single row of $N$ boxes under $SU(N)$. More intuitively, it is given electric charge entirely under a single $U(1) \subset U(1)^k$. Now to render this neutral, we must dress it with $N$ fermionic excitations, and moreover they must all come with the same gauge group index (since they carry charge under only one of the $k$ possible $U(1)$s). Hence we must come up with an non-vanishing operator by multiplying together a single fermionic field $N$ times. But just as with the bosonic problem, this requires using derivatives to prevent the operator vanishing (albeit now due to the fermionic statistics rather than the explicit antisymmetrization). Hence the counting is also the same, giving $\Delta \sim 2 N^{3/2}/3$.

In this way, the naivest possible matchings one could imagine powering these dualities are straightforwardly realized at leading order in the large $N$ limit. (In both calculations, the subleading terms are harder to calculate, of course, but in principle this can be checked order by order.) There are plenty more checks one can carry out, and they all go through well for Aharony's dualities. (Some have also been carried out for the master duality.)


\chapter{A Duality in Condensed Matter}
\labelChapter{condmat}

\begin{introduction}
  In condensed matter, dualities crop up because we are interested in many different effective descriptions low-energy physics of strongly interacting systems. In this section, we see how one of the dualities that we derived earlier on crops up in Quantum Hall physics.
\end{introduction}

\section{Quantum Hall Physics}

\lettrine{W}{e are going to turn} to one of the most famous physical effects to be uncovered last century -- the quantum Hall effect (QHE). This may seem a bit of a sudden change of direction, but it illustrates how helpful it is to think in the language of dualities when attacking practical experimentally-motivated questions.

This is part of an enormous story which we certainly will not have time to discuss in detail. See \cite{tongqhe} for an accessible introduction to many of the key ideas of the QHE.

The central idea is as follows: we study the physics of electrons confined to a plane, subject to a large magnetic field. The physics of this system is quite different from the sort of systems traditionally studied in high-energy physics. Firstly, it is non-relativistic: we are working in regimes far below the energy that would be required for electron particle-anti-particle creation to be relevant. Secondly, the presence of the background magnetic field drastically changes the spectrum, separating it out into Landau levels.

Classically, one predicts that the electron drift velocity $\textbf{v}$ in an electric field should satisfy a force balance equation
\be \textbf{F} = -e(\textbf{E} + \textbf{v} \times \textbf{B}) - \frac{m}{\tau} \textbf{v} = 0 \label{drude} \ee
where $\tau$ models the average scattering time of the system (the usual origin of resistance in this model, which you might know as the Drude model). The current is proportional to $\textbf{v}$, of course; it is given by
\be \textbf{J} = - n e \textbf{v} \ee
where $n$ is the density of the conducting particles. Rewriting the above equation in terms of $J$ and the the cyclotron frequency $\omega_B = eB/m$, we find
\be
\begin{pmatrix} 1 & \omega_B \tau \\ -\omega_B \tau & 1 \end{pmatrix} \textbf{J} = \frac{e^2 n\tau}{m} \textbf{E}
\ee
which one would conventionally write as Ohm's law, $J = \sigma E$, where $\sigma$ is the \textit{conductivity} of the system. One can also define the \textit{resistivity} of the system, using
\be \rho = \sigma^{-1} = \begin{pmatrix}
	\rho_{xx} & \rho_{xy} \\
	-\rho_{xy} & \rho_{xx}
\end{pmatrix}
 = \frac{m}{n e^2 \tau}  \begin{pmatrix} 1 & \omega_B \tau \\ -\omega_B \tau & 1 \end{pmatrix}
 \ee
where we have both the conventional longitudinal resistivity $\rho_{xx}$ and the novel Hall resistivity $\rho_{xy}$ with
\be \rho_{xx} = \frac{m}{ne^2\tau} \quad \mbox{and} \quad \rho_{xy} = \frac{B}{ne} \fstp \ee

This is indeed observed at low $B$. However, as we increase $B$ such that the number of filled Landau levels
\be \nu = \frac{n}{eB/2\pi\hbar} \ee
becomes small (on the order of $\nu \lesssim 10$) we start to get deviations due to quantum effects. (We have counted the 2 spin polarizations of a physical electron as filling 2 levels. Practically, this means working with films on the scale of nanometres and magnetic fields on the order of a few Tesla.)

In particular, the Integer Quantum Hall Effect (IQHE) refers to plateaus developed around filling fractions $\nu \in \mathbb{Z}$ where
\be \rho_{xx} = 0 \quad \mbox{and} \quad \rho_{xy} = \frac{2\pi \hbar}{e^2} \frac{1}{\nu} \fstp \ee
over a wide range of magnetic fields. (This quantization is exact to 1 in a hundred million.)  We also have the Fractional Quantum Hall Effect (FQHE), where a similar phenomenon is observed at
\be \nu = \frac{1}{3}, \frac{2}{3}, \frac{2}{5}, \ldots \ee
for various $\nu \in \mathbb{Q}$. 

Each one of these plateau refers to a distinct phase of matter! They are naturally parametrized by the levels of Chern-Simons terms in effective field theories that govern them. Notice that e.g. the effective field theory
\be S = \frac{e^2}{\hbar} \int \frac{k}{4\pi} A \rmd A \ee 
has the property that the current is
\be J_i = \frac{\delta S}{\delta A_i} = -\frac{k e^2}{2\pi\hbar} \epsilon_{ij} E_j \ee 
corresponding to
\be \rho_{xy} = \frac{2\pi \hbar}{e^2} \frac{1}{k} \ee
which is precisely the right resistivity to describe the $\nu = k$ filling state. Hence this is a natural guess for the effective theory of the IQHE, where $k \in \mathbb{Z}$.

Similarly -- and now setting $e^2 = \hbar = 1$ again -- we find that
\be S = \int  -\frac{k}{4\pi}  a \rmd a + \frac{1}{2\pi} a \rmd A \ee 
gives the correct description of the ``Laughlin states'' which have filling fraction $\nu = 1/k$. To see this quickly, ignoring subtleties on non-trivial manifolds, one solves $k a = A$ for $a$ to find an effective Lagrangian $\lag \sim \frac{1/k}{4\pi} A \rmd A$.
The Laughlin states only exist for odd $k$ for fermions, and even $k$ for bosons.

In general, there are many theories with the same filling fraction, however; some are equivalent and some are distinct. Their physical differences are obtained by investigating other things like their spectrum of excitations. This can be analyzed by adding matter fields to the Chern-Simons theory, giving rise to an anyonic spectrum for general Chern-Simons theories. (Typically these matter fields are non-relativistic.)

\section{Half-Filling and Composite Fermions}

However, there is something very interesting that happens as $\nu \to \frac{1}{2}$. In this regime, we no longer observe plateaus, and at first sight it seems like we simply recover the classical physics,
\be \rho_{xx} = \mbox{const.} \quad \mbox{and} \quad \rho_{xy} = \frac{B}{ne} \fstp \ee
Yet on closer inspection, things look \textit{wildly} different from the classical physics of electrons in a strong magnetic field. One can do an experiment that reveals that the excitations of the system behave exactly as if there was no magnetic field whatsoever! The electrical excitations are in a metallic phase, complete with a Fermi surface; they are not organized into Landau levels at all.

It is clear from this that these fundamental excitations are not electrons; at least, not as we know them. There are various approaches one can take to understanding this sort of state, but we will follow one of the cleanest approaches, known as the \textit{parton} or \textit{slave particle} construction. Specifically, we follow the presentation of \cite{ssww}.

The idea is to make a minimal modification of our naive electron-based understanding of the system. We propose that the physical electron $\Psi$ can be represented as
\be \Psi = \psi \phi \ee
where there fermionic field $\psi$ is some sort of dressed electron, whilst the bosonic $\phi$ somehow encodes the dressing. We will take the fermion $\psi$ to have the same charge as the electron $\Psi$ under the physical electromagnetic field $A$, whilst $\phi$ is neutral.

Now there is some inherent redundancy in the relative phase of $\psi$ and $\phi$. Let us account for this by gauging this $U(1)$ redundancy, giving $\psi$ charge $1$ and $\phi$ charge $-1$. This leads to a proposal for an effective field theory description of the state:
\be L_{\mathrm{parton}} = i \bar{\psi} \slashed{D}_{A+a}\psi + |D_{-a}\phi|^2 + \cdots \ee
where we didn't write any interactions, but we imagine they may be important! Note that this  action is trivially equivalent to
\be L_{\mathrm{parton}} = i \bar{\psi} \slashed{D}_{a}\psi + |D_{A-a}\phi|^2 + \cdots \ee
since we can shift $a \to a-A$.

We now further assume that it is productive to think of $\psi$ as the dynamical field in the theory, whilst $\phi$ can be treated by some sort of mean-field approach. That means solving for their lowest-energy state, and integrating them out by expanding around that state. Of course, without understanding the details of the dynamics, it is hard to know what this state is.\footnote{In the course so far, we have been confident that we are expanding around the $\phi = 0$ or $\phi = \textbf{const}$ states. But if, for example, the true state depended upon the gauge field $a-A$, we might find a different effective description. This will be the case for us.} But let us make some observations.

Firstly, notice that gauge invariance implies the number of $\phi$ and $\psi$ excitations are both equal to the number of $\Psi$ excitations, which we expect to be fixed to the number of physical electrons in the system. Therefore, we don't expect the $\phi$ state to be $\phi=0$. (One possibility is that $\phi$ condenses so as to Higgs $a$. This seems possible, but boring: it leaves only $\psi$ as a fermion in the background field $A$, and we haven't really gained anything from our parton construction.)

Secondly, $\phi$ is coupled to a electromagnetic field $A-a$, and there is a strong magnetic field $\rmd A$ in the system. It is plausible that the physics of this magnetic field might dominate the behaviour of $\phi$. Then $\phi$ could enter a \textit{bosonic} quantum Hall state! Since we are interested in the $\nu = 1/2$ state of the original theory, and the number of $\phi$ excitations is the same as the number of $\Psi$ excitations, it is only natural to speculate that $\phi$ might form the standard Laughlin $\nu = 1/2$ state described above! (This is a valid state for bosonic particles.) We generally expect this to be a low energy state, so it is reasonable to imagine that we might expand around such a vacuum.

Doing so, we obtain a proposed description
\be L_{\text{parton Laughlin}} = i \bar{\psi} \slashed{D}_{a}\psi + \frac{1}{2\pi} (a-A) \rmd b - \frac{2}{4\pi} b \rmd b + \cdots \fstp \ee

Since our effective field theory was constructed by assuming that $\phi$ is at half-filling, and the number of $\phi$s is such that they half-fill the original set of Landau levels, the magnetic field $\phi$ experiences should be exactly of size $A$. Hence $a$ should not really carry any magnetic field. Therefore, we will expect this to describe a number of fermions which do not experience a magnetic field -- it has been soaked up by the $\phi$ variables. This is something else which gets called "flux attachment". In the condensed matter literature, one commonly refers to $\psi$ as a \textit{composite fermion}, and says that we have "attached two units of flux" to the electron $\Psi$ to form $\psi$.

This is the fundamental picture which reveals the way that the emergent particle $\psi$ can effectively feel no background field: the electromagnetic field is experiences is screened by the filled Landau level of $\phi$ excitations. Instead, it forms a Fermi surface, with all the usual physics that goes along with that.

\section{The Application of Fermionic Particle-Vortex Duality}

We might have proceeded rather differently, of course. Suppose we wanted to write down a description in terms of the electrons directly. We might have done this by considering a theory like
\be \lag = i \bar{\Psi} \slashed{D}_{A} \Psi \ee
for example. This directly described a state at half-filling in terms of an electron field $\Psi$.

\begin{exercise}[subtitle=Landau Levels for Dirac Fermions]
	Prove that the Dirac equation
	\be i (\slashed{\partial} - i\slashed{A}) \psi = E\psi \ee
	in a background with $(\rmd A)_{12}=B$ has solutions
	\be E^2 = (2n+1)B \pm B \ee
	with $n=0,1,2,\ldots$ indexing the Landau levels. The last term here is Zeeman splitting of the two spin degrees of freedom.
\end{exercise}

How can both of these descriptions possibly be related? Well, in a course on 2+1d dualities, there is one obvious possibility...

Recall our first fermionic particle-vortex duality,
\be
i \bar{\psi}\slashed{D}_a \psi + \frac{1}{2\pi} b \rmd a - \frac{2}{4\pi} b \rmd b + \frac{1}{2\pi} b \rmd C
\qquad \longleftrightarrow \qquad
i \bar{\tilde{\psi}} \slashed{D}_C \tilde{\psi} + \frac{1}{4\pi} C \rmd C
\relabeleq{fermion-fermion}
\ee
and observe that the left-hand side precisely describes the partonic theory we just wrote down with $C=-A$! Now letting $\Psi$ be the time-reversal, charge-conjugated version of $\tilde{\psi}$, we conclude
\be L_{\text{parton Laughlin}} \bigdual i \bar{\Psi} \slashed{D}_{A} \Psi + \cdots \fstp \label{parton_dual} \ee
so that the theory is dual to an ungauged Dirac fermion, plus any interactions terms we have missed.

We learn that our proposed descriptions of the $\nu = \frac{1}{2}$ state of fermions is equivalent to an electron directly coupled to the electromagnetic field strength! Of course, one has to add interactions in principle to understand the details of what is going on, but the idea is that it may be possible to treat those interactions perturbatively in the partonic picture. If we had reason to believe this was the case (and it can be motivated from mean-field reasoning), then we would indeed \textit{predict} that there is a useful description in terms of an emergent fermionic field $\psi$ which experiences no magnetic field at half-filling!

(This picture makes clear the so-called \textit{particle-hole} symmetry of the lowest Landau-level, in which one simply switches which states are occupied and which are empty. It is natural that a $\nu=\frac{1}{2}$ state of fermions should have this symmetry. This symmetry is the hidden symmetry which we observed back in equation \eqref{fermion-fermion}.)

This is a rather beautiful story, and it opens the door to various nice ways of understanding the full spectrum of quantum Hall states in terms of $\psi$ instead of the original $\Psi$. We will not explore this any further, but will simply note that thinking in terms of dualities gives a rather more solid foundation upon which to build the ideas of mean-field theory and parton constructions which come so easily to condensed matter experts!


\begin{asidebox}[A Long Story]%
	Quantum Hall physics is amazingly rich, despite being entirely about the basic problem of the low-energy physics of electrons in 2+1 dimensions. We have barely scratched the surface of the physics here.
	
	Let's mention one interesting application of the above dualities. We said that an emergent $U(1)_k$ Chern-Simons theory describes the Laughlin state. The lightest excitations in this theory are observed to be anyons, carrying charge $1/k$ times that on the electron. We could add these excitations to our effective field theory by including a bosonic field with charge $1$ under the emergent $U(1)_k$ theory. But this theory is dual to some $SU(k)$ theory coupled to a fermion, a theory in which the gauge-invariant object is the $SU(k)$ baryon, also a fermion for $k$ odd. This can be thought of as the electron. Then emergent matter field is then seen to be another partonic description of the electron in which it is broken into $k$ pieces, $\Psi = \psi_1 \cdots \psi_k$. This is then reflected in the structure of the wavefunctions proposed by Laughlin.
	
	Secondly, we emphasized that the filling fraction does not uniquely determine the phase of matter we observe. Let's just mention one example of that, at $\nu = \frac{1}{2}$ filling. In fact, at $\nu=\frac{5}{2}$ (so there are two filled Landau levels, and then one half-filled Landau level, which naively is equivalent to just the half-filled Landau level) we observe different physics to the $\nu=\frac{1}{2}$ system. For example, it seems like the excitations carry different charges.
	
	The \textit{Moore-Read} or \textit{Pfaffian state} proposed to describe this can be thought of as the result of a deformation of the $\nu = \frac{1}{2}$ Fermi liquid state. In particular, recall that Fermi liquids are unstable to weakly attractive interactions between electrons: they can enter a BCS superconducting state! Indeed, it is thought that the physics of the $\nu=\frac{5}{2}$ state is that of a superconducting state of composite fermions.
\end{asidebox}


\part{Conclusion}
\labelPart{conclusion_part}
\chapter{Conclusion}
\labelChapter{conclusion}

\begin{introduction}
  A few closing words to try and draw some general lessons from what we have learned.
\end{introduction}

\section{Summary}

\lettrine{L}{et's take stock} of what we have learned. We started by quickly reviewing some standard exact dualities in 2 and 3 dimensions. These introduced a few ideas (magnetic symmetries, defect operators, the significance of topological terms, the potential for symmetry enhancement at self-dual points, and so on) which crop up again and again in studying dualities.

Then we started getting our teeth into the concept of an IR duality, emphasizing the roles of RG flows and CFTs in understanding low-energy physics. We illustrated this first with the well-established example of particle-vortex duality, seeing how the phase diagram and operators matched up (and in particular saw how understanding defect operators can play a crucial roles in a duality). We then moved on to Abelian bosonization, taking a little time to appreciate the central role of anomalies in understanding field theories. We also saw some of the subtleties involves in Chern-Simons theory, from statistical transmutation and spin structure sensitivity to topological degeneracy.

Then we studied the interrelations between these dualities, and then how they can be used to propose new and interesting dualities. We emphasized that we cannot prove the existence of CFTs using these techniques, but that this needn't dispirit us even if there is no CFT realizing the IR duality. We also discussed various non-Abelian generalizations of these dualities, including some applications to the phase diagram of QCD.

Finally, we briefly reviewed connections to supersymmetry, discretized (and in particular lattice) models and large $N$ physics, before spending a little more time giving an application of the duality to condensed matter physics in the context of the quantum Hall effect.

\section{The Space of Quantum Field Theories}\labelSection{conclusion_bootstrap}

At the end of this course, one might be left wondering \textit{why} any such dualities exist. We should not be satisfied with a simplistic understanding of it as a coincidence -- at the very least, having some sort of intuitive justification would be nice.


We attempted to give some sense of this back in \refChapterOnly{particle_vortex}: the idea was to seek a change of variables in the path integral which preserves locality of the action. Then one can consider the natural UV theories corresponding to the field content before and after the change of variables. One might reasonably expect that a CFT accessible from one of the theories ought to be accessible from the other.

This may be more practical in a lattice regularization, as discussed in \refChapterOnly{more_evidence}, where we saw that a lattice realization of the XY model can be written as a gauged XY model. One can then try to take the continuum limit of both theories, thereby constructing a continuum version of the whole flow from two distinct simple UV theories to a common interacting IR theory. This is somewhat hit-and-miss, although the word has seen plenty of ingenious lattice constructions. Let us try and attack the problem from a different angle.

Back in the introduction, and again on page~\pageref{walking-lattice}, we mentioned the bootstrap program. We will leave a detailed review to \cite{Poland}, but it is useful to impart a general sense of the significance of this approach. At a very high level, this is to push towards a classification of all CFTs (and ideally all QFTs). This means imposing the constraints arising from the fundamental definitions of field theory and seeing what can possibly satisfy those constraints.

At a more practical level, most progress has been made by restricting further to field theories whose spectrum satisfies certain additional requirements. For example, we might impose that there is a $SU(N_f)$ symmetry, and that there is a relevant operator transforming in the fundamental representation of that symmetry. Typically, even a very small number of constraints seem to heavily constrain the possible CFTs, even using a relatively small proportion of all possible constraints. It seems plausible that supplying some very minimal data along the lines of the above may uniquely identify CFTs. The general moral is that it is \textit{hard} to satisfy the axioms of conformal field theory! If this is right, we should think of CFTs are rare and typically isolated.

This gives a rather more mathematical and satisfying way to think about why dualities exist. If we are given two distinct UV theories flowing to a CFT which is constrained by some minimal data (the number of parameters we tune, their representations under the symmetry group, and anomalies) which agrees, then in all likelihood there are not two possible CFTs that we \textit{could} hit!

And why would we expect there to be distinct UV theories flowing to CFTs with the same constraints? The key is to identify matter content which mediates some particular transition, perhaps between two TQFTs for example. For us, this was made possible by level-rank duality. Then we tune the minimal set of parameters to reach that transition, checking that this requires the same amount of tuning on both sides (so that the putative set of relevant operators is the same). Then the only remaining question is whether we hit a CFT. This is hard to predict. We are saved in some cases where there is a parameter like $N$, and there is some approximation in which there clearly appear to be some RG fixed points. But in general, we can say very little.

Nonetheless, with the perspective that CFTs are rare, the existence of IR dualities seems a little less mysterious. But we are left with the big question: what CFTs \textit{do} exist? That is a question we are not yet quite ready to answer.
\chapter*{Acknowledgements}
\addcontentsline{toc}{chapter}{Acknowledgements}

Thank you to the hosts of the Modave school for the invitation to lecture there, and their excellent hospitality. It was a pleasure to interact with the other participants as well, and I thank them for their help in shaping the course. I also appreciate \href{https://tex.stackexchange.com/a/46243}{Gabriel Blindell's \LaTeX{} template}.

The papers which these lectures notes draw so heavily upon are generally unusually well written, so I think we all owe a debt of thanks to the many physicists involved for their lucidity! I recommend reading the original papers wholeheartedly. I also thank my various collaborators for helping me learn about these theories over the last couple of years, especially David Tong for his supervision during my PhD.

And, Alice: thank you.

\appendix
\part{Appendices}
\labelPart{appendices}
\chapter{Chern-Simons Theories}
\labelAppendix{chern-simons}

\begin{introduction}
  In this section, we review some key facts about three-dimensional Chern-Simons theories.
\end{introduction}

\section{Gauge Invariance}

\lettrine{C}{hern-Simons theories} -- which exist in all odd dimensions -- have many interesting properties. The first, and most important, is that they are in fact gauge-invariant, despite being expressed with explicit dependence on the gauge potential $a$. We'll offer two ways to see the theory is gauge-invariant: one direct computational method, and one more abstract approach. We will focus on $U(N)$ gauge theory, for which the level $k$ Chern-Simons action is
\be
k S_{\mathrm{CS}}[a] = \frac{k}{4\pi} \int_M \tr \left( a \wedge \rmd a - \frac{2i}{3} a \wedge a \wedge a \right)
\ee
where we write out wedge products for clarity. One can set a different level for the $U(1)$ and $SU(N)$ parts of $U(N)$ by adding $n S_\mathrm{CS} [ \tr a ]$. Writing $a = b + c 1_N$ where $b \in su(N)$ and $c \in u(1)$, we have
\be
U(N)_{k,k'} \text{: } \quad \frac{k}{4\pi} \tr \left( b \wedge \rmd b - \frac{2i}{3} b \wedge b \wedge b \right) + \frac{k'N}{4\pi} c \wedge \rmd c
\ee
where we take $k' = k + nN$. We will argue below that $k$ and $n$ should be taken to be integers. The slightly surprising $N$ dependence arises from the quotient in 
\be
U(N) = \frac{SU(N) \times U(1)}{\mathbb{Z}_N} \fstp
\ee

One direct way to analyze the gauge invariance of the theory is to simply implement a gauge transformation $a \to g^{-1} a g  + i g^{-1} \rmd g $ in the action. Under this transformation (essentially because of the Polyakov-Wiegmann property \cite{Polyakov:1983tt})
\be \label{cs_direct_variation} kS_{\mathrm{CS}} \to kS_{\mathrm{CS}} - \frac{ik}{4\pi} \rmd \tr (\rmd g g^{-1} \wedge a) + \frac{k}{12\pi} \tr (g^{-1} \rmd g \wedge g^{-1} \rmd g \wedge g^{-1} \rmd g) \ee
changes in a rather elaborate way. The first term is a total derivative. If we impose suitable boundary conditions, we can set it to zero. The second term is interesting; it is present only for non-Abelian groups. It turns out that 
\be w(g) = \frac{1}{24\pi^2} \int \tr (g^{-1} \rmd g \wedge g^{-1} \rmd g \wedge g^{-1} \rmd g) \in \mathbb{Z} \ee
is an integer-valued \textit{winding number} provided $g \to 1 \in SU(N)$ at infinity.\footnote{Technically, this arises from the non-trivial homotopy group $\pi_3(SU(N)) = \mathbb{Z}$ -- it is a topological invariant of maps from $S^3 \to SU(N)$. This $S^3$ is the one-point compactification of an $\mathbb{R}^3$ spacetime, which is possible because of the boundary condition on $g$.}
Therefore, if $\exp(ikS_{\mathrm{CS}})$ is to be well-defined under arbitrary so-called \textit{large gauge transformations}, we need only take $k \in \mathbb{Z}$.

The conclusion is that an $SU(N)$ level is quantized to an integer due to the winding of gauge transformations; however, the $U(1)$ level is apparently unconstrained, subject to mild boundary conditions.

It turns out that the $U(1)$ levels should also be quantized in many circumstances. Simply connected spacetimes do not have large $U(1)$ gauge transformations. But if, for example, we insist that the theory is well-defined on a thermal circle -- when we compactify the time direction -- then there are gauge transformations which wrap that circle. If one considers a bundle corresponding to a monopole with unit charge $\frac{1}{2\pi}\int_{\text{space}} f = 1$, then it follows once more that $k \in \mathbf{Z}$. See \cite{tongqhe} for an accessible discussion of this issue.

\subsubsection*{A Four-Dimensional Embedding}

It is perhaps more insightful to obtain the $SU(N)$ result from a more mathematical approach \cite{chernsimons,donaldson_furuta_kotschick_2002} that actually goes back to the discovery of Chern-Simons theory. The key observation is that the 3-dimensional Chern-Simons action is really the action of a 4-dimensional theory. We will take our spacetime to be a closed, oriented 3-manifold $M$. Then there exists a 4-manifold $X$ whose boundary is $\partial X = M$ -- in fact, there are many such manifolds, which we will return to in just a moment. Now consider an arbitrary connection $a$ in a principal $SU(N)$ bundle $P \to X$; again, both the bundle and the connection may be extended with bulk (4-dimensional) continuations $P_X,b$. Then there is a natural quantity which we can compute by integrating over the 4 manifold, namely the integral of the Pontryagin density
\be F_X[b] = \frac{1}{8\pi^2} \int_X \tr f \wedge f \ee
which is again clearly topological (independent of the metric). It is also gauge-invariant. If $X$ was a closed manifold, with no boundary, then $F_X[b]$ is the Chern-Weil formula for the \textit{second Chern number} $c_{2}(b;X)$ of the connection $b$, which is always a half-integer -- and if $X$ has a spin structure, it is an integer. However, we know that $X$ has a boundary, and that boundary is our spacetime $M$.

But this still tells us something useful about our various extensions. Suppose we looked at two distinct extensions $(X,P_X,b)$ and $(X',P'_{X'},b')$. Then we could consistently glue them together along their common boundary $\partial X = \partial X' = M$, forming the connection $\bar{b}$. But then (accounting for their relative orientation)
\be F_X[b] - F_{X'}[b'] = F_{X \cup X'} = c_{2}(\bar{b};X\cup X') \in \mathbb{Z} \ee
which establishes that the difference between the possible values of $F_X[b]$ depends on the particular extension only by an overall half-integer or integer shift. In the presence of a spin structure, the fractional part is well-defined; hence $F_M : a \mapsto \mathbb{R}/\mathbb{Z}$ is a good functional given only 3 dimensional data.

Now since we require only that the path integral (and hence $e^{iS}$) is well-defined, we can in fact use $2\pi k F_M[a]$ in the action for an arbitrary integer $k \in \mathbb{Z}$. All that remains is to find a 3 dimensional way to compute this quantity, which is not too hard. We take $X = M \times [0,1]$, parametrizing the interval with $t \in [0,1]$. The manifold is taken to shrink to a point at $t=0$, whilst the boundary of $M$ lies at $t=1$. Since $SU(N)$ is simply connected, we can just assume that all bundles are trivial. Therefore we can also take the continuation of the connection to be $b = t a$. Now the formula for $F_X$ is easy to compute. Exactly one of the four derivatives must be a $t$ derivative, and hence (using the language of 3-dimensional differential forms)
\begin{align}
	2\pi k F_X[a] &= \frac{k}{4\pi} \int_M \int_0^1 \rmd t \ \tr 2 \partial_t (ta) \wedge \left[ \rmd (ta) - i (ta) \wedge (ta) \right] \nn \\
	&= \frac{k}{4\pi} \int_M \tr a \wedge \rmd a \ \left( \int_0^1 \rmd t \  2 t \right) - i a \wedge a \wedge a \ \left( \int_0^1 \rmd t \  2 t^2 \right) \nn \\
	&= \frac{k}{4\pi} \int_M \tr a \wedge \rmd a - \frac{2i}{3} a \wedge a \wedge a \nn \\
	&\equiv k S_{\mathrm{CS}}[a]
\end{align}
is a perfectly well-defined term to include in the action.

Notice that the direct computation \eqref{cs_direct_variation} is of course very closely related with what we have done here; in fact, one can show directly that the Pontryagin density is a total derivative of a CS term, $\tr f \wedge f \propto \rmd (a \wedge \rmd a - 2i/3 a \wedge a \wedge a)$.

We emphasize that the choice of a spin structure is essential to guarantee gauge invariance of the Chern-Simons term with an odd coefficient -- without a spin structure, the theory is only gauge-invariant with an even coefficient.

This changes if $a$ is made into a \spinc field; in this case, there is a combination of the usual $U(1)_1$ action and a gravitational Chern-Simons term which is well-defined modulo $2\pi$ \cite{Seiberg:2016rsg}.


\section{Degrees of Freedom}

Having established that this theory is gauge invariant, we should try to understand how it changes the field theory. A conventional approach to understanding the effect of the Chern-Simons term -- which is quadratic in the fields -- is to analyze the classical equations of motion; a slightly more sophisticated approach is to compute the propagator. We will summarize these computations for the Abelian Maxwell-Chern-Simons theory, $U(1)_k$; see \cite{aspectscs} for a longer discussion of these issues.

Consider the Lagrangian
\be \lag = -\frac{1}{4g^2} f_{\mu\nu}f^{\mu\nu} + \frac{k}{4\pi} \epsilon^{\mu\nu\rho} a_\mu \partial_{\nu} a_\rho \fstp \ee
The classical field equations are
\be \partial_\mu f^{\mu\nu} + \frac{k g^2}{4\pi} \epsilon^{\nu\rho\sigma}f_{\rho\sigma} = 0 \ee
which it is convenient to rewrite in terms of $v^\mu = \frac{1}{2} \epsilon^{\mu\nu\rho} f_{\nu\rho}$ as
\be \left[ \partial_\mu \partial^\mu + \left(\frac{k g^2}{2\pi} \right)^2 \right] v^\nu = 0 \ee
where we should remember that (up to magnetic charges) $\partial_\mu v^\mu = 0$. We have eliminated any gauge-dependence by expressing the theory in terms of $v^\mu$. Therefore, the theory contains a single transverse degree of freedom, with a mass
\be \boxed{m_{\mathrm{top}} = \frac{k g^2}{2\pi}} \ee
and the theory is said to be \textit{topologically massive}. (One can also check that the spin of the excitation is $\sign k = \pm 1$.)

Alternatively, the propagator may be computed by adding a gauge fixing term $\lag_{\mathrm{gf}} = -\frac{1}{2g^2 \xi} (\partial_\mu a^\mu)^2$ and inverting the quadratic form in Fourier space. One finds
\be \Delta_{\mu\nu} = g^2 \left( \frac{p^2 \eta_{\mu\nu} - p_\mu p_\nu - i m_{\mathrm{top}}\epsilon_{\mu\nu\rho}p^\rho}{p^2\left(p^2 - m_{\mathrm{top}}^2\right)} + \xi \frac{p_\mu p_\nu}{p^4} \right) \ee
and the pole clearly identifies the physical mass. (Taking $\xi \to 0$ eliminates the spurious pole $p=0$; the term containing an $\epsilon$ does not lead to a propagating degree of freedom.)

Note that if the relevant parameter $g^2 \to \infty$ in the IR, the topological mass $m_\mathrm{top}$ also grows. Hence we tend, in the IR, to a pure Chern-Simons theory with no propagating degrees of freedom at all.

We should also note that, although the Higgs mechanism still ``works'' in pure Chern-Simons theory (in the sense that a charged field $\phi$ gaining a VEV gives a mass to all excitations in the theory), the way this happens is slightly different. In the unbroken phase, there is no photon, only the two real degrees of freedom of $\phi$. In the Higgs phase, we keep one massive degree of freedom in $\phi$, but the would-be Goldstone boson is eaten by the non-propagating longitudinal mode of the gauge field, leaving a massive gauge mode. The key point is that there is still no light Goldstone mode.

\backmatter
\RaggedRight

\bibliography{bibliography}{}

\begin{thebibliography}{10}
\providecommand{\url}[1]{\texttt{#1}}
\providecommand{\urlprefix}{URL }
\providecommand{\eprint}[2][]{\url{#2}}

\bibitem{yellow}
P.~Di~Francesco, P.~Mathieu and D.~Senechal, \emph{{Conformal Field Theory}},
  Graduate Texts in Contemporary Physics, Springer-Verlag, New York (1997).

\bibitem{tonggt}
D.~Tong, \emph{Gauge Theory (Lecture Notes)} (2018),
  \urlprefix\url{www.damtp.cam.ac.uk/user/tong/gaugetheory.html}.

\bibitem{senthiltalk}
T.~Senthil, \emph{(Effective) Field Theory and Emergence in Condensed Matter
  (Talk)} (2014), \urlprefix\url{http://web.mit.edu/~senthil/www/bu0514.pdf}.

\bibitem{Karch:2019lnn}
A.~Karch, D.~Tong and C.~Turner, \emph{{A Web of 2d Dualities: ${\mathbb{Z}}_2$
  Gauge Fields and Arf Invariants}}  (2019), \eprint{1902.05550}.

\bibitem{Coleman:1974bu}
S.~R. Coleman, \emph{{The Quantum Sine-Gordon Equation as the Massive Thirring
  Model}}, Phys. Rev. \textbf{D11}, p. 2088 (1975), [,128(1974)].

\bibitem{Mandelstam:1975hb}
S.~Mandelstam, \emph{{Soliton Operators for the Quantized Sine-Gordon
  Equation}}, Phys. Rev. \textbf{D11}, p. 3026 (1975), [,138(1975)].

\bibitem{Witten-nonab}
E.~Witten, \emph{{Nonabelian Bosonization in Two-Dimensions}}, Commun. Math.
  Phys. \textbf{92}, pp. 455 (1984).

\bibitem{simontopquant}
S.~Simon, \emph{Topological Quantum (Lecture Notes)} (2016),
  \urlprefix\url{http://www-thphys.physics.ox.ac.uk/people/SteveSimon/topological2016/TopoBook.pdf}.

\bibitem{lerda}
A.~Lerda, \emph{Quantum Mechanics of Particles with Fractional Statistics},
  Springer (1992).

\bibitem{jackiwpi}
R.~Jackiw and S.-Y. Pi, \emph{Classical and quantal nonrelativistic
  Chern-Simons theory}, Phys. Rev. D \textbf{42}, pp. 3500 (1990).

\bibitem{Nakahara}
M.~Nakahara, \emph{{Geometry, topology and physics}}, Taylor \& Francis (2003).

\bibitem{Peskin:1977kp}
M.~E. Peskin, \emph{{Mandelstam 't Hooft Duality in Abelian Lattice Models}},
  Annals Phys. \textbf{113}, p. 122 (1978).

\bibitem{Dasgupta:1981zz}
C.~Dasgupta and B.~I. Halperin, \emph{{Phase Transition in a Lattice Model of
  Superconductivity}}, Phys. Rev. Lett. \textbf{47}, pp. 1556 (1981).

\bibitem{Poland}
D.~Poland, S.~Rychkov and A.~Vichi, \emph{{The Conformal Bootstrap: Theory,
  Numerical Techniques, and Applications}}  (2018), \eprint{1805.04405v2}.

\bibitem{Polyakov:1988md}
A.~M. Polyakov, \emph{{Fermi-Bose Transmutations Induced by Gauge Fields}},
  Mod. Phys. Lett. \textbf{A3}, p. 325 (1988), [,214(1988)].

\bibitem{Metlitski:2015eka}
M.~A. Metlitski and A.~Vishwanath, \emph{{Particle-vortex duality of
  two-dimensional Dirac fermion from electric-magnetic duality of
  three-dimensional topological insulators}}, Phys. Rev. \textbf{B93}, no.~24,
  p. 245151 (2016), \eprint{1505.05142}.

\bibitem{Wang:2015qmt}
C.~Wang and T.~Senthil, \emph{{Dual Dirac Liquid on the Surface of the Electron
  Topological Insulator}}, Phys. Rev. \textbf{X5}, no.~4, p. 041031 (2015),
  \eprint{1505.05141}.

\bibitem{2016PhRvL.117a6802M}
D.~F. {Mross}, J.~{Alicea} and O.~I. {Motrunich}, \emph{{Explicit Derivation of
  Duality between a Free Dirac Cone and Quantum Electrodynamics in (2 +1 )
  Dimensions}}, Physical Review Letters \textbf{117}, no.~1, 016802 (2016),
  \eprint{1510.08455}.

\bibitem{Redlich:1983dv}
A.~N. Redlich, \emph{{Parity Violation and Gauge Noninvariance of the Effective
  Gauge Field Action in Three-Dimensions}}, Phys. Rev. \textbf{D29}, pp. 2366
  (1984), [,2366(1983)].

\bibitem{Wu:1976ge}
T.~T. Wu and C.~N. Yang, \emph{{Dirac Monopole Without Strings: Monopole
  Harmonics}}, Nucl. Phys. \textbf{B107}, p. 365 (1976).

\bibitem{Dray:1984gy}
T.~Dray, \emph{{The Relationship Between Monopole Harmonics and Spin Weighted
  Spherical Harmonics}}, J. Math. Phys. \textbf{26}, p. 1030 (1985).

\bibitem{aspectscs}
G.~Dunne, \emph{{Aspects of chern-simons theory}}, Topological Aspects of Low
  Dimensional Systems , no.~i (1999).

\bibitem{Witten:1988hf}
E.~Witten, \emph{{Quantum Field Theory and the Jones Polynomial}}, Commun.
  Math. Phys. \textbf{121}, pp. 351 (1989), [,233(1988)].

\bibitem{tongqhe}
D.~Tong, \emph{The Quantum Hall Effect (Lecture Notes)} (2016),
  \urlprefix\url{http://www.damtp.cam.ac.uk/user/tong/qhe.html}.

\bibitem{ssww}
N.~Seiberg, T.~Senthil, C.~Wang and E.~Witten, \emph{{A Duality Web in 2+1
  Dimensions and Condensed Matter Physics}}, Annals Phys. \textbf{374}, pp. 395
  (2016), \eprint{1606.01989}.

\bibitem{hsinseiberg}
P.-S. Hsin and N.~Seiberg, \emph{{Level/rank Duality and Chern-Simons-Matter
  Theories}}, JHEP \textbf{09}, p. 095 (2016), \eprint{1607.07457}.

\bibitem{Gaiotto:2014kfa}
D.~Gaiotto, A.~Kapustin, N.~Seiberg and B.~Willett, \emph{{Generalized Global
  Symmetries}}, JHEP \textbf{02}, p. 172 (2015), \eprint{1412.5148}.

\bibitem{fermionpathintegrals}
E.~Witten, \emph{Fermion path integrals and topological phases}, Rev. Mod.
  Phys. \textbf{88}, p. 035001 (2016).

\bibitem{anomaliesodddimensions}
L.~Alvarez-Gaumé, S.~D. Pietra and G.~Moore, \emph{Anomalies and odd
  dimensions}, Annals of Physics \textbf{163}, no.~2, pp. 288  (1985), ISSN
  0003-4916.

\bibitem{karchtong}
A.~Karch and D.~Tong, \emph{{Particle-Vortex Duality from 3d Bosonization}},
  Phys. Rev. \textbf{X6}, no.~3, p. 031043 (2016), \eprint{1606.01893}.

\bibitem{Gaiotto:2008ak}
D.~Gaiotto and E.~Witten, \emph{{S-Duality of Boundary Conditions In N=4 Super
  Yang-Mills Theory}}, Adv. Theor. Math. Phys. \textbf{13}, no.~3, pp. 721
  (2009), \eprint{0807.3720}.

\bibitem{Witten:2003ya}
E.~Witten, \emph{{SL(2,Z) action on three-dimensional conformal field theories
  with Abelian symmetry}} pp. 1173--1200 (2003), \eprint{hep-th/0307041}.

\bibitem{Son:2015xqa}
D.~T. Son, \emph{{Is the Composite Fermion a Dirac Particle?}}, Phys. Rev.
  \textbf{X5}, no.~3, p. 031027 (2015), \eprint{1502.03446}.

\bibitem{Wang:2016gqj}
C.~Wang and T.~Senthil, \emph{{Composite fermi liquids in the lowest Landau
  level}}, Phys. Rev. \textbf{B94}, no.~24, p. 245107 (2016),
  \eprint{1604.06807}.

\bibitem{Cordova:2017kue}
C.~Córdova, P.-S. Hsin and N.~Seiberg, \emph{{Time-Reversal Symmetry,
  Anomalies, and Dualities in (2+1)$d$}}, SciPost Phys. \textbf{5}, p. 006
  (2018), \eprint{1712.08639}.

\bibitem{2016PhRvB..94u4415C}
M.~{Cheng} and C.~{Xu}, \emph{{Series of (2+1)-dimensional stable self-dual
  interacting conformal field theories}}, Phys. Rev. B \textbf{94}, no.~21,
  214415 (2016), \eprint{1609.02560}.

\bibitem{PhysRevB.92.220416}
C.~Xu and Y.-Z. You, \emph{Self-dual quantum electrodynamics as boundary state
  of the three-dimensional bosonic topological insulator}, Phys. Rev. B
  \textbf{92}, p. 220416 (2015).

\bibitem{Benini:2017dus}
F.~Benini, P.-S. Hsin and N.~Seiberg, \emph{{Comments on global symmetries,
  anomalies, and duality in (2 + 1)d}}, JHEP \textbf{04}, p. 135 (2017),
  \eprint{1702.07035}.

\bibitem{Nahum:2015vka}
A.~Nahum, P.~Serna, J.~T. Chalker, M.~Ortuño and A.~M. Somoza, \emph{{Emergent
  SO(5) Symmetry at the Néel to Valence-Bond-Solid Transition}}, Phys. Rev.
  Lett. \textbf{115}, no.~26, p. 267203 (2015), \eprint{1508.06668}.

\bibitem{PhysRevX.7.031052}
Y.~Q. Qin, Y.-Y. He, Y.-Z. You, Z.-Y. Lu, A.~Sen, A.~W. Sandvik, C.~Xu and
  Z.~Y. Meng, \emph{Duality between the Deconfined Quantum-Critical Point and
  the Bosonic Topological Transition}, Phys. Rev. X \textbf{7}, p. 031052
  (2017).

\bibitem{Gorbenko:2018ncu}
V.~Gorbenko, S.~Rychkov and B.~Zan, \emph{{Walking, Weak first-order
  transitions, and Complex CFTs}}, JHEP \textbf{10}, p. 108 (2018),
  \eprint{1807.11512}.

\bibitem{Karch:2016aux}
A.~Karch, B.~Robinson and D.~Tong, \emph{{More Abelian Dualities in 2+1
  Dimensions}}, JHEP \textbf{01}, p. 017 (2017), \eprint{1609.04012}.

\bibitem{Intriligator:1996ex}
K.~A. Intriligator and N.~Seiberg, \emph{{Mirror symmetry in three-dimensional
  gauge theories}}, Phys. Lett. \textbf{B387}, pp. 513 (1996),
  \eprint{hep-th/9607207}.

\bibitem{Aitken:2018joz}
K.~Aitken, A.~Karch and B.~Robinson, \emph{{Deconstructing S-Duality}}, SciPost
  Phys. \textbf{4}, p. 032 (2018), \eprint{1802.01592}.

\bibitem{Blau:1993tv}
M.~Blau and G.~Thompson, \emph{{Derivation of the Verlinde formula from
  Chern-Simons theory and the G/G model}}, Nucl. Phys. \textbf{B408}, pp. 345
  (1993), \eprint{hep-th/9305010}.

\bibitem{Aharony:2015mjs}
O.~Aharony, \emph{{Baryons, monopoles and dualities in Chern-Simons-matter
  theories}}, JHEP \textbf{02}, p. 093 (2016), \eprint{1512.00161}.

\bibitem{Radicevic:2016wqn}
{\DJ. Radičević}, D.~Tong and C.~Turner, \emph{{Non-Abelian 3d Bosonization
  and Quantum Hall States}}, JHEP \textbf{12}, p. 067 (2016),
  \eprint{1608.04732}.

\bibitem{Radicevic:2015yla}
{\DJ. Radičević}, \emph{{Disorder Operators in Chern-Simons-Fermion
  Theories}}, JHEP \textbf{03}, p. 131 (2016), \eprint{1511.01902}.

\bibitem{Aharony:2016jvv}
O.~Aharony, F.~Benini, P.-S. Hsin and N.~Seiberg, \emph{{Chern-Simons-matter
  dualities with $SO$ and $USp$ gauge groups}}, JHEP \textbf{02}, p. 072
  (2017), \eprint{1611.07874}.

\bibitem{Benini:2017aed}
F.~Benini, \emph{{Three-dimensional dualities with bosons and fermions}}, JHEP
  \textbf{02}, p. 068 (2018), \eprint{1712.00020}.

\bibitem{Komargodski:2017keh}
Z.~Komargodski and N.~Seiberg, \emph{{A symmetry breaking scenario for
  QCD$_{3}$}}, JHEP \textbf{01}, p. 109 (2018), \eprint{1706.08755}.

\bibitem{Freed:2006mx}
D.~S. Freed, \emph{{Pions and Generalized Cohomology}}, J. Diff. Geom.
  \textbf{80}, no.~1, pp.~45 (2008), \eprint{hep-th/0607134}.

\bibitem{Vafa:1984xg}
C.~Vafa and E.~Witten, \emph{{Parity Conservation in QCD}}, Phys. Rev. Lett.
  \textbf{53}, p. 535 (1984).

\bibitem{Vafa:1984xh}
C.~Vafa and E.~Witten, \emph{{Eigenvalue Inequalities for Fermions in Gauge
  Theories}}, Commun. Math. Phys. \textbf{95}, p. 257 (1984).

\bibitem{Vafa:1983tf}
C.~Vafa and E.~Witten, \emph{{Restrictions on Symmetry Breaking in Vector-Like
  Gauge Theories}}, Nucl. Phys. \textbf{B234}, pp. 173 (1984).

\bibitem{Appelquist:1989tc}
T.~Appelquist and D.~Nash, \emph{{Critical Behavior in (2+1)-dimensional
  {QCD}}}, Phys. Rev. Lett. \textbf{64}, p. 721 (1990).

\bibitem{Appelquist:1988sr}
T.~Appelquist, D.~Nash and L.~C.~R. Wijewardhana, \emph{{Critical Behavior in
  (2+1)-Dimensional QED}}, Phys. Rev. Lett. \textbf{60}, p. 2575 (1988).

\bibitem{Jensen:2017bjo}
K.~Jensen, \emph{{A master bosonization duality}}, JHEP \textbf{01}, p. 031
  (2018), \eprint{1712.04933}.

\bibitem{Jain:2013gza}
S.~Jain, S.~Minwalla and S.~Yokoyama, \emph{{Chern Simons duality with a
  fundamental boson and fermion}}, JHEP \textbf{11}, p. 037 (2013),
  \eprint{1305.7235}.

\bibitem{Seiberg:2016rsg}
N.~Seiberg and E.~Witten, \emph{{Gapped Boundary Phases of Topological
  Insulators via Weak Coupling}}, PTEP \textbf{2016}, no.~12, p. 12C101 (2016),
  \eprint{1602.04251}.

\bibitem{Aharony:2013kma}
O.~Aharony, S.~S. Razamat, N.~Seiberg and B.~Willett, \emph{{3$d$ dualities
  from 4$d$ dualities for orthogonal groups}}, JHEP \textbf{08}, p. 099 (2013),
  \eprint{1307.0511}.

\bibitem{Gomis:2017ixy}
J.~Gomis, Z.~Komargodski and N.~Seiberg, \emph{{Phases Of Adjoint QCD$_3$ And
  Dualities}}, SciPost Phys. \textbf{5}, p. 007 (2018), \eprint{1710.03258}.

\bibitem{Atiyah:1975jf}
M.~F. Atiyah, V.~K. Patodi and I.~M. Singer, \emph{{Spectral asymmetry and
  Riemannian Geometry 1}}, Math. Proc. Cambridge Phil. Soc. \textbf{77}, p.~43
  (1975).

\bibitem{Witten:1999ds}
E.~Witten, \emph{{Supersymmetric index of three-dimensional gauge theory}} pp.
  156--184 (1999), \eprint{hep-th/9903005}.

\bibitem{Jensen:2017dso}
K.~Jensen and A.~Karch, \emph{{Bosonizing three-dimensional quiver gauge
  theories}}, JHEP \textbf{11}, p. 018 (2017), \eprint{1709.01083}.

\bibitem{Aitken:2018cvh}
K.~Aitken, A.~Baumgartner and A.~Karch, \emph{{Novel 3d bosonic dualities from
  bosonization and holography}}, JHEP \textbf{09}, p. 003 (2018),
  \eprint{1807.01321}.

\bibitem{Gaiotto:2017tne}
D.~Gaiotto, Z.~Komargodski and N.~Seiberg, \emph{{Time-reversal breaking in
  QCD$_{4}$, walls, and dualities in 2 + 1 dimensions}}, JHEP \textbf{01}, p.
  110 (2018), \eprint{1708.06806}.

\bibitem{Kachru:2016aon}
S.~Kachru, M.~Mulligan, G.~Torroba and H.~Wang, \emph{{Nonsupersymmetric
  dualities from mirror symmetry}}, Phys. Rev. Lett. \textbf{118}, no.~1, p.
  011602 (2017), \eprint{1609.02149}.

\bibitem{Tong:2000ky}
D.~Tong, \emph{{Dynamics of N=2 supersymmetric Chern-Simons theories}}, JHEP
  \textbf{07}, p. 019 (2000), \eprint{hep-th/0005186}.

\bibitem{KarchTongTurner1}
A.~Karch, D.~Tong and C.~Turner, \emph{{Mirror Symmetry and Bosonization in 2d
  and 3d}}  (2018), \eprint{1805.00941v1}.

\bibitem{WireCon}
D.~Mross, J.~Alicea and O.~Motrunich, \emph{{Explicit derivation of duality
  between a free Dirac cone and quantum electrodynamics in (2+1) dimensions}}
  (2015), \eprint{1510.08455}.

\bibitem{Polyakov:1983tt}
A.~M. Polyakov and P.~B. Wiegmann, \emph{{Theory of Nonabelian Goldstone
  Bosons}}, Phys. Lett. \textbf{B131}, pp. 121 (1983), [,195(1983)].

\bibitem{chernsimons}
S.-S. Chern and J.~Simons, \emph{Characteristic Forms and Geometric
  Invariants}, Annals of Mathematics \textbf{99}, no.~1, pp.~48 (1974), ISSN
  0003486X.

\bibitem{donaldson_furuta_kotschick_2002}
S.~K. Donaldson, M.~Furuta and D.~Kotschick, \emph{Floer Homology Groups in
  Yang-Mills Theory}, Cambridge Tracts in Mathematics, Cambridge University
  Press (2002).

\end{thebibliography}

\end{document}